\documentclass[a4paper,11pt]{article}
\usepackage{jheppub}
\usepackage[T1]{fontenc}
\usepackage{lmodern}

\usepackage{amstext}
\usepackage{amsfonts}
\usepackage{mathrsfs}
\usepackage{amsmath}
\usepackage{amssymb}
\usepackage{cases}
\usepackage{bm}
\usepackage{extarrows}
\usepackage{tensor}
\usepackage{xcolor}
\usepackage{float}
\usepackage{bbm, dsfont}
\usepackage{stmaryrd}
\usepackage{hyperref}
\usepackage[low-sup]{subdepth}

\newcommand{\red}{\color{red}}
\newcommand{\blu}{\color{blue}}
\definecolor{Green}{RGB}{199,238,206}
\makeatletter
\newsavebox\myboxA
\newsavebox\myboxB
\newlength\mylenA
\newcommand*\xoverline[2][0.82]{%
\sbox{\myboxA}{$\m@th#2$}%
\setbox\myboxB\null
\ht\myboxB=\ht\myboxA%
\dp\myboxB=\dp\myboxA%
\wd\myboxB=#1\wd\myboxA
\sbox\myboxB{$\m@th\overline{\copy\myboxB}$}
\setlength\mylenA{\the\wd\myboxA}
\addtolength\mylenA{-\the\wd\myboxB}%
\ifdim\wd\myboxB<\wd\myboxA%
\rlap{\hskip 1.2\mylenA\usebox\myboxB}{\usebox\myboxA}%
\else
\hskip -0.5\mylenA\rlap{\usebox\myboxA}{\hskip 0.5\mylenA\usebox\myboxB}%
\fi}
\newcommand{\eqrefe}{Eq.\eqref}
\newcommand{\beq}{\begin{equation}}
\newcommand{\eeq}{\end{equation}}
\newcommand{\ba}{\begin{array}}
\newcommand{\ea}{\end{array}}
\newcommand{\beqa}{\begin{eqnarray}}
\newcommand{\eeqa}{\end{eqnarray}}
\newcommand{\beqs}{\begin{subequations}}
\newcommand{\eeqs}{\end{subequations}}
\def\dis{\displaystyle}
\newcommand{\la}{\langle}
\newcommand{\ra}{\rangle}

\newcommand{\fr}[2]{\frac{{#1}}{#2}}
\newcommand{\Fr}[2]{\mbox{$\frac{\,{#1}\,}{#2}$}}
\renewcommand{\rm}{\mathrm}
\def\over{\xoverline}

\def\PT{\text{PT}}
\def\Pf{\text{Pf}}

\def\diag{\text{diag}}

\def\leqq{\leqslant}
\def\geqq{\geqslant}
\def\({\left(}
\def\){\right)}
\def\[{\left[}
\def\]{\right]}
\def\LB{\left\{}
\def\RB{\right\}}
\def\nn{\nonumber}
\def\pd{\partial}
\def\d{\rm{d}}
\def\pp{\prime}
\def\to{\rightarrow}
\def\ito{\!\rightarrow\!}

\def\dd{\mathbf{d}}
\def\shat{\hat{s}}
\def\ba{\bar{a}}

\def\whA{\widehat{A}}
\def\A{\mathcal{A}}

\def\whcA{\widehat{\mathcal{A}}}
\def\whcM{\widehat{\mathcal{M}}}

\def\CC{\mathcal{C}}
\def\mC{\mathbb{C}}
\def\D{\mathcal{D}}
\def\td{\text{d}}

\def\hg{\hat{g}}

\def\th{\tilde{h}}
\def\th{\tensor{h}}

\def\hh{\hat{h}}
\def\ii{\text{i}}

\def\whI{\widehat{\mathcal{I}}}

\def\KK{\mathcal{K}}

\def\La{\mathcal{L}}

\def\M{\mathcal{M}}

\def\dM{\delta\mathcal{M}}

\def\NN{\mathcal{N}}
\def\dNN{\delta\mathcal{N}}

\def\whN{\widehat{\mathcal{N}}}
\def\bN{\xoverline{N}}
\def\bsN{\scriptsize{\xoverline{N}}}

\def\bn{\mathfrak{n}}
\def\mO{\mathcal{O}}
\def\hp{\hat{p}}

\def\bR{\mathbb{R}}
\def\bs{\bar{s}}

\def\dT{\delta\mathcal{T}}
\def\TT{\mathcal{T}}

\def\Z{\mathbb{Z}}
\def\ZZ{\mathbb{Z}_2^{}}

\def\hPhi{\widehat{\Phi}}

\def\al{\alpha}
\def\alp{\alpha^\prime}
\def\be{\beta}
\def\ga{\gamma}

\def\ka{\kappa}

\def\mn{\mu\nu}

\def\da{\delta}

\def\lam{\lambda}
\def\hlam{\hat{\lambda}}

\def\hka{\hat{\kappa}}
\def\si{\sigma}
\def\hsi{\hat{\sigma}}
\def\ct{c_\theta^{}}
\def\st{s_\theta^{}}
\def\cct{c_\theta^2}
\def\sst{s_\theta^2}

\def\ctt{c_{2\theta}^{}}
\def\cttt{c_{3\theta}^{}}
\def\ctf{c_{4\theta}^{}}
\def\ctfv{c_{5\theta}^{}}
\def\cts{c_{6\theta}^{}}

\def\qb{\bar{q}}
\def\qqbp{\bar{q}^{\prime 2}}

\def\Mn{M_n}

\def\sgn{\text{sgn}}
\def\hze{\hat{\zeta}}
\def\sh{\hat{s}}
\def\th{\hat{t}}
\def\uh{\hat{u}}

\def\fA{\mathsf{A}}
\def\fB{\mathsf{B}}
\def\fC{\mathsf{C}}
\def\fP{\mathsf{P}}
\def\ax{\textbf{a}}
\def\bx{\textbf{b}}
\def\cx{\textbf{c}}
\def\ma{M_{\textbf{a}}^{}}
\def\mb{M_{\textbf{b}}^{}}
\def\mc{M_{\textbf{c}}^{}}

\def\hs{\hspace*{0.3mm}}
\def\hsx{\hspace*{0.5mm}}
\def\hsm{\hspace*{-0.3mm}}
\def\hsmx{\hspace*{-0.5mm}}

\allowdisplaybreaks[4]
\def\End{\end{document}}

\title{Massive Color-Kinematics Duality and Double-Copy\\
for Kaluza-Klein Scattering Amplitudes}

\author[a]{Yao Li,}
\emailAdd{neolee@sjtu.edu.cn}

\author[a]{~Yan-Feng Hang,}
\emailAdd{yfhang@sjtu.edu.cn}

\author[a,b]{~Hong-Jian He\,\footnote{corresponding author.}}
\emailAdd{hjhe@sjtu.edu.cn}

\affiliation[a]{Tsung-Dao~Lee Institute \&  School of Physics and Astronomy,\\
Key Laboratory for Particle Astrophysics and Cosmology (MOE),\\
Shanghai Key Laboratory for Particle Physics and Cosmology,\\
Shanghai Jiao Tong University, Shanghai, China}
\affiliation[b]{Physics Department \& Institute of Modern Physics,
Tsinghua University, Beijing, China; 
\\
Center for High Energy Physics, Peking University,
Beijing, China}

\abstract{\\
We study the structure of scattering amplitudes of massive
Kaluza-Klein (KK) states under toroidal compactification.\
We present a shifting method to quantitatively derive the
scattering amplitudes of massive KK gauge bosons and KK gravitons
from the corresponding massless amplitudes in the noncompactified
higher dimensional theories.\ With these we construct the massive
KK scattering amplitudes by extending the double-copy relations
of massless scattering amplitudes within the field theory
framework, including both the BCJ and CHY methods, and build up
their connections to the massive KK KLT relations.\
We present the massive BCJ-type double-copy construction of
the $N$-point KK gauge boson/graviton scattering amplitudes,
and as the applications we derive explicitly the four-point
KK scattering amplitudes as well as the five-point
KK scattering amplitudes.\
We further study the nonrelativistic limit of these massive
scattering amplitudes with the heavy external KK states and
discuss the impact of the compactified extra dimensions on
the low energy gravitational potential.\
Finally, we analyze the four-point and $N$-point
mass spectral conditions and newly propose a novel 
group theory approach to prove that {\it only the KK theories}
under toroidal compactification can satisfy
these conditions for directly realizing massive double-copy
in the field theory framework.
\\[3mm]
JHEP (in press), [arXiv:2209.11191 [hep-th]].
}

\begin{document}
\maketitle
\flushbottom

\setcounter{page}{2}

\vspace*{5mm}
\section{\hspace*{-2mm}Introduction}
\label{sec:1}

Scattering amplitude is an important means for studying
fundamental forces in nature,
and can bridge theories with experiments.\
One of the greatest challenges in modern physics
is the unification of gravitational force with the gauge forces
of electromagnetic, weak and strong interactions of
all elementary particles.\
This leads to the explorations of a higher dimensional
spacetime structure, with extra spatial dimensions
compactified on the boundaries and being smaller
than the existing observational limits.\ Such attempts started
a century ago from the Kaluza-Klein (KK) theory for
unifying the gravitational and electromagnetic forces with
a compactified fifth dimension (5d)\,\cite{KK}.\
This opened up a truly fundamental direction,
and has been seriously pursued and extensively developed
in various contexts, including
the string/M theories\,\cite{string} and
the extra dimensional field theories with large or small
extra dimensions\,\cite{Exd0}\cite{Exd}\cite{ExdRS}.\
The KK compactification predicts an infinite tower of
massive KK excitations for each known particle of the
Standard Model (SM).\
These have intrigued substantial phenomenological and experimental
efforts over the past two decades to search for the low-lying KK states
of the extra dimensional KK theories\,\cite{exdpheno}\cite{exd-rev},
which could produce the first signatures of new physics
beyond the SM, including the KK states of
gravitons and of the SM particles as well as the possible
dark matter candidate.
\\[-4.3mm]


The longstanding difficulty of unifying the gravitational force
with the gauge forces arises from the intricate nonlinearity and
perturbative nonrenormalizability of the Einstein theory
of General Relativity (GR),
in contrast to the modern renormalizable gauge theories
of the electroweak and strong
interactions in the SM of particle physics.\
However, the scattering amplitudes of gravitons and gauge bosons
are connected through the conjectured deep relation of double-copy,
$\hs\text{GR}\! =\! (\text{Gauge~Theory})^2\hs$,
even though it does not manifest at the Lagrangian (Hamiltonian)
level.\  The double-copy relation points to the common fundamental root
of both the gravity force and gauge forces.\
It has also become a powerful tool for efficiently computing
the highly intricate scattering amplitudes of spin-2 gravitons.\
The massive KK graviton scattering amplitudes are 
especially involved (with large energy cancellations)\,\cite{Chivukula:2020}\cite{Kurt-2019}\cite{Hang:2021fmp} 
and the $N$-point longitudinal KK graviton amplitudes have 
their leading energy dependence nontrivially cancelled down 
by a large power factor $\hs E^{2N}\!\!$ ($N\!\geqq\!4$) 
up to any loop order\,\cite{Hang:2021fmp}.\
The first double-copy was constructed by
Kawai-Lewellen-Tye (KLT)\,\cite{KLT} to link the 
scattering amplitudes of massless closed strings to the products of
scattering amplitudes of massless open strings at tree level.\
The KLT relation leads to the connection between the scattering
amplitudes of massless gravitons and the products of
the color-ordered amplitudes of massless gauge bosons
in the low energy field-theory limit.\
Then, Bern-Carrasco-Johansson (BCJ) constructed the double-copy
within the field theory formalism through the color-kinematics
duality\,\cite{BCJ:2008}\cite{BCJ:2019}, which connects
the (squared) scattering amplitudes of massless gauge bosons
to the corresponding massless graviton amplitudes.\
The massless double-copy relations of BCJ may be proven
and refined by analyzing the tree-level scattering amplitudes
of the massless heterotic strings and open strings\,\cite{Tye-2010}.\
They may also be proven\,\cite{BCJ:2019} by using the on-shell
recursion relation of Britto-Cachazo-Feng-Witten (BCFW)\,\cite{BCFW}
and be extended to loop levels in the field theory framework.\
Afterwards, the worldsheet Cachazo-He-Yuan (CHY) method\,\cite{CHY}
was motivated by Witten's twistor string
theory\,\cite{Witten:2003nn}, it further shows that
the KLT kernel can be interpreted as the inverse amplitudes
of bi-adjoint scalars; it can also extend the double-copy
relations to other field theories than the gauge/gravity
theories\,\cite{CHY14}.\
But, all these works were to formulate and test
double-copy for the scattering amplitudes
of massless gauge bosons/gravitons\,\cite{BCJ:2019}.\
Our recent works extended the massless double-copy method to
the massive double-copy constructions for  
the 5d KK gauge/gravity field theories
(with orbifold compactification)\,\cite{Hang:2021fmp}, 
for the compactified 26d KK bosonic string theory
plus its field theory limit\,\cite{KKString}, 
and for the 3d topologically massive Chern-Simons (CS) gauge/gravity 
theories\,\cite{Hang:2021oso}.\ 
In the literatue, some other recent works attempted to generalize 
the massless double-copy method to the case of 
massive double-copies, 
including the 4d massive Yang-Mills (YM) theory versus Fierz-Pauli-like
massive gravity\,\cite{dRGT}\cite{DC-4dx1},
the spontaneously broken YM-Einstein supergravity models
with adjoint Higgs fields\,\cite{SUGRAhiggs},
the KK-inspired effective gauge theory with extra global
U(1)\,\cite{Momeni:2020hmc},
the 3d Chern-Simons (CS) gauge/gravity theories
with or without supersymmetry\,\cite{3dCS-susy}\cite{3dCS1}\cite{Gonzalez:2021bes},
and certain massive scalar theories\,\cite{DC-scalar}.\

\vspace*{0.5mm}

Extensions of the conventional massless double-copy
formalism to the case of massive gauge/gravity theories are generally
difficult, because many theories of this kind violate gauge symmetry
and diffeomorphism invariance (which include the massive
YM theory and the massive Fierz-Pauli gravity\,\cite{FP}).\
By adding extra non-linear polynomial interaction terms in
the literature\,\cite{dRGT}\cite{DC-4dx1},
one could realize the double-copy between the
massive YM and Fierz-Pauli gravity theories,
but the high energy behavior
of the four-point massive graviton amplitudes could be improved
to no better than $E^6$ \cite{Cheung2016}\cite{Kurt-E6}
and is still much worse than the final energy-dependence of
$O(E^2)$ in the tree-level massive KK graviton scattering amplitudes\,\cite{Chivukula:2020}\cite{Kurt-2019}\cite{Hang:2021fmp}.\
The KK compactification can realize a geometric ``Higgs'' mechanism
for mass generation of KK gravitons both at
the Lagrangian level\,\cite{GHiggs}\cite{Hang:2021fmp} and
at the scattering $S$-matrix level\,\cite{Hang:2021fmp}.\
This is shown\,\cite{Hang:2021fmp} to make the massive KK GR theory
free from the van-Dam-Veltman-Zakharov (vDVZ) 
discontinuity\,\cite{vDVZ} and exhibit much better
high energy behaviors (as good as that of the massless Einstein gravity).\
Hence the KK GR theories provide a truly consistent realization
of the massive gravity in the effective field theory formulation.\
Another important realization of consistent massive double-copy
is given by the 3d topologically massive Chern-Simons gauge and
gravity theories\,\cite{TMG} because they have
topological mass-generations for the gauge bosons
and gravitons in a gauge-invariant manner and guarantee the
good high energy behaviors of the massive gauge boson/graviton
scattering amplitudes\,\cite{Hang:2021oso}.\

\vspace*{0.5mm}

Our recent works studied the massive double-copy
constructions for the KK gauge/gravity field theories
under the 5d compactification with the $S^1\!/\ZZ$
orbifold\,\cite{Hang:2021fmp},
and for the 26d KK bosonic string theory
under the compactification
of $\bR^{1,24}\hsm \otimes\hsm S^1$
\cite{KKString}.\footnote{%
In passing, a recent paper\,\cite{wKLT}
studied the general KLT factorization of winding string amplitudes
in the bosonic string theory and computed explicitly
the four-point tachyon amplitudes, which do not have  
the low energy field-theory limit.}\ 
We found\,\cite{Hang:2021fmp}
that the massive double-copy construction
for the 5d KK theories with $S^1\!/\ZZ$ orbifold compactification
is highly involved because the KK gauge boson scattering amplitudes
exhibit double-pole structures and the naive extension of the
conventional massless BCJ method does not work.\
By making the high energy expansion, we proved\,\cite{Hang:2021fmp}
that the leading order (LO) KK gauge boson amplitudes
are mass-independent and obey the color-kinematics duality.\
So the massive double-copy works at the LO
[which is enough for our double-copy
construction of the KK Gravitational Equivalence Theorem
(GRET)\,\cite{Hang:2021fmp}], but it does not exactly work at the
next-to-leading-order (NLO) and beyond,
unless certain special treatment is made.\
Then, by using the first principle approach of KK string theory,
we realized\,\cite{KKString} that the massive extension of
the KLT-type double-copy construction could
exactly work for the 5d $S^1\hsm$ compactification 
without orbifold$\hs$\footnote{%
We note that the twisted states of the KK bosonic strings under 
orbifold compactification such as $S^1/\mathbb{Z}_2$ will lift the 
vacuum energy on the worldsheet and increase the masses of 
KK open (closed) strings 
by a large amount $\frac{1}{\,16\hs \alpha'\,}$ 
($\hs\frac{1}{\,4\hs \alpha'\,}$)
which fully decouple in the field theory limit 
with string tension $\alpha'\hsm\ito 0\hs$ \cite{KKString}.\
Besides, the vertex operators of twisted KK states with 
orbifold compactification are not as simple as those for winding states and there is no such explicit formula as the exponential of a free field\,\cite{Polchinski1}.\
This makes it hard to directly realize 
the massive KLT relations under the
orbifold compactification even within the KK string theory.}
under which the KK amplitudes have single-pole structure,
exhibit the massive color-kinematics duality,
and obey the mass spectral condition.\ 
As the resolution for the nontrivial case of
the orbifold compactification, we found\,\cite{KKString} that
the correct double-copied KK graviton amplitudes can be constructed
in terms of proper combinations of the KK graviton amplitudes
(derived under the compactification without orbifold).\
We obtained these insights and the exact double-copy
construction of KK gauge boson/graviton amplitudes by first
deriving the massive KLT relations (connecting the KK closed
string amplitudes to the products of the KK open string amplitudes)
for the bosonic string theory under the compactification of $\bR^{1,24}\otimes S^1$.\
With these, we took the low energy field-theory limit
(with the string tension $\alpha'\!\ito 0$) and derived the
exact double-copied KK graviton amplitudes.\
Then, we found that the KK scattering amplitudes under the
$S^1\!/\ZZ\hs$ orbifold compactification can be constructed in terms of
proper combinations of the corresponding KK amplitudes
(computed without the $\ZZ$ orbifold).\
We note that the massive KK KLT-like relations derived in
Ref.\,\cite{KKString}
rely on the color-ordered amplitudes of KK gauge bosons.\
It is thus desirable to further construct in the present work
the BCJ-type massive double-copy with extended color-kinematics duality
for the KK gauge boson/graviton scattering amplitudes 
within the field theory framework and build its
quantitative connection to the massive KLT-like relations.\ 
Especially, the BCJ double-copy approach
has its own advantages\,\cite{BCJ:2019},
hence studying its massive extension to
the compactified KK gauge/gravity theories is valuable.\
It is also desirable to extend the
conventional massless CHY method to the massive KK double-copy
construction, and build up its connections to the massive
KLT-type relations and the massive BCJ-type construction.\

\vspace*{0.7mm}

In this work, we study the structure of 
scattering amplitudes of massive KK
states under toroidal compactification,
and analyze the realization of massive color-kinematics duality
and KK gauge/gravity double-copy construction.\
We will present a shifting method to quantitatively derive the
scattering amplitudes of massive KK gauge bosons and KK gravitons
from the corresponding massless amplitudes in the noncompactified
higher dimensional theories.\
The massive KK amplitudes derived in this way correspond to
the toroidal compactification without orbifold and can serve as
the basis KK amplitudes for further constructing other types of KK
amplitudes under the orbifold toroidal compactification.\
With these we construct the massive KK scattering amplitudes
by extending the double-copy relations
of massless scattering amplitudes within the field theory formulations,
including both the BCJ and CHY methods, and build up
their connections to the massive KK KLT relations.\
We present the massive BCJ-type double-copy construction of
the $N$-point KK gauge boson/graviton scattering amplitudes.\
As the applications, we derive explicitly the four-point KK
scattering amplitudes and the five-point KK scattering amplitudes.\
Under the KK compactification without orbifold and
using the generalized gauge invariance, we will derive a
mass spectral condition for the four-point
massive KK graviton amplitudes.\ We also use an extended
fundamental BCJ relation for the massive KK theories and
prove that the four-point KK scattering amplitudes have to
obey the same mass spectral condition.\
We further study the nonrelativistic limit of these massive
scattering amplitudes with the heavy external KK states and
discuss the impact of the compactified extra dimensions on
the low energy gravitational potential.\
We demonstrate that the elastic scattering amplitudes of the
heavy KK states can induce a leading-order behavior of the
classical potential which scales as $1\hsm /r$ at low energies.
Finally, we study the possible solutions to
the four-point KK mass spectral condition which is a necessary
and sufficient condition for realizing the massive KK double-copy.\
We newly propose a novel group theory approach to prove that it gives
a unique consistent solution of the mass spectrum
which could be realized only by the KK theories
under the toroidal compactifications without orbifold.\
We further extend this four-point spectral condition
to the general $N$-point spectral condition.\ We prove that the
KK mass spectrum is also the solution to the $N$-point spectral
condition, so it is a truly consistent solution.
In passing, we note that the usually generalized massive KLT relations 
and massive double-copy in the literature suffer from 
the problem of spurious poles\,\cite{DC-4dx1}\cite{Momeni:2020hmc}.\ 
But our approach propose to use the shifting method and derive
massive KK amplitudes from their higher dimensional {\it massless} counterparts which are free from the spurious poles.\  
It was also suggested\,\cite{DC-4dx1} 
that imposing a mass spectral 
condition can remove the spurious poles.\  
As we will demonstrate in Section\,\ref{sec:5},
our massive KK double-copy under toroidal
compactification not only obeys this spectral condition, 
but also serves as its unique solution.

\vspace*{1mm}

This paper is organized as follows.\ 
The main purpose of this paper is to study the structure of 
the massive KK gauge-boson/graviton scattering amplitudes and construct 
their double-copies via the extended massive BCJ and CHY methods 
within the pure quantum field theory (QFT) framework.\ 
In Section\,\ref{sec:2},
we establish an extended double-copy approach for scattering
amplitudes of massive KK gauge bosons and KK gravitons
under the toroidal compactification within
the QFT formulation.\ 
We propose a shifting method 
to construct the massive KK amplitudes from 
their massless counterparts in the 
noncompactified higher dimensional theories, 
with which we build up a correspondence 
from the conventional massless BCJ double-copy
to the extended massive KK double-copy.\
In Section\,\ref{sec:3}, we use this shifting method 
to construct the extended four-point massive BCJ-type
double-copy of the KK gauge-boson/graviton amplitudes 
under the 5d toroidal compactification without or with orbifold.\
Under the toroidal compactification without orbifold and
by using either the generalized massive gauge invariance 
or the extended massive fundamental BCJ relation,
we will derive a mass spectral condition for consistent
double-copy construction of the four-point KK graviton amplitudes.\
We further construct the five-point KK graviton scattering 
amplitudes from the double-copy construction.\
As an application, we also derive the nonrelativistic 
KK scattering amplitudes
for the heavy KK gauge bosons, heavy KK gravitons,
and heavy KK scalars, respectively.\
In Section\,\ref{sec:4}, using our shifting method 
we generalize the conventional massless CHY approach  
and present an extended massive CHY formulation of
the KK gauge-boson/graviton scattering amplitudes.\
We will derive a massive scattering equation
for the KK scattering amplitudes and construct the KK
bi-adjoint scalar amplitudes.\
We use this extended massive CHY approach to further construct
the scattering amplitudes of KK gauge bosons and of KK gravitons,
and derive their relations to the extended
BCJ-type KK amplitudes (Section\,\ref{sec:3}) and to the extended
KLT-type KK amplitudes (given by Ref.\,\cite{KKString}).\
In Section\,\ref{sec:5}, we study the possible solutions to the
mass spectral conditions for the four-point KK scattering amplitudes
and for the general $N$-point KK scattering amplitudes.\
For this we propose a novel group theory approach 
to prove that the four-point mass spectral condition 
can uniquely determine the allowed mass
spectrum to be that of the KK theories 
under toroidal compactification.\
Finally, we summarize and conclude in Section\,\ref{sec:6}.\
For making the analyses in the main text, we also define 
in Appendix\,\ref{App:A}
the kinematics of the four-point scattering for both the massless
5d theories and the compactified massive 4d KK theories.\ 
The kinematic numerators for KK gauge
boson scattering amplitudes are given in Appendix\,\ref{App:B},\
and the double-copied full scattering amplitudes of KK gravitons
are presented in Appendix\,\ref{App:C}.\

\section{\hspace*{-2mm}KK Scattering Amplitudes under Toroidal Compactification}
\label{sec:2}

In the recent work\,\cite{KKString},
we studied the scattering amplitudes of massive Kaluza-Klein (KK) states
of open and closed bosonic strings under toroidal compactification.\
We demonstrated that the $N$-point scattering amplitudes of
the massive KK gauge bosons and KK gravitons can be derived by taking the
field-theory limit $\hs\alp\!\ito 0\,$ for the corresponding
open and closed amplitudes of the massive KK bosonic strings.\
We demonstrated that the extended massive KLT-like relations can
realize the exact double-copy construction
under the toroidal compactification
(without orbifold) which conserves the KK numbers and ensures the
massive KK amplitudes to have single-pole structure
in each kinematic channel.\
For the toroidal compactification with $\ZZ$ orbifold,
we showed that any $N$-point KK amplitude
(with external states being $\ZZ$ even or odd)
can be decomposed into a sum of sub-amplitudes
which belong to the toroidal compactification without orbifold.\

\vspace*{0.6mm}

In this section, we will show that the eigenfunctions of
Laplace operator on a compact manifold can be chosen as
the exponential functions
with which the massive KK scattering amplitudes
of a higher dimensional theory under toroidal compactification
can be obtained by replacing the extra-dimensional
momentum-components in the corresponding massless amplitudes of
the noncompactified theory
by their discretized values (given by the KK compactification).\
The physical scattering amplitudes are independent of which basis of
eigenfunctions is chosen.\ Thus, the amplitudes defined under other
eigenfunction bases (such as the trigonometric functions)
can be obtained by proper transformations of
the external states.\
Using such a ``shifting'' method, we can directly construct the massive
KK scattering amplitudes from the corresponding massless amplitudes
of the noncompactified higher dimensional theory.\
Thus, we can establish an extended BCJ-type double-copy approach
for scattering amplitudes of the massive KK gauge bosons
and KK gravitons in the QFT formulation.\
We also stress {\it the importance of using the toroidal compactification
without orbifold as the base construction}
of the massive KK double-copy, with which the double-copy constructions
in other KK theories under the orbifold compactification (such as $S^1\!/\ZZ$)
can be formulated by proper transformations.\

\subsection{\hspace*{-2mm}Toroidal Compactification with Different Eigenbases}
\vspace*{1.5mm}
\label{sec:2.1}

Consider a generic extra-dimensional model defined on the manifold
$\,\mathbb{M}^{1,3}\hsm\otimes \mathbb{N}^{\delta}$,
where $\,\mathbb{M}^{1,3}$ denotes the (1+3)-dimensional
Minkowski spacetime and $\mathbb{N}^{\da}\!$ ($\hs\da\!\geqq\!1$)
denotes the extra $\da$-dimensional space
under toroidal compactification.\
The extra-dimensional mass operator $\Delta$ is a
Laplacian defined on $\mathbb{N}^{\delta}\hs$ and has
the following KK eigenvalue equation:
%
\begin{equation}
\Delta \mO_I^{} \,=\, M_I^2 \hs \mO_I^{} \,.
\end{equation}
where $\{\mO_I^{}\}$ denote the eigenfunctions with KK index $I$
and $\{M_I^{}\}$ are the corresponding
mass-eigenvalues\,\cite{Kurt-2019}.\
According to Ref.\,\cite{Nakahara},
we can express the Laplacian $\Delta$ as follows:
\begin{equation}
\Delta =(\dd\hsm + \dd^\dagger)^2
=\hs \dd\dd^\dagger\hsm + \dd^\dagger\dd\,,
\end{equation}
where $\hs\dd\hs$ denotes the exterior derivative operator for
the extra dimensional space and $\hs\dd^\dagger$
is the adjoint exterior derivative operator.
It can be shown\,\cite{Nakahara} that the nilpotency holds,
$\,\dd\dd =\dd^\dagger\dd^\dagger =0\,$.
Thus we can prove $M_I^2$ to be positive definite:
\begin{align}
M_I^2 \,= \int_{\mathbb{N}}^{} \! \mO_I^{} \Delta \mO_I^{}
\hs =\hs (\mO_I^{},\Delta \mO_I^{}\hsm )
\hs =\hs (\dd\mO_I^{}, \dd\mO_I^{}\hsm ) +
      (\dd^\dagger \mO_I^{}, \dd^\dagger \mO_I^{}\hsm )
\geqq 0 \,,
\end{align}
for $\mathbb{N}$ being a Riemann manifold,
where the integration is performed over the extra dimensional
coordinates.

\vspace*{1mm}

For the case without degenerate KK states,
the KK eigenfunctions $\{\mO_I^{}\}$
with different eigenvalues should be orthonormal to each other:
\begin{equation}
\int_{\mathbb{N}}^{} \!\mO_I^{} \hs \mO_J^{}  \,=\, \delta_{IJ}^{} \,.
\end{equation}
For the case with degenerate KK states,
the same condition can be retained by using the
Schmidt orthogonalization.
We can further define the general $N$-point coupling constants
among the KK states as follows:
\beqs
\begin{align}
\label{eq:Orth-1}
\int_{\mathbb{N}}^{}   \mO_{I_1^{}}^{} \mO_{I_2^{}}^{} \cdots \mO_{I_N}^{}
&=\, C^{}_{I_1^{} \cdots I^{}_N} \,,
\\[1mm]
\label{eq:Orth-2}
\int_{\mathbb{N}}^{}  \pd \hs \mO_{J_1^{}}^{} \mO_{I_2^{}}^{} \cdots \mO_{I_N^{}}^{}
&= \, D^{}_{J_1^{} I_2^{} \cdots I_N^{}} \,,
\end{align}
\eeqs
and so on.\ In Eq.\eqref{eq:Orth-2}, the symbol
$\hs\pd\hs\mO_{J_1^{}}^{}\hsm$ denotes the possible
extra-dimensional derivative(s) in a given interaction vertex.

\vspace*{1mm}

Under the toroidal compactification, there are two common choices
of the eigenfunctions $\,\mO_I^{}\,$, one is based on the
Fourier expansion with trigonometric functions, and another is the
choice of exponential functions:
\beqs
\label{eq:OI}
\begin{numcases}
{\{\mO_I^{} \}\,=\, }
\LB\hsx 
1,\, \sqrt{2\hs}\cos (n_i^{}\hs y_i^{}/\!R_i^{}), \,
     \sqrt{2\hs}\sin (n_i^{}\hs y_i^{}/\!R_i^{})
\RB \hsmx , ~~& $n_i^{} \in \mathbb{Z}^+$\hsm , \hspace*{5mm}
\label{eq:OI-1}
\\[2mm]
\label{eq:OI-2}
\LB \exp(\ii\hs n_i^{}y_i^{}/\!R_i^{}) \RB \hsm ,
~~& $n_i^{} \in \Z$\,,
\end{numcases}
\eeqs
where $\,y_i^{}\!\in\! [0,2\pi R_i^{}]\,$ represents the
$i$-th coordinate of the extra dimensional space
$\mathbb{N}^{\da}$
and $R_i^{}$ denotes the radius of the $i$-th dimension
of $\mathbb{N}^{\da}$.\
In addition, we note that the eigenfunctions \eqref{eq:OI-2}
will lead to Feynman rules which are analogous to that of the
non-compactified flat space, but withe the continuous momenta
replaced by the corresponding discrete values, namely,
$\,q_i^{}\ito n_i^{}/\!R_i^{}$\,.\
In fact, as we will show, it is really advantageous to first
analyze the KK scattering amplitudes
under the toroidal compactification without orbifold
and using the exponential eigenfunctions \eqref{eq:OI-2};
and then we can use these to derive the corresponding
KK scattering amplitudes in another given basis of
eigenfunctions such as Eq.\eqref{eq:OI-1}.

\vspace*{1mm}

We note that for a massless theory in the $d$-dimensional spacetime
$\,\mathbb{M}^{1,d-1}$,
the general $N$-point amplitude,$\!$\footnote{%
In the following text, the variables with an extra ``hat''
symbol denote the quantities defined in
the higher $d$-dimensional spacetime with
$\hs d\hsm >\hsm 4\hs$.}
$\,\whcA_N(\{\hp_i^{},\hze^{}_i\})$,\,
is an analytical function of the momenta and polarizations\footnote{%
If we consider the scalar fields, then the polarization
$\,\hze =1\,$ and has no effect.}
living in the full $d$-dimensional spacetime.
We can decompose the momentum $\hp_i^{}\hsm =\hsm (p_i^{},\,q_i^{})$
into a $(1+3)$-dimensional momenta $p_i^{}$ and
an extra $\da$-dimensional momentum  $q_i^{}$.
Thus, we can obtain the expression of an $N$-point KK
scattering amplitude (in the KK theory with the
compactified extra-dimensional space $\mathbb{N}^{\da}$)
from the corresponding massless amplitude (in the non-compactified
massless $d$-dimensional theory with $d=4\hsm +\hsm\da\hs$),
\begin{equation}
\label{eq:general-KK}
\A_N^{}\hsm\big[\{p_i^{},\hat\zeta_i^{},I_i^{}\}\big] \,=\,
\sum_{q_j^{}}\!\whcA_N^{}\hsm\big[\{p_i^{},\hze^{}_i, q_i^{}\}\big]\hsm
\prod_j\!\int_{\mathbb{N}}^{}\!\!\td y_j^{}\hs
e^{\ii\hs q_j^{} y_j^{}}  \hs \mO_{I_j}^{}\hsm (y_j^{}) \,,
\end{equation}
where $\,y_j^{}\,$ denotes the
coordinates of the extra-dimensional space
$\mathbb{N}^{\da}$ in the eigenfunction
$\hs\mO_{I_j}^{}\hsm (y_j^{})\hs$.\
Since the final physical KK amplitude should not depend on
choosing which base of extra-dimensional eigenfunctions,
the KK amplitudes defined by using the base $X$ should be connected
to the KK amplitudes defined under another base $Y$ through
the linear transformations of external KK states,
\begin{equation}
\label{eq:XY-states}
\Big|\mO^X_i \Big\rangle \,=\,
\int\!  \mO^X_i \hs \mO^Y_j \Big|\mO^Y_j \Big\rangle \,.
\end{equation}
The KK scattering amplitude may be expressed in the following form,
\begin{equation}
\A_N^{} \,\sim\, \la 0 | S | \mO_1^{} \mO_2^{} \ldots \mO_N^{} \ra \,,
\end{equation}
where we choose all the external particles to be incoming.\
Thus, the scattering amplitudes in the compactified KK theories
defined under different bases of KK eigenfunctions
can be transformed into each other under the transformation
\eqref{eq:XY-states} of their external states.\
We note that only the amplitudes under the toroidal compactification
can be connected to amplitudes in flat spacetime and obeys such
simple rules, as shown in \eqrefe{eq:general-KK}.

\vspace*{1mm}

For instance, consider an $N$-point amplitude
in $\,\mathbb{M}^{1,3} \otimes S^1$ spacetime
under the $S^1$ toroidal compactification.\
Substituting an eigenfunction
$\sqrt{2}\cos(n_j^{}y_j^{}/\!R)$
(with $\,n_j^{}\!\in \mathbb{Z}^+$)
of Eq.\eqref{eq:OI-1}
into \eqrefe{eq:general-KK} for each external state,
we derive:
\begin{align}
\hspace*{-6mm}
\A_N^{}[\{p_i^{} , q_i^{}, \hat\zeta^{}_i\}]
&=\, \sum_{q_j^{}}^{}\!
\whcA_N^{}[\{p_i^{},\hze^{}_i,q_i^{}\}]
\prod_j\! \int^{2\pi R}_0\!\!\!\td y_j^{}
\exp(\ii\hs q_j^{} y_j^{})\!
\(\!\!\sqrt{2} \cos \frac{\hs n_j^{} y_j^{}\,}{R} \!\!\)
\nn \\
&=\, 2^{-N/2} \sum_{q_j^{}}\!
\whcA_N^{}[\{p_i^{},\hze^{}_i, q_j^{}\}] \!
\prod_j\!\[\delta\!\(\!q_j^{},\hsm\frac{\,n_j^{}\,}{R}\!\)
\!+\delta\!\(\!q_j^{},\hsm -\frac{\,n_j^{}\,}{R}\!\)\hsm\]
\nn\\
&=\, 2^{-N/2}\!\!\sum_{\{\sgn_i^{}\}}\!\!
\whcA_N^{}[\{p_i^{},\hze^{}_i,\sgn_i^{}\!\times\!n_i^{}/\!R\}]
\hs,
\label{KK-s1z2}
\end{align}
where in the last line we have defined
$\,\sgn_i^{}\!=\text{sign}(n_i^{})=\!\pm 1$\,
and the sum of $\{\sgn_i^{}\}$ runs over all possible
independent combinations of the $N$-external KK states.\
In the second line of the above formula, the delta function
is defined as
$\,\da (a,\hs b)\hsm\equiv \da_{ab}^{}\hs$.\
In Eq.\eqref{KK-s1z2},
we have used the conservation of the extra-dimenstional momenta
$\,\sum_i q_i^{}\!=\hsm 0\,$
for the amplitude
$\whcA_N(\{p_i^{},\hze_i^{},q_i^{}\})$
because the flat non-compactified extra-dimensional space holds
the translation symmetry.\
Under the periodic boundary conditions,
the toroidal compactification of extra spatial dimensions
also holds the translation symmetry
and thus the conservation of KK numbers.\
Eq.\eqref{KK-s1z2} explicitly demonstrates that
under the toroidal compactification of the extra-dimensional space
$\mathbb{N}^{\da}$, a given $N$-point massive KK amplitude
with external particles being the $\ZZ$-even eigenfunctions
of Eq.\eqref{eq:OI-1}
can be obtained from a sum of the corresponding
{\it massless amplitudes defined in a higher dimensional
spacetime}  $\mathbb{M}^{1,3+\da}$
with each extra-dimensional momentum $q_i^{}$ replaced by the
discretized one
($\hs q_i^{}\!=\!+n_i^{}/\!R\,$, or,
 $\hs q_i^{}\!=\!-n_i^{}/\!R\,$),
which we will call the sub-amplitudes hereafter.\
For the external states being $\ZZ$-odd or certain more complicated
combination, their KK amplitudes can be derived in the similar
fashion by using \eqrefe{eq:general-KK}.\
We stress that
the above presentation gives a fairly general approach for deriving
the massive KK scattering amplitudes
(under the toroidal compactification)
from the corresponding massless scattering amplitudes
of the non-compactified higher dimensional theory.\
This construction can be applied to any physical KK fields
and is not limited to the KK gauge bosons and KK gravitons.\
Because the double-copy method has been well established
for the massless field theories, our current extended
double-copy approach for the massive KK scattering amplitudes
in the following section can be formulated within pure field-theory
framework without relying on the KK string theory construction
given in our previous work\,\cite{KKString}.

\vspace*{1mm}

As a final remark, we note that the polarization
$\hs\hze_i^{}\hs$ of each external state as appeared
on the left-hand-side (LHS) of
Eq.\eqref{eq:general-KK} or Eq.\eqref{KK-s1z2}
is still the one defined in the full $\hs d=4+\da\hs$ dimensions.
In the next subsection,
we will further derive the exact formulation of the
KK scattering amplitudes with the polarization $\hze_i^{}$
reduced to that of the compactified KK theory
in the 4-dimensional spacetime.

\vspace*{2mm}
\subsection{\hspace*{-2mm}Connecting Amplitudes Before and After  Toroidal Compactification}
\vspace*{2mm}
\label{sec:2.2}

In Eqs.\eqref{eq:general-KK} and \eqref{KK-s1z2},
the scattering amplitudes on their LHS still have the
polarization $\hze$ defined in the full
$\, d\hsm =\hsm 4 +\da\,$ dimensions.\
To obtain the $4$-dimensional KK scattering amplitudes,
we first note that the compactified components and
non-compactified components of $\,\hze\,$ transform
in different representations
of the Lorentz group for the non-compactified spacetime
$\mathbb{M}^{1,3}$.\
Hence, the polarization vector
$\hze$ can be separated into two parts in terms of
the representations of Lorentz group.\
One part of the ploarization $\hze$ is restricted to
(1+3)-dimensions,
$\hs \hze^{}\!=\hsm \{\zeta^{}, 0,0,\cdots\!,0\}
 \!=\!\{\zeta^{},\hs \vec{0}\hs\}\hs$,
whereas the other part of the ploarization vector $\hze^{}$
only has compactified components
$\hs\hze^{}\!=\!\{0,\cdots\!,0, \vec{\eta}\hs\}\hs$.

\vspace*{1mm}

In the following we analyze the polarization vector
$\hs\hze^{}\! =\hsm \{\zeta^{}, \vec{0}\hs\}\hs$.\
We consider the compactification from
the $(4\hsm +\hsm\da)$-dimensional Minkowski spacetime
to the $4$-dimensional spacetime.\
We denote the momenta and polarization vectors
as $\hp_i^{}$ and $\hze_i^{}$
in $(4 +\da)$-dimensions, and those in
the $4$-dimensional Minkowski spacetime
by $\hs p_i^{}\hs$ and $\hs\zeta_i^{}\hs$.\
Under the toroidal compactification,
the momentum and polarization vector
in the full $(4\hsm +\hsm\da)$-dimensional
spacetime are connected to those in the $4$-dimensional spacetime:
\begin{equation}
\label{eq:pzeD}
\hp^M \!= (p^\mu, \,
n_1^{}/\!R_1^{}, \cdots\hsm, n_{\da}^{}/\!R_{\da}^{}) \,, \qquad
\hze^{M} \!= (\zeta^\mu, 0, \cdots\hsm, 0) \,,
\end{equation}
where the integer $\hs n \!\in\! \mathbb{Z}$\, is the KK number.\
The tree-level scattering amplitudes should be rational functions
of the Lorentz invariants, such as
$\,\hp_i^{}\hsm\cdot\hp_j^{}\hs$,
$\,\hp^{}_i \hsm\cdot\hze^{}_j\hs$, and
$\,\hze^{}_i \hsm\cdot\hze^{}_j\,$.\
Using \eqrefe{eq:pzeD}, we can rewrite these Lorentz invariants
as follows:
\beqs
\label{eq:sij-pol-shift}
\begin{align}
\label{eq:sij-shift}
&\sh_{ij}^{} = s_{ij}^{} -
\sum_{k=1}^{\da}\!
\frac{\,(n_{k,i}^{} \!+\hsm n_{k,j}^{}\hsm )^2\,}{R_k^2}
\hs =\,  s_{ij}^{} -
\sum_{k=1}^{\da}\!
\(\hsm\sgn_i^{}M_{k,i}^{}\!+\sgn_j^{}M_{k,j}^{}\hsm\)^{\hsm\!2}
,
\\[1mm]
\label{eq:pol-shift}
&\hze^{}_i \cdot \hat{p}_j =\zeta^{}_i \cdot p_j
\,,  \qquad
\hze^{}_i \cdot \hze^{}_j = \zeta^{}_i \cdot \zeta^{}_j \,,
\end{align}
\eeqs
where
$\hs\sgn_i^{}\!\equiv\hsm\rm{sign}(n_{k,i}^{})\!=\!\hsm \pm 1\hs$
and
$M_{k,i}^{}\!=\!|n_{k,i}^{}|/\!R_k^{}\hs$.\
According to the above, we can derive the massive KK scattering amplitude
$\whcA_N(\{\hze^{}_i\},\hsm\{\hp_i^{}\})$
from the corresponding massless zero-mode scattering amplitude
$\A_N(\{\zeta^{}_i\},\{p_i^{}\})\hs$
by shifting its Mandelstam variables:\
$\,s_{ij}^{}\!\ito \hat{s}_{ij}^{}\,$,
where
$\,\hat{s}_{ij}^{}$ is defined in \eqrefe{eq:sij-pol-shift}.
We note that since the physical degrees of freedom
for each external state are conserved before and after
the toroidal compactification,
the number of physical polarization vectors
will remain the same after the compactification,
except that each massive KK gauge boson acquires a
physical longitudinal component with polarization vector
$\hs\zeta_L^\mu\hs$
which is ensured by the geometric Higgs mechanism\,\cite{5DYM2002}
of KK compactification
and charaterized by the KK gauge boson equivalence theorem
(GAET)\,\cite{5DYM2002}\cite{5DYM2002-2}\cite{KK-ET-He2004}
at the $S$-matrix level.\
This shifting method provides a powerful tool to efficiently
compute both the KK gauge boson amplitudes and
KK graviton amplitudes under toroidal compactification.

\vspace*{1mm}

We first consider an $N$-point scattering amplitude for
massless gauge boson in 5d.\ It can be generally expressed
in the following form:
\begin{align}
\label{eq:T5d}
\widehat{\TT}\Big[\whA^{a_1}_{\hlam_1} \whA^{a_2}_{\hlam_2} \ldots \whA^{a_N}_{\hlam_N}\Big]
\,=\, \hg^{N-2} \sum_{j}^{} \frac{~\CC_j^{}\, \whN_j^{}(\{\hlam\}\hsm )~}
 {\llbracket\widehat{\D}_j^{}\rrbracket} \,,
\end{align}
where on the left-hand-side (LHS) the subscripts
$\hlam_i \!=\! \{\pm 1, 0\}\hs$ denote the helicity states
for each 5d gauge boson,
and on the right-hand-side (RHS) the sum runs over all
$\,(2N \!-\hsm 5)!!\,$ distinct trivalent diagrams
and $\,\llbracket\widehat{\D}_{\hsm j}^{}\rrbracket\,$
denotes the product of denominators of the Feynman propagators.\
In each numerator, $\{\CC_j^{}\}$ and $\{\whN_j^{}\}$
are the color and kinematic factors, obeying the color
and kinematic Jacobi identities respectively:
\begin{equation}
\label{eq:Jacobi-Npt}
\CC_\al^{} + \CC_\be^{} + \CC_\ga^{} = 0 \,,
\hspace*{10mm}
\whN_\al^{} + \whN_\be^{} + \whN_\ga^{} = 0 \,,
\end{equation}
which contain only $(2N \!-\hsm 5)!!-(N\!-\hsm2)!\,$
independent equations for each group of equations
in \eqrefe{eq:Jacobi-Npt} \cite{BCJ:2008}\cite{BCJ:2019}.
Thus, according to \eqrefe{eq:sij-pol-shift},
the $N$-point scattering amplitude for the massive KK gauge bosons
can be derived from the corresponding non-compactified
higher dimensional massless scattering amplitude \eqref{eq:T5d}:
\begin{equation}
\label{eq:T4dS1}
\TT\!\[\hsm A_{\lam_1}^{a_1\hs \bn_1}\hsmx
A_{\lam_2}^{a_2\hs \bn_2}\hsmx
\cdots A_{\lam_N}^{a_N\hs \bn_N}\hsm\]
=\, g^{N-2} \sum_{j}^{}\!
\frac{\CC_j^{} \hs\NN_j^{\fP}(\{\lam\})}
{~\llbracket\D_j^{}\!-\hsm M_{\bn\bn_j}^2\rrbracket~} \,,
\end{equation}
where on the left side 
the subscripts
$\lam_i^{} \!=\! \{\pm 1, 0\}\!\equiv\!\{\pm 1, L\}$
denote the helicities
of each external KK gauge boson state, including
two transverse polarizations
and one longitudinal polarization.
On the right side of \eqrefe{eq:T4dS1},
the 4d gauge coupling $\hs g\hs$
is connected to the 5d gauge coupling $\hs\hg\hs$ via
$\,g\!=\!\hg/\sqrt{2\pi R\,}\,$.
The superscript $\fP$ labels every possible combination of
the signs of the KK indices $\{\bn_i^{}\}$
of the external gauge bosons obeying the $N$-point
neutral condition $\sum_{i=1}^N \bn_i^{}\!=\! 0$
\cite{KKString}.\ 
According to Eq.\eqref{eq:sij-pol-shift},
the numerator $\NN_j^{\fP}(\{\lam\})$ can be obtained from the 
corresponding 5d massless numerator $\whN_j^{}$ by the
shifting method:
%
\beqa 
\NN_j^{\fP}(\{\lam\}) &=  
\whN_j^{}\Big|_{\hat{s}_{ij}^{}\to s_{ij}^{}\hsm -M_{ij}^2}^{}\,,
\eeqa 
%
where $M_{ij}^2\!=\!
(\sgn_i^{}M_{i}^{}\!+\sgn_j^{}M_{j}^{})^{2}$,
and we also make the replacements for products of
the external-state polarizations and momenta
$\hze^{}_i \!\cdot \hat{p}_j \!=\zeta^{}_i\hsm\cdot\hsm p_j$ and 
$\hze^{}_i \cdot \hze^{}_j = \zeta^{}_i \cdot \zeta^{}_j\,$.\
In the denominator of Eq.\eqref{eq:T4dS1}, the symbol
$\hs\llbracket\D_j^{}\hsmx -\hsmx M_{\bn\bn_j}^2\rrbracket\hs$
denotes the product of the denominators of the propagators
with $M_{\bn\bn_j}^{}$
being the relevant KK mass.\ 
For instance, considering the four-point scattering process,
we have
$\hs\D_j^{} \!\in\! \{s,t,u\}$ and
$\hs\bn\bn_j \!\in\! \{\bn_1^{}\!+\bn_2^{},\,
\bn_1^{}\hsmx+\bn_4^{}, \, \bn_1^{}\!+\bn_3^{}\}\hs$,
where $\,j\!=\!\hsm 1,2,3$\,.\
We note that the massive KK gauge boson amplitude \eqref{eq:T4dS1}
obtained by the shifting method just corresponds to the
5d toroidal compactification without orbifold.\

\vspace*{1mm}

Then, we consider the 5d compactification under $S^1\!/\ZZ\hs$
orbifold.\ The KK gauge boson
$A_{\lam}^{a\hs n}\!=\hsm\zeta^\mu_{\lam}A^{an}_{\mu}$
is even under $\ZZ$ and is defined as
\beqa
\label{eq:An=Anp+Anm}
A_{\lam}^{a\hs n} \,=\, \frac{1}{\sqrt{2\,}\,}\hsmx
\[A_{\lam}^{a(+n)}\!+A_{\lam}^{a(-n)}\]\!.
\eeqa
Also a $\ZZ$-odd state $A_{\lam (-)}^{a\hs n}$ can be defined by
flipping the plus sign in the center of the brackets
of Eq.\eqref{eq:An=Anp+Anm} into minus sign.\
Thus, we can derive the $N$-point KK gauge boson
scattering amplitude as follows:
\begin{equation}
\label{eq:T4dS1Z2}
\TT\big[A_{\lam_1}^{a\hs n_1}A_{\lam_2}^{b\hs n_2} \cdots
A_{\lam_N}^{a_N \hs n_N}\big]
= \frac{~g^{N-2}~}{2^{{\bsN}/2}}
\sum_\fP\hsm \sum_{j} 
\frac{\CC_j^{} \hs\NN_j^{\fP}(\lam)}
{~\llbracket \D_j^{}\hsm -\hsm M_{\bn\bn_j}^2 \rrbracket~}
\hs,
\hspace*{6mm}
\end{equation}
where $\bN$ denotes the number of the external KK states
(with $n_j^{}\hsm\!\neq\! 0$)
and the difference $(N\hsmx -\hsm\bN)$ equals the
number of possible external zero-mode states.

\vspace*{1mm}
Next, using the color-kinematics duality
of the BCJ double-copy method,
we can construct the $N$-point 5d massless graviton amplitude
from the corresponding $N$-point massless gauge boson amplitude
\eqref{eq:T5d}:
\begin{equation}
\label{eq:5dGravitonAmp-massless}
\widehat{\M}\big[\hh_{\hsi_1} \hh_{\hsi_2} \cdots \hh_{\hsi_N}\big]
= (-)^{N+1}\!\hsm\(\frac{\hka}{4}\)^{\!N-2}
\sum_j\!\!\sum_{\hlam^{}_k,\hlam'_k}
\hsm\!\!\Big(\!\prod_k\!
C_{\hsm\hlam_k^{}\hlam'_k}^{\hs\hsi_{\hsm k}^{}}\Big)
\frac{~\whN_j(\hlam_k^{})\hs\whN_j(\hlam'_k)~}
 {\llbracket\widehat{\D}_j^{}\rrbracket}  \,,
\end{equation}
where the helicity index $\,\hsi_k^{}\!=\!\{\pm 2,\pm 1,0\}\,$
labels the five helicity states of each external 5d massless graviton.\
The 5d polarization tensor of the $k$-th external graviton state
is given by the following formulas:
\begin{equation}
\label{eq:zeta-munu-5d}
\hze^{\mn}_{\hs\hsi_{\hsm k}^{}}\hs =
\sum_{\hlam^{}_k,\hlam'_k}\!\!
C_{\hsm\hlam_k^{}\hlam'_k}^{\hs\hsi_{\hsm k}^{}}\hs
\hze_{\hlam_k^{}}^\mu \hs \hze_{\hlam'_k}^\nu \,,
\end{equation}
where on the right-hand-side
summations over the repeated helicity indices
$(\hlam_k^{},\hlam'_k)$ are implied.\
In Eq.\eqref{eq:zeta-munu-5d},
the normalization coefficients in each polarization tensor of graviton
are defined as follows:
\\[-6mm]
\beqs
\label{eq:Ckk'}
\begin{align}
\hspace*{-10mm}
\hh_{\pm 2}^{} = \hze^{\mn}_{\pm 2}\hs\hh_{\mn}\!:
&\hspace*{4mm}
C_{\hsm\hlam_k^{}\hlam'_k}^{\hs\hsi_{\hsm k}^{}}\!\hsm
= C_{\pm 1,\pm 1}^{\pm 2}\hsm =\hsm 1 \hs,
\\
\hspace*{-10mm}
\hh_{\pm 1}^{} = \hze^{\mn}_{\pm 1}\hs\hh_{\mn}\!:
&\hspace*{4mm}
C_{\hsm\hlam_k^{}\hlam'_k}^{\hs\hsi_{\hsm k}^{}}\!\hsm
= C_{\pm 1,0}^{\pm 1}\hsm = C_{0,\pm 1}^{\pm 1}\hsm
=\hsm \sqrt{\hsm\Fr{1}{2}\hs} \hs,
\\[0.5mm]
\hspace*{-10mm}
\hh_0^{} = \hze^{\mn}_{0}\hs\hh_{\mn}\!:
&\hspace*{4mm}
C_{\hsm\hlam_k^{}\hlam'_k}^{\hs\hsi_{\hsm k}^{}}\!\hsm
= C_{\pm 1,\mp 1}^{\hs 0}\hsm =\!\sqrt{\hsm\Fr{1}{6}\hs}\hs,~~~
C_{0,0}^{\hs 0}\hsm =\! \sqrt{\hsm\Fr{2}{3}\hs} \hs,
\end{align}
\eeqs
where the polarization tensors of the 5d massless graviton are
defined by ``doubling up'' the three transverse polarization vectors
of the 5d massless gauge boson via
\beqs
\label{eq:Gpol-massless}
\begin{align}
\label{eq:Gpol-21}
\hze^{\mn}_{\pm 2} &\hs =\hs
\hze_{\pm 1}^{\mu} \hze_{\pm 1}^{\nu}\,,\quad
\hze_{\pm 1}^{\mn} \hs =
\Fr{1}{\sqrt{2\,}\,}\!\hsm\(\hze_{\pm 1}^{\mu}\hze_0^{\nu}
\hsm +\hze_0^{\mu}\hze_{\pm 1}^{\nu}\) \!,
\\[1mm]
\label{eq:Gpol-L}
\hze^{\mn}_0 & = \Fr{1}{\sqrt{6\,}\,}\!\hsm
\(\hze^\mu_{+1}\hze^\nu_{-1}\hsmx + \hze^\mu_{-1}\hze^\nu_{+1}
\hsmx +2\hs\hze^\mu_0 \hze^\nu_0\)\!.
\end{align}
\eeqs
%
In the double-copy formula \eqref{eq:5dGravitonAmp-massless},
we have made the following replacement
between the gauge couplings
and gravitational couplings\,\cite{KKString}:
\begin{equation}
\label{eq:g2-kappa}
{\hat{g}^{N-2} }\,\longrightarrow~
(-)^{N+1}\hsm (\hka/4)^{N-2} \,.
\end{equation}
With the 5d compactification we have relations between the 5d and 4d
couplings, $(\hat{g},\hs\hat{\kappa})=(g,\hs\kappa)\sqrt{L\hs}\hs$,
where $L$ denotes the length of 5d.\

\vspace*{1mm}

Then, under the $S^1$ compactification, we can use
our shifting method in Eq.\eqref{eq:sij-pol-shift}
to derive the 4d $N$-point massive KK graviton scattering
amplitude from the corresponding double-copied 5d
massless graviton amplitude \eqref{eq:5dGravitonAmp-massless}
as follows:
\begin{equation}
\label{eq:MS1-Npt}
\hspace*{-6mm}
\M \big[h_{\si_1}^{\bn_1} h_{\si_2}^{\bn_2} \cdots
h_{\si_N}^{\bn_N}\big]
=\, (-)^{N+1}\!\hsm\(\!\frac{\hs\ka\,}{4}\!\)^{\!\!N-2}\!\!
\sum_j\!\!\sum_{\lam^{}_k,\lam'_k}\!\!\!
\Big(\!\prod_k\!\hsm C_{\hsm\lam_k^{}\lam'_k}^{\hs\si_{\hsm k}^{}}\Big)
\hsm\frac{~\NN_j^\fP(\lam_k^{}) \hs \NN^{\hs\fP}_j(\lam'_k)~}
{\llbracket\D_j^{}\!-\!M_{\bn\bn_j}^2\rrbracket}\hs,
\end{equation}
where the helicity index
$\,\si_k^{}\!=\!\{\pm 2,\pm 1,0\}
 \!\equiv\!\{\pm 2,\pm 1,L\}\,$
labels the five helicity states of each external 4d massive
KK graviton.\
Equivalently, we can also construct the above $N$-point
massive KK graviton scattering amplitude by directly applying the
color-kinematics duality to the corresponding $N$-point
massive KK gauge boson amplitude \eqref{eq:T4dS1}.
In Eq.\eqref{eq:MS1-Npt}, the polarization tensors of the external
KK graviton states are defined as
\beqs
\label{eq:zeta-munu-4d}
\begin{align}
& \zeta^{\mn}_{\hs\si_{\hsm k}^{}} =
\sum_{\lam^{}_k,\lam'_k} \!\!
C_{\hsm\lam_k^{}\lam'_k}^{\hs\si_{\hsm k}^{}}\hs
\zeta_{\lam_k^{}}^\mu\hs\zeta_{\lam'_k}^\nu \,,
\\
& C_{\pm 1,\pm 1}^{\pm 2} \!=\hsm 1 \hs,~~
C_{\pm 1,0}^{\pm 1}\hsm = C_{0,\pm 1}^{\pm 1}\hsm
=\hsm \sqrt{\hsm\Fr{1}{2}\hs} \hs,~~
C_{\pm 1,\mp 1}^{\hs 0}\hsm =\!\sqrt{\hsm\Fr{1}{6}\hs}\hs,~~
C_{0,0}^{\hs 0}\hsm =\! \sqrt{\hsm\Fr{2}{3}\hs} \hs.
\end{align}
\eeqs
Thus, the polarization tensors of a massive KK graviton
take the following forms:
\beqs
\label{eq:Gpol-massive-KK}
\begin{align}
\zeta^{\mn}_{\pm 2} &=\hs
\zeta_{\pm 1}^{\mu} \zeta_{\pm 1}^{\nu}\hs,\quad
\zeta_{\pm 1}^{\mn} =
\Fr{1}{\sqrt{2\,}\,}\big(\zeta_{\pm 1}^{\mu}\zeta_L^{\nu}
\!+\zeta_L^{\mu}\zeta_{\pm 1}^{\nu}\big) \hs,
\label{eq:Gpol-massive-KKT}
\\
\zeta^{\mn}_L & = \Fr{1}{\sqrt{6\,}\,}
\big(\zeta^\mu_{+1}\zeta^\nu_{-1}\hsmx
+\zeta^\mu_{-1}\zeta^\nu_{+1}\hsmx
+2\hs\zeta^\mu_L \zeta^\nu_L\big)\hs,
\label{eq:Gpol-massive-KKL}
\end{align}
\eeqs
where we denote
$\hs\zeta^\mu_L\hsm\equiv\hsm\zeta_{\hs 0}^\mu\hs$
and $\hs\zeta^{\mu\nu}_L\hsm\equiv\hsm\zeta_{\hs 0}^{\mu\nu}$.

\vspace*{1mm}

Finally, for the KK compactification under the orbifold $S^1\!/\ZZ$,
we can apply the color-kinematics duality to
the corresponding $N$-point massive KK gauge boson amplitude
\eqref{eq:T4dS1Z2} and derive the following
$N$-point massive KK graviton scattering amplitude:
\begin{equation}
\label{eq:MS1Z2-Npt}
\hspace*{-5mm}
\M\big[h_{\si_1}^{n_1} h_{\si_2}^{n_2} \ldots h_{\si_N}^{n_N}\big]
= \frac{\,(-)^{N+1}}{2^{{\bsN}/2}}\!
\(\!\frac{\hs\ka\,}{4}\!\)^{\!\!N-2}\!\!
\sum_\fP\!\sum_j\!\!\sum_{\lam^{}_k,\lam'_k}
\!\!\Big(\!\prod_k\! C_{\lam_k^{}\lam'_k}^{\,\si_k^{}}
\Big)\hsm
\frac{~\NN_j^\fP(\lam_k^{}) \NN^{\hs\fP}_j(\lam'_k)~}
{\llbracket\D_j^{}\!-\!M_{\bn\bn_j}^2\rrbracket}\,,
\end{equation}
where $\bN$ denotes the number of $\ZZ$-even external KK states
and the $(N\!-\!\bN)$ equals the number of possible external
zero-mode states.

\vspace*{2mm}
\section{\hspace*{-2mm}Extended Color-Kinematics Duality
for Massive KK Double-Copy}
\label{sec:3}
\vspace*{1.5mm}

In the previous section, we have established the general
correspondence between the massless scattering amplitudes in
the noncompactified higher dimensional theory
and the massive KK scattering amplitudes in the
compactified 4d KK theory.\
With these, we present systematically in this section
the explicit massive double-copy construction
for the four-point and five-point scattering amplitudes of
the massive KK gauge bosons and KK gravitons.\
In Section\,\ref{sec:3.1}, we present the general four-point
massive KK gauge boson amplitudes and the double-copied KK
graviton amplitudes from extending the corresponding massless
5d scattering amplitudes.\
Using the generalized massive gauge invariance 
on the double-copied KK graviton amplitudes
or imposing the massive fundamental BCJ relation, 
we will derive a mass spectral condition.\
Then, we will derive the same mass spectral condition
by using the massive fundamental BCJ relation.\
In Section\,\ref{sec:3.2}, we present the explicit double-copy
constructions of the four-point KK graviton amplitudes from
the corresponding KK gauge boson amplitudes under the
5d toroidal compactification of $S^1$.\
Then, in Section\,\ref{sec:3.3} we extend
this analysis to the case of the 5d orbifold compactification
of $S^1\!/\ZZ$.\ In Section\,\ref{sec:3.4}, we further present
the five-point KK graviton scattering amplitudes from double-copy.\
Finally, in Section\,\ref{sec:5} we will derive the nonrelativistic
KK scattering amplitudes for the heavy KK gauge bosons, heavy
KK gravitons, and heavy KK scalars respectively.

\vspace*{2mm}
\subsection{\hspace*{-2mm}KK Amplitudes and Double-Copy
under 
Toroidal Compactification}
\vspace*{1.5mm}
\label{sec:3.1}

Consider the massless gauge boson amplitude at tree level
in the 5d Yang-Mills theory.\
The four-point scattering amplitudes can be generally expressed
as the sum of the three kinematic channels with distinct color
structures, corresponding to the three pole-diagrams plus the
contributions of the contact diagram (which are absorbed into
the pole terms):
\begin{equation}
\label{eq:T5d-4pt}
\widehat{\TT}\!\[\hsm\whA^{a}_{\hlam_1} \whA^{b}_{\hlam_2}
\whA^{c}_{\hlam_3} \whA^{d}_{\hlam_4}\]
\,=\, \hg^{2} \sum_{j}
\frac{~\CC_j^{} \hs\whN_j^{}(\hlam)~}{\sh_j^{}} \,.
\end{equation}
On the left side of Eq.\eqref{eq:T5d-4pt},
the subscript $\,\hlam_i^{} \!=\!\{\pm 1, 0\}\,$
denotes the three transverse helicity states
for each 5d gauge boson, whereas on the right side
the subscript $\,j\!\in\!\{s,t,u\}\,$ runs over
the three kinematic channels
and $\,\hg\,$ is the 5d gauge coupling constant.\
In the numerator of \eqrefe{eq:T5d-4pt}, the color factors
$\CC_j^{}$ are defined as follows:
\begin{equation}
\label{eq:Cj-def}
(\CC_s,\hs \CC_t,\hs \CC_u)
\hs =\hs (f^{abe}f^{cde},\hs f^{ade}f^{bce},\hs f^{ace}f^{dbe})\hs,
\end{equation}
and the kinematic factors $\whN_j^{}$ are given
in \eqrefe{eq:Nstu-5d} of Appendix\,\ref{App:B1}.\
They obey the color and kinematic Jacobi identities
respectively\,\cite{BCJ:2008}\cite{BCJ:2019}:
\begin{equation}
\label{eq:Jacobi-C-N}
\CC_s^{}\hsm +\CC_t^{}\hsm +\CC_u^{} = 0\,, \hspace*{8mm}
\whN_s^{}+\whN_t^{}+\whN_u^{} = 0\,.
\end{equation}
In the denominator of \eqrefe{eq:T5d-4pt},
$\sh_j^{}$ denotes the 5d Mandelstam variables
and has the correspondence with the general denominator
of the $N$-point amplitude \eqref{eq:T5d}:
$\llbracket\widehat{\D}_j^{}\rrbracket\ito\sh_j^{}\,$,
their definitions in the center-of-mass frame are given by
\eqrefe{eq:stu-massless} of Appendix\,\ref{App:A}.\
We note that because of the color Jacobi identity
as given by the first formula of Eq.\eqref{eq:Jacobi-C-N},
the gauge boson amplitude \eqref{eq:T5d-4pt}
is invariant under the following generalized gauge
transformations\,\cite{BCJ:2008}:
\beq
\label{eq:GGT-Nj}
\whN_j^{\hs\prime} \,=\, \whN_j^{} +
\shat_j^{}\!\times\!\widehat\Delta \,,
\eeq
where the gauge parameter $\widehat\Delta\hs$
is an arbitrary function of kinematic variables.\

\vspace*{1mm}

Then, we compactify the 5d Yang-Mills theory on a flat $S^1$ space
with the fifth coordinate
$\,0 \!\leqslant\hsmx y \hsmx\leqslant\! 2\pi R$\,.\
Under such compactification, the 5d massless gauge bosons
$\,\whA^{a}_M\,$ acquire masses through the
geometric Higgs mechanism\,\cite{Hang:2021fmp}\cite{5DYM2002}
and result in a tower of massive KK gauge boson states,
whereas the zero-mode gauge bosons remain massless.\
As generally shown in Eq.\eqref{eq:T4dS1},
we can further derive the 4d massive KK gauge boson
scattering amplitude from the corresponding 5d massless
amplitude \eqref{eq:T5d-4pt} as follows:
\begin{equation}
\label{eq:T4dS1-4pt}
\TT\!\[A_{\lam_1}^{a\hs \bn_1^{}}A_{\lam_2}^{b\hs \bn_2^{}}
A_{\lam_3}^{c\hs \bn_3^{}}A_{\lam_4}^{d\hs \bn_4^{}}\]
=\, g^2 \sum_{j} \frac{~\CC_j^{} \hs\NN_j^{\fP}(\lam)~}
{~s_{\hsm j}^{}-M^2_{\bn\bn_j}~} \,,
\end{equation}
where on the left side
the index $\lam_i^{}\,$ denote the helicities of each external
gauge boson state which corresponds to two transverse polarizations
and one longitudinal polarization,
$\lam_i \!=\! \{\pm 1, 0\}\!\equiv\!\{\pm 1, L\}$.\
In Eq.\eqref{eq:T4dS1-4pt},
the KK number of each external gauge boson is
$\,\bn_i^{} \!=\hsmx \pm n_i^{}\,$ with
$\,n_i^{} \!\in\hsm \mathbb{Z}^+\hsm$,
where the sum of the KK numbers of all external states
should satisfy the neutral condition: 
\begin{equation}
\label{eq:NeutralCond}
\bn_1^{} + \bn_2^{} + \bn_3^{} + \bn_4^{}= 0 \,.
\end{equation}
On the right-hand side of \eqrefe{eq:T4dS1-4pt},
the mass pole is defined as
$\hs M_{\bn\bn_j}^{} \!\! =\!\bn\bn_j/R$\,, with
$\bn\bn_j \!\in\!
\{\bn_1^{}\! +\hsmx\bn_2^{},\, \bn_1^{}\hsm\!+\hsmx\bn_4^{},\,
\bn_1^{}\hsm\!+\hsmx\bn_3^{}\}\,$
and $\,j\!\!\in\!\!\{s,t,u\}$,
whereas the symbol $\fP$ labels each allowed combination
of the KK numbers of the external gauge bosons
obeying the neutral condition \eqref{eq:NeutralCond}.\
An essential property of the amplitude \eqref{eq:T4dS1-4pt}
is that it contains only the \textit{simple poles}
in the denominator.\
Thus, by setting all the internal momenta be on-shell,
we can factorize the higher-point amplitude into the products
of lower-point amplitudes at the presence of simple poles.\
We further note that the massive KK gauge boson amplitude
\eqref{eq:T4dS1-4pt} is invariant under the following
generalized gauge transformation:
\beq
\label{eq:GGT-Nj-KK}
\NN_j^{\fP\hs\prime} \,=\,
\NN_j^{\fP} + \(\hsm s_j^{}\hsm -\hsm M^2_{\bn\bn_j}\hsm\)
\!\times\!\Delta \,,
\eeq
where $\Delta$ denotes an arbitrary gauge parameter
which is a function of kinematic variables.\

\vspace*{1mm}

Finally, using the shifting method of Ref.\,\cite{KKString},
we also can directly derive
the massive KK gauge boson scattering amplitude \eqref{eq:T4dS1-4pt}
from the corresponding scattering amplitude of the 4d massless
zero-mode gauge bosons:
\begin{equation}
\label{eq:T4dS1-4pt-0mode}
\TT \big[A_{\lam_1}^{a\hs \bn_1}A_{\lam_2}^{b\hs\bn_2}
A_{\lam_3}^{c\hs \bn_3}A_{\lam_4}^{d\hs \bn_4}\big]
\,=\, \TT\big[A_{\lam_1}^{a\hs0}A_{\lam_2}^{b\hs 0}
A_{\lam_3}^{c\hs0}A_{\lam_4}^{d\hs0}\big]
\Big|_{
\substack{
s_j^{} \to\hs s_j^{} - M_{\bn\bn_j}^2}} ,
\end{equation}
which agrees to Eq.\eqref{eq:T4dS1-4pt}.\
We note that the massive KK amplitude \eqref{eq:T4dS1-4pt} is
deduced from the 5d massless gauge boson amplitude \eqref{eq:T5d-4pt}
by using the shifting method of Eq.\eqref{eq:sij-pol-shift},
whereas the massive KK amplitude
\eqref{eq:T4dS1-4pt-0mode} is obtained from the 4d
massless gauge boson amplitude by using of the
shifting method of Ref.\,\cite{KKString}.
In fact, the two methods of deriving Eq.\eqref{eq:T4dS1-4pt}
and Eq.\eqref{eq:T4dS1-4pt-0mode} give the same result
and are thus practically equivalent.

\vspace*{1mm}

Next, we study how to explicitly construct the four-point
massive KK graviton scattering amplitudes.\
We note that the four-point massless graviton ampliutde in 5d
can be obtained by applying the color-kinematics (CK)
duality\,\cite{BCJ:2008}\cite{BCJ:2019} to the corresponding
5d massless gauge boson amplitude \eqref{eq:T5d-4pt}.\
Because the color factors and kinematic factors of the
gauge boson amplitude \eqref{eq:T5d-4pt} satisfies
the Jacobi identities of Eq.\eqref{eq:Jacobi-C-N},
this reflects an exchangeability between
those two kinds of factors
$\,\CC_j^{}\!\leftrightarrow\!\whN_j^{}\hs$.\
Hence, we can replace the color factors by the kinematic factors
and replace gauge coupling $\,\hg\,$
with the gravity coupling $\,\hka/4\,$ simultaneously,
which lead to the following 5d massless graviton amplitude:
\begin{align}
\widehat{\M}\big[
\hh_{\hsi_1} \hh_{\hsi_2}\hh_{\hsi_3}\hh_{\hsi_4}\big]
&=\, -\, \widehat{\TT}\!\[\whA^{a}_{\hlam_1} \whA^{b}_{\hlam_2}
\whA^{c}_{\hlam_3} \whA^{d}_{\hlam_4}\]\!
\bigg|_{\vspace*{-10mm}
\substack{\hspace*{0mm} \CC_j \,\to\, \whN_j \\
{\hspace*{0.5mm} \hg \,\to\, \hka/4 }
\\{} \\ {} \\ {}}}
\nn\\[-5mm]
&=\, -\frac{~\hka^2~}{16} \sum_j\!\!\sum_{\hlam^{}_k,\hlam'_k}
\!\!\!\Big(\!\prod_{k=1}^4\!\hsm
C^{\hs\hsi_k^{}}_{\hlam_k^{}\hlam'_k}\Big)\hsm
\frac{~\whN_j^{}(\hlam_k)  \hs
\whN^{}_j (\hlam'_k)~}{\sh_j^{}}  \,,
\label{eq:M4pt-5d}
\end{align}
where helicity index $\,\hsi_k^{} = \{\pm2,\pm1,0\}\,$
denotes the five helicity states for each 5d massless 
graviton.\footnote{%
Here the overall gravitational coupling coefficient  
(including its minus sign) on the right-hand side of
\eqrefe{eq:M4pt-5d} cannot be predicted by the BCJ method itself;
instead it can be determined from the KLT counterpart 
as derived from the massless string amplitudes 
in the field-theory limit\,\cite{BCJ:2019}\cite{Tye-2010}
(where this overall minus sign comes from the contour integral 
on the complex plane for the closed
string\,\cite{string}\cite{KLT}\cite{Bjerrum-Bohr:2010pnr}).}\
It can be shown that the above massless BCJ formula
\eqref{eq:M4pt-5d} agrees to the graviton amplitudes
as derived from the field-theory limit of the 
conventional KLT relations for the massless closed/open 
string amplitudes\,\cite{BCJ:2019}\cite{Tye-2010}.\ 
In Eq.\eqref{eq:M4pt-5d},
the polarization tensor of each external graviton state 
is defined in Eqs.\eqref{eq:zeta-munu-5d}-\eqref{eq:Ckk'}.

\vspace*{1mm}

Then, we apply the generalized gauge transformation
\eqref{eq:GGT-Nj} to the double-copied
5d massless graviton scattering amplitude
\eqref{eq:M4pt-5d} and require it to be gauge-invariant.\
This leads to the conditions
$\hs\sum_j^{}\hsm\whN_j^{}\hsm\!=\!0\hs$ and
$\hs\sum_j\hsmx\shat_j^{}\!=\hsm 0\hs$ (with $j\!=\hsm s,t,u$),
where the first condition is just the kinematic Jacobi identity in
Eq.\eqref{eq:Jacobi-C-N}, and the second condition always holds
for the massless external states as shown by the first formula of
Eq.\eqref{eq:Sum-stu}.\

\vspace*{1mm}

In addition, we can substitute the numerators of \eqrefe{eq:Nstu-5d}
into \eqrefe{eq:M4pt-5d} and re-express the four-point graviton
scattering amplitude \eqref{eq:M4pt-5d} in a Lorentz-invariant form
which is a function of all the relevant kinematic variables:
\begin{equation}
\label{key}
\widehat{\M}
\big[\hh_{\hsi_1} \hh_{\hsi_2}\hh_{\hsi_3}\hh_{\hsi_4}\big]
=
\frac{\,\hka^2\,}{\,\sh \hs \th \hs \uh~}\!\!
\sum_{\hlam^{}_k,\hlam'_k}  \!\!\!
\Big(\!\prod_{k=1}^4\!\hsm C^{\hs\hsi_k^{}}_{\hlam_k^{}\hlam'_k}\Big)
K_{1234}^{} \!\times\! K_{1'2'3'4'}^{} \hs,
\end{equation}
where the factor $K_{1234}^{}\equiv K_{\lam_1\lam_2\lam_3\lam_4}^{}$
is given by
\begin{align}
\label{eq:K-5d}
\hspace*{-4mm}
K =&\   \sh\hs\th \hs (\hze^{}_1 \!\cdot \hze^{}_3)\hs
(\hze^{}_2 \!\cdot \hze^{}_4)
+ \sh\hs\uh \hs (\hze^{}_1 \!\cdot \hze^{}_4)\hs
(\hze^{}_2 \!\cdot \hze^{}_3)
+ \th\hs\uh\hs (\hze^{}_1 \!\cdot \hze^{}_2)\hs
(\hze^{}_3 \!\cdot \hze^{}_4)
\nn\\
&\ -2\hs\sh \, [(\hat{p}^{}_1\!\cdot \hze^{}_4)
(\hat{p}^{}_3\!\cdot \hze^{}_2) (\hze^{}_1\!\cdot \hze^{}_3)
+ (\hat{p}^{}_1\!\cdot \hze^{}_3) (\hat{p}^{}_4\!\cdot \hze^{}_2)
(\hze^{}_1\!\cdot \hze^{}_4)
+ (\hat{p}^{}_2\!\cdot \hze^{}_3) (\hat{p}^{}_4\!\cdot \hze^{}_1)
(\hze^{}_2\!\cdot \hze^{}_4)
\nn\\
&\ + (\hat{p}^{}_2\!\cdot \hze^{}_4)(\hat{p}^{}_3\!\cdot \hze^{}_1)
(\hze^{}_2\!\cdot \hze^{}_3)]
-2\hs\th\,[(\hat{p}^{}_1\!\cdot \hze^{}_2)
(\hat{p}^{}_3\!\cdot \hze^{}_4) (\hze^{}_1\!\cdot \hze^{}_3)
+ (\hat{p}^{}_1\!\cdot \hze^{}_3) (\hat{p}^{}_2\!\cdot
\hze^{}_4) (\hze^{}_1\!\cdot \hze^{}_2)
\nn\\
&\ +(\hat{p}^{}_2\!\cdot \hze^{}_1)(\hat{p}^{}_4\!\cdot \hze^{}_3)
(\hze^{}_2\!\cdot \hze^{}_4)
+ (\hat{p}^{}_3\!\cdot \hze^{}_1)(\hat{p}^{}_4\!\cdot \hze^{}_2)
(\hze^{}_3\!\cdot \hze^{}_4)]
-2\hs\uh \big[(\hat{p}^{}_1\!\cdot \hze^{}_2) (\hat{p}^{}_4\!\cdot
\hze^{}_3) (\hze^{}_1\!\cdot \hze^{}_4)
\nn\\
&\ + (\hat{p}^{}_1\!\cdot \hze^{}_4) (\hat{p}^{}_2\!\cdot \hze^{}_3)
(\hze^{}_1\!\cdot \hze^{}_2)
+(\hat{p}^{}_3\!\cdot \hze^{}_2) (\hat{p}^{}_4\!\cdot \hze^{}_1)
(\hze^{}_3\!\cdot \hze^{}_4) + (\hat{p}^{}_3\!\cdot \hze^{}_4)
(\hat{p}^{}_2\!\cdot \hze^{}_1)
(\hze^{}_2\!\cdot \hze^{}_3)\big]\hs ,
\end{align}
and the factor
$K_{1'2'3'4'}^{}\hsm\equiv\hsm K_{\lam_1'\lam_2'\lam_3'
\lam_4'}^{}$ takes the same form as Eq.\eqref{eq:K-5d}.\
As can be checked, the four-point 5d massless graviton
scattering amplitude
given by \eqrefe{eq:M4pt-5d} or Eqs.\eqref{key}-\eqref{eq:K-5d}
agrees with that of the direct Feynman diagram calculation.\
It also agrees with the field theory limit of
the four-point genus-zero closed string amplitude\,\cite{Schwarz:1982}.

\vspace*{1mm}

Then, under the toroidal compactification of $S^1$ and
from the $N$-point massive double-copy formula \eqref{eq:MS1-Npt},
we can deduce the four-point massive KK graviton
scattering amplitude as follows:
\begin{equation}
\label{eq:Mh-4pt-BCJ}
\hspace*{-10mm}
\M\!\[h_{\si_1}^{\bn_1} h_{\si_2}^{\bn_2} h_{\si_3}^{\bn_3}h_{\si_4}^{\bn_4}\]
=\, -\frac{~\ka^2~}{16}  \sum_j\!\!\sum_{\lam^{}_k,\lam'_k}
\!\!\!\Big(\!\prod_{k=1}^4\!\hsm
C_{\lam_k^{}\lam'_k}^{\hs\si_{\hsm k}^{}}\Big)\hsm
\frac{~\NN_j^\fP(\lam_k) \hs \NN^{\fP}_j(\lam'_k)~}
{s_j^{}-M^2_{\bn\bn_j}} \,,
\end{equation}
where $\{C_{\lam_k^{}\lam'_k}^{\hs\si_{\hsm k}^{}}\}$
denote the coefficients of the polarization tensor
of the $k$-th external KK graviton state as
given by Eqs.\eqref{eq:zeta-munu-4d}-\eqref{eq:Gpol-massive-KK}.
We can apply the generalized gauge transformation
\eqref{eq:GGT-Nj-KK} to the double-copied KK graviton scattering amplitude
\eqref{eq:Mh-4pt-BCJ} and require it to be gauge-invariant.\
From this we deduce the following two conditions:
\beq
\label{eq:CondKK-Nj-sj}
\dis \sum_j^{}\hsm\NN_j^{\fP}\hsm = 0\hs,~~~~~
\sum_j\! \(s_j^{}\hsm -\hsm M^2_{\bn\bn_j}\hsm\)\hsm = 0\hs.
\eeq
where $\hs j\hsmx\in\!\{s,t,u\}\hs$.\
The first condition in Eq.\eqref{eq:CondKK-Nj-sj} is the massive
kinematic Jacobi identity for the numerators of the four-point
KK gauge boson amplitude \eqref{eq:T4dS1-4pt}
and can be inferred from the massless
kinematic Jacobi identity in the second formula of
Eq.\eqref{eq:Jacobi-C-N} by using the shifting method
of Sec.\,\ref{sec:2.2}.\
The second condition in Eq.\eqref{eq:CondKK-Nj-sj} is also important
and can be derived from the 5d massless kinematic condition
$\hs\sum_j\hsmx\shat_j^{}\!=\hsm 0\hs$ in
Eq.\eqref{eq:Sum-stu} by using the shifting formula
\eqref{eq:sij-shift}.\
Thus, according to Eq.\eqref{eq:Sum-stu},
we have the sum of Mandelstam variables for the
four KK gauge boson scattering,
$\hs\sum_j\hsm s_j^{}\!=\!\sum_{i=1}^4\! M_{\bn_i^{}}^2\hs$.\
Hence, from the second condition in Eq.\eqref{eq:CondKK-Nj-sj},
we derive the following four-point mass spectral condition:
\beq
\label{eq:MSCond-4KK}
\sum_{i=1}^4\! M_{\bn_i^{}}^2 =\hs
M^2_{\bn\bn_s}\!+ M^2_{\bn\bn_t}\! + M^2_{\bn\bn_u}
\hs,
\eeq
where the left-hand side sums up the squared-mass $M_{\bn_i^{}}^2$
of each external KK state and the right-hand side is the sum
of the internal pole-mass-squared of all three kinematic channels.\
This condition requires that the sum of the squared-masses of 
all the external KK states equals the sum of the 
internal squared-mass-poles in the $(s,t,u)$ channels.\
It should hold for any four-point scattering amplitudes
of the KK gauge bosons and of the KK gravitons
in the KK gauge/gravity theories
under the 5d toroidal $S^1$ compactification (without orbifold).\


For the 5d toroidal $S^1$ compactification, the KK mass
$\hs M_{\bn_i}^{}\!\!=\hsm\bn_i^{}/R\,$ and the KK numbers are
conserved in each interaction vertex.\
Thus, the pole mass in each of the $(s,t,u)$ channels
is given by
\beq
M^{}_{\bn\bn_s}\!=M^{}_{\bn_1\!+\bn_2}\hs,~~~~
M^{}_{\bn\bn_t}\!=M^{}_{\bn_1\!+\bn_4}\hs,~~~~
M^{}_{\bn\bn_u}\!=M^{}_{\bn_1\!+\bn_3}\hs.
\eeq
With these we can re-express the mass spectral condition
\eqref{eq:MSCond-4KK} as follows:
\beq
\bn_1^2\hsm +\bn_2^2\hsm +\bn_3^2\hsm +\bn_4^2 \,=\,
(\bn_1^{}\!+\hsm\bn_2^{})^2 \!+
(\bn_1^{}\!+\hsm\bn_4^{})^2 \!+
(\bn_1^{}\!+\hsm\bn_3^{})^2\hs.
\eeq
This relation can be further reduced to
$\bn_1^{}\!+ \bn_2^{}\!+ \bn_3^{}\!+ \bn_4^{}\!=\!0\hs$,
which is just the same as the neutral condition \eqref{eq:NeutralCond}
and is thus guaranteed by the KK number conservation
under the toroidal compactification of $S^1$.\
The neutral condition \eqref{eq:NeutralCond} is again a direct outcome
of the conservation of KK indices under the 5d toroidal
$S^1$ compactification.\
Hence, the mass spectral condition \eqref{eq:MSCond-4KK} does hold
for the 5d toroidal $S^1$ compactification.\

\vspace*{1mm}

Regarding the mass spectral condition \eqref{eq:MSCond-4KK} we have
some comments in order.\
First, we stress that Eq.\eqref{eq:MSCond-4KK} is both a
{\it necessary and sufficient} condition
for {\it directly} realizing the massive double-copy construction
in any KK gauge/gravity theories under the 5d toroidal compactification.\
As we will explicitly demonstrate in Sec.\,\ref{sec:3.3},
under the 5d compactification with $S^1\!/\ZZ$ orbifold,
the condition \eqref{eq:MSCond-4KK} is violated for the four-point
KK scattering amplitudes and the direct double-copy is impossible.\
Instead we could realize the exact double-copy by decomposing each
KK graviton amplitude into a sum of the partial KK graviton amplitudes
which are constructed from the double-copy of the KK gauge boson
amplitudes under the $S^1$ compactification
{\it without} $\hs\ZZ\!$ orbifold.\
Second, for the gauge/gravity theories other than the
KK gauge/gravity theories under the toroidal compactification,
the BCJ-type double-copy could be realized even without obeying
the mass spectral condition \eqref{eq:MSCond-4KK}.\
An important example is the double-copy construction
for the 3d topological Chern-Simons gauge/gravity theories\,\cite{TMG},
where we find that the BCJ-type color-kinematics duality and massive
double-copy can be realized\,\cite{Hang:2021oso}, but the
spectral condition \eqref{eq:MSCond-4KK} is not obeyed
because all physical gauge bosons (gravitons) in the
pure non-Abelian Chern-Simons gauge (gravity) theories
have the same mass $M$ and thus the condition \eqref{eq:MSCond-4KK}
becomes $\hs 4M^2\!\neq\! 3M^2$.\
Finally, we will further analyze the mass spectral condition
\eqref{eq:MSCond-4KK} in Section\,\ref{sec:5} and
demonstrate that without assuming an underlying theory
a priori, solving this spectral condition \eqref{eq:MSCond-4KK}
can uniquely determine the mass spectrum to be that
of the KK theories under the toroidal compactification.

\vspace*{1mm}

In passing, from the KK gauge boson amplitude \eqref{eq:T4dS1-4pt}
and the KK graviton amplitude \eqref{eq:Mh-4pt-BCJ}, we can further
derive the double-copy formulation under the high energy expansion
of $\,s_j^{}\!\gg\!M^2_{\bn\bn_j}$.\
We expand the KK gauge boson amplitude \eqref{eq:T4dS1-4pt} as follows:
\beqs
\begin{align}
\label{eq:T4A-4dS1}
\hspace*{-4mm}
\TT\big[A_{\lam_1}^{a\hs\bn_1^{}}\hsm A_{\lam_2}^{b\hs \bn_2^{}}
\hsm A_{\lam_3}^{c\hs \bn_3^{}}\hsm A_{\lam_4}^{d\hs \bn_4^{}}\big]
\hsm & =\hs g^2\hsm \sum_{j}\hsm\!\frac{\,\CC_j^{}\hs\NN_j^{\fP}
\hsmx (\lam)\,}{\,s_{\hsm j}^{}\!-\hsm M^2_{\bn\bn_j}\,}
\nn\\
& =\hsx g^2\hsm\sum_{j}\!
\frac{\,\CC_j^{}\big[\NN_j^{0,\fP}\hsmx (\lam)\!+\!\dNN_j^{\fP}
\hsmx (\lam)\big]\,}{\,s_{\hsm j}^{}\,}
=\TT_0^{}+\dT  \hs,
\\[1mm]
\frac{\NN_j^{\fP}\hsmx (\lam)}{~1\!-M^2_{\bn\bn_j}/s_j^{}\,}
&=\hsx 
\NN_j^{0,\fP}\hsmx (\lam)\hsm +\hsm\dNN_j^{\fP}\hsmx (\lam)\hs.
\end{align}
\eeqs
Then, we can expand the double-copied four-point
KK graviton scattering amplitude \eqref{eq:Mh-4pt-BCJ} for
$h_{\si_1}^{\bn_1} h_{\si_2}^{\bn_2}\ito h_{\si_3}^{\bn_3}h_{\si_4}^{\bn_4}$
as follows:
\begin{align}
\label{eq:M4h-4dS1}
\hspace*{-4mm}
\M &= -\frac{\,\ka^2\hs}{16}\!\sum_j\!\!\sum_{\lam^{}_k,\lam'_k}\!
\prod_{k=1}^4\!\hsmx C_{\lam_k^{}\lam'_k}^{\hs\si_{\hsm k}^{}}\!
\frac{\,1\!-\! M^2_{\bn\bn_j}\hsm /\hsm s_j^{}\,}{\,s_{\hsm j}^{}\,}
\hsm\Big[\NN_j^{0,\fP}\hsmx (\lam_k)\!+\hsm\dNN^{\fP}_j\hsmx (\lam_k)\Big]\!
\Big[\NN_j^{0,\fP}\hsmx (\lam_k')\!+\hsm\dNN^{\fP}_j\hsmx (\lam_k')
\Big]
\nn\\[1mm]
&= \,\M_0^{}+\dM\,,
\end{align}
where $\M_0^{}$ and $\dM$ denote the expanded KK graviton amplitudes
at the leading order (LO) and next-to-leading order (NLO)
respectively,
\\[-3mm]
{\small
\beqs
\label{eq:M4h-LO-NLO}
\begin{align}
\label{eq:M4h-LO}
\hspace*{-2mm}
\M_0^{} &\hsmx =\hsmx  -\frac{\ka^2}{\,16\,}
\sum_{j} \!\!\sum_{\lam^{}_k,\lam'_k}
\prod_{k} C_{\lam_k^{}\lam'_k}^{\hs\si_{\hsm k}^{}}
\frac{\,\NN_{j}^{0,\fP}\! (\lam^{}_k)\hs \NN_{j}^{0,\fP}\!
(\lam'_k)~}{s_j^{}} \,,
\\
\label{eq:M4h-NLO}
\hspace*{-2mm}
\dM &\hsmx =\hsmx  -\frac{\ka^2}{\,16\,}\!
\sum_{j} \!\!\sum_{\lam^{}_k,\lam'_k}\!\!
\prod_{k} \!\hsm C_{\lam_k^{}\lam'_k}^{\hs\si_{\hsm k}^{}}\!
\frac{\,\NN_{j}^{0,\fP}\! (\lam^{}_k)\da \NN_{j}^{\fP}\!(\lam'_k)
\!+\! \NN_{j}^{0,\fP}\! (\lam'_k)\da\NN_{j}^{\fP}\!(\lam^{}_k)
\!-\!\NN_{j}^{0,\fP}\! (\lam^{}_k)\NN_{j}^{0,\fP}\!(\lam'_k)
M^2_{\bn\bn_j}/s_j^{}\,}{s_j^{}} .
\end{align}
\eeqs
}
\\[-4.5mm]
Using the general power counting method for the KK
theories\,\cite{Hang:2021fmp},
we can estimate that the above four-point LO
KK graviton amplitude $\hs\M_0^{}\!=\!O(E^2M_n^0)\hs$ and the
NLO KK graviton amplitude $\dM\!=\!O(E^0M_n^2)\hs$.\
The above expanded massive double-copy formulas
under the high energy expansion are
important for the formulation of the
KK gravitational equivalence theorem (GRET)\,\cite{Hang:2021fmp}
which quantitatively connects the scattering amplitudes
of the longitudinal KK gravitons to the scattering
amplitudes of the corresponding KK Goldstone bosons
in the high energy limit.

\vspace*{1mm}

We derived the extended KLT relations
for the massive KK open/closed string amplitudes in Ref.\,\cite{KKString}
by using the compactified bosonic string theory.\
We note that this is closely related to the massive KK BCJ
construction as generally discussed in the previous section.\
For the four KK graviton scattering,
we use the massive KLT relation\,\cite{KKString} to
express the tree-level KK graviton scattering amplitude
as the products of the two color-ordered KK gauge boson amplitudes:
\begin{equation}
\label{eq:Mhbar-4pt}
\M\!\hsm\[h_{\si_1}^{\bn_1^{}}h_{\si_2}^{\bn_2^{}}
h_{\si_3}^{\bn_3^{}}h_{\si_4}^{\bn_4^{}}\]
\hs=\hs \frac{~\ka^2~}{16}\!\! \sum_{\lam^{}_k,\lam'_k}
\!\!\!\(\prod_{k=1}^4\!\!C_{\lam_k^{}\lam'_k}^{\hs\si_{\hsm k}^{}}\!\)\!
(t \hsm -\hsm M_{\bn\bn_t}^2 )\hs
\TT_{\lam_k^{}}^\fP\hsm [1234] \hsx
\TT_{\lam'_k}^\fP\hsm [1324] \hs,
\end{equation}
where the superscript $\fP$
on the right-hand-side denotes each given choice of
the KK indices of the external KK gravitons
$\{\bn_1^{},\hs \bn_2^{},\hs \bn_3^{},\hs \bn_4^{}\}$.\
For the case of elastic scattering, there are three types of
independent combinations of the KK numbers for the external state,
$\{\bn_1^{},\hs \bn_2^{},\hs \bn_3^{},\hs \bn_4^{}\} \!=\!
\{\pm n,\pm n,\mp n,\mp n\},\{\pm n,\mp n,\mp n,\pm n\},
\{\pm n,\mp n,\pm n,\mp n\}$.\
In the above Eq.\eqref{eq:Mhbar-4pt},
we have used the shorthand notations
for the color-ordered amplitudes
$\hs\TT[1^{\bn_1}_{\lam_1}2^{\bn_2}_{\lam_2}
 3^{\bn_3}_{\lam_3}4^{\bn_4}_{\lam_4}]
\hsm\equiv\hsm\TT_{\lam_k^{}}^\fP[1234]\,$ and
$\hs\TT[1^{\bn_1}_{\lam_1}3^{\bn_3}_{\lam_3}
 2^{\bn_2}_{\lam_2}4^{\bn_4}_{\lam_4}]
\hsm\equiv\hsm\TT_{\lam_k^{}}^\fP[1324]\hs$.\

\vspace*{1mm}

Then, we can express the above two color-ordered KK gauge boson
scattering amplitudes as follows:
\begin{alignat}{3}
\label{eq:T1234-T1243}
\(\!\!
\begin{aligned}
\TT_{}^\fP\hsm [1234]
\\[1mm]
-\TT_{}^\fP\hsm [1324]
\end{aligned}\)\!
&= \begin{pmatrix}\!
- \frac{1}{\,s-M_{\bn\bn_s}^2\,} \hsm -\hsm
  \frac{1}{\,t-M_{\bn\bn_t}^2\,}~~   ~&
-\frac{1}{t -M_{\bn\bn_t}^2}
\\[2mm]
-\frac{1}{\,t-M_{\bn\bn_t}^2\,} \!
~&
-\frac{1}{\,t-M_{\bn\bn_t}^2\,}\hsm -\hsm
\frac{1}{\,u -M_{\bn\bn_u}^2\,}	
\!\end{pmatrix}
\!\!\!\(\! \begin{aligned}
\NN_s^\fP
\\[1mm]
\NN_u^\fP
\end{aligned} \)
&&\!\equiv \Theta\! \(\! \begin{aligned}
\NN_s^\fP
\\[1mm]
\NN_u^\fP
\end{aligned} \) \hsm\!,
\end{alignat}
where we have used the notations
$\hs\TT[1^{\bn_1}_{\lam_1}2^{\bn_2}_{\lam_2}
 3^{\bn_3}_{\lam_3}4^{\bn_4}_{\lam_4}]
\hsm\!\equiv\!\hsm\TT^\fP\hsm [1234]\,$ and
$\hs\TT[1^{\bn_1}_{\lam_1}3^{\bn_3}_{\lam_3}
 2^{\bn_2}_{\lam_2}4^{\bn_4}_{\lam_4}]
\hsm\!\equiv\!\hsm\TT^\fP\hsm [1324]\hs$
with the helicity indices suppressed for simplicity.\
On the right-hand side of Eq.\eqref{eq:T1234-T1243},
the kinematic numerators have encoded the helicity indices of
the external KK gauge bosons,
$\NN^\fP_s \!=\!\NN^\fP_s(\lam )$ and
 $\NN^\fP_u \!=\!\NN^\fP_u(\lam )$.\
In the derivation of Eq.\eqref{eq:T1234-T1243},
we have used a massive fundamental BCJ identity:
\begin{equation}
\label{eq:fBCJ-relation}
(s-\!M_{\bn\bn_s}^2\hsm )\hs\TT_{}^\fP[1234]
\,=\, (u-\!M_{\bn\bn_u}^2\hsm )\hs\TT_{}^\fP[1324] \hs,
\end{equation}
which is deduced by applying the shifting method to the
conventional massless fundamental BCJ relation
$\,s\hsx\TT[1234]\hsm =\hsm u\hsx\TT[1324]\hs$ \cite{BCJ:2008}
or by taking the field theory limit of the corresponding
relation for the KK open string amplitudes\,\cite{KKString}.\
Then, we can substitute the color-ordered KK gauge boson amplitudes
in Eq.\eqref{eq:T1234-T1243} into \eqrefe{eq:Mhbar-4pt}, and derive
the following BCJ-type KK graviton scattering amplitude:
\begin{equation}
\label{eq:MhbarL-4pt-BCJ}
\hspace*{-6mm}
\M\!\hsm\[h_{\si_1}^{\bn_1^{}}h_{\si_2}^{\bn_2^{}}
h_{\si_3}^{\bn_3^{}}h_{\si_4}^{\bn_4^{}}\]
=\, - \frac{~\ka^2~}{16}\!\sum_j\!\!\sum_{\lam^{}_k,\lam'_k}
\!\!\!\(\prod_{k=1}^4\!\!C_{\lam_k^{}\lam'_k}^{\hs\si_{\hsm k}^{}}\!\)\!
\frac{~\NN_j^\fP\!(\lam_k^{})\hs\NN_j^\fP\!(\lam'_k)~}
{~s_j^{}\hsm -\hsm M_{\bn\bn_j}^2~}\,.
\end{equation}
The above fully agrees with the extended massive BCJ-type double-copy
formula \eqref{eq:Mh-4pt-BCJ} [which was derived from Eq.\eqref{eq:MS1-Npt}
by using the shifting method].\
This full agreement demonstrates that
the massive BCJ-type double-copy formula
\eqref{eq:MhbarL-4pt-BCJ} [or \eqref{eq:Mh-4pt-BCJ}] can be also
derived from the massive KK KLT relation \eqref{eq:Mhbar-4pt}
(which was based on our previous derivation of the KK KLT relations
from analyzing the KK open/closed string amplitudes\,\cite{KKString}).\
Hence, the two types of the massive double-copy constructions
of the four-point KK graviton amplitudes are shown
to be equivalent at tree level.\
We will systematically present the explicit results of the four-point
full scattering amplitudes of the massive longitudinal KK gravitons
in Appendix\,\ref{App:B2}.
\\[-3mm]


Next, we further analyze the structure of the color-ordered
KK gauge boson amplitudes \eqref{eq:T1234-T1243}
in connection to the massive fundamental BCJ relation
\eqref{eq:fBCJ-relation}.\
This relation \eqref{eq:fBCJ-relation} shows that the two
color-ordered KK gauge boson amplitudes
$\TT_{}^\fP[1234]$ and $\TT_{}^\fP[1324]$ are not independent,
so the kernel matrix $\hs\Theta$ of Eq.\eqref{eq:T1234-T1243}
has rank $1\hs$ and leads to a vanishing determinant
$\det\Theta \!=\!0\hs$.\
Using Eq.\eqref{eq:T1234-T1243}, we directly compute this determinant:
\begin{equation}
\label{eq:Theta}
\det \Theta \,=\,
-\frac{~s + t +u \hsm -\hsm
(M_{\bn\bn_s}^2\! +\hsm M_{\bn\bn_t}^2\! +\hsm M_{\bn\bn_u}^2)~}
{(s\hsm -\hsm M_{\bn\bn_s}^2)(t\hsm -\hsm M_{\bn\bn_t}^2)
(u\hsm -\hsm M_{\bn\bn_u}^2)}\,,
\end{equation}
which should vanish once we impose a kinematic condition
$\hsm s + t + u \hsm =
M_{\bn\bn_s}^2\hsm\!+\hsm M_{\bn\bn_t}^2\! + M_{\bn\bn_u}^2\hsm$.
Since Eq.\eqref{eq:Sum-stu} gives the sum of Mandelstam variables
$\hs s + t + u \!=\!\sum_{i=1}^{4}\! M_{\bn_i}^2$,
we can express this kinematic condition as follows:
\begin{equation}
\label{eq:MassCond-4pt}
\sum_{i=1}^{4}\hsm M_{\bn_i}^2 =
M_{\bn\bn_s}^2\!+\hsm M_{\bn\bn_t}^2\!+\hsm M_{\bn\bn_u}^2 \hs.
\end{equation}
This just gives the same four-point mass spectral condition as
Eq.\eqref{eq:MSCond-4KK} which we derived earlier by requiring the
double-copied KK graviton scattering amplitude \eqref{eq:Mh-4pt-BCJ}
to be invariant under the generalized KK gauge transformation
\eqref{eq:GGT-Nj-KK}.\
The above proof of the condition \eqref{eq:MassCond-4pt} relies on
using the massive KK fundamental BCJ relation \eqref{eq:fBCJ-relation}
for the rank reduction of the kernel matrix $\hs\Theta\hs$,
where we deduced Eq.\eqref{eq:fBCJ-relation} to hold for
the KK gauge theories under the 5d toroidal $S^1$ compactification.\
The exact agreement between
the conditions \eqref{eq:MSCond-4KK} and \eqref{eq:MassCond-4pt}
means that the generalized KK gauge transformation
\eqref{eq:GGT-Nj-KK} and the massive KK fundamental BCJ relation
\eqref{eq:fBCJ-relation} are fully compatible.\

\vspace*{1mm}

In passing, we note that a recent literature\,\cite{DC-4dx1} studied
the constraints on a massive double-copy between the massive YM theory and
the dRGT gravity model\,\cite{dRGT}; they found that the massive kernel
matrix of the four-point bi-adjoint scalar amplitudes could take
the minimal rank if a spectral condition is met; their spectral condition
turns out to be similar to our above KK spectral condition
\eqref{eq:MassCond-4pt}.\
But, the literature\,\cite{dRGT} essentially differs from our
present study because we focus on the KK gauge/gravity theories
under the 5d toroidal compactification in which both the
massive KK fundamental BCJ relation \eqref{eq:fBCJ-relation}
and the massive KK double-copy construction
\eqref{eq:MhbarL-4pt-BCJ} or [\eqref{eq:Mh-4pt-BCJ}]
can be proven without additional assumption.\
Our mass spectral condition \eqref{eq:MassCond-4pt}
is a {\it prediction} of the KK gauge/gravity theories
under the 5d toroidal $S^1\hsm$ compactification.\
Moreover, our proof of the same condition \eqref{eq:MSCond-4KK}
only relies on the fact that the double-copied four-point KK graviton
scattering amplitudes \eqref{eq:Mh-4pt-BCJ}
should be gauge-invariant under the generalized
KK gauge transformation \eqref{eq:GGT-Nj-KK}.\
This independently gives the simplest direct proof of
the spectral condition \eqref{eq:MSCond-4KK}.\

\vspace*{2mm}
\subsection{\hspace*{-2mm}Explicit Double-Copy Construction of KK Graviton Amplitudes}
\vspace*{2mm}
\label{sec:3.2}

In this subsection, we compute explicitly the four-point
elastic and inelastic scattering amplitudes
of the KK gauge bosons.\
Then, using the extended BCJ-type massive double-copy construction
of Section\,\ref{sec:3.1},
we derive the corresponding four-point KK graviton
scattering amplitudes under the toroidal compactification of $S^1$.

\vspace*{1mm}

For the convenience and simplicity of explicit demonstration of
our extended massive KK BCJ construction
of the four-point KK graviton amplitudes
in a compact analytical form, we will consider
the leading scattering amplitudes of the longitudinal KK gravitons,
which are defined by using an effective leading-order
polarization tensor
$\hs\bar{\zeta}^{\mn}_L\!\!=\!\zeta^\mu_L \hs \zeta^\nu_L\hs$
of the KK graviton.\footnote{%
We will present the double-copied full scattering amplitudes of
longitudinal KK gravitons using the exact
longitudinal polarization tensor \eqref{eq:Gpol-massive-KKL}
in Appendix\,\ref{App:B2}.}\
Thus, we define the corresponding on-shell
longitudinal KK graviton field as follows:
\begin{equation}
\label{eq:LO-Gpol}
\bar{h}_L = \bar{\zeta}^{\mn}_L h_{\mn} \,, \qquad
\bar{\zeta}^{\mn}_L = \zeta^\mu_L \hs \zeta^\nu_L \,.
\end{equation}
As we will demonstrate in this subsection, the effective longitudinal KK
graviton amplitudes defined by this effective leading-order
polarization tensor
$\hs\bar{\zeta}^{\mn}_L\!\!=\!\zeta^\mu_L \hs \zeta^\nu_L\hs$
can always give the correct leading-order results of
the exact longitudinal KK graviton scattering amplitudes
under the high energy expansion, and thus have a physical meaning.

\vspace*{1.5mm}
\subsubsection*{$\blacklozenge$~Inelastic KK Scattering Amplitudes of
\boldmath{$\,\{0, 0, n, n$\}}:}
\vspace*{1.5mm}

Since the structures of the inelastic KK scattering amplitudes
are simpler than those of the elastic KK
scattering amplitudes, we will start with analyzing the
inelastic KK amplitudes.\
Thus, we first consider the simplest nontrivial example
of the inelastic scattering process.\
This is the ``mixed'' four-particle scattering process
of gauge bosons with the initial states being zero-mode
gauge bosons (gravitons) and the final states being
longitudinal KK gauge bosons (KK gravitons).\
The neutral condition \eqref{eq:NeutralCond} allows
two possible choices of the external KK states
under the $S^1$ toroidal compactification:
\begin{equation}
\label{eq:KK00nn}
\{\bn_1^{},\hs\bn_2^{},\hs \bn_3^{},\hs \bn_4^{}\} =
\{0, 0, \pm n, \mp n\}
\equiv \fA \,,
\end{equation}
where only one of the combinations gives the
independent amplitude.\
Then, we derive the following four-point gauge boson
scattering amplitude for this reaction:
\begin{equation}
\label{eq:T00nn}
\TT\big[\hsm A^{a\hs0}_{\pm 1} A^{b\hs0}_{\mp 1}
A^{c \hs\pm n}_L\hsm A^{d\hs\mp n}_L\big]
=\, g^2 \!\(\!\frac{\,\CC_s^{}\hs\NN_s^{\fA}\,}{s} + \frac{\CC_t^{}\hs\NN_t^{\fA}}{\,t-\!\Mn^2\,} + \frac{\CC_u^{}\hs\NN_u^{\fA}}{\,u-\!\Mn^2\,} \!\) \!,
\end{equation}
where 
the kinematic numerators $\,\{\NN_j^{\fA}\}\,$ are
given in \eqrefe{eq:NA-00nn} of Appendix\,\ref{App:B1}.
With this, we impose the CK duality and derive the
corresponding four-point scattering amplitude
of the longitudinal KK gravitons with the
external states chosen as in Eq.\eqref{eq:KK00nn}:
\begin{align}
\label{eq:M00nn-LO}
\over{\M}\big[h_{\pm 2}^{0} h_{\mp 2}^{0}
\bar{h}_L^{\mp n}\bar{h}_L^{\pm n}\big]
&=\, -\frac{~\ka^2~}{16} \bigg[
\frac{\,(\NN_s^\fA)^2\,}{s}
+ \frac{\,(\NN_t^\fA)^2\,}{\,t\!-\!\Mn^2\,}
+ \frac{\,(\NN_u^\fA)^2\,}{\,u\!-\!\Mn^2~} \bigg]
\nn\\[1mm]
&=\, \frac{~\ka^2\hsm M_n^2\hs (\bs \!+\!4)^2
\hs s^4_\theta~}{~{8}\hs
[(\bs\hsm +\hsm 4)\hsmx -\hsmx (\bs \hsm -\hsm 4)\hs\ctt\hs]\,} \,,
\end{align}
where $\,\bs \!=\! s/\Mn^2\,$ and
$\,(s_{k \theta}^{},\, c_{k \theta}^{}) \!\equiv\!
(\sin \hsm k\theta,\, \cos \hsm k \theta)\hs$.\
In Eq.\eqref{eq:M00nn-LO},
we have made the replacement of couplings
$\,g^2\ito -(\ka/4)^2\,$
based on Eq.\eqref{eq:g2-kappa}.\
We see that the above double-copied inelastic graviton amplitude
with the leading-order longitudinal polarization tensor
\eqref{eq:LO-Gpol} takes a fairly compact form.\
Then, we make the high energy expansion of the amplitude
\eqref{eq:M00nn-LO} and derive the LO amplitude as follows:
\begin{equation}
\label{eq:M00nn-expdLO}
\over{\M}_0^{}\big[h_{\pm 2}^{0} h_{\mp 2}^{0}
\bar{h}_L^{\mp n}\bar{h}_L^{\pm n}\big] \hs =\hs
\frac{~\ka^2\hs s^2_\theta\,s~}{\,{16}\,}
= \frac{~\ka^2\hs (1\!-\hsm \ctt)\hs s~}{\,{32}\,} \hs.
\end{equation}
We further present the full graviton amplitude in
\eqrefe{eq:M00nn-full} of Appendix\,\ref{App:B2}
by using the exact longitudinal polarization tensor
\eqref{eq:Gpol-massive-KKL} of the KK gravitons.\
Under the high energy expansion, the LO and NLO parts
of the full graviton amplitude \eqref{eq:M00nn-full} are
derived in Eqs.\eqref{eq:h00nnFull-LO} and \eqref{eq:h00nnFull-NLO}
respectively and we have
${\M}\big[h_{\pm 2}^{0} h_{\mp 2}^{0}
\bar{h}_L^{\mp n}\bar{h}_L^{\pm n}\big]
\hsm\!=\! {\M}_0^{} +{\dM}\hs$.\
Comparing Eqs.\ \eqref{eq:M00nn-expdLO} and \eqref{eq:h00nnFull-LO},
we can deduce:
\beq
\label{eq:00nnLO-M0bar=M0}
\over{\M}_0^{}\big[h_{\pm 2}^{0} h_{\mp 2}^{0}
\bar{h}_L^{\mp n}\bar{h}_L^{\pm n}\big] =
{\M}_0^{}\big[h_{\pm 2}^{0} h_{\mp 2}^{0}
{h}_L^{\mp n}{h}_L^{\pm n}\big].
\eeq
%
Hence, the scattering amplitude \eqref{eq:M00nn-LO}
which we have computed using the leading polarization tensor
\eqref{eq:LO-Gpol} of longitudinal KK gravitions {\it does have
an important physical meaning,} namely, its expanded LO amplitude
\eqref{eq:M00nn-expdLO} can give the correct LO contribution
\eqref{eq:h00nnFull-LO}
of the full graviton amplitude \eqref{eq:M00nn-full}.\
This is a striking feature because using the leading polarization tensor
\eqref{eq:LO-Gpol} of longitudinal KK gravitions we only need to
compute the corresponding gauge boson amplitude including the external
longitudinal polarization state alone (for the double-copy
construction), but in the full graviton amplitude
its external longitudinal KK graviton state contains the exact
longitudinal polarization tensor
\eqref{eq:Gpol-massive-KKL}
which includes the products of the gauge boson polarization vectors
of both the helicities $\pm 1$ and $0\hs$, and is thus much
more complicated.\
In the following, we will demonstrate explicitly that this
feature holds for all the four-point KK graviton scattering amplitudes
studied in this section.


\vspace*{1.5mm}
\subsubsection*{$\blacklozenge$~Inelastic KK Scattering Amplitudes of
\boldmath{$\{n, 2n, 3n, 4n\}$}:}
\vspace*{1.5mm}

As the second example, we study the inelastic KK scattering
process containing the following KK indices
of the external states:
\begin{equation}
\label{eq:KK-1234n}
\{\bn_1^{},\hs\bn_2^{},\hs \bn_3^{},\hs \bn_4^{}\} =
\{\pm n, \mp 2n, \mp 3n, \pm 4n\}
\equiv \fA \,.
\end{equation}
Thus, we derive the following four-point leading
scattering amplitude of longitudinal KK gauge bosons:
\begin{equation}
\label{eq:TAL-1234}
\TT\!\[A^{a\hs\pm n}_L A^{b\hs\mp 2n}_L
A^{c\hs\mp 3n}_L A^{d\hs\pm 4n}_L\]
=\, g^2 \!\( \frac{\CC_s^{}\hs\NN_s^{\fA}}{\,s-\Mn^2\,}
+ \frac{\CC_t^{}\hs\NN_t^{\fA}}{\,t-25\Mn^2\,}
+ \frac{\CC_u^{}\hs\NN_u^{\fA}}{\,u-4\Mn^2\,} \) \!,
\end{equation}
where 
the kinematic numerators $\,\{\NN_j^{\fA}\}\,$
are given by \eqrefe{eq:NA-00nn}.\
With the longitudinal KK gauge boson amplitude
\eqref{eq:TAL-1234} and using the CK duality, we compute
the corresponding four-point leading longitudinal KK
graviton amplitude as follows:
\begin{align}
\label{eq:M1234n-LO}
& \over{\M}\!\[\bar{h}^{\pm n}_L \bar{h}^{\mp 2n}_L
\bar{h}^{\mp 3n}_L \bar{h}^{\pm 4n}_L\]
= - \frac{~\ka^2~}{16}\!
\[\hsm\frac{(\NN_s^{\fA})^2}{~s\hsm -\hsm\Mn^2\,} +
\frac{(\NN_t^{\fA})^2}{\,t\hsm -\hsm 25\Mn^2\,} +
\frac{(\NN_u^{\fA})^2}{\,u\hsm -\hsm 4\Mn^2~}\]
\hspace*{8mm}
\nn\\[1.5mm]
&\hspace*{5mm}
= \frac{~\ka^2 M_n^2(49\hs\bs'\!-\!1)
[(343\hs\bs^{\prime 2} \! -\hsm 174\hs\bs' \! +\hsm 27)
\hsm +\hsm 12\hs{\omega}_-^{}\ct
\hsm +\hsm {\omega}_+^{2}\ctt\hs ]^2~}
{3136\hs\bs^{\prime 2} \hs [(7\hs\bs'\!+\! 3)
\!+\hsm{\omega}_-^{}\ct]
[(7\hs\bs'\!-\! 3) \!-\! {\omega}_-^{}\ct\hs ]} \,,
\hspace*{8mm}
\end{align}
where we have defined $\,\bs\!=\!s/\Mn^2\hs$,
$\,\bs'\!=\!\bs/49\hs$, and
$\hs{\omega}_\pm \hsm\!=\!\hsm
 \sqrt{49\hs\bs^{\prime\hs 2}\!\pm 58\hs\bs'\! +\hsm 9\,}\hs$.\hs\
Then, making the high energy expansion for Eq.\eqref{eq:M1234n-LO},
we derive the following LO KK graviton scattering amplitude:
\begin{align}
\label{eq:M1234nLO-LO}
& \over{\M}_0
\big[\bar{h}^{\pm n}_L \bar{h}^{\mp 2n}_L
\bar{h}^{\mp 3n}_L \bar{h}^{\pm 4n}_L\big]
= \frac{~\ka^2\hs s~}{\,64\,}\hs (7\hsmx +\hsm\ctt)^2\csc^2\!\theta \,,
\hspace*{8mm}
\end{align}
which have $O(E^2)$ and are mass-independent, as expected.


Moreover, we have used the exact longitudinal polarization
tensor \eqref{eq:Gpol-massive-KKL} to further derive
the corresponding full inelastic KK graviton amplitudes
$\M\!\[{h}^{\pm n}_L{h}^{\mp 2n}_L {h}^{\mp 3n}_L{h}^{\pm 4n}_L\]$
in \eqrefe{eq:M1234-full} of Appendix\,\ref{App:B2}.\
Under high energy expansion,
the LO and NLO contributions of the full amplitude
\eqref{eq:M1234-full} are given by Eqs.\eqref{eq:M1234-full-LO}
and \eqref{eq:M1234-full-NLO}, respectively.\
Inspecting the two LO amplitudes \eqref{eq:M1234nLO-LO} and
\eqref{eq:M1234-full-LO}, we find that they are equal:
\beq
\label{eq:1234nLO-M0bar=M0}
\over{\M}_0^{}
\big[\bar{h}^{\pm n}_L \bar{h}^{\mp 2n}_L
\bar{h}^{\mp 3n}_L \bar{h}^{\pm 4n}_L\big]
=
{\M}_0^{}
\big[{h}^{\pm n}_L {h}^{\mp 2n}_L {h}^{\mp 3n}_L {h}^{\pm 4n}_L\big]
\hs.
\eeq
Hence, as we observed earlier, the scattering amplitude \eqref{eq:M1234n-LO}
which we have computed using the leading polarization tensor
\eqref{eq:LO-Gpol} of longitudinal KK gravitions does have
an important physical meaning, because its expanded LO amplitude
\eqref{eq:M1234nLO-LO} can give the correct LO contribution
\eqref{eq:M1234-full-LO} of the full KK graviton amplitude \eqref{eq:M1234-full}.

\subsubsection*{$\blacklozenge$~Inelastic KK Scattering Amplitudes of
\boldmath{$\{n, n, m, m\}$}:}
\vspace*{1.5mm}

As the third example of the inelastic KK scattering process,
we study the four longitudinal KK gauge boson scattering
amplitude whose external states have the following KK indices:
\begin{equation}
\begin{aligned}
\label{eq:KK-num-2n2m}
\{\bn_1^{},\hs\bn_2^{},\hs \bn_3^{},\hs \bn_4^{}\} \,=\,
\left\{ \begin{array}{r}
\{\pm n,\mp n, \mp m,\pm m\} \equiv \fA \,,
\\[0.7mm]
\{\pm n,\mp n,\pm m,\mp m\} \equiv \fB \,,
\end{array}\right.
\end{aligned}
\end{equation}
where the positive integers $(n,\hs m)$ are unequal
($\hs n\!\neq\! m\hs$), and the inelastic amplitudes
of type-$\fA$ and type-$\fB$ are independent.\
Thus, we compute the corresponding leading longitudinal
KK gauge boson scattering amplitudes as follows:
\beqs
\label{eq:T-nnmm}
\begin{align}
\label{eq:Tnnmm+--+}
\TT\!\[A^{a\hs\pm n}_{L} A^{b\hs\mp n}_{L}
A^{c\hs\mp m}_L A^{\hs\pm m}_L\]
&=\, g^2 \!\(\frac{\,\CC_s^{}\hs\NN_s^{\fA}\,}{s} +
\frac{\,\CC_t^{}\hs\NN_t^{\fA}\,}{t\hsm -\hsm M_{n+m}^2}
+ \frac{\,\CC_u^{}\hs\NN_u^{\fA}\,}{u\hsm -\hsm M_{n-m}^2} \) \!,
\\[1mm]
\label{eq:Tnnmm+-+-}
\TT\!\[A^{a\hs\pm n}_{L} A^{b\hs\mp n}_{L}
A^{c\hs\pm m}_L A^{d\hs\mp m}_L\]
&=\, g^2 \!\( \frac{\,\CC_s^{}\hs\NN_s^{\fB}\,}{s}
+ \frac{\CC_t^{}\hs\NN_t^{\fB}}{\,t\hsm -\hsm M_{n-m}^2\,}
+ \frac{\CC_u^{}\hs\NN_u^{\fB}}{\,u\hsm -\hsm M_{n+m}^2\,} \) \!,
\end{align}
\eeqs
where
the kinematic numerators
$\{\NN_j^{\fA},\hs\NN_j^{\fB}\}$ are presented
in Eqs.\eqref{eq:NA-2n2m} and \eqref{eq:NB-2n2m}
of Appendix\,\ref{App:B1}.

\vspace*{1mm}

Using the inelastic KK gauge boson scattering amplitudes
\eqref{eq:Tnnmm+--+}-\eqref{eq:Tnnmm+-+-}
and the extended massive CK duality,
we can derive the following leading scattering amplitudes
for longitudinal KK gravitons:
\beqs
\begin{align}
\label{eq:nnmm+--+}
\hspace*{-5mm}
&\over{\M}\!\[\bar{h}_L^{\pm n}\bar{h}_L^{\mp n}
\bar{h}_L^{\mp m}\bar{h}_L^{\pm m}\]
= -\frac{\ka^2}{\,16\,} \bigg[\,
\frac{(\NN_s^\fA)^2}{s} +
\frac{(\NN_t^\fA)^2}{\,t \hsm -\hsm M_{n+m}^2\,} +
\frac{(\NN_u^\fA)^2}{\,u \hsm -\hsm M_{n-m}^2\,} \,\bigg]
\hspace*{12mm}
\nn\\[1mm]
& 
=\frac{~\ka^2 \Mn^2\hs
[(7 \bs^2 \!-\! 12\hs\bs\hs r_+^2 \!+\! 48\hs r^2) \!+\!
64\hs\qb\hs\qb' r \ct \!+\! (\bs \!+\! 4)(\bs \!+\! 4\hs r^2)
\ctt]^2~}
{64\hs\bs\hs (\bs \!-\! 4\hs r \!-\! 4\hs\qb\hs\qb'\ct)
(\bs \!+\! 4\hs r \!+\! 4\hs\qb\hs\qb' \ct)} \,,
\\[2mm]
\label{eq:nnmm+-+-}
\hspace*{-5mm}
& \over{\M}\!\[\bar{h}_L^{\pm n}\bar{h}_L^{\mp n}
\bar{h}_L^{\pm m}\bar{h}_L^{\mp m}\] =
-\frac{\ka^2}{\,16\,} \bigg[\,
\frac{(\NN_s^\fB)^2}{s} +
\frac{(\NN_t^\fB)^2}{\,t \hsm -\hsm M_{n-m}^2\,} +
\frac{(\NN_u^\fB)^2}{\,u\hsm -\hsm M_{n+m}^2\,} \,\bigg]
\nn\\[1mm]
& 
=\frac{~\ka^2 \Mn^2\hs
[(7\hs\bs^2 \!-\! 12\hs\bs\hs r_+^2 \!+\! 48\hs r^2 ) \!-\!
64\hs\qb\hs\qb' r \ct \!+\! (\bs \!+\! 4) (\bs \!+\! 4 r^2) \ctt]^2~}
{64\hs\bs\hs (\bs \!-\! 4\hs r \!+\! 4\hs\qb\hs\qb'\ct)
(\bs \!+\! 4\hs r \!-\! 4\hs\qb\hs\qb' \ct)} \,,
\end{align}
\eeqs
where we have used the notations,
\begin{align}
\label{eq:qsr}
& r =\hsm M_m^{}/\Mn \hs ,
\hspace*{4mm}
r_+^2 = 1+r^2 ,
\hspace*{4.mm}
q \!=\hsm \sqrt{E^2 \!-\! M_n^2\,}\,,
\hspace*{4.mm}
q' \!=\hsm \sqrt{E^2 \!-\! M_m^2\,}\,,
\nn\\[-3.5mm]
\\[-2mm]
&\bar{q}^2\!= q^2/M_n^2 = \bs/4 \hsm -\hsm 1 \hs,
\hspace*{3mm}
\bar{q}^{\pp 2}\!= q^{\pp 2}\hsm / M_n^2= \bs/4\hsm -\hsm r^2 ,
\hspace*{3mm}
\qb^2\qqbp \hsm =\hsm (\bs\hsm -\hsm 4)
(\bs\hsm -\hsm 4\hs r^2)/16 \hs.~~~
\hspace*{3mm}
\nn
\end{align}
Making the high energy expansion, we can derive the LO amplitudes
of Eqs.\eqref{eq:nnmm+--+}-\eqref{eq:nnmm+-+-} as follows:
\begin{align}
\label{eq:MnnmmLO-LO}
\over{\M}_0^{}\big[\bar{h}_L^{\pm n}\bar{h}_L^{\mp n}
\bar{h}_L^{\mp m}\bar{h}_L^{\pm m}\big] =
\over{\M}_0^{}\big[\bar{h}_L^{\pm n}\bar{h}_L^{\mp n}
\bar{h}_L^{\pm m}\bar{h}_L^{\mp m}\big]
= \frac{~\ka^2\hs s~}{\,64\,}\hs (7\hsmx +\hsm\ctt)^2\csc^2\!\theta \,,
\end{align}
%
which are of $O(E^2)$ and mass-independent, as expected.
We note that the LO amplitudes \eqref{eq:M1234nLO-LO} and
\eqref{eq:MnnmmLO-LO} coincide.

\vspace*{1mm}

In addition, using the exact longitudinal polarization tensor
\eqref{eq:Gpol-massive-KKL} for each external KK graviton state,
we have further derived the full inelastic KK graviton
scattering amplitudes
$\M\!\[{h}_L^{\pm n}{h}_L^{\mp n}{h}_L^{\mp m}{h}_L^{\pm m}\]$
and
$\M\!\[{h}_L^{\pm n}{h}_L^{\mp n}{h}_L^{\pm m}{h}_L^{\mp m}\]$
in Eqs.\eqref{eq:Amp-4hL-nnmm}-\eqref{eq:Xj-nnmm}
of Appendix\,\ref{App:B2}.\
Under the high energy expansion, their LO and NLO scattering
amplitudes are given in Eq.\eqref{eq:Mnnmm-LONLO}.\
Comparing their LO amplitudes \eqref{eq:Mnnmm-LO} with
our above LO amplitudes \eqref{eq:MnnmmLO-LO},
we find that they are equal:
%
\begin{align}
&\over{\M}_0^{}\big[\bar{h}_L^{\pm n}\bar{h}_L^{\mp n}
\bar{h}_L^{\mp m}\bar{h}_L^{\pm m}\big] =
\over{\M}_0^{}\big[\bar{h}_L^{\pm n}\bar{h}_L^{\mp n}
\bar{h}_L^{\pm m}\bar{h}_L^{\mp m}\big]
\nn\\[1mm]
=\,\,&
{\M}_0^{}\big[{h}_L^{\pm n}{h}_L^{\mp n}{h}_L^{\mp m}{h}_L^{\pm m}\big]
={\M}_0^{}\big[{h}_L^{\pm n}{h}_L^{\mp n}{h}_L^{\pm m}{h}_L^{\mp m}\big].
\label{eq:nnmmLO-M0bar=M0}
\end{align}
This is similar to Eqs.\eqref{eq:00nnLO-M0bar=M0} and \eqref{eq:1234nLO-M0bar=M0},
which we have demonstrated for other four-point KK graviton
scattering amplitudes at the LO.

\vspace*{2mm}
\subsubsection*{$\blacklozenge$~Elastic KK Scattering Amplitudes of
\boldmath{$\{n, n, n, n\}$}:}
\vspace*{1.5mm}

Next, we study the elastic scattering processes for the
four longitudinal KK gauge bosons and KK gravitons.\
The external states are allowed to have the following combinations
of KK indices:
\begin{equation}
\begin{aligned}
\label{eq:KK-num-4-point}
\{\bn_1^{},\hs\bn_2^{},\hs \bn_3^{},\hs \bn_4^{}\} =
\left\{ \begin{array}{r}
\{\pm n,\pm n,\mp n,\mp n\} \equiv \fA \,,
\\[0.8mm]
\{\pm n,\mp n, \mp n,\pm n\} \equiv \fB \,,
\\[0.8mm]
\{\pm n,\mp n,\pm n,\mp n\} \equiv \fC \,,
\end{array}\right.
\end{aligned}
\end{equation}
where only three types of them, denoted by $(\fA,\hs\fB,\hs\fC)$,
are independent.\
Then, we compute the four-point elastic scattering amplitudes
of longitudinal KK gauge bosons as follows:
\beqs
\label{eq:Amp-4AL-nnnn}
\begin{align}
\label{eq:Amp-4AL-nnnn++--}
\TT\!\[A^{a\hs\pm n}_{L} A^{b\hs\pm n}_{L}
A^{c\hs\mp n}_L A^{d\hs\mp n}_L\]
&=\, g^2\!\(\!
\frac{\CC_s^{}\hs \NN_s^{\fA}}
{\,s\hsm -\hsm 4\Mn^2\,}+ \frac{\,\CC_t^{}\hs\NN_t^{\fA}\,}{t} +
\frac{\,\CC_u^{}\hs\NN_u^{\fA}\,}{u} \!\) \!,
\\[1mm]
\label{eq:Amp-4AL-nnnn+--+}
\TT\!\[A^{a\hs\pm n}_{L} A^{b\hs\mp n}_{L} A^{c\hs\mp n}_L A^{d\hs\pm n}_L\]
&=\, g^2\!\(\!\frac{\,\CC_s^{}\hs\NN_s^{\fB}\,}{s}
+ \frac{\CC_t^{}\hs\NN_t^{\fB}}{\,t\hsm -\hsm 4\Mn^2\,}
+ \frac{\,\CC_u^{}\hs\NN_u^{\fB}\,}{u} \!\) \!,
\\[1mm]
\label{eq:Amp-4AL-nnnn+-+-}
\TT\!\[A^{a\hs\pm n}_{L} A^{b\hs\mp n}_{L}
A^{c\hs\pm n}_L A^{d\hs\mp n}_L\]
&=\, g^2 \!\(\! \frac{\,\CC_s^{}\hs\NN_s^{\fC}\,}{s}
+ \frac{\,\CC_t^{}\hs\NN_t^{\fC}\,}{t} +
\frac{\CC_u^{}\hs\NN_u^{\fC}}{\,u\hsm -\hsm 4\Mn^2\,} \!\) \!,
\end{align}
\eeqs
where the kinematic numerators
$\{\NN_j^{\fA},\hs\NN_j^{\fB},\hs\NN_j^\fC\}$
are given by Eqs.\eqref{eq:NA-4n}-\eqref{eq:NC-4n}
of Appendix\,\ref{App:B1}.

\vspace*{1mm}

Using the above KK gauge boson scattering amplitudes
\eqref{eq:Amp-4AL-nnnn++--}-\eqref{eq:Amp-4AL-nnnn+-+-}
and the extended massive CK duality,
we derive the following leading elastic scattering amplitudes
of longitudinal KK gravitons:
\beqs
\label{eq:Mbar-4hL-nnnn}
\begin{align}
\label{eq:Mbar-4hL-nnnn++--}
&\over{\M}\big[\bar{h}_L^{\pm n}\bar{h}_L^{\pm n} \bar{h}_L^{\mp n}\bar{h}_L^{\mp n}\big]
=- \frac{~\ka^2~}{16}
\bigg[
\frac{(\NN_s^{\fA})^2}{\,s\hsm -\hsm 4\Mn^2\,} +
\frac{\,(\NN_t^{\fA})^2\,}{t} +
\frac{\,(\NN_u^{\fA})^2\,}{u} \bigg],
\hspace*{5mm}
\\[1.5mm]
\label{eq:Mbar-4hL-nnnn+--+}
& \over{\M}\!\[\bar{h}_L^{\pm n}\bar{h}_L^{\mp n} \bar{h}_L^{\mp n}\bar{h}_L^{\pm n}\] =
-\frac{~\ka^2~}{16} \bigg[\hsm
\frac{\,(\NN_s^{\fB})^2\,}{s} +
\frac{(\NN_t^{\fB})^2}{\,t\hsm -\hsm 4\Mn^2\,} +
\frac{\,(\NN_u^{\fB})^2\,}{u} \bigg] ,
\hspace*{5mm}
\\[1.5mm]
\label{eq:Mbar-4hL-nnnn+-+-}
& \over{\M}\!\[\bar{h}_L^{\pm n}\bar{h}_L^{\mp n} \bar{h}_L^{\pm n}\bar{h}_L^{\mp n}\] =
-\frac{~\ka^2~}{16} \bigg[\hsm
\frac{\,(\NN_s^{\fC})^2\,}{s} +
\frac{\,(\NN_t^{\fC})^2\,}{t} +
\frac{(\NN_u^{\fC})^2}{\,u\hsm -\hsm 4\Mn^2~} \bigg].
\hspace*{5mm}
\end{align}
\eeqs
After substituting the expressions of the numerators
into the above formulas, we derive the explicit forms
of these scattering amplitudes:
\beqs
\begin{align}
\hspace*{-2mm}
\over{\M}\big[\bar{h}_L^{\pm n}\bar{h}_L^{\pm n} \bar{h}_L^{\mp n}\bar{h}_L^{\mp n}\big]
&
= \frac{~\ka^2\Mn^2~}{64}
(\bs\!-\!4)(7\!+\!\ctt)^2 \csc^2\hsmx \theta \,,
\\[1.5mm]
\hspace*{-2mm}
\over{\M}\!\[\bar{h}_L^{\pm n}\bar{h}_L^{\mp n} \bar{h}_L^{\mp n}\bar{h}_L^{\pm n}\]
&
= \frac{~\ka^2 \Mn^2\hs
[(7\hs\bs^2\!-\!24\hs\bs \!+\! 48)
\!+\! 16\hs (\bs\hsm -\hsm 4) \ct
\!+\! (\bs\hsm +\hsm 4)^2 \ctt\hs ]^2 \csc^2\!\fr{\theta}{2}~}
{128\hs\bs\hs (\bs\!-\!4)[(\bs\hsm +\hsm 4) \hsm +\hsm
(\bs\hsm -\hsm 4)\ct\hs ]} \hs ,
\\[1.5mm]
\hspace*{-2mm}
\over{\M}\!\[\bar{h}_L^{\pm n}\bar{h}_L^{\mp n} \bar{h}_L^{\pm n}\bar{h}_L^{\mp n}\]
&
= \frac{~\ka^2 \Mn^2\hs
[(7\hs\bs^2\!-\!24\hs\bs \hsm +\hsm 48)
\!-\! 16\hs (\bs\hsm -\hsm 4) \ct
\hsm +\hsm (\bs\hsm +\hsm 4)^2 \ctt\hs ]^2
\sec^2\!\fr{\theta}{2}~}
{128\hs\bs\hs (\bs\hsm -\hsm 4)[(\bs\hsm +\hsm 4)
\hsm -\hsm (\bs\hsm -\hsm 4)\ct\hs ]} \hs.
\end{align}
\eeqs
Then, making the high energy expansion, we derive the following
LO scattering amplitudes:
\begin{align}
\label{eq:MnnnnLO-LO}
\hspace*{-4mm}
&\over{\M}_0^{}\big[\bar{h}_L^{\pm n}\bar{h}_L^{\pm n} \bar{h}_L^{\mp n}\bar{h}_L^{\mp n}\big]
= \over{\M}_0^{}\big[\bar{h}_L^{\pm n}\bar{h}_L^{\mp n} \bar{h}_L^{\mp n}\bar{h}_L^{\pm n}\big]
= \over{\M}_0^{}\big[\bar{h}_L^{\pm n}\bar{h}_L^{\mp n} \bar{h}_L^{\pm n}\bar{h}_L^{\mp n}\big]
\nn\\[1.5mm]
&
= \frac{~\ka^2\hs s~}{64}
(7\!+\!\ctt)^2\hsm\csc^2\hsmx \theta \,.
\end{align}
We see that the LO elastic KK graviton amplitudes
of the above three helicity combinations
are equal and are mass-independent, as expected.

\vspace*{1mm}

Moreover, using the exact longitudinal polarization tensor
\eqref{eq:Gpol-massive-KKL} for each external KK graviton state,
we have further derived the full elastic KK graviton
scattering amplitudes
in Eqs.\eqref{eq:Amp-4hL-nnnn}-\eqref{eq:Xj-nnnn}
of Appendix\,\ref{App:B2}.\
The LO and NLO contributions of these KK graviton scattering amplitudes
are given in \eqrefe{eq:Amp-4hL-nnnn}.\
Comparing Eqs.\eqref{eq:MnnnnLO-LO} and \eqref{eq:MnnnnFull-LO},
we find that they are equal:
%
\begin{align}
&
\over{\M}_0^{}\big[\bar{h}_L^{\pm n}\bar{h}_L^{\pm n} \bar{h}_L^{\mp n}\bar{h}_L^{\mp n}\big]
= \over{\M}_0^{}\big[\bar{h}_L^{\pm n}\bar{h}_L^{\mp n} \bar{h}_L^{\mp n}\bar{h}_L^{\pm n}\big]
= \over{\M}_0^{}\big[\bar{h}_L^{\pm n}\bar{h}_L^{\mp n} \bar{h}_L^{\pm n}\bar{h}_L^{\mp n}\big]
\hspace*{12mm}
\nn\\[0.6mm]
=\hs\,&
{\M}_0^{}\big[h_L^{\pm n}h_L^{\pm n} h_L^{\mp n}h_L^{\mp n}\big] =
{\M}_0^{}\big[h_L^{\pm n}h_L^{\mp n} h_L^{\mp n}h_L^{\pm n}\big] =
{\M}_0^{}\big[h_L^{\pm n}h_L^{\mp n} h_L^{\pm n}h_L^{\mp n}\big].
\label{eq:nnnnLO-M0bar=M0}
\end{align}
%

In general, we expect that the equality between the two types of the
LO KK graviton amplitudes, such as those shown in
Eqs.\eqref{eq:00nnLO-M0bar=M0}, \eqref{eq:1234nLO-M0bar=M0}, \eqref{eq:nnmmLO-M0bar=M0}, and \eqref{eq:nnnnLO-M0bar=M0},
can hold for any such $N$-point KK graviton
scattering amplitudes at the LO (with some or all external KK graviton
states being longitudinally polarized), as long as
these KK amplitudes respect the
KK GAET\,\cite{5DYM2002}\cite{KK-ET-He2004}
and KK GRET\,\cite{Hang:2021fmp}
in which the residual terms can be supressed and belong to the NLO.\
For the current study, considering the four-point KK
scattering amplitudes, we can use the KK GAET and GRET to prove
this equality.\ For any four longitudinal KK gauge boson scattering
amplitude, the KK GAET states\,\cite{5DYM2002}\cite{KK-ET-He2004}
that under the high energy expansion its LO scattering amplitude
of longitudinal KK gauge bosons ($A_L^{a\bn}$)
equals that of the corresponding
KK Goldstone bosons ($A_5^{a\bn}$) at the LO:
\begin{equation}
\label{eq:GAET-ALLLL-LO}
\TT_0^{}\big[A_L^{a\bn_1^{}}\! A_L^{b\bn_2^{}}\!A_L^{c\bn_3^{}}\!A_L^{d\bn_4^{}}\big]
= \TT_0^{}\big[A_5^{a\bn_1^{}}\!A_5^{b\bn_2^{}}\!A_5^{c\bn_3^{}}\!A_5^{d\bn_4^{}}
\big],
\end{equation}
which are of $O(g^2\hsm E^0)$ and 
are mass-independent.$\!$\footnote{%
This KK GAET\,\cite{5DYM2002}\cite{KK-ET-He2004} formulates the
geometric ``Higgs'' mechanism for the KK gauge boson mass-generation
at the scattering $S$-matrix level.\ 
It differs from the usual 4d equivalence theorem (ET)\,\cite{SM-ET} 
of the SM (describing the conventional Higgs mechanism 
at the $S$-matrix level) because the geometric KK mass-generation 
and the KK GAET do not invoke any Higgs boson.}\
As we have demonstrated in this work and \cite{Hang:2021fmp},
we can make double-copy construction on both sides of
Eq.\eqref{eq:GAET-ALLLL-LO} and thus deduce the corresponding
GRET formula which connects the LO scattering amplitude
of longitudinal KK gravitons ($\bar{h}_L^{\bn}$)
to that of the corresponding
KK Goldstone bosons ($h_{55}^{\bn}$) at the LO:
\begin{equation}
\label{eq:DC-hLh55-LO}
\xoverline{\M}_0^{}\big[\bar{h}_L^{\bn_1^{}}\bar{h}_L^{\bn_2^{}}
\bar{h}_L^{\bn_3^{}}\bar{h}_L^{\bn_4^{}}\big]
= \M_0^{}\big[h_{55}^{\bn_1^{}}h_{55}^{\bn_2^{}}
h_{55}^{\bn_3^{}}h_{55}^{\bn_4^{}}\big],
\end{equation}
where on the left-hand-side of the equality each external
longitudinal KK graviton state
$\bar{h}_L^\bn\!=\bar{\zeta}_{\mn}^{}h_\bn^{\mn}$
has its leading-order longitudinal polarization tensor defined
in Eq.\eqref{eq:LO-Gpol}.\footnote{%
We note that the double-copy of the LO scattering amplitudes
of the KK Goldstone bosons, $\,\TT_0^{}\big[A_5^{a\bn_1^{}}\!A_5^{b\bn_2^{}}\!
A_5^{c\bn_3^{}}\!A_5^{d\bn_4^{}}\big]\!\ito\! \M_0^{}\big[h_{55}^{\bn_1^{}}h_{55}^{\bn_2^{}}
h_{55}^{\bn_3^{}}h_{55}^{\bn_4^{}}\big]$,
is well justified because at the LO of the high energy expansion
the KK Goldstone bosons $A_5^{a\bn}$ and $h_{55}^\bn$
behave as massless states and their LO amplitudes are
mass-independent\,\cite{Hang:2021fmp}.\ Hence, the double-copy
of the LO KK Goldstone amplitudes should be guaranteed by the
double-copy of the corresponding massless scattering amplitudes,
$\TT\big[\hat{A}_5^{a}\hat{A}_5^{b}\hat{A}_5^{c}\hat{A}_5^{d}\big]
\!\ito \M\big[\hat{h}_{55}^{}\hat{h}_{55}^{}
\hat{h}_{55}^{}\hat{h}_{55}^{}\big]$,
in the non-compactified 5d gauge/gravity theories.\
This conclusion also holds for the general $N$-point massless
Goldstone boson amplitudes as is obvious.}\
On the other hand, the KK GRET\,\cite{Hang:2021fmp}
gives the following LO equality:
\begin{equation}
\label{eq:GAET-hLh55-LO}
{\M}_0^{}\big[{h}_L^{\bn_1^{}}{h}_L^{\bn_2^{}}{h}_L^{\bn_3^{}}{h}_L^{\bn_4^{}}
\big] = \M_0^{}\big[h_{55}^{\bn_1^{}}h_{55}^{\bn_2^{}}
h_{55}^{\bn_3^{}}h_{55}^{\bn_4^{}}
\big],
\end{equation}
where according to the original
GRET formulation\,\cite{Hang:2021fmp},
the LO scattering amplitude of longitudinal KK gravitons
on the left-hand-side of the equality is computed
by using the exact longitudinal polarization tensor
\eqref{eq:Gpol-massive-KKL} for each external KK graviton state
$\hs {h}_L^\bn\!={\zeta}_{\mn}^{}h_\bn^{\mn}$.\
Comparing Eqs.\eqref{eq:DC-hLh55-LO} and \eqref{eq:GAET-hLh55-LO},
we deduce the following identity for the two types of
four-point longitudinal KK graviton scattering amplitudes at the LO:
\begin{equation}
\label{eq:M0-hLbar-hL-LO} \xoverline{\M}_0^{}\big[\bar{h}_L^{\bn_1^{}}\bar{h}_L^{\bn_2^{}}\bar{h}_L^{\bn_3^{}}\bar{h}_L^{\bn_4^{}}\big] =
{\M}_0^{}\big[{h}_L^{\bn_1^{}}{h}_L^{\bn_2^{}}{h}_L^{\bn_3^{}}{h}_L^{\bn_4^{}}
\big] .
\end{equation}
We can extend the above identity to the general case of the
scattering amplitudes of $N$ longitudinal KK gravitons
($N\!\geqq\hsm 4\hs$):
\begin{equation}
\label{eq:N-M0-hLbar-hL-LO} \xoverline{\M}_0^{}\big[\bar{h}_L^{\bn_1^{}}\bar{h}_L^{\bn_2^{}}
\cdots\bar{h}_L^{\bn_N^{}}\big] =
{\M}_0^{}\big[{h}_L^{\bn_1^{}}{h}_L^{\bn_2^{}}
\cdots {h}_L^{\bn_N^{}}
\big] .
\end{equation}
We can readily prove the $N$-point identity \eqref{eq:N-M0-hLbar-hL-LO}
by repeating the same reasoning of the above proof for the four-point
scattering case, because the GRET generally holds
for $N$ longitudinal KK graviton
scattering amplitudes\,\cite{Hang:2021fmp}
and the double-copy also holds for
$N$-point massless scalar KK Goldstone boson amplitudes
(as noticed in footnote-5).\
%
We have explicitly verified this identity for a number of
typical four-point longitudinal KK graviton scattering amplitudes
as shown in Eqs.\eqref{eq:1234nLO-M0bar=M0},
\eqref{eq:nnmmLO-M0bar=M0}, and \eqref{eq:nnnnLO-M0bar=M0}.\
We can readily extend the above proof to other four-point
KK graviton scattering amplitudes
and to the case of $N$-point
KK graviton scattering amplitudes with $N\!\!>\!4\hsx$.\
%
For the extension to other cases, the scattering amplitudes
should contain more than one external longitudinal KK state.\
This is because
for a given scattering amplitude having only one
external longitudinal KK state, the the residual
term of the KK GAET or GRET has the same order as the
LO $A_5^\bn$ amplitude or the LO
$h_{55}^\bn$ amplitude\,\cite{Hang:2021fmp}; hence
the LO amplitudes of KK longitudinal gauge bosons and
of KK Goldstone bosons will not equal,
unlike the case of Eq.\eqref{eq:GAET-ALLLL-LO}
or Eq.\eqref{eq:GAET-hLh55-LO}.\
This means that the extension
of our conclusion of Eq.\eqref{eq:M0-hLbar-hL-LO}
or Eq.\eqref{eq:N-M0-hLbar-hL-LO} to
other KK graviton scattering amplitudes should contain
more than one external longitudinal KK state.\
In addition, we note that
the identities \eqref{eq:M0-hLbar-hL-LO}-\eqref{eq:N-M0-hLbar-hL-LO}
and their extension should hold for any consistent
5d KK compactifications including the $S^1$ compactification
and the orbifold compactification of $S^1\hsm\!/\ZZ\hs$.

\vspace*{1mm}

According to the LO double-copy identities
\eqref{eq:M4h-LO-NLO} and \eqref{eq:N-M0-hLbar-hL-LO},
we can translate the identity \eqref{eq:M0-hLbar-hL-LO} into
a nontrivial sum rule condition on the LO kinematic numerators
of KK gauge boson scattering amplitudes:
%
\beq
\label{eq:LO-Cond-N0L=N0}
\sum_{j}\! \frac{1}{\,s_j^{}\,}\!
\left\{\! \Big[\NN_{j}^{0,\fP}\! (L)\Big]^2 \!-
\!\sum_{\lam^{}_k,\lam'_k}\!
\prod_{k}\! C_{\lam_k^{}\lam'_k}^{\,\si_{\!k}^{}}\hs
{\NN_{j}^{0,\fP}\! (\lam^{}_k)\hs \NN_{j}^{0,\fP}\!(\lam'_k)}
\!\right\} =\, 0 \,.
\eeq

In summary, the above identities \eqref{eq:M0-hLbar-hL-LO} and \eqref{eq:N-M0-hLbar-hL-LO} demonstrate that
under high energy expansion,
each leading-order (LO) scattering amplitude of longitudinal
KK gravitions computed by using the simple effective leading
longitudinal polarization tensor \eqref{eq:LO-Gpol} always
equals the LO amplitude computed by using the exact longitudinal polarization tensor
\eqref{eq:zeta-munu-4d}-\eqref{eq:Gpol-massive-KK}.\
This leads to the above nontrivial sum rule condition
\eqref{eq:LO-Cond-N0L=N0},
from which we draw an important conclusion:\ each $N$-point
longitudinal KK graviton scattering amplitude (with two or more
external KK graviton states being longitudinally polarized) can be
constructed by double-copy of a single amplitude of KK gauge bosons
(in which the corresponding KK gauge bosons are longitudinally
polarized only) according to the effective leading
longitudinal polarization tensor \eqref{eq:LO-Gpol}.\

\vspace*{2.5mm}
\subsection{\hspace*{-2mm}{KK Amplitudes and Double-Copy under
Orbifold Compactification}}
\vspace*{1.5mm}
\label{sec:3.3}

In this subsection, we further study the massive double-copy
construction of the four-point scattering amplitudes of
KK gauge bosons and KK gravitons,
under the 5d orbifold compactification of $S^1\!/\ZZ\hs$.\
In this case, the KK gauge boson fields $A^{a\mu}_n$
and KK graviton fields $h^{\mu\nu}_n$
are defined as $\ZZ$ even\,\cite{Hang:2021fmp}.\
Thus, the KK amplitudes can be expressed
as the sum of the relevant subamplitudes under the $S^1$
toroidal compactification without orbifold.\
Similar to Eq.\eqref{eq:An=Anp+Anm},
the $\ZZ$ even states of KK gauge bosons and KK gravitons
can be defined as follows:
\beqa
\label{eq:Anhn-Z2even}
A_{\lam}^{a\hs n} = \frac{1}{\sqrt{2\,}\,}\!
\[A_{\lam}^{a(+n)}\!+\hsm A_{\lam}^{a(-n)}\]\!,
~~~~
h_{\si}^{n} = \frac{1}{\sqrt{2\,}\,}\!
\[h_{\si}^{(+n)}\!+\hsm h_{\si}^{(-n)}\]\!,
\eeqa
where we denote the helicity of KK gauge boson by
$\hs\lam\,(=\!0,\pm 1)\hs$ and the helicity of KK graviton by
$\hs\si\,(=\!0,\pm 1,\pm 2)\hs$.\
With these, we can derive the four-point scattering amplitudes
of the KK gauge bosons and KK gravitons:
\beqs
\begin{align}
\label{eq:T4dS1Z2-4pt}
\TT [A_{\lam_1}^{a n_1}\! A_{\lam_2}^{b n_2}\! A_{\lam_3}^{c n_3}\!A_{\lam_4}^{d n_4}]
&\,=\, \frac{~g^2~}{\,2^{\bsN\!/2\,}}
\sum_\fP \sum_{j} \, \frac{~\CC_j^{} \hs\NN_j^{\fP}(\lam)~}{~s_j^{}\!-\hsm M^2_{\bn\bn_j}~} \,,
\\
\label{eq:M4dS1Z2-4pt}
\M [h_{\si_1}^{n_1} \hs h_{\si_2}^{n_2} \hs h_{\si_3}^{n_3} \hs h_{\si_4}^{n_4}]
&\,=\, -\frac{~\ka^2~}{\,2^{\bsN\hsm/2+4}_{}\,}
\sum_\fP \sum_j\!\sum_{\lam^{}_k,\lam'_k} \prod_{k=1}^4\! C_{\lam_k^{}\lam'_k}^{\hs\si_{\hsm k}^{}}
\frac{~\NN^\fP_j(\lam_k^{})\hs\NN^{\hs \fP}_j(\lam'_k)~}
{s_j^{}\hsm -\hsm M^2_{\bn\bn_j}} \,,
\hspace*{7mm}
\end{align}
\eeqs
where the external KK indices are non-negative integers
$\hs n_i^{}\!\hsm\geqq\!0\hs$,
and the sum of $\fP$ runs over the possible combinations
of the external KK indices
which obey the neutral condition \eqref{eq:NeutralCond}.\
In the above formulas, the external states could contain possible
zero-mode gauge bosons (gravitons), so we denote the number of
external KK states (with $n_i^{}\!>\hsm 0\hs$) as $\bN\hs$,
and thus the number of the external zero-mode states
(with $n_i^{}\!=\hsm 0\hs$) equals $(4\hsm -\!\bN)$.\
The factor $\,2^{-\bsN /2}$
arises from the overall coefficient $\hs 1/\!\sqrt{2}\,$
of Eq.\eqref{eq:Anhn-Z2even}.
Substituting the kinematic numerators into
Eqs.\eqref{eq:T4dS1Z2-4pt}-\eqref{eq:M4dS1Z2-4pt},
we have verified that the scattering amplitudes
\eqref{eq:T4dS1Z2-4pt}-\eqref{eq:M4dS1Z2-4pt} agree with
our previous results\,\cite{KKString}.

\vspace*{1mm}

In addition, using the method of
Eqs.\eqref{eq:M4h-4dS1}-\eqref{eq:M4h-LO-NLO},
we can make high energy expansion for the double-copied KK
graviton amplitides \eqref{eq:M4dS1Z2-4pt}.\
Similar to Eqs.\eqref{eq:M4h-4dS1}-\eqref{eq:M4h-LO-NLO},
we derive the LO and NLO scattering amplitudes of four
KK gravitons from Eq.\eqref{eq:M4dS1Z2-4pt},
$\hs\M=\M_0^{}+\dM\hs$,
under the $S^1\!/\ZZ\hs$ orbifold compactification:
\\[-4mm]
{\small
\beqs
\label{eq:M4h-LO-NLO-Z2}
\begin{align}
\label{eq:M4h-LO-Z2}
\hspace*{-2mm}
\M_0^{} &\hsmx =\hsmx
-\frac{~\ka^2~}{\,2^{\bsN\hsm/2+4}_{}\,}
\sum_{\fP}^{}\! \sum_{j}
\!\!\sum_{\lam^{}_k,\lam'_k}\!
\prod_{k}\! C_{\lam_k^{}\lam'_k}^{\hs\si_{\hsm k}^{}}
\frac{~\NN_{j}^{0,\fP}\! (\lam^{}_k)\hs \NN_{j}^{0,\fP}\!(\lam'_k)~}{s_j^{}} \,,
\\[1mm]
\label{eq:M4h-NLO-Z2}
\hspace*{-2mm}
\dM &\hsmx =\hsmx
-\frac{~\ka^2~}{\,2^{\bsN\hsm/2+4}_{}\,}\! 
\!\sum_{\lam^{}_k,\lam'_k}^{\fP,\hs j}\!\!
\prod_{k} \! C_{\lam_k^{}\lam'_k}^{\hs\si_{\hsm k}^{}}\!
\frac{~\NN_{j}^{0,\fP}\! (\lam^{}_k)\da \NN_{j}^{\fP}\!(\lam'_k)
\!+\! \NN_{j}^{0,\fP}\! (\lam'_k)\da\NN_{j}^{\fP}\!(\lam^{}_k)
\!-\! N_{j}^{0,\fP}\! (\lam^{}_k)N_{j}^{0,\fP}\!(\lam'_k)
M^2_{\bn\bn_j}/s_j^{}\,}{s_j^{}} ,
\end{align}
\eeqs
}
\\[-6mm]
where the LO and NLO kinematic numerators
$(\NN_j^{0,\fP}\!,\hs \dNN_j^{\fP})$
are derived under the high energy expansion,
\\[-8mm]
\begin{align}
\frac{\NN_j^{\fP}\hsmx (\lam)}{~1\!-M^2_{\bn\bn_j}/s_j^{}\,}
&=\, \NN_j^{0,\fP}\hsmx (\lam)\hsm +\dNN_j^{\fP}\hsmx (\lam)\hs.
\end{align}
For the four-point elastic KK graviton scattering,
we verified\,\cite{Hang:2021fmp} explicitly
that the above BCJ-type massive double-copy
formulas give the correct results at both the LO and NLO,
and they agree with the results of the extended massive KLT
relations\,\cite{KKString}
derived from the KK string theory approach.

\vspace*{1mm}

Using Eq.\eqref{eq:Anhn-Z2even} and \eqref{eq:T4dS1Z2-4pt},
we can derive relations between
the KK scattering amplitudes under
the $\hs S^1\!/\ZZ\hs$ orbifold compactification
and the corresponding KK scattering amplitudes under
the $S^1$ compactification.\
For the inelastic KK scattering processes
$\hs (0,0)\hsm\ito (n,n)$ and $(n,2n)\hsm\ito (3n,4n)$
discussed in the example-I and example-II of Section\,\ref{sec:3.2},
we derive the following relations:
\beqs
\begin{align}
\label{eq:S1xZ2-00LL}
{\M}\!\[{h}_{\pm 2}^{0}{h}_{\mp 2}^{0}{h}_L^{n}{h}_L^{n}\]
&=\,
{\M}\!\[{h}_{\pm 2}^{0}{h}_{\mp 2}^{0}
{h}_L^{\pm n}{h}_L^{\mp n}\] \!,
\\[1mm]
\label{eq:S1xZ2-1234n}
{\M}\!\[{h}^{n}_L {h}^{2n}_L {h}^{3n}_L {h}^{4n}_L\]
&=\, \Fr{1}{\,2\,}
{\M}\!\[{h}^{\pm n}_L {h}^{\mp 2n}_L {h}^{\mp 3n}_L {h}^{\pm 4n}_L\]\!.
\end{align}
\eeqs
The above relations hold for the external longitudinal KK graviton
states with either the exact longitudinal polarization tensor
$\zeta_L^{\mn}\hsm$ of Eq.\eqref{eq:Gpol-massive-KKL}
or the leading-order longitudinal polarization tensor
$\hs\bar\zeta_L^{\mn}$ of Eq.\eqref{eq:LO-Gpol}.\
To derive the relation \eqref{eq:S1xZ2-00LL},
we have used the fact of
$\hs\M\big[{h}_{\pm 2}^{0}{h}_{\mp 2}^{0}
{h}_L^{\pm n}{h}_L^{\pm n}\big]\!=\!0\hs$
based upon Eq.\eqref{eq:KK00nn},
whereas for deriving the relation \eqref{eq:S1xZ2-1234n},
we have used the fact that Eq.\eqref{eq:KK-1234n} gives
all the allowed combinations of the external KK indices in
this scattering process.\

\vspace*{1mm}

For the inelastic KK scattering process
$(n,n)\hsm\ito (m,m)$ discussed in the example-III
of Section\,\ref{sec:3.2}, we can derive the KK scattering amplitude
under $S^1\!/\ZZ$ compactification:
\begin{align}
&\over{\M}\!\[\bar{h}_L^{n}\bar{h}_L^{n} \bar{h}_L^{m}\bar{h}_L^{m}\]
=\, \Fr{1}{\,2\,}\hsm
\left\{\over{\M}\!\[\bar{h}_L^{\pm n}\bar{h}_L^{\mp n}
\bar{h}_L^{\mp m}\bar{h}_L^{\pm m}\]\hsmx
+\hsm\over{\M}\!\[\bar{h}_L^{\pm n}\bar{h}_L^{\mp n}
\bar{h}_L^{\pm m}\bar{h}_L^{\mp m}\] \right\}
\nn\\[1mm]
&=\,
\frac{~\ka^2\Mn^2\hs (X_0^{} \!+\! X_2^{}\hs\ctt \!+\!
X_4^{}\hs\ctf \!+\! X_6^{}\hs\cts)~}
{64\hs\bs\hsx (Q_0^{} \!+\hsm Q_2^{}\hs\ctt \!+\hsm Q_4^{}\hs\ctf)}\,,
\label{eq:MnnmmLO-Z2}
\end{align}
where the polynomials $\,\{X_i^{},Q_j^{} \hs\}\,$ are given by
\begin{align}
X_0^{} &= 2\hs \big[\hs 85\hs \bs^6
+ 92\hs\bs^5 r_+^2
\hsm-\hsm  16\hs\bs^4 (49\hs r^4\hsm+\hsm  95\hs r^2\hsm+\hsm  49)
+ 64\hs\bs^3 (13\hs r^6\hsm+\hsm  45\hs r^4\hsm+\hsm  45\hs r^2\hsm+\hsm  13)
\nn\\
& \hspace*{4.5mm}
\hsm-\hsm  256\hs\bs^2 r^2 (29\hs r^4+ 19\hs r^2\hsm+\hsm  29)
+ 15360\hs\bs r^4 r_+^2\hsm-\hsm  28672\hs r^6 \hs\big] \hs,
\nn\\[1mm]
X_2^{} &= -\big[\hs 143\hs\bs^6 \!-\! 1796\hs\bs^5 r_+^2
\hsmx +\hsmx  16\hs\bs^4
(193\hs r^4\hsmx+\hsmx 335\hs r^2\hsmx+\hsmx  193)
\!-\hsmx 320\hs\bs^3 (3\hs r^6 \!-\hsm 29\hs r^4
\nn\\
&\hspace*{4.5mm}
\hsm-\hsm  29\hs r^2\!+\hsm  3)
\!-\hsm 256\hs\bs^2 r^2 (31\hs r^4\!+\hsm  273\hs r^2\!+\hsm 31)
\hsm+\hsm  64512\hs\bs\hs r^4 r_+^2
- 69632\hs r^6 \hs\big]\hs,
\nn\\[1mm]
X_4^{} &= -2\hs\big[ 13\hs\bs^6 \!-\hsm 36\hs\bs^5 r_+^2\!
-\!16\hs \bs^4 (17\hs r^4\!+\hsm 7\hs r^2\!+\hsm  17)
\hsm +  64\hs\bs^3(5\hs r^6\!-\hsmx  11\hs r^4\!-\hsm
11\hs r^2\hsm+\hsm  5)~~~
\\
&\hspace*{4.5mm}
+\hsm  256\hs\bs^2 r^2 (19\hs r^4\!+\hsm 53\hs r^2\!+\hsm  19)
\hsm-\!  17408\hs\bs\hs r^4 r_+^2\hsm+\hsm  4096\hs r^6\big] \hs,
\nn\\[1mm]
X_6^{} &= - (\bs^2\hsm+\hsm  4\hs\bs\hs r_+^2\hsm+\hsm 16\hs r^2 )^2
(\bs^2 \hsm-\hsm  4 \bs r_+^2\hsm+\hsm  16 r^2 ) \hs,
\nn\\[1mm]
Q_0^{} &= 3\hs\bs^4\!+\hsm 8\hs\bs^3 r_+^2
\hsm +\hsm 16\hs\bs^2(3\hs r^4 \!-\hsm 20\hs r^2 \!+\hsm 3)
\hsm +\hsm 128\hs\bs\hs r^2 r_+^2 \hsm + 768\hs r^4 \hs ,
\nn\\
Q_2^{} &= - 4\hs (\bs^2\!-\!16)(\bs^2 \!-\! 16\hs r^4)  \hs,
~~~~~
Q_4^{} = (\bs-4)^2 (\bs - 4\hs r^2)^2 \hs .
\nn
\end{align}
\\[-6mm]
In the above Eq.\eqref{eq:MnnmmLO-Z2} and hereafter,
the external on-shell
KK graviton states such as $\bar{h}_L^n$ or $\bar{h}_L^{m}$
are defined as in \eqrefe{eq:LO-Gpol} by using the leading-order
longitudinal polarization tensor $\hs\bar{\zeta}^{\mn}_L$.\
We also denote their amplitudes by $\over{\M}\hs$.
Making the high energy expansion for Eq.\eqref{eq:MnnmmLO-Z2}
and using the LO relation \eqref{eq:MnnmmLO-LO},
we derive the LO inelastic KK graviton scattering amplitude
as follows:
\begin{align}
\label{eq:MnnmmLO-LO-Z2}
\over{\M}_0^{}\big[\bar{h}_L^{n}\bar{h}_L^{n} \bar{h}_L^{m}\bar{h}_L^{m}\big] =
\over{\M}_0^{}\big[\bar{h}_L^{\pm n}\bar{h}_L^{\mp n}
\bar{h}_L^{\pm m}\bar{h}_L^{\mp m}\big]
= \frac{~\ka^2\hs s~}{\,64\,}\hs (7\hsmx +\hsm\ctt)^2
\hsm\csc^2\!\theta \,.
\end{align}
%

\vspace*{1mm}

Next, we inspect the elastic KK scattering amplitudes of
$(n,n)\hsm\ito (n,n)$ as discussed in the example-IV
of Section\,\ref{sec:3.2}.\ Under the $S^1\!/\ZZ$ compactification,
we derive the elastic longitudinal KK graviton scattering amplitude
in terms of the corresponding KK graviton amplitudes
under the $S^1$ compactification:
\begin{align}
\over{\M}\big[\bar{h}_L^{n}\bar{h}_L^{n}  \bar{h}_L^{n}\bar{h}_L^{n}\big]
&= \Fr{1}{\,2\,}\! \left\{
\over{\M}\big[\bar{h}_L^{\pm n}\bar{h}_L^{\pm n}
\bar{h}_L^{\mp n}\bar{h}_L^{\mp n}\big]
\!+\!\over{\M}\big[\bar{h}_L^{\pm n}\bar{h}_L^{\mp n}
\bar{h}_L^{\mp n}\bar{h}_L^{\pm n}\big]
\!+\!\over{\M}\big[\bar{h}_L^{\pm n}\bar{h}_L^{\mp n}
\bar{h}_L^{\pm n}\bar{h}_L^{\mp n}\big] \hsm \right\}
\nn\\
&= -\frac{~\ka^2 \Mn^2\hs (X_0^{} \hsm +\hsm X_2^{}\hs\ctt\hsm+\hsm
X_4^{}\hs\ctf \hsm+\hsm X_6^{}\hs\cts)\csc^2 \hsmx\theta~}
{512\hs\bs\hs (\bs \!-\! 4)[\hs
(\bs^2 \!+\! 24\hs\bs\!+\! 16) \!-\! (\bs \!-\! 4)^2\hs \ctt \hs]} \,,
\label{eq:MnnnnLO-Z2}
\end{align}
where the polynomials $\{X_i\}$ take the following forms,
\begin{equation}
\begin{aligned}
X_0^{} &= -2\hs (255\hs\bs^5\!+2856\hs\bs^4\!-\hsm 19168\hs\bs^3\!
+35840\hs\bs^2\!+9984 \bs+\hsm 14336) \hs ,
\\
X_2^{} &= 429\hs \bs^5\!-\hsm 10120\hs\bs^4\!+\hsm 26976\hs\bs^3
\!-\hsm 36352\hs\bs^2\!-\hsm 19200\hs\bs+34816 \hs ,
\\
X_4^{} &= 2\hs (39\hs\bs^5\!-\hsm 280\hs\bs^4\!+ 32\hs\bs^3\!
-\hsm 11264\hs\bs^2\!+\hsm 20224\hs\bs -\hsm 2048) \hs,
\\
X_6^{} &= 3\hs\bs^5\! + 8\hs\bs^4\!+160\hs\bs^3\!-512\hs\bs^2
\!-\hsm 1280\hs\bs -\hsm 2048 \hs.
\end{aligned}
\end{equation}
Then, making high energy expansion for the amplitude
\eqref{eq:MnnnnLO-Z2} and using Eq.\eqref{eq:MnnnnLO-LO},
we derive the following LO elastic KK graviton amplitude
of Eq.\eqref{eq:MnnnnLO-Z2}:
%
\begin{align}
\label{eq:M0bar-nnnn-S1Z2}
\over{\M}_0^{}\big[\bar{h}_L^{n}\bar{h}_L^{n}
\bar{h}_L^{n}\bar{h}_L^{n}\big]
=\fr{\,3\,}{2}\,\over{\M}_0^{}
\big[\bar{h}_L^{\pm n}\bar{h}_L^{\pm n} \bar{h}_L^{\mp n}\bar{h}_L^{\mp n}\big]
= \frac{~3\hs\ka^2\hs s~}{128}
(7\hsm +\hsm\ctt)^2 \hsm\csc^2\hsmx \theta \,.
\end{align}
%
This equals the LO elastic KK graviton amplitude
\eqref{eq:MnnnFull-LO-Z2} [as derived by using the full
elastic KK graviton amplitude \eqref{eq:M4hLZ2-nnnn}],
\beqa
\label{eq:M0bar=M0-nnnn}
\over{\M}_0^{}\big[\bar{h}_L^{n}\bar{h}_L^{n}
\bar{h}_L^{n}\bar{h}_L^{n}\big] =
{\M}_0^{}\big[{h}_L^{n}{h}_L^{n}
{h}_L^{n}{h}_L^{n}\big].
\eeqa
This agrees with the identity \eqref{eq:M0-hLbar-hL-LO}.\
We note that based upon the first equality of
Eq.\eqref{eq:M0bar-nnnn-S1Z2} together with
Eq.\eqref{eq:M0bar=M0-nnnn},
we can make a double-copy construction
using directly the KK gauge boson scattering amplitude
$\TT_0^{}\big[A_L^{an_1^{}}\! A_L^{bn_2^{}}\!A_L^{cn_3^{}}\!A_L^{dn_4^{}}\big]$
under $S^1\!/\ZZ$ as in \cite{Hang:2021fmp}
and deduce the correct LO KK graviton scattering amplitude
${\M}_0^{}\big[{h}_L^{n}{h}_L^{n}
{h}_L^{n}{h}_L^{n}\big]$
as in Eq.\eqref{eq:MnnnFull-LO-Z2}.

\vspace*{1mm}

In fact the identity \eqref{eq:M0-hLbar-hL-LO} generally
holds for the current KK compactification of $S^1\!/\ZZ$
as well.\
According to the LO double-copy formula \eqref{eq:M4h-LO-Z2},
we can express the LO KK graviton scattering amplitudes
on the two sides of Eq.\eqref{eq:M0-hLbar-hL-LO} as follows:
\beqs
\begin{align}
\label{eq:barM4h-LO-Z2}
\over{\M}_0^{}\big[\bar{h}_L^{n_1^{}}\bar{h}_L^{n_2^{}}
\bar{h}_L^{n_3^{}}\bar{h}_L^{n_4^{}}\big]
&=
-\frac{~\ka^2~}{\,64\,}
\sum_{\fP}^{}\!\sum_{j}\!\frac{1}{\,s_j^{}\,}\!\hsm
\[\NN_{j}^{0,\fP}\! (L)\]^{\hsm 2}
\!,
\\
{\M}_0^{}\big[{h}_L^{n_1^{}}{h}_L^{n_2^{}}{h}_L^{n_3^{}}{h}_L^{n_4^{}}
\big]
&=
-\frac{~\ka^2~}{\,64\,}\!\!
\sum_{\fP}^{}\!\sum_{j}\!\frac{1}{\,s_j^{}\,}
\hsm\bigg[\hsm\sum_{\lam^{}_k,\lam'_k}\!\!
\prod_{k}\! C_{\lam_k^{}\lam'_k}^{\hs\si_{\hsm k}^{}}
{\NN_{j}^{0,\fP}\! (\lam^{}_k)\hs \NN_{j}^{0,\fP}\!(\lam'_k)}\bigg]
,
\end{align}
\eeqs
Thus, we can translate the identity \eqref{eq:M0-hLbar-hL-LO} into
a spectral condition on the LO kinematic numerators
of KK gauge boson scattering amplitudes:
%
\beq
\sum_{\fP}^{}\!
\sum_{j}\! \frac{1}{\,s_j^{}\,}\!
\bigg\{\hsm \Big[\NN_{j}^{0,\fP}\! (L)\Big]^{\hsm 2} \!-
\!\sum_{\lam^{}_k,\lam'_k}\!
\prod_{k}\! C_{\lam_k^{}\lam'_k}^{\hs\si_{\hsm k}^{}}\hs
{\NN_{j}^{0,\fP}\! (\lam^{}_k)\hs \NN_{j}^{0,\fP}\!(\lam'_k)}
\hsm\bigg\} =\, 0 \,.
\eeq
We see that this condition is guaranteed by the stronger
condition \eqref{eq:LO-Cond-N0L=N0},
because Eq.\eqref{eq:LO-Cond-N0L=N0} holds for each
given combination ``$\fP$'' of KK indices of external states
and has no extra summation over all allowed combinations
``$\fP$''.\

\vspace*{2mm}
\subsection{\hspace*{-2mm}Five-Point Massive KK Graviton
Amplitudes from Double-Copy}
\label{sec:3.4}
\vspace*{2mm}

\begin{figure}[tp]
\centering
\includegraphics[width=14.5cm]{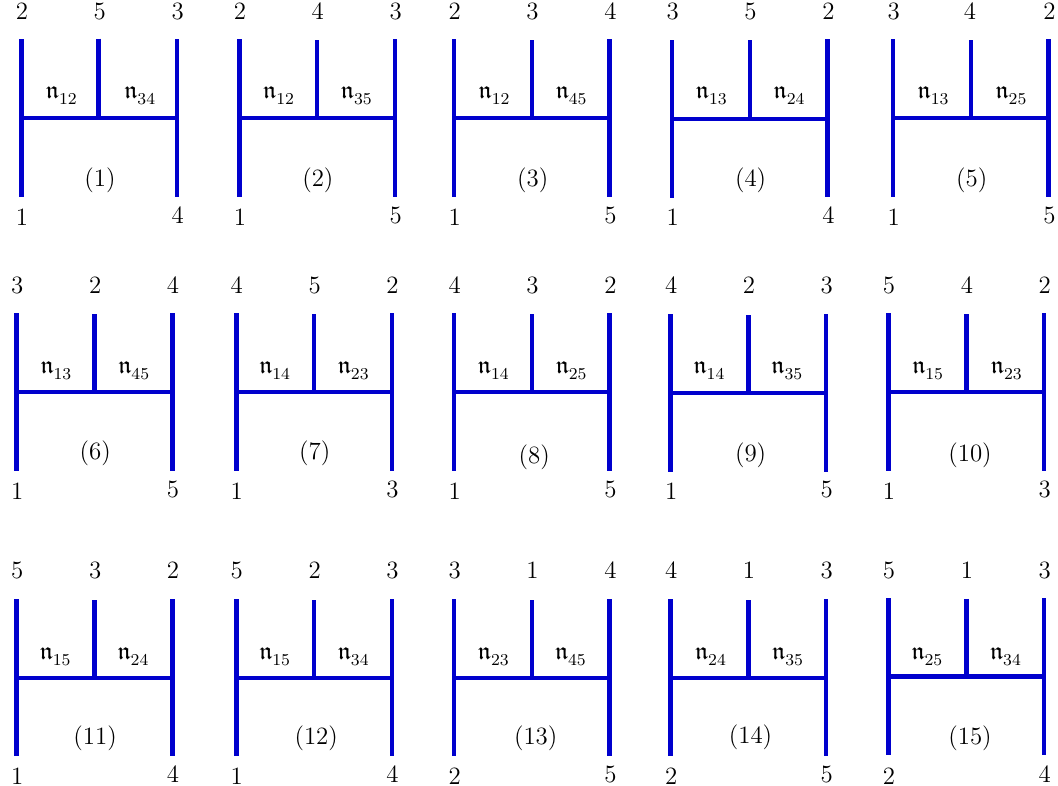}
\caption{Independent structures of the color-dressed
five-point scattering amplitudes of KK gauge bosons
or KK gravitons at tree level,
where the numbers $\{1,\cdots\hsm,5\}$ denote
the KK indices of the external KK states,
$\{1,\cdots\hsm,5\}\!\sim\hsm
 \{\bn_1^{},\cdots\hsm,\bn_5^{}\}$, and each internal line
has its KK index $\bn_{ij}^{}\!=\hsm\bn_i^{}\!+\hsm\bn_j^{}\hs$.
}
\label{fig:1}
\end{figure}

The above analyses of the four-point KK scattering amplitudes
can be further extended to the $N$-point scattering amplitudes
with $\hs N\!\geqq\hsm 5\hsx$.
In this subsection, we study the double-copy construction of
the five-point massive KK graviton scattering amplitudes
with external states having KK indices
$(\bn_1,\bn_2,\bn_3,\bn_4,\bn_5)$.\

\vspace*{1mm}
There are 15 independent structures of the five-point
scattering amplitudes of the KK gauge bosons or KK gravitons,
and each structure contains two kinematic poles in total,
which we present in Fig.\,\ref{fig:1}.\
Thus, using the extended massive BCJ-type double-copy formula
\eqref{eq:MS1-Npt} of the $N$-point KK graviton amplitudes,
we can derive the structure of the general five-point
scattering amplitudes of KK gravitons
under the $S^1$ compactification.\
For each given combination of KK indices
{$\{\bn_1^{},\cdots\!,\bn_5\}$} of external states,
we can construct the KK graviton scattering amplitude as follows:
\\[-7mm]
\begin{align}
\label{eq:MS1-5pt}
&\M\big[h_{\bn_1}^{\si_1}h_{\bn_2}^{\si_2}h_{\bn_3}^{\si_3}
h_{\bn_4}^{\si_4}h_{\bn_5}^{\si_5}\big]
=
\frac{~\ka^3\,}{64}\hsm\!\sum_{\lam^{}_k,\lam'_k}
\!\prod_{k=1}^5\! C_{\lam_k^{}\lam'_k}^{\hs\si_{\hsm k}^{}}
\Bigg[
\frac{\NN^{}_1\NN^{\pp}_1}{(s_{12}^{}\!-\! M_{12}^2) (s_{34}^{}\!-\! M_{34}^2)}
+\frac{\NN^{}_2\NN^{\pp}_2}{(s_{12}^{}\!-\! M_{12}^2) (s_{35}^{}\!-\! M_{35}^2)}
\hspace*{6mm}
\nn\\[-1mm]
&\qquad\hspace*{8mm}
+\frac{\NN^{}_3\NN^{\pp}_3}{(s_{12}^{}\!-\! M_{12}^2) (s_{45}^{}\!-\! M_{45}^2)}
+\frac{\NN^{}_4\NN^{\pp}_4}{(s_{13}^{}\!-\! M_{13}^2)(s_{24}^{}-M_{24}^2)}
+\frac{\NN^{}_5\NN^{\pp}_5}{(s_{13}^{}\!-\! M_{13}^2)(s_{25}^{}-M_{25}^2)}
\nn\\
&\qquad\hspace*{8mm}
+\frac{\NN^{}_6\NN^{\pp}_6}{(s_{13}^{}\!-\! M_{13}^2) (s_{45}^{}\!-\! M_{45}^2)}
+\frac{\NN^{}_7\NN^{\pp}_7}{(s_{14}^{}\!-\! M_{14}^2)(s_{23}^{}-M_{23}^2)}
+\frac{\NN^{}_8\NN^{\pp}_8}{(s_{14}^{}\!-\! M_{14}^2)(s_{25}^{}-M_{25}^2)}
\nn\\
&\qquad\hspace*{8mm}
+\frac{\NN^{}_9\NN^{\pp}_9}{(s_{14}^{}\!-\! M_{14}^2)(s_{35}^{}\!-\! M_{35}^2)}
+\frac{\NN^{}_{10}\NN^{\pp}_{10}}{(s_{15}^{}\!-\! M_{15}^2)(s_{23}^{}-M_{23}^2)}
+\frac{\NN^{}_{11}\NN^{\pp}_{11}}{(s_{15}^{}\!-\! M_{15}^2)(s_{24}^{}-M_{24}^2)}
\nn\\
&\qquad\hspace*{8mm}
+\frac{\NN^{}_{12}\NN^{\pp}_{12}}{(s_{15}^{}\!-\! M_{15}^2)(s_{34}^{}\!-\! M_{34}^2)}
+\frac{\NN^{}_{13}\NN^{\pp}_{13}}{(s_{23}^{}-M_{23}^2)(s_{45}^{}\!-\! M_{45}^2)}
+\frac{\NN^{}_{14}\NN^{\pp}_{14}}{(s_{24}^{}-M_{24}^2)(s_{35}^{}\!-\! M_{35}^2)}
\nn\\
&\qquad\hspace*{8mm}
+\frac{\NN^{}_{15}\NN^{\pp}_{15}}{(s_{25}^{}-M_{25}^2)(s_{34}^{}\!-\! M_{34}^2)} \Bigg] .
\end{align}
%
In the above amplitude,
for simplicity we have used the shorthand notations
of the shifted massive numerators which are given by
%
%
\begin{equation}
\label{eq:5pt-Numerator}
\{\NN_j^{},\,\NN_j^{\pp}\} \hs =\hs
\{\NN_j^{}(\lam_k^{}),\,\NN_j^{}(\lam'_k)\}
\Big|_{s_j^{}\to\hsx s_j^{}\hsm -\hs M_{\bn\bn_j}^2} ,
\end{equation}
where we also make the replacements for products of
the external-state polarizations and momenta
$\hze^{}_i \!\cdot \hat{p}_j \!=\zeta^{}_i\hsm\cdot\hsm p_j$ and 
$\hze^{}_i \cdot \hze^{}_j = \zeta^{}_i \cdot \zeta^{}_j\,$
according to Eq.\eqref{eq:sij-pol-shift}.\ 
On the right-hand side of Eq.\eqref{eq:5pt-Numerator},
the kinematic numerators
$\{\NN_j^{}(\lam_k^{}),\,\NN_j^{}(\lam'_k)\}$ are given by 
the corresponding five-point massless gauge boson amplitudes.\  
The kinematic numerators of the five-point massless gauge boson amplitudes were studied in many literatures\,\cite{BCJ:2019} and 
can be obtained by a number of methods, such as the direct Feynman 
diagram calculation, the CHY formula, and the low energy limit 
of open string amplitudes.\  To save space we will not list all 
the explicit formulas of these five-point massless numerators.\ 
As an example, we consider a sample five-point scattering amplitude
of the inelastic process $(2n,n)\ito (n,n,n)$, where
all the external KK states are chosen to be $\ZZ\hsmx$ even.\
Thus, given the condition of the KK number conservation
$\sum_{i=1}^5 \hsm\bn_i^{}\!=\! 0\hs$ for each KK sub-amplitude,
we find the following allowed sub-amplitudes with their external
KK gauge bosons (gravitons) having the KK-numbers:
\begin{align}
\label{eq:KK-num-5-point}
&\{+2n,+n,-n,-n,-n\} \hs , \hspace{2.5mm}
\{+2n,-n,+n,-n,-n\} \hs , \hspace{2.5mm}
\{+2n,-n,-n,+n,-n\} \hs ,
\nn\\
&\{+2n,-n,-n,-n,+n\} \hs , \hspace{2.5mm}
\{\rm{all \ permutations \ of\  \( +,-\)}\} ,
\end{align}
where for convenience we assign the external state-1 to have
the KK number $+2n\hs$.\  For instance, we can explicitly
construct the KK longitudinal graviton scattering amplitude
with the external KK-number combination $\{+2n,+n,-n,-n,-n\}$:
\\[-6mm]
%
\begin{align}
\label{eq:MS1-5pt-2}
&\M\big[h^{+2n}_L h^{+n}_L h^{-n}_L h^{-n}_L h^{-n}_L\big]
\hs =\hs
\frac{~\ka^3\,}{64}\!\sum_{\lam^{}_k,\lam'_k}
\!\prod_{k=1}^5\! C_{\lam_k^{}\lam'_k}^{\hs 0}
\Bigg[
\frac{\NN^{}_1\NN^{\pp}_1}{\,(s_{12}^{}\!-\! 9\Mn^2)
(s_{34}^{}\!-\! 4\Mn^2)\,}
\nn\\
&~
+\frac{\NN^{}_2\NN^{\pp}_2}{(s_{12}^{}\!-\! 9\Mn^2) (s_{35}^{}\!-\! 4\Mn^2)}
+\frac{\NN^{}_3\NN^{\pp}_3}{(s_{12}^{}\!-\! 9\Mn^2) (s_{45}^{}\!-\! 4\Mn^2)}
+\frac{\NN^{}_4\NN^{\pp}_4}{(s_{13}^{}\!-\! \Mn^2)s_{24}^{}}
+\frac{\NN^{}_5\NN^{\pp}_5}{(s_{13}^{}\!-\! \Mn^2)s_{25}^{}}
\nn\\
&~
+\frac{\NN^{}_6\NN^{\pp}_6}{(s_{13}^{}\!-\! \Mn^2) (s_{45}^{}\!-\! 4\Mn^2)}
+\frac{\NN^{}_7\NN^{\pp}_7}{(s_{14}^{}\!-\! \Mn^2)s_{23}^{}}
+\frac{\NN^{}_8\NN^{\pp}_8}{(s_{14}^{}\!-\! \Mn^2)s_{25}^{}}
+\frac{\NN^{}_9\NN^{\pp}_9}{(s_{14}^{}\!-\! \Mn^2)(s_{35}^{}\!-\! 4\Mn^2)}
\nn\\
&~
+\frac{\NN^{}_{10}\NN^{\pp}_{10}}{(s_{15}^{}\!-\! \Mn^2)s_{23}^{}}
+\frac{\NN^{}_{11}\NN^{\pp}_{11}}{(s_{15}^{}\!-\! \Mn^2)s_{24}^{}}
+\frac{\NN^{}_{12}\NN^{\pp}_{12}}{(s_{15}^{}\!-\! \Mn^2)(s_{34}^{}\!-\! 4\Mn^2)}
+\frac{\NN^{}_{13}\NN^{\pp}_{13}}{s_{23}^{}(s_{45}^{}\!-\! 4\Mn^2)}
\nn\\
&~
+\frac{\NN^{}_{14}\NN^{\pp}_{14}}{s_{24}^{}(s_{35}^{}\!-\! 4\Mn^2)}
+\frac{\NN^{}_{15}\NN^{\pp}_{15}}{s_{25}^{}(s_{34}^{}\!-\! 4\Mn^2)} \Bigg] \,.
\end{align}
The five-point KK graviton scattering amplitudes with other
combinations of the external KK indices given in \eqrefe{eq:KK-num-5-point} can be constructed in the
similar manner.\

\vspace*{2mm}
\subsection{\hspace*{-2mm}Nonrelativistic Scattering Amplitudes
of KK States}
\label{sec:3.5}
\vspace*{2mm}

The KK compactification predicts an infinite tower of KK states
for each type of particles
in the compactified 4d effective theory, which serve as the
KK excitation states of the corresponding zero-mode state.\
The KK mass spectrum always contains many heavy KK states,
so they can have KK masses much larger than
their kinetic energy and thus become nonrelativistic.\
Hence it is interesting to study how such nonrelativistic
scattering amplitudes of heavy KK states behave.\
In addition, there are recent studies to extract the
classical observables in the weak gravitational systems
by calculating scattering amplitude\,\cite{BernC1}\cite{CalAmp-Rev}.\
In these toy models, the macroscopic star or black holes are treated
as super massive particles.\  Their classical gravitational potential
is be obtained by computing the four-point scattering amplitudes,
while the testable gravitational wave signals can be calculated in
term of the five-point scattering amplitudes.\
This motivates us to study the scattering amplitudes of the
super massive KK states through the gravitational interactions
and examine their behavior in the nonrelativistic limit.\

\vspace*{5mm}
\noindent
{\bf 3.5.1.\,Nonrelativistic Scattering
Amplitudes of KK Gauge\,Bosons/Gravitons}
\vspace*{3mm}


We consider the four-point KK graviton amplitudes
in the nonrelativistic limit under low energy expansion
$M_n^2\!\gg\! q^2\hs$, where
$\hs q\hs$ 
denotes the magnitude of the 3-momentum of the
initial or final states of KK particles.\
For the inelastic channel $\{0,0,n,n\}$,
we have the magnitude of the 3-momentum of the
massless initial state $\hs q\!=\!E_n^{}\hs$
and the magnitude of the 3-momentum of the KK final state
$\hs q'\!=\!\hsm\sqrt{E_n^2\!-\!M_n^2\,}
 \!\ll\!M_n^2\hs$.\
Thus, with the full KK gauge boson scattering amplitude
in Eq.\eqref{eq:T00nn}  and Eq.\eqref{eq:NA-00nn}
(Appendix\,\ref{App:B}),  we derive the
following expanded nonrelativistic results at the LO and NLO:
\beqs
\begin{align}
\TT_0[A_{\pm1}^{a0}A_{\mp1}^{b0}A_L^{cn}A_L^{dn}] &=\hs
g^2(\CC_u^{}\hsm -\CC_t^{})(1\hsm -\hsm\ct) \hs,
\\[1mm]
\da\TT[A_{\pm1}^{a0}A_{\mp1}^{b0}A_L^{cn}A_L^{dn}] &=
-\frac{\,g^2\hs q'\,}{~2\Mn\,}\hs\CC_s^{}\hs (\ct-\cttt)\hs,
\end{align}
\eeqs
where we have applied the 5d orbifold compactification
of $S^1\!/\ZZ$.\
Then, using the full KK graviton scattering amplitude
\eqref{eq:M00nn-full} of Appendix\,\ref{App:C}, we derive the
following expanded nonrelativistic results at the LO and NLO:
\beqs
\begin{align}
\hspace*{-5mm}
\M_0\big[h^0_{\pm 2} h^0_{\mp 2} h^n_L h^n_L\big]
&= \M_0\big[h^0_{\pm 2} h^0_{\mp 2} h^{\pm n}_L h^{\mp n}_L\big]
=\frac{~3\hs\ka^2\hsm\Mn^2\,}{32}
(3\hsm -\hsm 4\hs\ctt\!+\hsm\ctf)\hs,
\\[1mm]
\hspace*{-5mm}
\dM\big[h^0_{\pm 2} h^0_{\mp 2} h^n_L h^n_L\big]
& =\dM\big[h^0_{\pm 2} h^0_{\mp 2} h^{\pm n}_L h^{\mp n}_L\big]
=\frac{~3\hs\ka^2q^{\prime\hs 2}\,}{128}
(2\hsm -\hsm \ctt\!-\hsm 2\hs\ctf\!+\cts)
\hs.
\end{align}
\eeqs
We see that for the inelastic channel of $\{0,0,n,n\}$, the LO
KK amplitude is of $O(M_n^2q^0)$ and the NLO KK amplitude has
$O(M_n^0q^2)\hs$.\

\vspace*{1mm}

Next, we consider the nonrelativistic limit of the
elastic KK gauge boson/graviton scattering channel
$\{n,n,n,n\}$.\
With the full KK gauge boson scattering amplitude
in Eq.\eqref{eq:Amp-4AL-nnnn} and
Eqs.\ \eqref{eq:NA-4n}-\eqref{eq:NC-4n}
(Appendix\,\ref{App:B}),  we derive the
following expanded nonrelativistic results at the LO and NLO:
{\small
\beqs
\begin{align}
\label{eq:4AL-nnnn-NR-LO}
\hspace*{-5mm}
\TT_{\hs 0}[A_L^{an}A_L^{bn}A_L^{cn}A_L^{dn}] &=
\frac{~g^2\Mn^2~}{q^2}\!
\[\CC_t^{}\frac{~1\hsm +\ctt~}{1\hsm +\ct}
  +\CC_u^{}\frac{\,-(1+\ctt)~}{1\hsm -\ct}\]\!,
\\[1mm]
\hspace*{-5mm}
\dT[A_L^{an}A_L^{bn}A_L^{cn}A_L^{dn}] &=
g^2\!\hsm\[\CC_s^{}\frac{\hs -9\hs\ct\,}{2} +
  \CC_t^{}\frac{\,4\!-\!19\hs\ct\!-\hsm 6\hs\ctt\!+\hsm\cttt}
{4\hs (1\hsm +\ct)}
+ \CC_u^{}\frac{\,-(4\!+\!19\hs\ct\!-\hsm 6\hs\ctt\!-\hsm\cttt)\,}
{4\hs (1\hsm -\ct)}\hsm\]\hsm\!.\hspace*{-2mm}
\end{align}
\eeqs
}
\hspace*{-3mm}
We note that the LO elastic KK gauge boson amplitude $\TT_0^{}$ has
$O(M_n^2\hs q^{-2})$ and exhibits a low energy behavior of
$\,1\hsm /q^2\,$ due to the exchange of massless zero modes
in the $t$ and $u$ channels with the momentum transfer
$Q^2\!=\hsm 2q^2(1\!\pm\hsm\ct)$,
where $Q^2\!=\!-t$ or $Q^2\!=\!-u\hs$.\
It is worth to note that after Fourier transformation,
this $\,1\hsm /Q^2\,$ behavior reproduces
the classical Coulomb potential $1/\hs r\,$.\

\vspace*{1mm}

Then, using the double-copied full KK graviton amplitudes
\eqref{eq:Amp-4hL-nnnn}-\eqref{eq:Xj4hLZ2-nnnn}
of Appendix\,\ref{App:C}, we derive the
following expanded nonrelativistic partial amplitudes at the LO:
\beqs
\begin{align}
{\M}_0^{}\big[h_L^{\pm n}h_L^{\pm n} h_L^{\mp n}h_L^{\mp n}\big]
& = 0 \,,
\\[1.5mm]
{\M}_0^{}\big[h_L^{\pm n}h_L^{\mp n} h_L^{\mp n}h_L^{\pm n}\big]
&= \frac{~\ka^2\Mn^4\hs
(11\hsm +\hsm 12\hs\ct\hsm +9\hs\ctt)~}
{16\hs (1\hsm -\ct)\hs q^2} \,,
\\[1.5mm]
{\M}_0^{}\big[h_L^{\pm n}h_L^{\mp n} h_L^{\pm n}h_L^{\mp n}\big]
& = \frac{~\ka^2\Mn^4\hs (11\hsm +\hsm 12\ct\hsm +9\ctt)~}
{16\hs (1\hsm +\ct)\hs q^2} \,.
\end{align}
\eeqs
Under the 5d orbifold compactification of $S^1\!/\ZZ$,
we further deduce the following LO amplitude with
the external KK states being $\ZZ\hsm$ even:
\begin{align}
\label{eq:4hL-nnnn-NR-LO}
{\M}_0^{}\big[h_L^{n}h_L^{n} h_L^{ n}h_L^{n}\big]\!
&=\hsm \Fr{1}{\,2\,}\hsm\!
\(\!\M_0^{}\big[h_L^{\pm n}h_L^{\pm n} h_L^{\mp n}h_L^{\mp n}\big] \!+\!
{\M}_0^{}\big[h_L^{\pm n}h_L^{\mp n} h_L^{\mp n}h_L^{\pm n}\big]
\!+\!
{\M}_0^{}\big[h_L^{\pm n}h_L^{\mp n} h_L^{\pm n}h_L^{\mp n}\big]\!\)
\nn\\[1.mm]
&= \frac{~\ka^2\Mn^4~}{8\hsx q^2}
(1\!+\hsm 3\hs\ctt)^2\hsm\csc^2\!\theta
\hs.
\end{align}
We see that these LO elastic KK amplitudes have
$O(M_n^4\hs q^{-2})$ and they exhibit a low energy behavior of
$\,1/q^2\,$.\ This is due to the exchange of massless zero modes
in the $t$ and/or $u$ channels with the momentum transfer
$Q^2\!=\hsm 2q^2(1\!\pm\hsm\ct)$,
where $Q^2\!=\!-t$ or $Q^2\!=\!-u\hs$.\
It is worth to note that after Fourier transformation,
this $\,1\hsm /Q^2\,$ behavior reproduces
the classical Newtonian gravitational potential $1/\hs r\,$.\

\vspace*{1mm}

Then, under the nonrelativistic expansion,
we derive the following NLO KK amplitude for
the elastic scattering channel $\{n,n,n,n\}$:
\beqs
\begin{align}
\hspace*{-5mm}
\da\M\big[h_L^{\pm n}h_L^{\pm n} h_L^{\mp n}h_L^{\mp n}\big]
& = 0 \,,
\\[1.5mm]
\hspace*{-5mm}
\dM\big[h_L^{\pm n}h_L^{\mp n} h_L^{\mp n}h_L^{\pm n}\big]
&=\frac{\ka^2\Mn^2}{\,64\hs (1\!-\hsm\ct)\,}
(102\hsm +\hsm 22\hs\ct\hsmx +\hsm 72\hs\ctt\hsmx +\hsm
51\hs\cttt\hsmx +\hsm 18\hs\ctf\hsmx -\hsm 9\hs\ctfv)\hs,
\\
\hspace*{-5mm}
\dM\big[h_L^{\pm n}h_L^{\mp n} h_L^{\pm n}h_L^{\mp n}\big]
&=\frac{\ka^2\Mn^2}{\,64\hs (1\!+\hsm\ct)\,}
(102\hsm -\hsm 22\hs\ct\hsmx +\hsm 72\hs\ctt\hsm
-\hsm 51\hs\cttt\hsm +\hsm 18\hs\ctf\hsm +\hsm 9\hs\ctfv)\hs,
\end{align}
\eeqs
which are of $O(M_n^2q^0)$ and energy-independent.\
For the 5d orbifold compactification of $S^1\!/\ZZ$,
we further derive the corresponding NLO elastic KK amplitude:
\beq
\dM \big[h_L^{n}h_L^{n} h_L^{ n}h_L^{n}\big]
= \frac{~\ka^2\Mn^2~}{128}\hs
(226\hsm +\hsm 217\hs\ct\hsm +\hsm 78\hs\ctf\hsm
 -\hsm 9\hs\cts)\hsm\csc^2\!\theta \,,
\eeq
which also has the magnitude of $O(M_n^2\hs q^0)$.

\vspace*{5mm}
\noindent
{\bf 3.5.2.\,Nonrelativistic KK Scattering Amplitudes
with KK Graviton Exchange}
\vspace*{3mm}

For comparison with the above nonrelativistic pure
KK gauge/gravity amplitudes, we consider the
Einstein gravity coupled to scalar fields or gauge fields
under the 5d KK compactification of $S^1\!/\ZZ$
and analyze the nonrelativistic scattering amplitudes of
KK scalars (KK gauge bosons) through the KK graviton exchanges.\

\vspace*{1mm}

We first consider the 5d Einstein-Scalar (ES) theory as a toy model,
which contains a massless real scalar field $\widehat{\Phi}$
coupled to gravity:
\begin{equation}
\label{eq:SEH}
S_{\rm{ES}}^{} \,=\, \int\!\!\rm{d}^4\hsm{x}\hs\d y \,
\sqrt{\!-\hat{g}\,}\!\(\!\frac{2}{~\hka^2\,} \widehat{R}
+\frac{1}{\,2\,}\hat{g}^{}_{\mn}\pd^\mu_{}\widehat\Phi\hs
 \pd^{\nu}\widehat\Phi
\hsm\)\!,
\end{equation}
Under the 5d compactification, we make the KK expansion
of $\widehat{\Phi}$ and derive the trilinear
KK scalar-graviton interaction vertex
$\hs\phi_n^{}$$\hsm -$$\phi_m^{}$$\hsm-$$h_k^{\mn}$
and the KK scalar-radion interaction vertex
$\hs\phi_n^{}$$\hsm -$$\phi_m^{}$$\hsm-$$h_0^{55}$.\
Thus, we can compute the four-point scalar scattering amplitudes
with zero-mode and KK graviton-exchanges as well as the
radion exchange.\
For the inelastic channel
$\phi^{}_0\hs\phi^{}_0\ito\phi^{}_n\hs\phi^{}_n$,
we derive the following scattering amplitude:
\begin{equation}
\label{eq:4phi-00nn}
\hspace*{-5mm}
\TT_{\hs h}^{}[\phi^{}_0\phi^{}_0\phi^{}_n\phi^{}_n]
\,=\,-\frac{~\ka^2\Mn^2\hs\big[(7\hs\bs\hsm+\hsm 4)
\hsm +(\bs\hsm +\hsm 4)\hs\ctt\hs\big]^2 ~}
{32\hs\big[(\bs\hsm +\hsm 4)\hsmx -\hsmx (\bs\hsm -\hsm 4)
\hs\ctt\big]}
\,,
\end{equation}
where
$\hs\bs\hsm =\hsm s/M_n^2\!
 =\hsm 4\hs (\bar{q}^{\hs\prime\hs 2}\!+\hsm 1)\hs$
and $\bar{q}^{\hs\prime}\!=\hsm {q}^{\hs\prime}\hsm /M_n^{}\hs$.\
We have verified the above amplitude by using the shifting method
of Sec.\,\ref{sec:2.2} to derive the massive KK amplitude from
computing the corresponding 5d massless amplitude.\
Then, from Eq.\eqref{eq:4phi-00nn}
we make the nonrelativistic expansion
$\hs q^{\prime\hs 2}\!\ll\!M_n^2\hs$,
and derive the following LO and NLO scalar amplitudes:
\beqs
\label{eq:4phi-00nn-NRexp}
\begin{align}
\label{eq:4phi-00nn-NRexpLO}
\TT_{\hs h}^{(0)}[\phi^{}_0\phi^{}_0\phi^{}_n\phi^{}_n]
&\hs =\hs -\frac{\,9\,}{\,4\,}{\ka^2\Mn^2} \,,
\\
\label{eq:4phi-00nn-NRexpNLO}
\da\TT_{\hs h}^{}[\phi^{}_0\phi^{}_0\phi^{}_n\phi^{}_n]
&\hs =\hs -\frac{\,3\,}{\,8\,}{\ka^2q^{\hs\prime\hs 2}}
(11\!+5\hs\ctt) \hs.
\end{align}
\eeqs
We see that the LO and NLO amplitudes are of
$O(\ka^2M_n^2)$ and
$O(\ka^2q^{\hs\prime\hs 2})$, respectively.

\vspace*{1mm}

For the elastic channel $\{n,n,n,n\}$,
we compute the KK scalar scattering amplitude as follows:
\begin{equation}
\TT_{\hs h}^{}[\phi^{}_n\phi^{}_n\phi^{}_n\phi^{}_n]
= -\frac{~\ka^2\Mn^2\hs (X_0^{} \!+\!X_2^{}\hs\ctt\!+\!X_4^{}\ctf\!+\!X_6^{}\hs\cts)
\hsm\csc^2\!\theta~}
{~512\hs\bs\hs (\bs\hsm -\hsm 4)
\big[(\bs^2+24\hs\bs + 16)\hsm -\hsm (\bs\hsm -\hsm 4)^2 \ctt \big]~} \,,
\end{equation}
where
$\hs\bs\hsm =\hsm s/M_n^2\!
 =\hsm 4\hs (\bar{q}^{2}\!+\! 1)\hs$
with $\bar{q}\hsm =\hsm {q}/M_n^{}\hs$,
and the polynomial functions $\{X_j^{}\}$ take the following form:
\begin{align}
X_0^{} &=510 \hs\bs^5\!+\hsm 8144\hs\bs^4\!-\hsm 27584\hs\bs^3\!
+61440\hs\bs^2\!+60928\hs\bs +\hsm 28672 \hs,
\nn\\
X_2^{} &=-429\hs\bs^5\!+6984\hs\bs^4\!-\hsm 30816\hs\bs^3\!
+\hsm 51712\hs\bs^2\!-\hsm 42240\hs\bs \hsm -\hsm 34816\hs,
\nn\\[-3.3mm]
\\[-3.3mm]
X_4^{} &=-78 \hs\bs^5\!+\hsm 1200 \hs\bs^4\!-\hsm 6720\hs\bs^3
\!+\hsm 16384\hs\bs^2\!-\!15872\hs\bs\hsm +\hsm 4096\hs,
\nn\\
X_6^{} &=-3\hs\bs^5\!+\hsm 56\hs\bs^4\!-\hsm 416\hs\bs^3\!
+\hsm 1536\hs\bs^2\!-\hsm 2816\hs\bs\hsm +\hsm 2048\,.
\nn
\end{align}
Thus, making the nonrelativistic expansion
$\hs q^{2}\!\ll\!M_n^2\hs$,
we derive the following KK scalar scattering amplitudes
at the LO and NLO:
\beqs
\label{eq:4phi/h-nnnn}
\begin{align}
\label{eq:4phi/h-nnnn-LO}
\TT_{\hs h}^{(0)}[\phi^{}_n\phi^{}_n\phi^{}_n\phi^{}_n]
&=\hs
-\frac{~2\hs\ka^2\Mn^4\csc^2\!\theta~}{q^2}
\hs,
\\
\label{eq:4phi/h-nnnn-NLO}
\da\TT_{\hs h}^{}[\phi^{}_n\phi^{}_n\phi^{}_n\phi^{}_n]
&=\hs -\frac{1}{\,2\,}\ka^2\Mn^2\hs{(7\hsm +\hsm\ctt)}
\hsm \csc^2\!\theta\hs,
\end{align}
\eeqs
whose magnitudes are of $O(\ka^2M_n^4/q^2)$ and
$O(\ka^2M_n^2)$, respectively.
Note that the LO elastic KK scalar amplitude
\eqref{eq:4phi/h-nnnn-LO} also exhibits a low energy behavior of
$\,1/q^2\,$ due to the exchange of massless zero modes
in the $t$ and $u$ channels with the momentum transfer
$Q^2\!=\hsm 2q^2(1\!\pm\hsm\ct)$,
where $Q^2\!=\!-t$ or $Q^2\!=\!-u\hs$.\
This is similar to the LO amplitudes of the pure KK gauge boson
(KK graviton) scattering in
Eqs.\eqref{eq:4AL-nnnn-NR-LO} and  \eqref{eq:4hL-nnnn-NR-LO}.\
After a Fourier transformation
such $\,1\hsm /Q^2\,$ behavior recovers the classical
Newtonian gravitational potential of $1\hsm /r$ at low energy.\
We also see that the NLO KK amplitude
\eqref{eq:4phi/h-nnnn-NLO} behaves as $q$-independent constant term.\

\vspace*{1mm}

Next, for comparison we further consider the 5d gauge theory
coupled to GR, which has the following
5d action:
\begin{equation}
\label{eq:EYM5}
S_{\rm{GRA}}^{} \,=\, \int\!\!\rm{d}^4\hsm{x}\hs\d y \,
\sqrt{\!-\hat{g}\,}\hsm\(\!\frac{2}{~\hka^2\,} \widehat{R}
-\frac{1}{\,4\,}\hat{g}^{\mu\al}\hat{g}^{\nu\be}
\widehat{F}^{\hs a}_{\mn}\widehat{F}^{\hs a}_{\al\be}
\)\!,
\end{equation}
where $\widehat{F}^{\hs a}_{\mn}$ is the 5d gauge field strength of a
non-Abelian gauge group of YM theory or
Abelian gauge group of Maxwell theory (with $a\!=\!1$).\
Under the 5d compactification, we derive the 4d KK theory
which includes the trilinear KK gauge-boson-graviton interaction
vertex $\hs A_{n}^{a\hs\mu}$$\hsm -$$A_m^{b\hs\nu}$$\hsm-$$h_k^{\mn}$.\
Then, we will compute the four-point KK gauge boson scattering
amplitudes with graviton exchange and derive the nonrelativistic
results at low energy, which should be compared to the
nonrelativistic KK scalar amplitudes we derived for the
Einstein-Scalar theory \eqref{eq:SEH}.\
For the case of non-Abelian gauge theory, the four-point KK gauge
boson amplitudes will also receive the contribution
by pure gauge interactions including the contact gauge vertex and the
gauge boson exchange via cubic gauge vertex.\ But such pure gauge
contribution is gauge-invariant by itself and has already been
computed in Sec.\,3.5.1 and Sec.\,\ref{sec:3.2}-\ref{sec:3.3}.\
So, in the following we will focus on analyzing the
nonrelativistic KK gauge boson amplitudes through the graviton
exchanges and compare them with the nonrelativistic KK scalar amplitudes
which we derived in the first part of Sec.\,3.5.2.\

\vspace*{1mm}

For the inelastic channel $\{0,0,n,n\}$, we compute the
four-point gauge boson scattering amplitude via graviton
and radion exchanges:
\begin{equation}
\TT_{\hs h}^{}\big[A_{\pm1}^{a0}\hsm A_{\pm1}^{a0}
\hsm A_L^{bn}\!A_L^{bn}\big]
\hs =\hs \frac{~\ka^2\Mn^2 \big[(3\hs\bs^2\!+\!18\hs\bs\hsm -\hsm 8)
\hsm -\hsm (3\hs\bs^2\!-\!14\hs\bs\hsm +\hsm 8)\hs\ctt\hs\big]~}
{8\hs\big[(\bs\hsm +\hsm 4)\hsm -\hsm (\bs\hsm -\hsm 4)\hs\ctt\hs\big]}
\,.
\end{equation}
Then, we make the nonrelativistic expansion and derive
the LO and NLO scattering amplitudes as follows:
\\[-6mm]
\beqs
\label{eq:4A-00nn-NRexp}
\begin{align}
\label{eq:4A-00nn-NRexpLO}
\TT_{\hs h}^{(0)}\hsm\big[A_{\pm1}^{a0}\hsm A_{\pm1}^{a0}
\hsm A_L^{bn}\!A_L^{bn}\big]
& = \frac{7}{\,4\,}\hs\ka^2\Mn^2\,,
\\
\label{eq:4A-00nn-NRexpNLO}
\dT_{\hs h}^{}\hsm\big[A_{\pm1}^{a0}\hsm A_{\pm1}^{a0}
\hsm A_L^{bn}\!A_L^{bn}\big]
& = \frac{1}{\,4\,}\ka^2q^{\hs\prime\hs 2}(7 +\ctt)
\hs.
\end{align}
\eeqs
The above LO and NLO amplitudes are of $O(\ka^2M_n^2)$ and
$O(\ka^2q^{\hs\prime\hs 2})$ respectively.\
These have the same structures as that of the corresponding
LO and NLO scalar amplitudes in Eq.\eqref{eq:4phi-00nn-NRexp},
except that the angular dependence of
the above NLO amplitude \eqref{eq:4A-00nn-NRexpNLO}
has different coefficients from that of
Eq.\eqref{eq:4phi-00nn-NRexpNLO}.
%

\vspace*{1mm}

For the elastic channel $\{n,n,n,n\}$, we first consider all
the external KK gauge bosons having transverse polarizations
with the helicity indices $(\pm 1,\pm 1,\pm 1,\pm 1)$.\
Thus, we compute their elastic scattering amplitude as follows:
%
\begin{equation}
\TT_{\hs h}^{}\big[A_{\pm1}^{an}A_{\pm1}^{an}\hsm
A_{\pm1}^{bn}\hsm A_{\pm1}^{bn}\big]
= -\frac{~\ka^2\Mn^2\big[
(3\hs\bs^2\!-\!11\hs\bs\hsm +\!12)\hsm -\hsm (\bs\!-\!4)\ctt\big]~}
{2\hs\bs\hs (\bs-4)} \hs,
\end{equation}
where
$\hs\bs\hsm =\hsm s/M_n^2\!
 =\hsm 4\hs (\bar{q}^{2}\!+\! 1)\hs$
with $\bar{q}\hsm =\hsm {q}/M_n^{}\hs$.\
Then, we make the nonrelativistic expansion and derive
the following LO and NLO elastic scattering amplitudes:
\beqs
\label{eq:4A-nnnn-NRexp}
\begin{align}
\label{eq:4A-nnnn-NRexpLO}
\TT_{\hs h}^{(0)}\hsm\big[A_{\pm1}^{an}A_{\pm1}^{an}\hsm
A_{\pm1}^{bn}\hsm A_{\pm1}^{bn}\big]
&=-\frac{~\ka^2\Mn^4~}{2\hs q^2} \,,
\\
\label{eq:4A-nnnn-NRexpNLO}
\dT_{\hs h}^{}\big[A_{\pm1}^{an}A_{\pm1}^{an}\hsm
A_{\pm1}^{bn}\hsm A_{\pm1}^{bn}\big]
&=-\frac{1}{\,8\,}\ka^2\hsm\Mn^2\hs (9\hsm -\hsm\ctt) \hs.
\end{align}
\eeqs
We see that the LO nonrelativistic KK amplitude
\eqref{eq:4A-nnnn-NRexpLO}
exhibits a low energy behavior
$\,1/q^2\,$ due to the exchange of massless zero modes
in the $t$ and $u$ channels
which is similar to the LO KK scalar amplitude
\eqref{eq:4phi/h-nnnn-LO}.\
But, the above LO and NLO KK gauge boson scattering amplitudes have
different angular structures from that of the
elastic KK scalar scattering amplitudes \eqref{eq:4phi/h-nnnn}.

\vspace*{1mm}


Finally, we compute the four-point elastic longitudinal
KK gauge boson scattering amplitude via the graviton
and radion exchanges:
\begin{equation}
\TT_{\hs h}^{}\big[A_L^{an}\!A_L^{an}\!A_L^{bn}\!A_L^{bn}\big]
= -\frac{~\ka^2\hsm\Mn^2\hs\big(X_0^{}\hsm +\hsm X_2^{}\hs\ctt \hsm +\!X_4^{}\hs\ctf\hsm +\!X_6^{}\hs\cts\big)\hsm\csc^2\!\theta~}
{~512\hs\bs\hs (\bs\hsm -\hsm 4)
\big[(\bs^2\!+\hsm 24\hs\bs\hsm +\!16)
\hsm -\hsm (\bs\hsm -\hsm 4)^2 \ctt\big]~} \,,
\end{equation}
where
$\hs\bs\! =\! s/M_n^2\!
 =\hsm 4\hs (\bar{q}^{2}\hsm +\hsm 1)\hs$
with $\bar{q}\! =\! {q}/M_n^{}\hs$,
and the polynomial functions $\{X_j^{}\}$ are expressed as follows:
\begin{align}
X_0^{}&= 510\hs\bs^5\!+\hsm 7696\hs\bs^4\!-\hsm 35520\hs\bs^3\!
+\hsm 39936\hs\bs^2\!+\hsm 40448\hs\bs\hsm +\hsm 28672\hs,
\nn\\
X_2^{}&= -429 \hs\bs^5\!+\hsm 7528\hs\bs^4\!-\hsm 25056\hs\bs^3\!
+77824\hs\bs^2\!-\! 11520 \hs\bs\hsm -\hsm 34816\hs,
\nn\\
X_4^{}&= -78\hs\bs^5\!+\! 1136\hs\bs^4\!-\hsm 4928\hs\bs^3\!
+\!13312\hs\bs^2\!-\hsm 28160\hs\bs\hsm +\hsm 4096\hs,
\nn\\
X_6^{}&= -3\hs\bs^5\!+\hsm 24\hs\bs^4\!-\hsm 32\hs\bs^3\!-\hsm
768\hs\bs\hsm +\hsm 2048\hs.
\end{align}
From the above, we make the nonrelativistic expansion and derive
the LO and NLO longitudinal KK scattering amplitudes as follows:
\beqs
\label{eq:4AL-nnnn-NRexp}
\begin{align}
\label{eq:4AL-nnnn-NRexpLO}
\TT_{\hs h}^{(0)}\big[A_L^{an}A_L^{an} A_L^{bn} A_L^{bn}\big]
&=-\frac{~2\hs\ka^2\Mn^4\cot^2\!\theta~}{q^2}
 =-\frac{~\ka^2\Mn^4~}{q^2}
(1\!+\hsm\ctt)\hsm\csc^2\!\theta \hs,
\\
\label{eq:4AL-nnnn-NRexpNLO}
\dT_{\hs h}^{}\big[A_L^{an}A_L^{an} A_L^{bn} A_L^{bn}\big]
&=-\frac{~\ka^2\Mn^2~}{8} (23\hsm +\hsm 8\hs\ctt\hsm +\hsm\ctf)\hsm
\csc^2\!\theta\hs.
\end{align}
\eeqs
The above can be compared to the LO and NLO amplitudes
\eqref{eq:4A-nnnn-NRexp} with all the external KK states being
transversely polarized, and they have different angular structures.\
Moreover, comparing the above nonrelativistic LO and NLO
longitudinal KK scattering amplitudes with
the scalar KK scattering amplitudes \eqref{eq:4phi/h-nnnn},
we find that their LO amplitudes have the magnitude of
$O(\ka^2M_n^4/q^2)$ and their NLO amplitudes are of
$O(\ka^2M_n^2)$, but they have rather different
angular structures at the LO and NLO respectively.\
These nontrivial differences of angular structures between
the nonrelativistic KK scalar scattering amplitudes
\eqref{eq:4phi/h-nnnn} and the KK gauge boson scattering amplitudes
\eqref{eq:4A-nnnn-NRexp} are due to the different spins of the
external KK states in the two cases and their different
cubic interaction vertices with the internal gravitons.\
Such differences remain even in the nonrelativistic KK scattering
amplitudes.\

\section{\hspace*{-2mm}Worldsheet Formulation of Massive KK Double-Copy}
\label{sec:4}
\vspace*{1mm}

The conventional Cachazo-He-Yuan (CHY) formalism\,\cite{CHY}
can be used to analyze the scattering amplitudes of
the massless scalar, gauge, and gravity theories
in terms of the localized integrals
in the moduli space $\mathfrak{M}_{0,N}^{}$ of
the $N$-punctured genus-zero Riemann sphere $\mathbb{CP}^1$.\
Its physical origin is based on the ambitwistor string theory\,\cite{Mason:2013sva,Geyer:2014fka,Adamo:2013tsa,Casali:2015vta}.
In this section, we generalize the conventional massless
CHY scattering equations to the case of the massive KK scattering
amplitudes under the 5d toroidal compactification.\
We derive an extended massive KK scattering equation
for the KK scattering amplitudes in Section\,\ref{sec:4.0}
and construct the KK bi-adjoint scalar amplitudes
in Section\,\ref{sec:4.1}.\
Then, in Section\,\ref{sec:4.2} we use this massive CHY approach to
further construct the scattering amplitudes of KK gauge bosons
and of KK gravitons, and derive their relations to the extended
BCJ-type mmasive KK amplitudes (given in Section\,\ref{sec:3}) and
to the extended KLT-type masive KK amplitudes
(formulated in Ref.\,\cite{KKString}).


\vspace*{2mm}
\subsection{\hspace*{-2mm}Extended CHY Formulation of
Massive KK Scattering Amplitudes}
\vspace*{2mm}
\label{sec:4.0}

Consider the $(4\hsm +\hsm\delta)$-dimensional spacetime with
$\delta$ extra spatial dimensions under the
toroidal compactification.\
For a momentum $\hp^{}_a$ living in the
$(4\hsm +\hsm\delta)$-dimensions,
we can decompose it in the following form:
\begin{equation}
\hat{p}_{a}^{M}=
\(\hsm p_{a}^{\mu},\hs
\fr{\bn_1^{}}{\,R_1^{}}\bigg|_a \hs,\cdots\hsm,
\fr{\bn_{\da}^{}}{\,R_{\da}^{}} \bigg|_a \)  \!, \quad~~
a \in \{1 , \cdots\hsm, N\}  \hs,
\end{equation}
where $\,p^{\mu}_a\,$ and
$\,(\bn_i^{}/R_i^{})\big|_a\,$ are the momenta defined in
$(3\!+\!1)$-dimensions and $\da$-dimensions respectively.\
For the simplicity of demonstration,
we set $\hs \da\!\!=\hsmx\!1\hs$
with the compactification radius
$\hs R_1^{}\!=\!R$\,.\
Hence, similar to \cite{Dolan}\cite{Naculich}
we can express the $\rm{SL}(2,\mC)$ invariant
scattering equation in the $(4\! +\! 1)$-dimensional
spacetime as follows:
\begin{equation}
\label{eq:SEq-Ea}
\widehat{E}_a \,=\,
\sum_{\substack{b=1 \\ (b \neq a)}}^N\!\!
\frac{~\hat{p}_a^{} \!\cdot \hat{p}_b^{}~}
{z_a^{} \!-\hsm z_b^{}} \,=\,
\sum_{\substack{b=1 \\ (b \neq a)}}^N\!\!
\frac{~p_a^{} \!\cdot p_b^{} +\! M^2_{ab}~}
{z_a^{} \!-\hsm z_b^{}} \,=\, 0 \,,
\end{equation}
where the $N$ holomorphic variables $\{z_a^{}\}$
parametrize the moduli space $\mathfrak{M}_{0,N}^{}$ of
the $N$-punctured Riemann sphere $\mathbb{CP}^1$,
and summation runs over the $\,(N\!-3)!\,$
solutions for $z_a^{}$.\
The mass parameter in Eq.\eqref{eq:SEq-Ea} is given by
\begin{equation}
\label{eq:Mab^2}
M_{ab}^{2} = M_{ba}^{2}
=\frac{~\bn_a^{} \bn_b^{}~}{R^2} \,.
\end{equation}

As analyzed in Section\,\ref{sec:2.1}, we find that
each massive KK amplitude can be decomposed into
a sum of higher dimensional sub-amplitudes in which
the components of momenta associated with the
compactified dimension are discretized.\
Hence, under the toroidal compactification,
we can generalize the conventional CHY formulation of
the gauge boson amplitude and graviton amplitude
to the following forms:
\beqs
\label{eq:general-KK-CHY}
\begin{align}
\label{eq:AN-KK-CHY}
\A_N^{}\big[\{p_i^{}, \hze^{}_i,I_i^{}\}\big] &\,=\,
\sum_{q_j^{}}\!\whcA_N^{}\big[\{p_i^{}, \hze_i, q_i^{}\}\big]\hsm
\prod_j\!\hsm \int_{\mathbb{N}}^{}\td y_j^{}\hs
e^{\ii q^{}_j y_j^{}}\mO_{I_j}^{}\hsm (y_j^{}) \hs,
\\
\label{eq:MN-KK-CHY}
\M_N^{}\big[\{p_i^{}, \hze^{}_i,I_i^{}\}\big] &\,=\,
\sum_{q_j^{}}\!\whcM_N^{}\big[\{p_i^{}, \hze_i, q_i^{}\}\big]\hsm
\prod_j\!\hsm \int_{\mathbb{N}}^{}\td y_j^{}\hs
e^{\ii q^{}_j y_j^{}}\mO_{I_j}^{}\hsm (y_j^{}) \hs.
\end{align}
\eeqs
On the right-hand side of \eqrefe{eq:general-KK-CHY},
the amplitude $\whcA_N^{}\hs$ is defined as
\begin{equation}
\label{eq:CHY}
\whcA_N^{}\big[\{\hp^{}, \hze\}\big]
=\, \int\!\!\hsm
\underbrace{\frac{\td^N \!z^{}}
{\,\rm {Vol}\hs[\hs \rm{SL}(2,\mC)]\,}
\sideset{}{'}\prod_{a=1}^N\!
\delta\big(\widehat{E}_a\big)}_{\td \mu_N^{}}
\whI_N^{}(\{\hp, \hze, z\})  \,,
\end{equation}
\\[-4mm]
where $\td \mu_N^{}$ is the measure and $\whI_N^{}$ is a theory-dependent integrand.\
The above product operation $\prod^{\prime}$
is defined by
\\[-3mm]
\beq
\sideset{}{'}\prod_{a=1}^N \!
\delta\big(\widehat{E}_a\big)
\,\equiv\, (z_i^{}\!-\hsm z_j^{})
(z_j^{}\!-\hsm z_k^{})
(z_k^{}\!-\hsm z_i^{})\hspace*{-3.1mm}
\prod_{\substack{a=1 \\ (a \neq i,j,k)}}^N\hspace*{-3.8mm}
\delta\big(\widehat{E}_a\big) \hs,
\eeq
\\[-3mm]
which excludes the choice of the three fixed points
$(z_i^{},z_j^{},z_k^{})\!=\!(0,1,\infty)$
on the Riemann sphere and is permutation invariant\,\cite{CHY}.\
The scattering amplitude \eqref{eq:CHY} is invariant
under $\rm{SL}(2,\mC)$ transformation acting on the coordinates
$\{z_a^{}\}$.\
Moreover, it is well established
that the conventional KLT relations
for the $(4\! +\! 1)$-dimensional
scattering amplitudes of the massless gauge bosons
($\whcA_N^{}$) and of the massless gavitons
($\widehat{\M}_N^{}$) can be given by the CHY formalism:
\begin{equation}
\label{eq:KLT-d+1}
\widehat{\M}_N^{}  \hs =\!
\sum_{\{\al,\be\}}\!\!
\whcA_N^{}[\al] \hsx \widehat{\KK}\hsx [\al|\be] \whcA_N^{}[\be] \,,
\end{equation}
where $\{\al,\be\} \!\in\! S_{\hsm N-3}\hs$
denote the linearly independent
color-orderings of the $N$-point gauge boson amplitudes,
and $\widehat{\KK}\hsx [\al|\be]$ is the KLT kernel.\
Thus, combining the $(4\hsm +\! 1)$-dimensional
massless KLT formula \eqref{eq:KLT-d+1} with
Eq.\eqref{eq:general-KK-CHY}, we derive the
extended massive KLT-type relation under the
toroidal compactification, which expresses the KK graviton
scattering amplitude ($\M_{\hsm N}^{}$) as products of two
color-ordered KK gauge boson amplitudes ($\A_N^{}$)
together with a kernel ($\hs\KK\hs$) as follows:
\begin{equation}
\label{eq:KK-CHY-DC}
\hspace*{-3mm}
\M_{\hsm N}^{}\big[
{\{p_k^{}, \zeta^{\mn}_{\si_k^{}}\}}
\big]
\hsm =\hs c_0^{}\hsm\sum_{\fP}
\!\!\sum_{\{\al,\hs\be\}}
\!\sum_{\lam_k^{}\lam_k'}\!\hsm\!
\Big(\!\prod_{k}\!
C^{\,\si_k^{}}_{\lambda_k^{}\lambda_k'}\Big)
\A_N^{}\hsm\big[\al,\hs\zeta^\mu_{\lambda_k^{}}\!,\{\bn_i^{}\hsm\}
\hsm\big]
\KK\!\([\al|\be],\!\{\bn_i^{}\hsm\}\hsm\)\hsm
\A_N^{}\hsm\big[\be,\hs\zeta^\nu_{\lambda_k'}\!,\hsmx \{\bn_i^{}\hsm\}\hsm\big] \hs ,
\end{equation}
where $\hs c_0^{}\!=\!(-)^{N+1}\hsm [\kappa/(4g)]^{N-2}$ is
the conversion constant between the gauge and gravity couplings
as given by Eq.\eqref{eq:g2-kappa}.\
In the above,
$\{\bn_i\}$ denote the KK indices of the external states
and $\fP$ labels every possible combination of the signs of
these KK indices $\{\bn_i\}$ which obeying the
$N$-point neutral condition
$\sum_{i=1}^N \bn_i^{}\!=\! 0$ \cite{KKString}.\
In Eq.\eqref{eq:KK-CHY-DC}, $k$ counts the external states
of each amplitude and
$C_{\lam_k^{}\lam_k'}^{\si_k^{}}$ denotes the coefficients
of the polarization tensor $\hs\zeta^{\mn}_{\si_{\hsm k}^{}}$
of the $k$-th external KK graviton as defined
in Eqs.\eqref{eq:Ckk'} and \eqref{eq:Mh-4pt-BCJ}.\
Finally, we note that
the above formulation can be readily extended to the
case of having two or more compactified extra dimensions.\

\vspace*{1mm}

In summary, our above analysis has extended
the conventional massless CHY formulation
of the KLT double-copy relations
to the compactified massive KK gauge/gravity theories.\
In this massive KK CHY construction, the KLT kernel
$\KK\!\([\al|\be],\!\{\bn\}\hsm\)$
is formulated as the inverse of amplitudes of the KK
bi-ajoint scalar theory under the toroidal compactification
and will be discussed in Section\,\ref{sec:4.2}.

\vspace*{2mm}
\subsection{\hspace*{-2mm}CHY Construction for Massive KK Bi-ajoint Scalar Amplitudes}
\vspace*{2mm}
\label{sec:4.1}

Consider a massless bi-adjoint scalar (BAS) theory
respecting the (global) color symmetry
{$\rm{U}(N)\!\otimes\!\rm{U}({N})\hs$}.\
The simplest Lagrangian of such a BAS theory
contains a cubic interaction term\,\cite{CHY}.\
We can write down its Lagrangian formulation in 5d as follows:
\begin{equation}
\label{eq:LagPhi3}
\La_{\rm{BAS}} \,=\,
\frac{1}{2} (\pd_{M} \hPhi^{a a'}\hsm )^2
+\frac{\hlam}{\,3!\,}\hsm f^{a b c} f^{a' b'\hsm c'}
\hPhi^{a a'} \hPhi^{b b'} \hPhi^{c c'}\!,
\end{equation}
where $\hs\hlam\hs$ is the 5d coupling for the cubic scalar interaction.\

\vspace*{1mm}

Then, under the 5d KK compactification of $S^1$,
we make the following KK expansion for the bi-adjoint scalar field
$\hPhi^{ab}$:
\begin{equation}
\label{eq:Phi-KK-exp}
\hPhi^{ab} (x^\mu, y) \,= \frac{1}{\sqrt{2\pi R\,}\,}\!\!
\sum_{\bn=-\infty}^{+\infty} \!\!\Phi_{\bn}^{ab} (x^\mu)\hs
e^{\ii\hs\bn\hs y/\!R}  \,.
\end{equation}
Substituting the expansion \eqref{eq:Phi-KK-exp}
into the BAS Lagrangian \eqref{eq:LagPhi3}
and integrating over the 5d coordinate $y\hs$, we can
derive the effective KK Lagrangian:
\begin{align}
\La_{\rm{BAS}}^{\rm{KK}} =&\,
\frac{1}{\,2\,} |\pd_\mu \Phi_0^{a a'}|^2
+\frac{\lam}{\,6\,} f^{a b c}\hsm f^{a'b'c'}
\Phi^{a a'}_0\Phi^{b b'}_0\Phi^{c c'}_0
+\sum_{\bn\neq0} \(\frac{1}{\,2\,}|\pd_\mu \Phi_\bn^{a a'}|^2
+\frac{1}{\,2\,}M_\bn^2|\Phi_\bn^{a a'}|^2 \)
\nn\\
&+ \frac{\lam}{\,6\,} f^{a b c}f^{a'b'c'}\!\!\!\!\!
\sum_{\bn_1,\bn_2,\bn_3\neq0}\!\!
\(\Phi^{a a'}_0 \Phi^{b b'}_{\bn_1} \Phi^{c c'}_{\bn_2}
\delta_{\bn_1+\bn_2,0} + \Phi^{a a'}_{\bn_1}\Phi^{b b'}_{\bn_2}
\Phi^{c c'}_{\bn_3}\delta_{\bn_1+\bn_2+\bn_3,0}\) \!,
\end{align}
where the effctive 4d KK scalar coupling
$\,\lam\!=\!\hlam /\hsmx\sqrt{2\pi R\,}$\,.

\vspace*{1mm}

From the CHY fomulation \eqref{eq:CHY}, the full scattering amplitude
in the massless 5d bi-adjoint scalar theory
takes the following form:
\begin{align}
\label{eq:Amp-BAS-1}
\whcA_N^{\rm{BAS}} = \int\!\!\td\mu_N^{}\!\! \sum_{\{\al, \be\}}
\!\!
\PT[\al]\,\CC[\al] \!\times\! \PT[\be] \,\CC[\be]\hs.
\end{align}
In the above, the integration measure $\td\mu_N^{}\hsm$ is defined
in Eq.\eqref{eq:CHY} and can be computed according to the massive
scattering equations \eqref{eq:SEq-Ea}-\eqref{eq:Mab^2}
which include the KK masses via discretized fifth momentum-components.\
In Eq.\eqref{eq:Amp-BAS-1},
the factor $\hs\CC[\al]$ is decomposed into
the Del\,Duca-Dixon-Maltoni (DDM) basis\,\cite{DDM}
which contains $(N\!-\hsm 2)!\hs$ elements:
\begin{equation}
\label{eq:DDMcolor}
\CC [\al] \,= \sum_{c_1,\cdots\hsm ,\hs c_{N-3}^{}}
\hspace*{-4mm}
f^{a_1 a_{\al(2)} c_1} f^{c_1 a_{\al(3)} c_2} \cdots
f^{c_{N-3}^{} a_{\al(N-1)} a_N^{}}
\,,
\end{equation}
where the ordering of the 1st and $N$-th labels are fixed
at the beginning and end in the ordering $\al\hs$, i.e.,
$\al\!=\hsm\![1,\al(2),\cdots\!,\al(N\hsm\!-\hsm\!1),N]
 \!\hsm\in\hsm\! S_{N-2}\hs$,
and the ordering $\be$ is defined in the similar way.\
We also use the same label $\al$ to denote the possible
orderings and the permutations.\
In Eq.\eqref{eq:Amp-BAS-1}, the Parke-Taylor factor
$\PT[\al]\hs$ is defined as
\begin{equation}
\PT[\al] \,=\,
\frac{1}{~z_{1\al(2)}^{}\cdots\hs z_{\al(N-1)N}^{}\hs z_{N1}^{}~}\,,
\end{equation}
where in the denominator each factor
$\,z_{ij}^{}\!\!=\! z_i^{}\!-\hsmx z_j^{}\hs$.\
As for a consistent check, consider the case of four-point scattering
amplitudes, which have two allowed color orderings:
$\al\! =\! [1234]$ and $\be\! =\! [1324]\hs$.\
Thus, the color factors defined in \eqrefe{eq:DDMcolor}
are connected to the definitions \eqref{eq:Cj-def} via
\begin{equation}
\label{eq:C1234-C1324}
\CC[1234]=\CC_s \,, \qquad \CC[1324]= -\CC_u \,,
\end{equation}
where the third color factor
$\hs\CC_t\!=\!-\hsx\CC_s\!-\hsm\CC_u$ is determined by
using the color Jacobi identity in Eq.\eqref{eq:Jacobi-C-N}.\
For demonstration, we first consider the elastic four-point
KK BAS scattering amplitude.\
To solve the scattering equation \eqref{eq:SEq-Ea} for
the present case, we fix
$(z_1^{},\hs z_2^{},\hs z_3^{})
 \!=\!(0,1,+\infty)$
and derive the following solutions of $\hs z_4^{}\hs$
for the three independent combinations of
KK indices of the external states respectively:
%
\beqs
\label{eq:sol-sigma4}
\begin{align}
\fP_1^{}=\{\pm n,\pm n,\mp n,\mp n\}\!:
&~~~~~
z_4^{}=-\frac{u}{~s\hsm -\hsm 4M_n^2~}\hs,
\\[0.5mm]
\fP_2^{}=\{\pm n,\mp n,\mp n,\pm n\}\!:
&~~~~~
z_4^{}=-\frac{~u\hsm -\hsm 4M_n^2~}{s}\hs,	
\\[1mm]
\fP_3^{}=\{\pm n,\mp n,\pm n,\mp n\}\!:
&~~~~~
z_4^{}=-\frac{\,u\,}{s}\,.
\end{align}
\eeqs
%
With these, we compute the double partial amplitudes
of the KK bi-adjoint scalars for each KK combination $\fP_i^{}$
as follows:
\beqs
\label{eq:BASamp4partial}
\begin{align}
\A_{\rm{S}}^{}[1^{\pm n}2^{\pm n}3^{\mp n} 4^{\mp n}
|1^{\pm n}2^{\pm n}3^{\mp n} 4^{\mp n}] &=
-2\hs \lam^2\!\(\!\frac{1}{\,s\!-\! 4\Mn^2\,}
+\!\frac{1}{\,t\,} \)
\hsm\equiv  {\A^{}_{\rm{S}}}
[\al^{}_{\fP_1^{}}\hsm |\al^{}_{\fP_1^{}}] \hs,
\\[1mm]
\A^{}_{\rm{S}}[1^{\pm n}2^{\mp n}3^{\mp n} 4^{\pm n}
|1^{\pm n}2^{\mp n}3^{\mp n} 4^{\pm n}]
&=
-2\hs \lam^2\!\(\!\frac{1}{\,s\,}\hsm +\hsm\frac{1}{\,t\!-\!4\Mn^2\,}\hsmx\)
\hsm\equiv {\A^{}_{\rm{S}}}
[\al_{\fP_2}^{}\hsm |\al_{\fP_2}^{}] \hs,
\\[1mm]
\A^{}_{\rm{S}}[1^{\pm n}2^{\mp n}3^{\pm n} 4^{\mp n}
|1^{\pm n}2^{\mp n}3^{\pm n} 4^{\mp n}]
&=
-2\hs \lam^2\!\(\!\frac{1}{\,s\,}\hsm +\hsm \frac{1}{\,t\,} \hsm\)
\hsm\equiv {\A^{}_{\rm{S}}}
[\al_{\fP_3}^{}|\al_{\fP_3}^{}]  \hs.
\end{align}
\eeqs
Then, summing up the contributions of the three combinations above,
we further derive the double partial KK-amplitude
under the $S^1\!/\ZZ$ compactification:
\begin{align}
\label{eq:BAS-1234-1234}
\A^{}_{\rm{S}}[1^n2^n3^n 4^n|1^n2^n3^n 4^n]
&= \frac{1}{\,2\,} \sum_i
{\A^{}_{\rm{S}}}[\al_{\fP_i}^{}|\al_{\fP_i}^{}]
\hs\equiv\hs \A^{}_{\rm{S}}[\al^n|\al^n]
\nn\\
&=-\lam^2\!\(\!\frac{1}{\,s\hsmx -\hsm 4\Mn^2\,}
\!+\! \frac{1}{\,t\hsmx -\hsm 4\Mn^2\,}
\!+\!\frac{2}{\,s\,} + \frac{2}{\,t\,}\hsm\) \!.
\end{align}
Similarly, for the other two double partial KK-amplitudes,
we derive the following:
\beqs
\begin{align}
\A^{}_{\rm{S}}[\al^n|\be^n]
&= \frac{1}{\,2\,} \sum_i  {\A^{}_{\rm{S}}}
[\al_{\fP_i}^{}|\be_{\fP_i}^{}]
= \lam^2\!\(\!
\frac{1}{~t\!-\hsm 4\Mn^2~}\hsm +\hsm \frac{2}{\,t\,}
\)\!,
\\[1mm]
\A^{}_{\rm{S}}[\be^n|\be^n]
&= \frac{1}{\,2\,} \sum_i  {\A^{}_{\rm{S}}}
[\be_{\fP_i}^{}|\be_{\fP_i}^{}]
=-\lam^2\!\(\!\frac{1}{\,t\!-\hsm 4\Mn^2}\!+\!
\frac{1}{\,u\!-\hsm 4\Mn^2\,} \!+\!\frac{2}{\,t\,}
\!+\!\frac{2}{\,u\,}\hsm\) \!,
\end{align}
\eeqs
where we have denoted
$[\alpha^n]\hsm =\hsm [1^n 2^n 3^n 4^n]$ and
$[\beta^n]\hsm =\hsm [1^n 3^n 2^n 4^n]\hs$.\

\vspace*{1mm}

From the above,
we can compute the four-point elastic KK BAS scattering amplitude
as follows:
\begin{align}
\A^{}_{\rm{S}}\big[\Phi_n^{aa'}\hsm\Phi_n^{bb'}\hsm
\Phi_n^{cc'}\hsm\Phi_n^{dd'}\big]
&=
\CC_s^{}\CC_s'\A^{}_{\rm{S}}[\al^n|\al^n]
- (\CC_s^{} \CC_u'+\CC_u^{} \CC_s')
\A^{}_{\rm{S}}[\al^n|\be^n] + \CC_u^{} \CC_u' \A^{}_{\rm{S}}[\be^n|\be^n]
\nn \\
\hspace*{-6mm}
&=
-\sum_{j}\hsm \lam^2\hs\CC_j^{}\hs\CC'_j \!\(\!\hsm
\frac{1}{~s_j^{} \!-\hsm 4\Mn^2~} +\frac{2\,}{\,s_j^{}\,}
\!\)  \!,
\label{eq:Amp-BAS-2}
\end{align}
where the color factors $\{\CC_j^{}\}$  ($\hs j\!=\!s,t,u\hs$)
are defined in Eq.\eqref{eq:Cj-def}, and
$\{\CC_j'\}$ are obtained from $\{\CC_j^{}\}$ by replacing the
hidden color indices $(a,b,c,d)\ito (a',b',c',d')\hs$.\
In the second line of Eq.\eqref{eq:Amp-BAS-2} we have used
a color identity
$\CC_s^{}\CC_u'+\CC_u^{}\CC_s'\hsm =
 \CC_t^{}\CC_t'-\CC_s^{}\CC_s'-\CC_u^{}\CC_u'$.\
Eq.\eqref{eq:Amp-BAS-2} shows that after simplification
the scattering amplitude of the KK bi-adjoint scalars
takes a BCJ-like form.
Inspecting the explicit formulas of the KK gauge boson amplitude
\eqref{eq:Amp-4AL-nnnn} and of the KK graviton amplitude
\eqref{eq:Mbar-4hL-nnnn} for the elastic scattering $\{n,n,n,n\}$,
we see that
apart from the group factors $\,\CC_j^{}$ (\hs$j\!=\!s,t,u\hs$)
and numerators $\NN_j^{\hs\fP}$ ($\fP\hsm\!=\!\fA,\fB,\fC$),
the summed KK amplitudes of Eq.\eqref{eq:Amp-4AL-nnnn} or
Eq.\eqref{eq:Mbar-4hL-nnnn} contain the massive kernels
$\,1/(s_j^{}\!-\!M_n^2)\,$ and massless kenels $\,1/s_j^{}\,$
in each channel of $(s,\hs t,\hs u)$, which match
the corresponding kernel of the above elastic KK BAS amplitude
\eqref{eq:Amp-BAS-2}, respectively.

\vspace*{1mm}

Next, we can further compute the inelastic KK BAS scattering
amplitude of $(0,0) \ito (n,n)\hs$.\
For this process, we derive the solution of the massive
scattering equation as follows:
\begin{equation}
\fP\hsm =\hsm\{0,0,\pm n,\mp n\}\!:~~~~
z_4^{}=-\frac{~u\hsm -\hsm\Mn^2~}{s} \,.
\end{equation}
Then, we derive the following color-ordered
inelastic double amplitudes:
\begin{align}
\label{eq:BASamp4-00nn}
\A_{\rm{S}}^{}[1^0 2^0 3^n 4^n|1^0 2^0 3^n 4^n]
& = -2\hs \lam^2\!\(\!\frac{1}{\,s\,} +\frac{1}{\,t\!-\!\Mn^2\,}\!\)
\hsm\equiv\hs
{\A^{}_{\rm{S}}}
[\al^{\hsm 0n}|\al^{\hsm 0n}] \hs,
\nn\\[1mm]
\A_{\rm{S}}^{}[1^0 2^0 3^n 4^n|1^0 3^0 2^n 4^n]
& = \frac{2\hs \lam^2}{~t\hsm -\!\Mn^2~}
\equiv\hs
{\A^{}_{\rm{S}}}
[\al^{\hsm 0n}|\be^{0n}] \hs,
\nn\\[1mm]
\A_{\rm{S}}^{}[1^0 3^0 2^n 4^n|1^0 3^0 2^n 4^n]
& = -2\hs \lam^2\!
\(\!\frac{1}{~t\!-\!\Mn^2~}+\frac{1}{~u\!-\!\Mn^2~}\!\)
\hsm\equiv\hs
{\A^{}_{\rm{S}}} [\be^{0n}|\be^{0n}] \hs.
\end{align}
With these, we can compute the massive KK BAS scattering amplitude
of the inelastic channel
$\Phi_0^{aa'}\hsm\Phi_0^{bb'}\!\ito
\Phi_n^{cc'}\hsm\Phi_n^{dd'}$ as follows:
\begin{align}
\A_{\rm{S}}^{}\hsm\big[\Phi_0^{aa'}\hsm\Phi_0^{bb'}\hsm
\Phi_n^{cc'}\hsm\Phi_n^{dd'}\big]
&=\CC_s^{}\CC_s'
{\A^{}_{\rm{S}}}
[\al^{\hsm 0n}|\al^{\hsm 0n}] \!-\!
(\CC_s^{}\CC_u'\!+\!\CC_u^{}\CC_s')
{\A^{}_{\rm{S}}}
[\al^{0n}|\be^{0n}] + \CC_u^{} \CC_u' {\A^{}_{\rm{S}}}
[\be^{0n}|\be^{0n}]
\nn\\[1mm]
&= -2\hs \lam^2\!\(\!\frac{\,\CC_s^{}\hs\CC_s'\,}{s}+
\frac{\CC_t^{}\hs\CC_t'}{\,t\!-\!\Mn^2\,}+
\frac{\CC_u^{}\hs\CC_u'}{\,u\!-\!\Mn^2~}\!\) \!,
\label{eq:AmpS-00nn}
\end{align}
where in the last step we have also applied the
color identity
$\CC_s^{}\CC_u'+\CC_u^{}\CC_s'\hsm =
 \CC_t^{}\CC_t'-\CC_s^{}\CC_s'-\CC_u^{}\CC_u'$.\
Inspecting our explicit formulas of the KK gauge boson amplitude
\eqref{eq:T00nn} and of the KK graviton amplitude \eqref{eq:M00nn-LO}
for the inelastic scattering $\{0,0,n,n\}$, we see that
apart from the group factors $\CC_j^{}$ and numerators
$\NN_j^{\fA}$, these KK amplitudes contain the kernels
$(1/s,\,1/(t\!-\!M_n^2),\,1/(u\!-\!M_n^2))$
in each channel of $(s,\hs t,\hs u)$, which equal
each kernel of the above KK BAS amplitude \eqref{eq:AmpS-00nn},
respectively.

\vspace*{1mm}

For the $N$-point ($N\!\hsmx\geqq\!5\hs$) KK bi-adjoint scalar amplitudes
of the BAS theory, it is much more complicated to directly calculate
the color-ordered scattering amplitude by \eqrefe{eq:CHY}.\
But according to the labeled tree method 
based on the Cayley functions\,\cite{Gao:2017dek},
the color-ordered BAS amplitude
$\mathcal A_{\rm{S}}^{}[\alpha|\beta]$
of massless BAS theory is given by the common part of
cubic trees respecting the ordering $\alpha$ and $\beta$ up to
a factor $(-1)^{(N-3)+\rm{flip}}$.\
For the 5d massless BAS theory, the five-point
BAS amplitudes have two linearly independent double amplitudes
with color-ordering $[\al ]\!=\![12345]$ and $[\be ]\!=\![13254]$:
\beqs
\begin{align}
\widehat{\A}_\rm{S}[12345|12345] &= \hlam^3\!\(\!
\frac{1}{\,\shat^{}_{12}\shat^{}_{34}\,}
+\frac{1}{\,\shat^{}_{34}\shat^{}_{15}\,}
+\frac{1}{\,\shat^{}_{15}\shat^{}_{23}\,}
+\frac{1}{\,\shat^{}_{23}\shat^{}_{45}\,}
+\frac{1}{\,\shat^{}_{45}\shat^{}_{12}\,} \!\) \!,
\\[1mm]
\widehat{\A}_\rm{S}[13254|13254] &= \hlam^3\!\(\!
\frac{1}{\,\shat_{13}^{}\shat_{25}^{}\,}
+ \frac{1}{\,\shat_{25}^{}\shat_{14}^{}\,}
+ \frac{1}{\,\shat_{14}^{}\shat_{23}^{}\,}
+ \frac{1}{\,\shat_{23}^{}\shat_{45}^{}\,}
+ \frac{1}{\,\shat_{45}^{}\shat_{13}^{}\,} \!\) \!,
\\[1mm]
\widehat{\A}_\rm{S}[12345|13254] &=
\frac{\hlam^3}{\,\shat^{}_{23}\shat^{}_{45}\,} \,.
\end{align}
\eeqs
From the above and using the shifting method established
in Section\,\ref{sec:2}, we can derive corresponding color-ordered
massive KK BAS amplitudes for the scattering process
$(\bn_1^{},\bn_2^{})\ito (\bn_3^{},\bn_4^{},\bn_5^{})$
with $\sum_{j=1}^{5}\hsm\bn_j^{}\!=\!0\hs$ under
5d compactification of $S^1$:
{\small
\beqs
\label{eq:BAS5pt}
\begin{align}
\label{eq:BAS5pt-12345}
\hspace*{-3mm}
\A_{\rm{S}}^{}[12345|12345] =\,& \lam^3\!\hsm\(\!\hsm
\frac{1}{\,(s^{}_{12}\!-\!M_{12}^2)(s^{}_{34}\!-\!M_{34}^2)}
+\frac{1}{\,(s^{}_{34}\!-\!M_{34}^2)(s^{}_{15}\!-\!M_{15}^2)\,}
+\frac{1}{\,(s^{}_{15}\!-\!M_{15}^2)(s^{}_{23}\!-\!M_{23}^2)\,}
\right.
\nn \\
& \left.
+\frac{1}{(s^{}_{23}\!-\!M_{23}^2)(s^{}_{45}\!-\!M_{45}^2)}
+\frac{1}{\,(s^{}_{45}\!-\!M_{45}^2)(s^{}_{12}\!-\!M_{12}^2)\,}
\!\)\!,
\\[1mm]
\label{eq:BAS5pt-13254}
\hspace*{-3mm}
\A_{\rm{S}}^{}[13254|13254] =\,& \lam^3\!\hsm\(\!\hsm
\frac{1}{\,(s^{}_{13}\!-\!M_{13}^2)(s^{}_{25}\!-\!M_{25}^2)}
+\frac{1}{\,(s^{}_{25}\!-\!M_{25}^2)(s^{}_{14}\!-\!M_{14}^2)\,}
+\frac{1}{\,(s^{}_{14}\!-\!M_{14}^2)(s^{}_{23}\!-\!M_{23}^2)\,}
\right.
\nn \\
& \left.
+\frac{1}{(s^{}_{23}\!-\!M_{23}^2)(s^{}_{45}\!-\!M_{45}^2)}
+\frac{1}{\,(s^{}_{45}\!-\!M_{45}^2)(s^{}_{13}\!-\!M_{13}^2)\,}
\!\)\!,
\\[1mm]
\hspace*{-3mm}
\A_{\rm{S}}^{}[12345|13254] =\,&
\frac{\lam^3}{\,(s^{}_{23}\!-\!M_{23}^2)(s^{}_{45}\!-\!M_{45}^2)\,}\,,
\end{align}
\eeqs
}
\hspace*{-3mm}
where for simplicity we have suppressed the labels of KK numbers
for the external KK states and have defined the mass formula
$\hs M_{ij}^{2}\!=\hsm (\bn_i^{}\hsm\!+\hsm\bn_j^{})^2\!/\hsm R^2\,$.

\vspace*{2mm}
\subsection{\hspace*{-2mm}CHY Construction of Massive
KK Gauge/Gravity Amplitudes}
\vspace*{2mm}
\label{sec:4.2}

In this subsection, we present the extended CHY construction of
the massive KK gauge boson amplitudes and KK graviton amplitudes
in the compactified 5d YM and GR theories.\
With these, we demonstrate the connection (equivalence) between 
the massive CHY construction of the KK graviton amplitudes
and that of the massive BCJ double-copy 
or the massive KLT relations at tree level.\

\vspace*{1mm}

According to the CHY formula \eqref{eq:CHY},
we express the integrand $\,\whI_N^{\rm{YM}}\,$
in the massless 5d YM theory and the integrand
$\,\whI_N^{\rm{GR}}\,$
in the massless 5d GR theory as follows:
\begin{equation}
\label{eq:I-YM-GR}
\whI_N^{\rm{YM}} =  \sum_{\{\al\}}\CC[\alpha]\PT^{}[\al]\hs
\Pf'(\widehat{\Psi}_{\hsm N}^{}) \hs, \qquad
\whI_N^{\rm{GR}} = \rm{ Pf}'(\widehat{\Psi}_{\hsm N}^{})\hs
\Pf'(\widehat{\Psi}_{\hsm N}^{})\hs,
\end{equation}
where $\rm{ Pf}'(\widehat{\Psi}_{\hsm N}^{})$ is the
reduced Pfaffian of a $2N \!\times\hsmx 2N$
anti-symmetric matrix consisting of four $N \!\times\! N$
building blocks:
\\[-3mm]
\begin{equation}
\rm{Pf}'(\widehat{\Psi}_{\hsm N}^{}) = \frac{~(-)^{i+j}~}{z_i^{} \!-\! z_j^{}}  \rm{Pf}(\widehat{\Psi}^{ij}_{N,ij}) \hs.
\end{equation}
The definition of Pfaffian $\Pf$ is given by
\begin{equation}
\Pf(X) \hs =\, \sum_{\al} \rm{sign}(\al)
X_{\al(1)\al(2)}^{}\hsm\cdots\hsm
X_{\al(N-1)\al(N)}^{} \hs,
\end{equation}
where $X$ is an arbitrary $N\!\times\!N$ matrix,
$\{\al\}$ denotes all the possible ways to decompose
{$\{1,\cdots\hsmx,N\}$} into pairs of indices, and
$\hs\rm{sign}(\al)\hs$ equals to $+1$ for even permutations
and equals $-1$ otherwise.\
In addition, $\widehat{\Psi}^{ij}_{Nij}$ is a
$2(N\!\!-\!\!1)\!\hsm\times\! 2(N\!\!-\!\!1)$ matrix,
so we can also express
$\Pf(\widehat{\Psi}^{ij}_{N,ij})\hsm\!=\!\!
\sqrt{\det(\widehat{\Psi}^{ij}_{N,ij})\,}\hs$.\
Under the 5d toroidal compactification, we derive the extended
anti-symmetric matrix  $\widehat{\Psi}^{}_{N}$
with four $N\!\times\!N$ blocks in the following form:
\\[-5mm]
\beqs \label{eq:psi-matrix}
\begin{alignat}{3}
\widehat{\Psi}_{\hsm N}^{} &=
\left(\!\begin{array}{cc}
A_{\hsm N}^{} & -C^{\rm{T}}_{\hsm N} \\[1mm]
C_{\hsm N}^{} & B_{\hsm N}^{}
\end{array}\!\right) \!,
\hspace*{10mm}
&&
(A_{\hsm N}^{})_{ab}^{} = \left\{\!\begin{aligned}
\frac{~p_a^{} \!\cdot p_b^{}\!+\!M_{ab}^2~}
{z_a^{} \!-\hsm z_b^{}}\hs,
& \hspace*{4mm} (a \neq b),
\\[1mm]
0 \hs, \hspace*{8.5mm}
& \hspace*{4mm}  (a=b),
\end{aligned}\right.
\\[1mm]
(B_{\hsm N}^{})_{ab}^{} &= \left\{\!\! \begin{aligned}
\frac{\zeta_a^{} \!\cdot\zeta_b^{}}
{~z_a^{} \!-\! z_b^{}~} \hs,
& \hspace*{4mm}  (a \neq b),
\\[1mm]
0 \hs, \hspace*{5.5mm}
& \hspace*{4mm}  (a=b),
\end{aligned}\right.  \hspace*{10mm}
&& (C_{\hsm N}^{})_{ab}^{} =
\left\{\! \begin{aligned}
\frac{\zeta_a^{} \!\cdot p_b^{}}
{~ z_a^{}\!-\hsm z_b^{}~} \hs,
\hspace*{4mm}
& \hspace*{6mm}  (a \neq b),
\\[0.5mm]
\,\sum_{\substack{c=1 \\ c \neq a}}^N\!
\frac{~\zeta_a^{}\!\cdot p_c^{}\,}
{~ z_a^{} \!-\hsm z_c^{}~} \hs,
& \hspace*{6mm} (a=b).
\end{aligned}\right.
\end{alignat}
\eeqs
In the formula of $A_N^{}$, the squared-mass
$\,M_{ab}^2\!=\hsm{\bn_a^{}\bn_b^{}}/\hsm{R^2}\hs$
corresponds to the product of discretized 5th components
of the 5-dimensional momenta as in Eq.\eqref{eq:Mab^2}.\
For the case of $N\!=\!4\hs$, we can explicitly solve the
KK scattering equations
and compute the integrands via \eqrefe{eq:CHY}, where
$(z_i^{},z_j^{},z_k^{})\!=\!(0,1,\infty)$
are fixed on the Riemann sphere.\
With these we can derive the four-point scattering amplitudes of
the KK gauge bosons and of the KK gravitons,
which agree with the results of Appendix\,\ref{App:C}
and Ref.\,\cite{KKString}.
\\[-3mm]


It is shown\,\cite{CHY} that in the massless
scalar, gauge and gravity theories
the reduced Pfaffian can be decomposed into a sum of
products of the PT factor and the kinematic factor:
\\[-5mm]
\begin{align}
\label{eq:Pf'=PT.N}
\rm{Pf}'(\Psi) &\,=\, \sum_{\al \hs\in S_{N-2}}\!\!\!
\PT[\alpha]\hs\NN[\alpha] \,.
\end{align}
\\[-4mm]
The factors $\NN[\al]$ denote the kinematic numerators
which are functions of the momenta and polarization vectors and
can be calculated by labeled tree method\,\cite{Gao:2017dek}.\
In the compactified massive KK theories,
each $\NN[\al]$ should be replaced by
${\NN}[\al,\hsm\{\bn_i^{}\}]$ which also depends on
KK indices (KK masses) of the external states, and can be
deduced from $\NN[\al]$ under the shifting
{$\hs s_{ij}^{}\hsm\ito s_{ij}^{}\hsm -\hsmx
(\bn_i^{}\!+\bn_j^{})^2\!/R^2$},
as shown in Eq.\eqref{eq:sij-pol-shift}.
\\[-3mm]


Thus, using Eqs.\eqref{eq:I-YM-GR}, \eqref{eq:Pf'=PT.N} and
\eqref{eq:Amp-BAS-1},
we can derive the following BCJ-type double-copy construction
of the KK graviton amplitude from
the product of KK gauge boson amplitudes including
a kernel of the KK BAS amplitude:
\beqs
\label{eq:CHY-DC1}
\begin{align}
\label{eq:CHY-DC1-A-BAS}
\A^{\rm{BAS}}_{N} &\,=\,
\sum_{\fP}\!\!\sum_{\{\al,\be\}} \!\hsm
\CC[\al]\hs \bar{\A}^{\hs\rm{BAS}}_{N}
\big([\al|\be],\hsmx\{\bn_i^{}\hsm\}\big)\hs
\CC[\beta] \hs,
\\
\label{eq:CHY-DC1-A-YM}
\TT^{\rm{YM}}_{N} &\,=\,
\sum_{\fP}\!\!\sum_{\{\al,\be\}}\!\hsm
\NN[\al,\hsmx\{\bn_i^{}\}]\hs\bar{\A}^{\hs\rm{BAS}}_{N}
\big([\al|\be],\hsmx\{\bn_i^{}\}\big)\hs\CC[\be] \hs,
\\
\label{eq:CHY-DC-GR}
\M^{\rm{GR}}_{N} &\,=\,
\sum_{\fP}\!\!\sum_{\{\al,\be\}}\!\hsm
\NN[\al,\hsmx\{\bn_i^{}\}] \hs\bar{\A}^{\hs\rm{BAS}}_{N}
\big([\al|\be],\hsmx\{\bn_i^{}\}\big)\hs
\NN[\be,\hsmx\{\bn_i^{}\}] \hs,
\end{align}
\eeqs
where $\{\al,\be\}\in S_{N-2}$ and for simplicity we have
suppressed the helicity indices and polarizations as well as
a conversion factor of the gauge-gravity coupling constants
in the last formula.\ The KK BAS kernel is defined as
\beq
\label{eq:BAS-kernel-PTPT}
\bar{\A}^{\hs\rm{BAS}}_{N}\big([\al|\be],\hsm\{\bn_i^{}\}\hsm\big)
= \int\!\!\td\mu_N^{}\hs\PT[\al]\hs\PT[\be]
\hs.
\eeq
In Eq.\eqref{eq:CHY-DC1},  the BAS kernel $\bar{\A}_N^{\hs\rm{BAS}}\hsm\big([\al|\be],\hsm\{\bn_i^{}\}\big)$
can be viewed as an $(N\!-\hsm 2)!\times\hsm (N\!-2)!$ matrix
and we denotes its inverse matrix as
$\KK\big([\alpha|\beta],\hsm\{\bn_i^{}\}\big)$.\
Then, inserting a unit matrix
$\hs I\hsm =\hsm
\sum_{\{\ga\}}\!
\bar{\A}_N^{\hs\rm{BAS}}\hsm\big([\al|\ga],\hsmx\{\bn_i^{}\}\big)
\KK\big([\ga|\be],\hsmx\{\bn_i^{}\}\big)
\hs$
into Eq.\eqref{eq:CHY-DC-GR}, we can derive the following
double-copy relation between the KK graviton amplitude
and the products of the KK gauge boson amplitudes:
\begin{equation}
\label{eq:CHY-DC-GR-BCJ}
\M^{\rm{GR}}_{N} \,=\,
\sum_{\fP}\!\!\!\sum_{\{\al,\be\}}\!\!
\TT^{\rm{YM}}_{N}\hsm\big[\al,\hsmx\{\bn_i^{}\}\big]
\hs\KK\big([\al|\be],\hsmx\{\bn_i^{}\}\hsm\big)\hs
\TT^{\rm{YM}}_{N}\hsm\big[\beta,\hsm\{\bn_i^{}\}\big] \hs,
\end{equation}
where the color-ordered KK gauge boson amplitude
$\TT^{\rm{YM}}_{N}\big[\alpha,\hsm\{\bn_i^{}\}\big]$ is given by\,\footnote{%
For notational simplicity, we will remove the extra superscripts 
``YM'' and ``GR'' on the amplitudes $\TT$ and $\M$ in the following 
text after Eq.\eqref{eq:A=N/K}.}
\begin{equation}
\label{eq:A=N/K}
\TT^{\rm{YM}}_{N}\big[\al,\hsmx\{\bn_i^{}\}\big] \hs =\,
\sum_{\{\ga\}}\NN\big[\ga,\hsmx\{\bn_i^{}\}]\hs
\bar{\A}^{\hs\rm{BAS}}_{N}\big([\ga|\al],\hsmx\{\bn_i^{}\}
\hsm\big)\hs.
\end{equation}
\\[-5.mm]
Comparing the above Eq.\eqref{eq:CHY-DC-GR-BCJ} with
Eq.\eqref{eq:KK-CHY-DC} in Section\,\ref{sec:4.0}
and keeping in mind the suppressed helicity indices
and polarizations mentioned below Eq.\eqref{eq:CHY-DC1},
we see that the KK double-copy relation
\eqref{eq:CHY-DC-GR-BCJ} just reproduces the
extended massive KLT-type relation \eqref{eq:KK-CHY-DC}
with the inverse matrix
$\KK\big([\alpha|\beta],\hsmx\{\bn_i^{}\}\hsm\big)$
of Eq.\eqref{eq:CHY-DC-GR-BCJ}
identified as the KLT kernel.\
In this way, the derivation of Eq.\eqref{eq:CHY-DC-GR-BCJ}
has proven {\it the interpretation of the kernel
$\,\KK\big([\alpha|\beta],\hsmx\{\bn_i^{}\}\hsm\big)$
on the right-hand side of Eq.\eqref{eq:KK-CHY-DC}
as the inverse of the color-ordered KK BAS amplitude
$\bar{\A}^{\hs\rm{BAS}}_{N}\big([\al|\be],\hsm\{\bn_i^{}\}\hsm\big)$
in Eq.\eqref{eq:CHY-DC1-A-BAS}.}\

\vspace*{1mm}

In the following, we consider the four-point KK scattering amplitudes
to demonstrate explicitly how
Eqs.\eqref{eq:CHY-DC1-A-BAS}-\eqref{eq:CHY-DC-GR}
are realized.\
From \eqref{eq:CHY-DC1-A-BAS}, we compute the four-point scattering
amplitude of KK bi-adjoint scalars:
\beqs
\label{eq:BAS-Amp4}
\begin{align}
\label{eq:BAS-Amp4a}
\A^{\rm{BAS}}_{4} &=
\big(\CC_s~\,\CC_u\big)\bar{\A}^{\hs\rm{BAS}}_{4}\!
\begin{pmatrix}
\mathcal C'_s   \\[1mm]
\mathcal C'_u	
\end{pmatrix}
= \,-\!\sum_{j}\!
\frac{\CC_j^{}\hs\CC'_j}{~s_j\hsm -\hsm M_{\bn\bn_j}^2~}\,,
\\[2mm]
\label{eq:BAS-Amp4b}
\bar{\A}^{\hs\rm{BAS}}_{4} &=
\begin{pmatrix}
- \frac{1}{\,s-M_{\bn\bn_s}^2\,}\hsm -
\hsm\frac{1}{\,t-M_{\bn\bn_t}^2\,}
~~&  -\frac{1}{\,t -M_{\bn\bn_t\,}^2}
\\[3mm]
-\frac{1}{\,t-M_{\bn\bn_t}^2\,} \!
~~&
-\frac{1}{\,t-M_{\bn\bn_t}^2\,}\hsm -\hsm
\frac{1}{\,u -M_{\bn\bn_u}^2\,}	
\!\end{pmatrix}\!,
\end{align}
\eeqs
where $\hs\CC_s \!=\!\CC [1234]\,$ and
$\,\mathcal C_u\!=\! -\CC [1324]\,$
according to Eq.\eqref{eq:C1234-C1324},
and the color factors $\{\CC_j'\}$ ($\hs j\!=\!s,t,u$)
are obtained from $\{\CC_j^{}\}$ by replacing the
hidden color indices
$(a,b,c,d)\hsm\ito\hsm (a'\hsm ,\hs b'\hsm,\hs c'\hsm ,\hs d')\hs$.\
Comparing the above Eq.\eqref{eq:BAS-Amp4} with Eq.\eqref{eq:T1234-T1243},
we find that the BAS amplitude kernel
$\hs\bar{\A}^{\hs\rm{BAS}}_{4}\!=\hsm \Theta\hs$.\
Hence, the BAS kernel $\hs\bar{\A}^{\hs\rm{BAS}}_{N}\hs$ should
have the minimal rank of $(N\!-\hsm 3)!\!=\!1\hs$
for $\hs N\!=\!4\hs$,
which leads to 
$\hs\det\!\big(\bar{\A}^{\hs\rm{BAS}}_{N}\big)\!=\!0\hs$
and thus the same mass spectral condition \eqref{eq:MassCond-4pt}.\
\\[-4mm]


There are correspondences between the kinematic numerators,
\begin{equation}
\label{eq:N1234-Nstu}
\NN[1234,\!\{\bn_i^{}\}] = \NN_s\,,\qquad
\NN[1324,\!\{\bn_i^{}\}] = -\NN_u \,,
\end{equation}
which are analogous to the definition of their color counterparts
in \eqrefe{eq:C1234-C1324}.\
Using the relations \eqref{eq:N1234-Nstu} and Eq.\eqref{eq:CHY-DC1-A-YM},
we derive the BCJ-type four-point massive KK gauge boson
scattering amplitude as follows:
\begin{equation}
\label{eq:T4-BAS-BCJ}
\TT_4^{}\hs =\hs 
\big(\mathcal N_s~\,\mathcal N_u\big)
\bar{\A}^{\hs\rm{BAS}}_{4}\!
\begin{pmatrix}
\mathcal C'_s   \\[1mm]
\mathcal C'_u	
\end{pmatrix}
= -\!\sum_{j} \!\frac{\NN_j^{}\hsx\CC'_j\,}
{~s_j^{}\hsm -\hsm M_{\bn\bn_j}^2~}\,,
\end{equation}
where we have suppressed an overall coefficient
including the gauge coupling factor as indicated
below Eq.\eqref{eq:CHY-DC1}.\
In Eq.\eqref{eq:T4-BAS-BCJ} and hereafter, for notational simplicity
we have removed the superscript ``YM'' on the color-ordered 
KK gauge boson amplitude \eqref{eq:A=N/K}.\
Then, substituting Eq.\eqref{eq:A=N/K} into Eq.\eqref{eq:CHY-DC1-A-YM},
we can express the $N$-point color-dressed KK gauge boson amplitude
in terms of the color-ordered KK gauge boson amplitudes:
\begin{equation}
\TT_4^{}\hsm\big[\{\bn_i^{}\}\big] 
=\,
\sum_{\{\alpha\}}\CC [\al]
\TT^{}_{N}\hsm\big[\al,\hsmx\{\bn_i^{}\}\big] ]\,.
\end{equation}
For the $N\!=\!4\hs$ case, we derive the following explicit
relations to decompose the four-point color-dressed KK gauge boson
amplitude into the color-ordered KK amplitudes:
%
\begin{equation}
\TT_4^{}\hsm\big[\{\bn_i^{}\}\big]
=\hs \CC_s\hs\TT^{}_{4}\big[1234,\hsm\{\bn_i^{}\}\big]
-\CC_u\hs\TT^{}_{4}\big[1324,\hsm\{\bn_i^{}\}\big] \hs ,
\end{equation}
where each color-ordered KK gauge boson amplitude is
gauge-invariant separately.\
Next, with Eq.\eqref{eq:N1234-Nstu}
we substitute the KK BAS kernel matrix \eqref{eq:BAS-Amp4b}
into Eq.\eqref{eq:CHY-DC-GR} and derive the following BCJ-type
double-copy construction of the four-point KK graviton
scattering amplitudes:
\begin{equation}
\M_4^{}\big[\{\bn_i\}\big]
=\hs\big(\NN_s~\, \NN_u\big)\hs \bar{\A}^{\hs\rm{BAS}}_{4}\!
\begin{pmatrix}
\NN_s   \\[1mm]
\NN_u	
\hsm\end{pmatrix}
=\,-\!\sum_{j} \!\frac{~\NN_j^{}(\lam)\NN_j^{}(\lam')~}
{~s_j\hsm -\hsm M_{\bn\bn_j}^2~}\,,
\end{equation}
where $\hs j\!\in\!\{s,t,u\}\hs$ and we have suppressed the overall
coupling coefficient as well as possible summations of the
external KK gauge boson polarizations
[with helicity indices indicated by $(\lam,\lam')$]
on the right-hand side according to the definition of
the KK graviton polarization tensors
\eqref{eq:zeta-munu-4d}.\
\\[-4mm]


We further note that the kernel in \eqrefe{eq:CHY-DC-GR-BCJ} is a
$(N\!-\hsm 2)!\hsmx\times\!(N\!-\hsm 2)!$ matrix,
while the kernel in \eqrefe{eq:KK-CHY-DC} is a
$(N\hsm\!-\hsmx 3)! \hsm\times\! (N\! -\hsm 3)!\hs$ matrix.\
Thus, we will eliminate the redundant variables
by making use of the extended fundamental BCJ relations
for KK gauge boson amplitudes.\
As an example, according to \eqrefe{eq:CHY-DC-GR}, the four-point
KK graviton scattering amplitude can be written as follows:
\begin{align}
\label{eq:Ex-4pt}
\M_4^{}\big[\{\bn_i\}\big]
&=\hs
\NN[1234,\!\{\bn_i^{}\}]\hs
\TT_4^{} [1234,\!\{\bn_i^{}\}]
+\NN[1324,\!\{\bn_i^{}\}]\hs
\TT_4^{} [1324,\!\{\bn_i^{}\}]
\nn \\[1mm]
&=\(\!\NN[1234,\!\{\bn_i^{}\}]+
\frac{\,s\!-\!M_{\bn\bn_s}^2\,}{\,u\!-\!M_{\bn\bn_u}^2\,}\hs
\NN[1324,\!\{\bn_i^{}\}]\!\)\!
\TT_4^{} [1234,\!\{\bn_i^{}\}]
\nn\\[1mm]
&=\hs (\hs t\!-\!M_{\bn\bn_t}^2) \hs
\TT_4^{} [1234,\!\{\bn_i^{}\}]\hs 
\TT_4^{} [1324,\!\{\bn_i^{}\}] \,,
\end{align}
where we have denoted the color-ordered KK gauge boson amplitudes
$\hs{\TT_4^{}} [1^{\bn_1}_{\lam_1} 2^{\bn_2}_{\lam_2}
 3^{\bn_3}_{\lam_3} 4^{\bn_4}_{\lam_4}]
 \hsm\equiv\hsm{\TT_4^{}}[1234,\!\{\bn_i\}]\,$ and
$\hs{\TT_4^{}} [1^{\bn_1}_{\lam_1} 3^{\bn_3}_{\lam_3}
 2^{\bn_2}_{\lam_2} 4^{\bn_4}_{\lam_4}]
\hsm\equiv\hsm{\TT_4^{}} [1324,\!\{\bn_i\}]\hs$, 
and have suppressed an overall factor 
$\hs\ka^2/16\hs$ as well as possible summations of the
external KK gauge boson polarizations on the right-hand side
according to the definition of the KK graviton polarization tensors
\eqref{eq:zeta-munu-4d}.\
In the second line of \eqrefe{eq:Ex-4pt},
we have used the extended fundamental BCJ relation
\eqref{eq:fBCJ-relation} and the
correspondences of kinematic numerators in Eq.\eqref{eq:N1234-Nstu}.\
%
%
The above formula \eqref{eq:Ex-4pt}
gives an extended KLT-like relation between the four-point KK gaviton
scattering amplitude and product of the two corresponding color-ordered
KK gauge boson amplitudes.\
Moreover, we note that the kernel
$\hs t\!-\!M_{\bn\bn_t}^2$
is given by the inverse of massive double color-ordered amplitude
of the KK bi-adjoint scalars:
\begin{equation}
\A_{\rm S}[1234|1324;\!\{\bn_i^{}\}]=\frac{1}{~t-M_{\bn\bn_t}^2~}\,,
\end{equation}
which can be derived based on the analysis in Sec.\,\ref{sec:4.1}
or given by the off-diagonal element $\{12\}$ of the
$\hs\Theta\hs$ matrix in \eqrefe{eq:T1234-T1243}.\
Then, comparing Eq.\eqref{eq:Ex-4pt} with Eq.\eqref{eq:Mhbar-4pt},
we see that Eq.\eqref{eq:Ex-4pt} gives the same form
as the four-point massive KLT relation \eqref{eq:Mhbar-4pt}
except that Eq.\eqref{eq:Mhbar-4pt} has made a special choice of
the external KK states to have longitudinal polarizations and
Eq.\eqref{eq:Ex-4pt} has suppressed an overall factor and
the possible summations of the external KK gauge boson polarizations on
its right-hand side [as mentioned below Eq.\eqref{eq:Ex-4pt}].\

\vspace*{1mm}

According to the above massive KK CHY formulation,
the KK graviton scattering amplitudes \eqref{eq:KK-CHY-DC}
can be constructed by inputting the kernel
$\KK\!\([\al|\be],\!\{\bn_i^{}\hsm\}\hsm\)$
(as provided by the inverse of the KK BAS amplitudes given by
Sec.\,\ref{sec:4.1}) and the color-ordered KK gauge boson
amplitudes {$\TT_N\big[\al,\hsm\{\bn_i^{}\hsm\}\big]$} 
[as further expressed in Eq.\eqref{eq:A=N/K}].\
For the four-point scattering, there is one independent
color-ordering of the KK gauge boson amplitudes after applying
the massive fundamental BCJ relation \eqref{eq:fBCJ-relation}.\
For instance, we may choose the color-ordering [1234] and
derive the corresponding massive KK gauge boson amplitude
as follows:
%
\begin{align}
&{\TT_4} \big[1^{\bn_1}_{\lam_1}2^{\bn_2}_{\lam_2}3^{\bn_3}_{\lam_3}
4^{\bn_4}_{\lam_4}\big]
\hsm \hs =\hs g^2\!\(\!
\frac{\over{\NN}^{}_1}{~s_{12}^{}\!-\!M_{12}^{2}~}
+\frac{\over{\NN}^{}_2}{~s_{23}^{}\!-\!M_{23}^{2}~}
\!\)\!,
\end{align}
where $\hs s_{12}^{}\!=\!-(p_1^{}\!+\hsm p_2^{})^2\!=\! s\,$
and $\hs s_{23}^{}\!=\!\hsm -(p_2^{}\!+\hsm p_3^{})^2\hsm\!=\! t\hs$.\
The above color-ordered numerators
$(\over{\NN}^{}_1,\hs\over{\NN}^{}_2)$
are expressed as
%
{
\beqs
\begin{align}
\over{\NN}^{}_1\! =&
-2\hs (s_{23}^{}\!-\!M^2_{23})[12][34]
\!-\hsm 4\hs\Big\{\!
(12)(13)[14]\!+\hsm (12)(23)[14]\!-\hsm (12)(31)[34]
\!+\hsm (12)(34)[13]
\nn\\
& -\hsm (13)(21)[24]\!+\hsm (13)(24)[12]
\!-\hsm (21)(23)[24]\!+\hsm (21)(32)[34]
\!-\hsm (21)(34)[23]
\nn\\
& +\hsm (23)(24)[12]\!+\hsm (23)(34)[12]\Big\} \hs,
\\ 
\over{\NN}^{}_2\! =&\,
2\hs (s_{13}^{}\!-\!M^2_{13})[14][23]
\!-\hsm 2\hs (s_{23}^{}\!-\!M^2_{23})
\Big\{\![12][34]\!-\![13][24] \!\Big\}
\!-\hsm 4\hs\Big\{\!(12)(23)[14]\!-\hsm (13)(32)[14]
\nn\\
& -\hsm (21)(23)[24]\!+\hsm (21)(32)[34]\!-\hsm (21)(34)[23]
\!+\hsm (23)(24)[12]\!-\hsm (23)(31)[24]\!+\hsm (23)(34)[12]
\nn\\
& +\hsm (24)(31)[23]\!-\hsm (24)(32)[13]
\!+\hsm (32)(31)[34]\!-\hsm (32)(34)[13] \hsm\Big\}\hs.
\end{align}
\eeqs
}
\hspace*{-3mm}
In the above we have defined abbreviations
$(ij)\hsm\equiv\hsm (p_i^{}\hsm\cdot\zeta_j^{})$,
$[ij]\hsm\equiv\hsm (\zeta_i^{}\hsm\cdot\zeta_j^{})\hs$, and
$\,\zeta_j^\mu \!\equiv\! \zeta_{\lam_j}^\mu$
to simplify the notations.\

\vspace*{1mm}

As another explicit example,
we take the five-point KK gauge boson amplitude
to show how the above massive KK CHY formulation works.\
The five-point gauge boson amplitudes have
{$\hs (5\hsm -3)!\hsm =\hsm 2\,$}
independent color-orderings, which may be chosen as
$[12345]$ and $[13254]$.\
The five-point KK BAS amplitudes are studied
in Section\,\ref{sec:4.1}.\
We deduced the extended BCJ-type five-point KK gauge boson
amplitude in Section\,\ref{sec:3.4}.\
The five-point color-ordered massless
KK gauge boson amplitude can be calculated by the CHY formula.\
Then, we use the shifting method of Section\,\ref{sec:2.2}
to derive the corresponding five-point massive KK amplitude
as follows:
{\small
\beqs
\label{eq:T12345-13254}
\begin{align}
\label{eq:T12345-sub}
\hspace*{-3mm}
&{\TT_5^{}} \big[1^{\bn_1}_{\lam_1}2^{\bn_2}_{\lam_2}3^{\bn_3}_{\lam_3}
4^{\bn_4}_{\lam_4}5^{\bn_5}_{\lam_5}\big]
\hsm = g^3\!\left\{\!
\frac{\over{\NN}^{}_{\hsmx 1}}{\,(s_{12}^{}\hsm\!-\hsm\!M_{12}^{2})
(s_{34}^{}\hsm\!-\hsm\!M_{34}^{2})\,}
+\frac{\over{\NN}^{}_{\hsmx 2}}{\,(s_{34}^{}\hsm\!-\hsm\!M_{34}^{2})
(s_{15}^{}\hsm\!-\hsm\!M_{15}^{2})\,}
\right.
\nn\\
&\hspace*{10mm}\left.
+\frac{\over{\NN}^{}_{\hsmx 3}}{\,(s_{15}^{}\hsm\!-\hsm\!M_{15}^{2})
(s_{23}^{}\hsm\!-\hsm\!M_{23}^{2})\,}
+\frac{\over{\NN}^{}_{\hsmx 4}}{\,(s_{23}^{}\hsm\!-\hsm\!M_{23}^{2})
(s_{45}^{}\hsm\!-\hsm\!M_{45}^{2})\,}
+\frac{\over{\NN}^{}_{\hsmx 5}}{\,(s_{45}^{}\hsm\!-\hsm\!M_{45}^{2})
(s_{12}^{}\hsm\!-\hsm\!M_{12}^{2})\,}\!\right\} \!,
\\[1mm]
\label{eq:T3254-sub}
&\TT_5^{}\big[1^{\bn_1}_{\lam_1}3^{\bn_3}_{\lam_3}2^{\bn_2}_{\lam_2}
 5^{\bn_5}_{\lam_5}4^{\bn_4}_{\lam_4}\big]=g^3 \bigg\{\frac{\over{\NN}^{\pp}_{\hsmx 1}}{\,(s^{}_{13}\!-\!M_{13}^2)(s^{}_{25}\!-\!M_{25}^2)}
+\frac{\over{\NN}^{\pp}_{\hsmx 2}}{\,(s^{}_{25}\!-\!M_{25}^2)(s^{}_{14}\!-\!M_{14}^2)\,}
\nn\\
&\hspace*{10mm}+\frac{\over{\NN}^{\pp}_{\hsmx 3}}{\,(s^{}_{14}\!-\!M_{14}^2)(s^{}_{23}\!-\!M_{23}^2)\,}
+\frac{\over{\NN}^{\pp}_{\hsmx 4}}{(s^{}_{23}\!-\!M_{23}^2)(s^{}_{45}\!-\!M_{45}^2)}
+\frac{\over{\NN}^{\pp}_{\hsmx 5}}
{\,(s^{}_{45}\!-\!M_{45}^2)(s^{}_{13}\!-\!M_{13}^2)} \bigg\} \,.
\end{align}
\eeqs
}
\hspace*{-3mm}
We note that in the above color-ordered KK gauge amplitudes
contain the product of the numerator
$\over{\NN}^{}_{\hsmx {\ell}}$ 
(or $\over{\NN}^{\pp}_{\hsmx {\ell}}\hs$)
and the fraction $\,1/[(\cdots)(\cdots)]\,$ in each term,
where the fraction $\,1/[(\cdots)(\cdots)]\,$ just equals
the corresponding BAS kernel amplitude given in
Eqs.\eqref{eq:BAS5pt-12345} and \eqref{eq:BAS5pt-13254}.\
Moreover, we can derive each numerator
$\over{\NN}_{\hsmx {\ell}}^{}\,$ or
$\hs\over{\NN}^{\pp}_{\hsmx {\ell}}\hs$
(${\ell} \!=\!1,2,\cdots\!,5$)
in \eqrefe{eq:T12345-13254} by shifting the massless numerators
(computed previously\,\cite{CHY}),
$\,\over{\NN}_{\hsmx {\ell}}^{}\!=\hsmx
{\over{\NN}_{\hsmx \ell}^{\,\bn_j= 0}
\big|}_{\hat{s}_{ij}\to\hs  s_{ij}^{}\hsm -M_{ij}^2}^{}\hs$.\
For instance, we derive the shifted massive numerator
$\over{\NN}_{\hsmx 1}^{}$ as follows:
{\small
\begin{align}
\label{eq:N1-12345-sub}
\hspace*{-3mm}
\over{\NN}_{\hsmx 1}^{}=&\,
2\hs\hsm (s_{23}^{}\!-\!M_{23}^2)[12]
\Big\{\hsm [34](45)\!+\hsmx [35](14)\!+\hsmx [35](24)\!+\hsmx
 [35](34)\!-\hsmx [45](43)\!\Big\}
\nn\\
&+\hsm 2\hs (s_{24}^{}\!-\!M_{24}^2)[12][45]
\Big\{\!(13)\!+\hsm (23)\!\Big\}
\!+\hsm 2\hs (s_{34}^{}\!-\!M_{34}^2)[45]\Big\{\![12](23)+[13](12)
\!-\![23](21)\!\Big\}
\nn\\
&
+\hsm 4\hs\Big\{\!
(12)(13)\hsm\Big[\hsm (45)[14]\!-\hsm (41)[45]\!-\hsm (54)[15]\Big]
\hsm\!+\hsm (12)(14)\hsm
\Big[\hsm (23)[15]\!-\hsm (31)[35]\!+\hsm (35)[13]\hsm\Big]
\nn\\
&
+\hsm (12)(23)\hsm
\Big[\hsm (24)[15]\!+\hsm (34)[15]\!-\hsm (41)[45]
    \!+\hsm (45)[14]\Big]
\!-\hsm (12)(24)\hsm\Big[\hsm (31)[35]\!-\hsm (35)[13]\Big]
\nn\\
&
-\hsm (12)(31)\hsm\Big[\hsm (34)[35]\!-\hsm (43)[45]\!+\hsm (45)[34]\Big]
\!+\hsm (12)(34)\hsm\Big[(35)[13]\!+\hsm (45)[13]\Big]
\\
&
-\hsm (13)(14)\hsm\Big[(21)[25]\!-\hsm (25)[12]\Big]
\!-\hsm (13)(21)\hsm\Big[\hsm (24)[25]\!+\hsm (34)[25]
\!-\hsm (42)[45]\!+\hsm (45)[24]\Big]
\nn\\
&
+\hsm (13)(24)\hsm\Big[\hsm (25)[12]\!+\hsm (45)[12]\Big]\!\hsm
+\! (13)(25)(34)[12]\!
-\hsm (14)(21)\hsm
\Big[\hsm (23)[25]\!-\hsm (32)[35]\!+\hsm (35)[23]\hsm\Big]
\nn\\
&
+\hsm (14)(23)[12]\hsm\Big[\hsm (25)\hsm +\hsm (35)\Big]
\!-\hsm (21)(23)\hsm\Big[\hsm (24)[25]\hsm +\hsm (34)[25]
 \hsm -\hsm (42)[45]\hsm +\hsm (45)[24]\Big]
\nn\\
&
+\hsm (21)(24)\hsm\Big[(32)[35]\hsm -\hsm (35)[23]\Big]\hsm
\!+\hsm (21)(32)\Big[(34)[35]\!-\hsm (43)[45]\!+\hsm (45)[34]\Big]
\nn\\
&
-\hsm (21)(34)[23]\hsm\Big[\hsmx (35)\hsmx +\hsmx (45)\Big]\hsm\!
-\hsm (23)[12]\hsm\Big[\hsm (24)(15)\hsmx -\hsmx (25)(34)\Big]
\hsm\!+\hsm (23)(34)[12]\hsm
\Big[\hsmx (35)\hsmx +\hsmx (45)\hsm\Big]\Big\},
\nn
\end{align}
}
\hspace*{-3mm}
where 
$(ij)\!\equiv\!(p_i^{}\hsmx\cdot\hsm\zeta_j^{})$,
$[ij]\!\equiv\!(\zeta_i^{}\hsmx\cdot\zeta_j^{})$, and
$\,\zeta_j^\mu \!\equiv\! \zeta_{\lam_j}^\mu$.\
The other massive numerators 
$\hs\over{\NN}_{\hsmx {\ell}}^{}\hs$ and
$\hs\over{\NN}^{\pp}_{\hsmx {\ell}}\hs$
can be derived similarly.\
Thus, to construct the five-point KK graviton scattering amplitudes
from the extended massive CHY formula \eqref{eq:KK-CHY-DC}
[or, Eq.\eqref{eq:CHY-DC-GR-BCJ}],
we can use the above color-ordered massive KK gauge boson amplitudes
and express the kernel of \eqrefe{eq:KK-CHY-DC}
in terms of the inverse of the KK bi-adjoint scalar amplitudes
\eqref{eq:BAS5pt}.

\vspace*{2mm}
\section{\hspace*{-2mm}Determining KK Structure from Mass Spectral Condition}
\label{sec:5}
\vspace*{1.5mm}

We proved in Section\,\ref{sec:3.1} that
under the toroidal compactification (without orbifold)
the four-point scattering amplitudes of KK gauge bosons
and KK gravitons satisfy a mass spectral condition
\eqref{eq:MSCond-4KK} [or \eqref{eq:MassCond-4pt}].\
We may rewrite it in the following generic form:
\begin{equation}
\label{eq:mass-spectrum}
M_1^2+M_2^2+M_3^2+M_4^2 \,=\, M_{12}^2+M_{13}^2+M_{14}^2\,,
\end{equation}
where $M_j^{}$ 
denotes the mass of each external state
and $(M_{12}^2,M_{14}^2,M_{13}^2)$ equal respectively the
squared pole-mass in each of the $(s,t,u)$ channels.\
Due to the momentum conservation for the four-particle scattering,
we also have relations
$M_{12}^2\!=\!M_{34}^2\hs$,
$M_{13}^2\!=\!M_{24}^2$\hs, and
$M_{14}^2\!=\!M_{23}^2\hs$.\
It is natural to ask: whether there exist any different theories
(other than such compactified KK theories) which could satisfy
this mass spectral condition?

\vspace*{1mm}

We inspect Eq.\eqref{eq:mass-spectrum} and find that it
implies three {\it necessary} conditions.\
The first one is that the massive theory should contain
at least two types of particles with unequal masses.\
This is because for a theory having only
one type of particles with the same mass $\hs M\hs$,
the two sides of Eq.\eqref{eq:mass-spectrum} would become
$\,4\hs M^2\!\neq\! 3\hs M^2\hs$
(based on the cubic interaction vertex to be defined below).\
The second condition is that there exists only a simple pole
in each of the $(s,t,u)$ channels.\
The third condition is that the scattering amplitude should
include the contributions from all three
kinematic channels of $(s,t,u)$.\
The second condition means that, for a generic cubic vertex
$V_{\textbf{abc}}^{}$ involving three particles
$(\ax,\bx,\cx)$,
the mass $\hs\mc\hs$ of the third particle $\cx$
is uniquely determined once we specify the particles
$\hs\ax\hs$ and $\hs\bx\hs$ (having masses $\hs\ma\hs$
and $\mb$ respectively).\
In a given theory, we denote the particles of type
${\bf a}_j^{}$ if they all have the same mass
$\hs M_{\textbf{a}_j}^{}$ and same spin $\textsf{s}_{\ax_j}^{}\hsm$.\
Each type of particles (say, type ${\bf a}_j^{}$)
can be viewed as a set of states (having the same mass and spin)
in Hilbert space $\mathcal{H}\hs$.\ For a given cubic vertex
$V_{\textbf{abc}}^{}$, once the states $\ax$ and $\bx$ are chosen,
then the state $\cx$ in the vertex $V_{\textbf{abc}}^{}$ is also fixed
due to conservations of the momentum and angular momentum.\
Hence, if a given four-point scattering amplitude
has single-pole structure in each kinematic channel,
we can always realize such an
amplitude by using the cubic vertex $V_{\textbf{abc}}^{}$
where once the states $\ax$ and $\bx$ are chosen as the
external states, then the third state $\cx$ is uniquely determined.\
Applying such cubic interactions of $V_{\textbf{abc}}^{}$
to the states in the Hilbert space,
we have the following operation $\mathcal{F}\,$:
%
\beq
\label{eq:V3-operation}
\mathcal{F}(\ax,\bx ) ~\mapsto~ \cx\,,
\eeq
%
where $\,\ax,\bx,\cx\!\in\! \mathcal{H}\,$.\
%
For instance, in the 4d massless QCD the cubic gluon vertex provides
the simplest example of the operation \eqref{eq:V3-operation},
where the three gluons in the cubic vertex have different color indices
due to the anti-symmetric structure constant $f^{abc}$ of the gauge
group.\ In this case, the four-point massless gluon
scattering amplitudes obey the  mass-spectral condition
\eqref{eq:mass-spectrum} trivially.\footnote{%
In the Yang-Mills type of non-Abelian gauge theories, there also exist
quartic gauge interaction vertices, which contribute to the four-point
contact diagrams.\ But the contributions of such contact diagrams can
be ``blown up'' into the $(s,t,u)$ channel pole diagrams (induced by
cubic vertices)\,\cite{BCJ:2008}\cite{BCJ:2019}.\
Hence, for our current analysis only the cubic
interactions (denoted by $V_{\textbf{abc}}^{}$) need to be concerned.\
From the BCJ-type double-copy constructions, the GR type of four-point
graviton amplitudes will exhibit similar structures with the
$(s,t,u)$ channel pole diagrams (induced by cubic gravitational vertices).\
It is known that all the nonlinear gravitational self-interactions
can be induced precisely as having
cubic graviton interactions\,\cite{Deser}.}

\vspace*{1mm}

Then, we denote a set of particles (having the same mass $M_j^{}$
and joining the same type of cubic interactions
$V_{\textbf{abc}}^{}\hs$) by $\hs\mathfrak{M}_{j}^{}\hs$.\
All the particles in the given theory can form a set denoted as
$\mathbb{P}$ which is divided into the smaller sub-sets
in terms of their masses and types of interactions:
\begin{equation}
\mathbb{P} \,=\,
\mathfrak{M}_0^{} \cup
\mathfrak{M}_1^{} \cup\mathfrak{M}_2^{} \cup \cdots \,.
\end{equation}
Then we can define a new group
$\mathcal{G}=\{\mathfrak{M}_0^{},
\mathfrak{M}_1,\mathfrak{M}_2,\cdots\}$,
where each sub-set $\hs\mathfrak{M}_{j}^{}\hs$
serves as a group element.\
We may use $\mathfrak{M}_0^{}$ to denote the set
collecting the massless particles such as the massless gravitons
or gauge bosons, which can be further divided into several
subsets $\mathfrak{M}_{0\texttt{s}}^{}$
according to the different spin-$\texttt{s}$
of each type of the massless particles.\
Formally, we can define a multiplicative operation
$\hs *\hs$ for the elements of $\mathcal{G}\hs$:
%
\beq
\label{eq:Mi*Mj=Mk}
\mathfrak{M}_{ij}^{} \equiv \mathfrak{M}_i^{} *
\mathfrak{M}_j^{} ~\mapsto~ \mathfrak{M}_k^{}\,,
\eeq
%
where $\hs\mathfrak{M}_i^{},\mathfrak{M}_j^{},\mathfrak{M}_k^{}
\!\in \mathcal{G}\hs$.\
This means that for any states $\ax\!\in\!\mathfrak{M}_i^{}$
and $\bx\!\in\!\mathfrak{M}_j^{}$, the above multiplicative operation
\eqref{eq:Mi*Mj=Mk}
is realized by the cubic interaction vertex
$V_{\textbf{abc}}^{}$, namely,
$\,\ax * \bx \mapsto \cx\,$ with
$\hs\cx\hsm\in\hsm\mathfrak{M}_k^{}\hs$.\
The crossing symmetry of the scattering amplitudes guarantees
$\,\mathfrak{M}_{ij}^{}\!=\mathfrak{M}_{ji}^{}\,$,
which realizes the commutative law of multiplication.\
We also note that all the particles in a given set
$\hs\mathfrak{M}_i^{} \!\in\! \mathcal{G}\hs$
share the same mass $\hs m_i^{}\hs$,
but different types of particles in different sets may
have same masses.\ For instance, the KK gluons and KK gravitons
of the same KK-level $n$ can belong to two different sets of
$\hs\mathfrak{M}_{n_i}^{}$ and $\hs\mathfrak{M}_{n_j}^{}$.

\vspace*{1mm}

The second hidden property implied by the mass-spectral condition
\eqref{eq:mass-spectrum} is the existence of all the three
kinematic channels $(s,t,u)$.\ This ensures the following
associative law of multiplication for the group $\mathcal{G}\hs$:
\begin{equation}
\label{eq:Mi*Mj*Mk}
(\mathfrak{M}_i^{} * \mathfrak{M}_j^{})*\mathfrak{M}_k^{}
=(\mathfrak{M}_k^{}*\mathfrak{M}_i^{})*\mathfrak{M}_j^{}
=(\mathfrak{M}_j^{}*\mathfrak{M}_k^{})*\mathfrak{M}_i^{}\,.
\end{equation}
Hence the multiplication $\hs *\hs$
is an associative binary operation on the group $\mathcal{G}\hs$.

\vspace*{1mm}

We note that the massless gravitons can form a set
(as denoted by $\mathfrak{M}_{0h}^{}\!=\!\{h^{\mn}\}$)
and deserve a special attention.\
In general, any given QFT model can naturally contain
massless gravitons or be combined with the GR (in the
effective field theory formulation\,\cite{EFT})
because gravitation is universal and the
massless gravitons of the GR interact with all partices living
in the spacetime.\
For the generic graviton-matter coupling via
$h_{\mn}^{}T_i^{\mn}$ with $T^{\mn}$ being the
energy-momentum tensor, it always conatins the cubic vertex
$\,h_{\mn}^{}$-$\hs\phi_i^{}\hs$-$\hs\phi_i^{}\,$,
where $\phi_i^{}$ denotes any matter field in the theory.\
Consider, for instance, the scattering amplitude of
$\hs h\,\phi_1^{} \hsm\ito \phi_2^{}\hs\phi_3^{}\,$
at tree level, where the intermediate particle can only be one of
$(\phi_1^{},\hs \phi_2^{},\hs \phi_3^{})$ for each channel of
$(s,t,u)$.\
As we can readily check, this scattering amplitude does obey
the mass spectral condition \eqref{eq:mass-spectrum}:\
$\,0+ M_1^2\hsm +\hsm M_2^2\hsm +\hsm M_3^2\hsm
 =\!M_1^2\!+\!M_2^2\!+\!M_3^2\hs$.\
Moreover, it is clear that the tree-level
four-point massless graviton scattering amplitude
$\M[h_{\si_1}^{} h_{\si_2}^{} h_{\si_3}^{}h_{\si_4}^{}]$
also satisfies the condition \eqref{eq:mass-spectrum}
because all the external and internal graviton states are massless.\
Then, we consider a more general theory containing additional
massive gravitons and massive gauge bosons.\ 
Hence, for the set
$\hs\mathfrak{M}_{0h}^{}\hsm\!=\!\{h^{\mn}\}$
of massless gravitons,
we can use the cubic vertex
$\,V_3^{}(h^{\mn}\hsm\phi_i^{}\phi_i^{})\,$
to define the multiplicative operations on the group $\mathcal{G}\,$:
\\[-6mm]
\beqs
\label{eq:M0-Mi-antiMi}
\begin{align}
\label{eq:M0*Mi=Mi}
\mathfrak{M}_{0h}^{}* \hs\mathfrak{M}_i^{}
&\hs =\, \mathfrak{M}_i^{}\,,
\\
\label{eq:Mi*antiM=M0}
\mathfrak{M}_i^{}* \hs\over{\mathfrak{M}}_i^{}
&\hs =\, \mathfrak{M}_{0h}^{} \,,
\end{align}
\eeqs
\\[-6mm]
where in the second equation
$\hs\over{\mathfrak{M}}_i^{}\!=\!\{\bar\phi_i^{}\}$
denotes the set of anti-particles
which have the same mass $\hs m_i^{}$ as those particles in
the set $\mathfrak{M}_i^{}\!=\!\{\phi_i^{}\}\hs$.\
Thus, the set $\mathfrak{M}_{0h}^{}$ serves as the identity element
of the group $\mathcal{G}$.\
In the case that the particles
$\phi_j^{}\!\hsm\in\!\mathfrak{M}_j^{}$
are real fields and carry no charge, we have
$\,\mathfrak{M}_j^{}\hsm\!=\hsm\over{\mathfrak{M}}_j^{}\hs$.\
In general, because the commutative law of multiplication
requires $\,\mathfrak{M}_{ij}^{}\!=\mathfrak{M}_{ji}^{}\,$
as discussed below Eq.\eqref{eq:Mi*Mj=Mk},
we deduce that the group $\mathcal{G}\hs$ forms
a Abelian group and $*$ is the multiplicative operation
defined for its elements.\
Moreover, in a physical system
all the particle masses should be described
by a finite number of physical parameters.\
It means that this Abelian group $\mathcal{G}\hs$
should be a finitely generated Abelian group\,\cite{abelian-group}
and thus has a finite number of generating elements (generators)
which compose the generating set of $\mathcal{G}\hs$.

\vspace*{1mm}

The fundamental theorem of the finitely generated Abelian group
states\,\cite{abelian-group} that
every finitely generated Abelian group is isomorphic to a
direct sum of the primary cyclic groups ($\hs \Z_{p}\hs$)
and the free Abelian group (containing copies of the infinite
cyclic group which is just the integer group $\Z\hs$).\
Each primary cyclic group $\Z_p^{}$ contains natural numbers
modulo $\hs p\hs$, i.e.,
$\Z_p^{}\!=\!\{0,1,2,\cdots\hsm,p\!-\!1\}$,
with $p$ being a prime number or a power of prime number.\footnote{%
The cyclic group $\Z_p^{}$ is isomorphic to the Gaussian residue class
group $\hs\Z/p\hs \Z\hs$ \cite{abelian-group}.}\
Thus, the finitely generated Abelian group $\mathcal{G}$
can be decomposed as follows:
\begin{equation}
\label{eq:G-decompose}
\mathcal{G} \hs\cong\, \Z^r \oplus \Z_{p_1^{}} \oplus
\Z_{p_2^{}} \!\oplus\! \cdots \oplus \Z_{p_s^{}} ,
\hspace*{10mm}
(\hs r,s \!\in\! \mathbb{N}\hs),
\end{equation}
where the natural number $\hs r\hs$ is the rank of
group $\mathcal{G}$ and $(p_1^{},p_2^{},\cdots,p_s^{})$
are powers of (not necessarily distinct) prime numbers.\
The values of $(r,\hs p_1^{},\hs p_2^{},\cdots\hsm,\hs p_s^{})$
are uniquely determined by $\mathcal{G}\hs$.\
The primary cyclic groups
$(\Z_{p_1^{}},\hs \Z_{p_2^{}},\cdots\hsm,\Z_{p_s^{}})$
have finite orders
$(p_1^{},\hs p_2^{},\cdots\hsm,\hs p_s^{})$
which are the torsion coefficients of $\mathcal{G}\hs$.
Hence, we can assign any particle in this theory by
a set of integers $\{n_1^{}, n_2^{},\cdots\hsm,n_{r+s}^{}\}$
(which belong to the cyclic groups of 
$\hs\mathcal{G}\hs$ respectively)
to characterize its mass
$\hs M_{\{n_1^{},\hsx n_2^{},\cdots\hsm,\hsx n_{r+s}^{}\}}^{}$,
where each integer $\hs n_k^{}\hs$ is additively conserved mod
$p_k^{}\hs$.\   Thus, the identity elements
$\{n_1^{},\hs n_2^{},\cdots\hsm,n_{r+s}^{}\}
\! =\!\{0,\cdots\hsm,\hs 0\}$
should correspond to the mass
$\hs M_{\{0,\hs\cdots\hsm,\hsx 0\}}^{}\!=0\hs$.\
For the convenience of demonstration, let us consider
for instance a nontrivial cyclic group $\hs\Z_{n}$
(with $n\!\geqq\! 1\hs$)
in Eq.\eqref{eq:G-decompose}.\
With the above formulation, we can compute both sides of the
mass spectral condition \eqref{eq:mass-spectrum} for the
external particles having indices
$\{n_1^{},\hs n_2^{},n_3^{},n_{4}^{}\}
\hsmx =\hsmx\{1,\hs n\!-\!1,\hs 1,\hs n\!-\!1\}$
and obtain the following:
\begin{equation}
\label{eq:mCond-check}
M_{\{1\}}^2 + M_{\{n-1\}}^2 +  M_{\{1\}}^2+  M_{\{n-1\}}^2
=\,  0 + 0 + M_{\{2\}}^2 \hs,
\end{equation}
where on the right-hand side of the equality
we have used the fact of
$\hs M_{\{n\}}^{}\!\!=\hsm\!M_{\{0\}}^{}\!\!=\!0\hs$
because the elements of $\hs\Z_{n}$ are mod $\hs n\hs$
and thus $\hs\{n\}\!=\!\{0\}\hs$.\
Moreover, applying the relation
$\hs M_{\{1\}}^2\hsm\!=\hsm M_{\{n-1\}}^2$
to Eq.\eqref{eq:mCond-check},
we deduce from Eq.\eqref{eq:mCond-check} a condition
$\,M_{\{2\}}^2\hsm\!=\! 4\hs M_{\{1\}}^2$.
Then, we will use induction method to prove the mass formula
$\hs M_{\{k\}}^2\hsm\!=\hsm k^2 M_{\{1\}}^2$.\
This formula obviously holds for $k=0,1\hs$,\
and we already explicitly verified that it holds for $k\!=\!2$
as shown above.\
Let us suppose that the formula
$\hs M_{\{\ell\}}^2\hsm\!=\hsm\ell^2 M_{\{1\}}^2\hs$
holds for $\hs\ell\hsm\leqq\hsm k\hs$.\
Then, the remaining task is to prove that
this formula can hold for $\hs\ell \!=\!k\hsm +\! 1\hs$.\
Thus, we further consider a choice of the mass indices of
the external particles
$\{n_1^{},\hs n_2^{},n_3^{},n_{4}^{}\}
\hsmx =\hsmx\{1,\hs k,\hs n\!-\!1,\hs n\!-\!k\}$,
and compute both sides of the spectral condition
\eqref{eq:mass-spectrum} as follows:
\begin{equation}
\label{eq:mCond-check3}
M_{\{1\}}^2\hsm + M_{\{k\}}^2\hsm +  M_{\{n-1\}}^2\hsm +  M_{\{n-k\}}^2=\hs  M_{\{1+k\}}^2\hsm +M_{\{1-k\}}^2\hsm + 0 \,.	
\end{equation}
Because these mass indices are additively conserved mod $n$ of
$\hs\Z_n$, we can simplify Eq.\eqref{eq:mCond-check3} as,
%
$\hs M_{\{k+1\}}^2 \!=\!
2\hs \big(M_{\{k\}}^2\hsm\! +\hsm M_{\{1\}}^2\big)\hsmx
-\! M_{\{k-1\}}^2\hs$.\
%
Here we have made use of the fact that a particle and its
anti-particle must share the same mass, leading to
$\hs M_{\{k\}}^2\hsm\!=\! M_{\{-k\}}^2$ for any integer $k\hs$.\
Using the formula
$\hs M_{\{\ell\}}^2\!\!=\!\hsm\ell^2 M_{\{1\}}^2$
for $\hs\ell\!\leqq\! k\hs$, we can thus deduce:\
$M_{\{k+1\}}^2 \!\!=\!(k\!+\!1)^2 M_{\{1\}}^2$.\
This completes our proof by induction, and we can conclude:
\begin{equation}
\label{eq:mk=k*m1}
M_{\{k\}}^2=\hs k^2 M_{\{1\}}^2\hs , \hspace*{8mm}
(\hs\text{for}~ 0\leqq k \leqq n\hsm -\! 1\hs)\hs.
\end{equation}
But we must have the relation
$\hs M_{\{n-1\}}^{2}\!\!=\hsm\! M_{\{1\}}^2\hs$
because the mass indices are additively conserved mod $n$ of
$\hs\Z_n\hs$.\ Together with Eq.\eqref{eq:mk=k*m1} for
$\hs k\!=\!n\!-\!1\hs$, this leads to the condition:
\beq
\label{eq:n(n-2)=0}
n(n\!-\!2)\hs M_{\{1\}}^2 =\,0\,.
\eeq
It has a nontrivial solution $\hs n\hsm\!=\! 2\hsx$,
or, we can avoid this condition \eqref{eq:n(n-2)=0} altogether
by removing the constraint of the cyclic relation
$\hs M_{\{n-1\}}^{2}\!\!=\hsm\! M_{\{1\}}^2$
which could happen {\it only if}
$\hs n\!=\!\infty\hs$.\
Because of $\hs\Z_{\infty}\hsm\!=\!\Z^+\!\!\subset\!\Z\hsx$,
the case of $\,\Z_n \hsm\!=\!\Z_{\infty}\hs$ just goes back to
a subgroup of the integer groups $\Z^r$
(the free Abelian group) of the finitely generated group
$\mathcal{G}$ as defined in Eq.\eqref{eq:G-decompose}.\

\vspace*{1mm}

In the following we will first check the solution
$\hs n\hsm\!=\! 2\hsx$ and show that it is excluded by
the mass spectral condition \eqref{eq:mass-spectrum}.\
For this, using the group $\Z_2\hs$, we construct the simplest
realization including the massless graviton field $h^{\mn}$
and a massive real field $\hs\phi\hs$
(which has no self-interaction and describes a type of
particles equal to their own anti-particles).\
The field $\phi$ only interacts with the graviton field $h^{\mn}$
through the cubic vertex of $\phi$$-$$\phi$$-$$h\hs$,
denoted as $V_3^{}(h\hs\phi\hs\phi)$.\
For example, $\phi$ can be a real massive scalar field, or a
massive photon, or a massive Majorana fermion.\
This model contains two types of cubic interaction vertices
$\phi$$-$$\phi$$-$$h$ and $h$$-$$h$$-$$h\hs$.\
We present the multiplication table of this model
on the left, in comparison
with the $\Z_2$ multiplication table on the right:
\\[-1mm]
\begin{equation}
\label{eq:V3(hpp)-Z2}
\begin{tabular}{|c||c|c|}
\hline
$V_3^{}(h\hs\phi\hs\phi)$ & $\blu h^{\mn}$ & $\blu\phi$  \\
\hline
\hline
$\blu h^{\mn}$ & $\red h^{\mn}$ &  $\red\phi$ \\
		\hline
		$\blu\phi$ &  $\red\phi$ & $\red h^{\mn}$  \\
		\hline
	\end{tabular} \hspace*{10mm}
    	\begin{tabular}{|c||c|c|}
    	\hline
    	$\Z_2$& $\blu 0$ & $\blu 1$  \\
    	\hline
    	\hline
    	$\blu 0$ & $\red 0$ &  $\red 1$ \\
    	\hline
    	$\blu 1$ &  $\red 1$ & $\red 0$  \\
    	\hline
\end{tabular}
\end{equation}
\\[-1mm]
The above tables demonstrate that the defined operations for the
$\{ \phi,\hs h^{\mn}\}$ model do satisfy the multiplication rule
(addition) of the $\Z_2$ group.\
Consider for instance the scattering amplitude of
$\,\phi\,\phi \!\to\! \phi\,\phi\,$,
which is contributed by exchanging
the massless graviton fields $h^{\mn}$.\
It is clear that this process violates
the mass spectral condition \eqref{eq:mass-spectrum}
unless $\hs\phi\hs$ is also set as massless which
however becomes a trivial case.\
Thus, the condition \eqref{eq:mass-spectrum}
excludes $\hs n\!=\hsm 2\hsx$ as a nontrivial solution
of the physical massive theory.\
Hence, we conclude that
{\it the integer groups $\Z^r$ give the unique realization of the
mass spectral condition \eqref{eq:mass-spectrum}.}

\vspace*{1mm}

Next, taking one of the integer group $\Z$ of $\Z^r$,
we will prove that the mass spectral condition \eqref{eq:mass-spectrum}
has the unique solution:
\beq
\label{eq:Mk=k*M1f}
M_{\{k\}}^2\hsm =\hs k^2 M_{\{1\}}^2\hs , \hspace*{8mm}
(\hs\text{for}~ k \in \Z\hs )\hs.
\eeq
This can be proven by induction method.\
Choosing the indices of the external states as
$\{n_1^{},\hs n_2^{},n_3^{},n_{4}^{}\}
\hsmx =\hsmx\{0,\hs 0,\hs 0,\hs 0\}$, we deduce
from the spectral condition \eqref{eq:mass-spectrum}:\
$4M_{\{0\}}^2\!\!=\!3M_{\{0\}}^2\hs$ and thus
$\hs M_{\{0\}}^{}\!\!=\!0\hs$.
Then, choosing $\{n_1^{},\hs n_2^{},n_3^{},n_{4}^{}\}
\hsmx =\hsmx\{n,\hs -n,\hs 0,\hs 0\}$ and using
$\hs M_{\{0\}}^{}\!=\!0\hs$, we deduce from the condition
\eqref{eq:mass-spectrum}:\
$M_{\{-n\}}^2\!=\!M_{\{n\}}^2\hs$.\
We further choose  $\{n_1^{},\hs n_2^{},n_3^{},n_{4}^{}\}
\hsmx =\hsmx\{n,\hs n,\hs -n,\hs -n\}$
and deduce from the condition
\eqref{eq:mass-spectrum}:\
$M_{\{2n\}}^2\!\!=\!4M_{\{n\}}^2\hs$
and thus $\hs M_{\{2\}}^2\!\!=\!4M_{\{1\}}^2\hs$.\
For induction proof, we suppose that Eq.\eqref{eq:Mk=k*M1f} holds
for $\hs k\!=\!n\hs$.\
Then, choosing $\{n_1^{},\hs n_2^{},n_3^{},n_{4}^{}\}
\hsmx =\hsmx\{n,\hs -n,\hs 1,\hs -1\}$, we
deduce from the condition \eqref{eq:mass-spectrum}:\
\begin{equation}
\label{eq:mCond-check}
M_{\{n\}}^2 \!+\hsm M_{\{-n\}}^2 \!+\hsm  M_{\{1\}}^2\!+\hsm  M_{\{-1\}}^2
=\,  0 +\hsm M_{\{n+1\}}^2\hsm +\hsm M_{\{n-1\}}^2 \hs,
\end{equation}
which can be further simplified as
$M_{\{n+1\}}^2\!\!=\!(n\!+\!1)^2 M_{\{1\}}^2$.\
This completes our proof of the mass formula \eqref{eq:Mk=k*M1f}
by induction.\
Since the physical mass $M_{\{k\}}^{}$ should be defined
as nonnegative, we can solve Eq.\eqref{eq:Mk=k*M1f} as follows:
\beq
\label{eq:TMk=k*M1f}
\hs M_{\{k\}}^{}\!=|\widetilde{M}_{\{k\}}^{}|\hs, ~~~~~
\widetilde{M}_{\{k\}}^{}\!= k\hs M_{\{1\}}^{} \hs,\
\eeq
where $\hs k\!\in\! \Z\hs$.\
Thus, under the integer group $\Z$,
we have the following multiplication (additive) rule
for the mass parameters:
\beq
\label{eq:Mk1+Mk2}
\widetilde{M}_{\{k_1\}}^{} \!+\hsm \widetilde{M}_{\{k_2\}}^{} =
\widetilde{M}_{\{k_1\!+k_2\}}^{}\hs.
\eeq
For the cubic vertex $V_{{\bf abc}}^{}$ under the multiplication
operation of the group $\Z$, its group indices should obey
$\hs k_{\bf a}^{}\!+\hsm k_{\bf b}^{}\!=\hsm k_{\bf c}^{}\hs$,
and the masses for the three fields in this cubic vertex
hold the group relation
$\widetilde{M}_{\{k_{\bf a}\}}^{} \!+\hsm
 \widetilde{M}_{\{k_{\bf b}\}}^{} =
\widetilde{M}_{\{k_{\bf c}\}}^{}\hs$.

\vspace*{1mm}

Then, we can substitute Eqs.\eqref{eq:TMk=k*M1f} and \eqref{eq:Mk1+Mk2}
into the original spectral condition \eqref{eq:mass-spectrum} and
arrive at
\beq
k_1^2+k_2^2+k_3^2+k_4^2 \,=\hs
(k_1^{}\!+\hsm k_2^{})^2\!+\! (k_1^{}\!+\hsm k_3^{})^2
\!+\hsm (k_1^{}\!+\hsm k_4^{})^2\hs ,
\eeq
where $\hs k_j\!\in\!\Z\hs$ with $j\!=\!1,2,3,4\hs$.\
This can be further simplified to give the following condition:
\beq
\label{eq:k1234}
k_1^{}\!+k_2^{}\!+k_3^{}\!+k_4^{} =\hs 0 \hs .
\eeq
It holds for any finite integers
$\hs k_j\!\in\!\Z\hs$ and its important physical implication
will be discussed shortly.\

\vspace*{1mm}

From the above proof, we deduce that
the nontrivial realization of the group \eqref{eq:G-decompose}
for a physical theory [holding the mass spectral
condition \eqref{eq:mass-spectrum}] is isomorphic to
a direct product of integer groups $\Z^r$.\
Thus, for a given type of particles in this theory, their mass
is characterized by the indices corresponding to each of the
subgroup $\Z\hs$, namely,
$\hs M_{\{n_1^{},\hsx n_2^{},\cdots\hsm,\hsx n_{r}^{}\}}^{}$,
where $\hs r\!\!\geqq\!\! 1\hs$.\
Extending Eq.\eqref{eq:Mk=k*M1f} under a single integer group
$\Z$ to the general case of the product group
$\,\Z^r\hs$ and defining an index vector
$\hs\mathbf{n}\!=\!
 (n_1^{},\hsx n_2^{},\cdots\hsm,\hsx n_{r}^{})\hs$,
we can express the (mass)$^2$ for the particles of
type-$\mathbf{n}$ by the following $\,r\times r\,$
real symmetric matrix:
\begin{equation}
\label{eq:Mn2-Zr}
\mathbf{M}_{\mathbf{n}}^2
\,=\, \mathbf{n}\hsx \overline{\bf M}^2\hsx \mathbf{n}^T ,
\end{equation}
where $\overline{\bf M}^2$ is a positive-definite matrix coefficient.\ 
For the mass matrix \eqref{eq:Mn2-Zr},
the corresponding four-point scattering amplitudes
will obey the mass spectral condition \eqref{eq:mass-spectrum}.\
It is clear that among all the known consistent field theory models
{\it only the KK theories with $\delta\,(=\!r)$ extra dimensions
under the toroidal compactification could have a mass spectrum
behave exactly as in} Eq.\eqref{eq:Mn2-Zr}.\
The matrix coefficient $\overline{\bf M}^2$ in Eq.\eqref{eq:Mn2-Zr}
depends on detail of the KK compactification.\ 
For the case of a single extra dimension  
$\delta\!=\!r\!=\!1$, this matrix 
$\overline{\bf M}^2$ will reduce to 
a non-matrix quantity $\overline{\bf M}^2\!=\!1/R^2$ 
with $R\hs$ being the 5d radius.\
We further note that for the $\delta\!=\!1\hs$ case, 
the condition \eqref{eq:k1234} just means the requirement of the
KK number conservation, namely, the discretized
5th component of the momentum is conserved in the 5d space.\
This imposes a nontrivial condition requiring
that the 5d extra dimensional space should be
properly compactified (without $\ZZ$ orbifold).\
For the general case of $\hs\delta\hs$ extra dimensions,
we can substitute Eq.\eqref{eq:Mn2-Zr}
into the spectral condition \eqref{eq:mass-spectrum} and derive
the following conservation condition on the KK index vectors:
\beq
\label{eq:k1234-r}
\mathbf{n}_1^{}+\mathbf{n}_2^{} + \mathbf{n}_3^{} +\mathbf{n}_4^{}
=\hs 0 \hs,
\eeq
with each index vector
$\hs\mathbf{n}_j^{}\!=\hsm 
 (n_{j1}^{},\hsx n_{j2}^{},\cdots\hsm,\hsx n_{j\delta}^{})\hs$
and $\hs j\!=\!1,2,3,4\hs$.\

\vspace*{1mm}

We note that in the above proof [which leads to the unique
KK mass-spectrum \eqref{eq:Mn2-Zr} from solving the spectral
condition \eqref{eq:mass-spectrum}], we have only made two
modest assumptions:
{\bf (i).}\,the particle masses in a massive field theory
are characterized by a finite number of basic parameters
(as required by a sensible physical theory);\footnote{%
We note that the validity of the spectral condition \eqref{eq:mass-spectrum}
also requires the theory to contain at least two types of particles
with unequal masses.}\
{\bf (ii).}\,the cubic interaction vertex $V_{\textbf{abc}}^{}$
can unqiuely fix the mass $\hs\mc\hs$
of the third particle $\cx$
once the other two particles 
$\hs\ax\hs$ and $\hs\bx\hs$ are specified
(which have masses $\hs\ma\hs$ and $\mb$ respectively).\
With these we can solve the spectral condition \eqref{eq:mass-spectrum}
and uniquely deduce that the mass-spectrum of such massive theories
has to be given by Eq.\eqref{eq:Mn2-Zr}
which is same as that of the KK theories
under the toroidal compactification.\
But, this does not imply that there would exist no other double-copy
constructions if the condition \eqref{eq:mass-spectrum} does not hold.\
For instance, the double-copy could be realized in a more
complicated form such as the case of the double-copy construction
for the KK gauge/gravity theories under the 5d compactification
of the $S^1\!/\ZZ$ orbifold (cf.\ Section\,\ref{sec:3.3}).\
The KK theories under such $S^1\!/\ZZ$ orbifold compactification
do not obey the above assumption-(ii) and lead to double-pole
structure in the KK scattering amplitudes which violate the
spectral condition \eqref{eq:mass-spectrum}; so their double-copy
construction has to be realized in a nontrivial way, as we
analyzed in Section\,\ref{sec:3.3}.\footnote{%
As we clarified in the footnote-2 of Section\,\ref{sec:1}, 
it is hard to directly realize the massive KLT relations 
under the orbifold compactification even within the KK string 
theory\,\cite{KKString}, where the masses of KK open (closed) strings 
are lifted by a large amount 
$\frac{1}{\,16\hs \alpha'\,}$ ($\hs\frac{1}{\,4\hs \alpha'\,}$) 
and the KK states fully decouple in the field theory limit 
of $\alpha'\hsm\ito 0\hs$.}\
We also note that in
the 3d topologically massive Chern-Simons
gauge/gravity theories\,\cite{TMG},
all particles share {\it the same mass} and
thus directly violate the spectral condition \eqref{eq:mass-spectrum}
because of $\,4\hs M^2\!\neq\! 3\hs M^2\hs$.\
Although the scattering amplitudes of
such massive Chern-Simons theories violate the
spectral condition \eqref{eq:mass-spectrum}, it can still
hold the color-kinematics duality in a proper massive BCJ-representation 
and realize the double-copy in a different manner\,\cite{Hang:2021oso}\cite{Gonzalez:2021bes}.

\vspace*{1mm}

In addition, we note that the RS model\,\cite{ExdRS} has warped 5d geometry and its compactification at IR/UV boundaries leads to the following consequences:\ (i).\,it must have a nonlinear KK mass spectrum [different from our mass-spectrum solution \eqref{eq:Mk=k*M1f}]; (ii).\,its four-point KK amplitude contains exchanges of infinite number of KK particles.\ (iii).\,it violates the KK number conservation condition \eqref{eq:k1234} because the RS compactification with warped 5d geometry excludes the periodic boundary conditions.\  Hence, the RS model violates the spectral condition \eqref{eq:mass-spectrum}, and thus cannot realize the double-copy by the existing methods.\

\vspace*{1mm}

In general, it is difficult to solve the mass spectral conditions
like Eq.\eqref{eq:mass-spectrum}.\
This difficult was also realized in
Ref.\,\cite{Momeni:2020hmc},
so these authors studied\,\cite{Momeni:2020hmc} certain 
KK-inspired massive gauge theories by assuming their mass-spectrum
to be identical to that of the 5d YM theory under the
$S^1$ compactification.\ By demanding the scattering amplitudes
to respect the color-kinematics duality for BCJ-like double-copy
it was found\,\cite{Momeni:2020hmc}
that the couplings of the 4d Lagrangian
are fixed to that of the 5d KK YM theory
together with higher derivative operators.\
Our above proof does not assume any KK mass-spectrum from the start,
instead we newly propose a group theory approach to
{\it derive} the KK mass-spectrum as in Eq.\eqref{eq:Mn2-Zr} under
fairly simple and modest assumptions (explained above).\
Hence our approach differs essentially from the
literature\,\cite{Momeni:2020hmc}.
\\[-3.6mm]


Finally, we note that 4-point spectral condition \eqref{eq:mass-spectrum}
is the minimal spectral condition because the case of 3-point amplitudes
has no internal propagator at tree level.\
Then, we discuss the general spectral condition for
the $N$-point massive scattering amplitudes including
$\hs N\!\hsm\geqq\hsm 5\hs$.\ For this, we factorize an $N$-point
scattering amplitude into a 4-point sub-amplitude plus an
$(N\hsm\!-\hsm 2)$-point sub-amplitude.\
This 4-point sub-amplitude has a new 4th external state
which originates from breaking an internal propagator.\
According to the properties of the generic cubic vertex
$V_{\textbf{abc}}^{}$ and of its associated group
\eqref{eq:G-decompose}, we deduce that the 4th external state of
this 4-point sub-amplitude has a pole-mass
$\hs M_{X_4}^{}\!$ with its mass index given by
$\hs X_4^{}=k_4^{}+\cdots +k_N^{}\hs$.\
Thus, according to the additive rule \eqref{eq:Mk1+Mk2}
[which has been proven by solving the
four-point spectral condition \eqref{eq:mass-spectrum}], we can deduce
\beq
\label{eq:MX4}
M_{X_4}^{} =\hs M_{k_4^{}}^{} +\cdots \hsm + M_{k_N^{}}^{}\hs .
\eeq
Because this four-point sub-amplitude does hold
the spectral condition \eqref{eq:mass-spectrum},
we can use Eq.\eqref{eq:mass-spectrum} together
with the relation \eqref{eq:MX4} to derive a new spectral condition:
\beq
\label{eq:mass-cond-N}
M_1^2\hsm +\! M_2^2\! +\! M_3^2\!+\!
\big(M_4^{}+\cdots\hsm +M_N^{}\big)^2
=\hs  M_{12}^2\hsm +\hsm M_{13}^2\hsm +\! M_{23}^2\,,
\eeq
where on the righ-hand side we make use of the relation
$M_{1X_4}^2\!=\!M_{23}^2\hs$ due to the conservation of the sum
of all the $N$ external momenta.\
Hence, Eq.\eqref{eq:mass-cond-N} just gives the new
$N$-point spectral condition as generalized from the four-point
spectral condition \eqref{eq:mass-spectrum}.\

\vspace*{1mm}

It is clear that the unique solution \eqref{eq:TMk=k*M1f}
to the four-point spectral condition \eqref{eq:mass-spectrum}
is also the solution to above $N$-point spectral condition
\eqref{eq:mass-cond-N}.\
We can substitute Eq.\eqref{eq:TMk=k*M1f}
into Eq.\eqref{eq:mass-cond-N} and arrive at:
\beq
k_1^2+k_2^2+k_3^2+(k_4^{}\hsm +\cdots +k_N^{})^2 \hs =\hs
(k_1^{}\!+\hsm k_2^{})^2\!+\! (k_1^{}\!+\hsm k_3^{})^2
\!+\hsm (k_2^{}\!+\hsm k_3^{})^2\hs .
\eeq
From this, we further derive the following condition:
\beq
\label{eq:k1234-1N}
k_1^{}\!+k_2^{}\!+k_3^{}\!+k_4^{}\!+\cdots +k_N^{} =\hs 0 \hs ,
\eeq
which requires that the sum of the mass indices $\{k_j^{}\}$ for all
$N$ external states must vanish and be conserved.\
This is just a statement of the KK number conservation for the
$N$-point scattering amplitudes, which is a generalization of
the conservation law \eqref{eq:k1234} for $N\!=\!4$ case.\
We note that the KK number conservation of
Eqs.\eqref{eq:k1234} and \eqref{eq:k1234-1N} can be realized under
the 5d toroidal $S^1$ compactification without the $\ZZ$ orbifold.\

\vspace*{1mm}

For the general case of $\hs\delta\hs$ extra dimensions, we
can substitute the mass formula \eqref{eq:Mn2-Zr} into
the spectral condition \eqref{eq:mass-cond-N} and derive
the conservation condition on the $\delta$-dimensional KK index
vectors:
\beq
\label{eq:k1234-rN}
\mathbf{n}_1^{}+\mathbf{n}_2^{} + \mathbf{n}_3^{} +\mathbf{n}_4^{}
+\cdots +\mathbf{n}_N^{} =\hs 0 \hs,
\eeq
with each index vector
$\hs\mathbf{n}_j^{}\!=\!
 (n_{j1}^{},\hsx n_{j2}^{},\cdots\hsm,\hsx n_{j\delta}^{})\hs$
and $\hs j\!=\!1,2,\cdots\!,N$.\
\\[-4mm]

In passing, we also note that the literature\,\cite{DC-4dx1}
discussed explicitly an extended spectral condition
for the $N\!\!=\!4,5$ cases
and the application for reducing the rank of the BAS kernel
for the BCJ double-copy; it found that the massive YM theory and
dRGT gravity model do not obey such spectral conditions;
so it has no real overlap with our study since
it did not solve the spectral condition for either $N\hsm\!=\!4$
or $N\hsmx\!=\!5$ to determine the allowed mass spectrum
[such as our general solutions
\eqref{eq:Mk=k*M1f} and \eqref{eq:Mn2-Zr}],
and also did not study the general constraints on the mass indices
[$\hsx$such as our conditions \eqref{eq:k1234} ($\hs\delta \!=\!1\hs$)
and \eqref{eq:k1234-r} ($\hs\delta \!\geqq\!1\hs$)
for $N\hsm\!=\!4\hs$,
and our conditions \eqref{eq:k1234-1N} ($\hs\delta \!=\!1\hs$) 
and \eqref{eq:k1234-rN} ($\hs\delta \!\geqq\!1\hs$) 
for any $N$, which
can impose crucial constraint on the KK compactifications of the
higher dimensional gauge/gravity theories
as we have demonstrated$\hsx$].\
It focused on the applications to the massive YM theory and
dRGT gravity model with possible constraints
by the spectral conditions of $N\!=\!4,5\hs$.\

\vspace*{1mm}
\section{\hspace*{-2mm}Conclusions}
\label{sec:6}
\vspace*{1mm}

The Kaluza-Klein (KK) compactification\,\cite{KK} of higher dimensional
gauge and gravity theories predicts an infinite tower of
massive KK excitations for each known particle of the
Standard Model (SM), and has been a major frontier for
new physics beyond the SM,
including the string/M theories\,\cite{string} and
the extra dimensional field theories with large or small
extra dimensions\,\cite{Exd0}\cite{Exd}\cite{ExdRS}.\
The double-copy construction between the scattering amplitudes
of the gauge bosons and gravitons has pointed to fundamental clues
to the deep gauge-gravity connection.\ It has also become a powerful
tool for efficiently computing the highly intricate 
scattering amplitudes of spin-2 gravitons.\  
Especially, the massive KK graviton scattering amplitudes are 
even more involved (with large energy cancellations)\,\cite{Chivukula:2020}\cite{Kurt-2019}\cite{Hang:2021fmp} 
and the $N$-point longitudinal KK graviton amplitudes have 
their leading energy dependence nontrivially cancelled down 
by a large power factor $\hs E^{2N}\!\!$ ($N\!\geqq\!4$) 
up to any loop order\,\cite{Hang:2021fmp}.\
Hence, it is compelling to further study the double-copy constructions
of the massive KK graviton scattering amplitudes from the massive
KK gauge boson scattering amplitudes.\

\vspace*{0.5mm}

In this work, we systematically investigated the structure 
of massive scattering amplitudes of KK gauge bosons
and KK gravitons under toroidal compactifications.\
For this, we proposed a shifting method 
to construct the massive KK amplitudes from 
their massless counterparts in the 
noncompactified higher dimensional theories 
and used this to establish a correspondence 
from the conventional massless BCJ double-copy
to the extended massive KK double-copy (Section\,\ref{sec:2}).\
Then, using this shifting method
we studied the massive KK double-copy constructions
via both the extended BCJ approach (Section\,\ref{sec:3})  
and the extended CHY approach (Section\,\ref{sec:4})  
as well as their quantitative connections
to the massive KLT relations.\ 
We further solved the four-point
and $N$-point mass spectral conditions, 
and demonstrated that KK theories under toroidal compactification
provide the unqiue solution to such mass spectral conditions
and thus guarantee successful realization of the extended 
massive gauge/gravity double-copies (Section\,\ref{sec:5}).\
We summarize our findings in more detail as follows.

\vspace*{0.5mm}

In section\,\ref{sec:2},
we constructed an extended massive double-copy approach for
the scattering amplitudes of KK gauge bosons and KK gravitons
under the toroidal compactification within the  
quantum field theory (QFT) framework.\
For this, using the exponential eigenfunctions of
the Laplace operator, we derived the massive KK scattering amplitudes
of a higher dimensional theory under toroidal compactification
by replacing the extra-dimensional
momentum-components in the corresponding massless amplitudes of
the noncompactified theory by their discretized values.\
With this shifting method, we built up a
correspondence between the conventional massless BCJ double-copy
formulation and the extended massive KK double-copy formulation.\
We gave in Eqs.\eqref{eq:T4dS1} and \eqref{eq:MS1-Npt}
the shifted $N$-point massive KK gauge boson/graviton
amplitudes under the toroidal $S^1\hsmx$ compactification.\
We stressed the importance of using the toroidal compactification
without orbifold as the {\it base construction}
of the massive KK double-copy, with which the double-copy constructions
in other KK theories under orbifold compactification (such as $S^1\!/\ZZ$) can be formulated by proper transformations of the external KK states.\
We gave in Eqs.\eqref{eq:T4dS1Z2} and \eqref{eq:MS1Z2-Npt}
the shifted $N$-point massive KK gauge boson/graviton amplitudes
under the orbifold compactification of $S^1\!/\ZZ$.\


In Section\,\ref{sec:3},
we presented the extended four-point massive BCJ-type
double-copy construction of scattering amplitudes
of the KK gauge bosons and KK gravitons under the 5d toroidal
compactification with or without orbifold.\
In Section\,\ref{sec:3.1},
under the toroidal compactification without orbifold and
by requiring the double-copied KK graviton amplitudes
\eqref{eq:Mh-4pt-BCJ} being invariant under the
generalized massive gauge transformation \eqref{eq:GGT-Nj-KK},
we derived a four-point mass spectral condition \eqref{eq:MSCond-4KK}.\
Then, we deduced a massive fundamental BCJ relation
\eqref{eq:fBCJ-relation} for the KK gauge boson amplitudes and
applied it to derive a spectral condition \eqref{eq:MassCond-4pt}
which fully agrees to what we derived in
Eq.\eqref{eq:MSCond-4KK} by using the generalized
massive gauge transformation \eqref{eq:GGT-Nj-KK}.\
We showed that the spectral condition \eqref{eq:MSCond-4KK}
can serve as a necessary and sufficient condition to ensure
a consistent double-copy of the KK gauge boson/graviton
scattering amplitudes.\
Using the massive fundamental BCJ relation
\eqref{eq:fBCJ-relation}, we also proved that
the four-point massive BCJ-type double-copied KK graviton amplitudes
\eqref{eq:MhbarL-4pt-BCJ} [\eqref{eq:Mh-4pt-BCJ}] can be
derived from the massive KK KLT relation \eqref{eq:Mhbar-4pt}
(which was based on our previous derivation of the KK
open/closed string amplitudes\,\cite{KKString}), and thus
they are equivalent within the QFT.\

\vspace*{1mm}

In Section\,\ref{sec:3.2}, we systematically derived the
explicit double-copy constructions for various
four-point elastic and inelastic KK graviton scattering amplitudes
at tree level from the corresponding KK gauge boson amplitudes
under the 5d toroidal compactification of $S^1$.\
Here we used the effective leading longitudinal polarization tensor
\eqref{eq:LO-Gpol} of the KK gravitons to derive the double-copied
massive KK graviton amplitudes which take rather compact forms
of the BCJ-type.\
Then, we used the exact longitudinal polarization tensor
\eqref{eq:zeta-munu-4d}-\eqref{eq:Gpol-massive-KK} to compute
the corresponding full four-point KK graviton scattering amplitudes
which are presented in Appendix\,\ref{App:C}.\
We demonstrated in
Eqs.\eqref{eq:M0-hLbar-hL-LO}-\eqref{eq:N-M0-hLbar-hL-LO}
that under high energy expansion,
each leading-order (LO) scattering amplitude of longitudinal
KK gravitions computed by using the simple effective leading
longitudinal polarization tensor \eqref{eq:LO-Gpol} always
equals the LO amplitude computed by using the exact longitudinal polarization tensor
in Eqs.\eqref{eq:zeta-munu-4d}-\eqref{eq:Gpol-massive-KK}.\
From Eqs.\eqref{eq:M0-hLbar-hL-LO}-\eqref{eq:N-M0-hLbar-hL-LO},
we derived a nontrivial sum rule condition \eqref{eq:LO-Cond-N0L=N0}
and made an important conclusion stating that each $N$-point 
longitudinal KK graviton amplitude at the LO (with two or more
external KK graviton states being longitudinally polarized)
can be constructed by double-copy of a single amplitude of KK
gauge bosons (in which the corresponding KK gauge bosons
are longitudinally polarized only) according to the effective leading
longitudinal polarization tensor \eqref{eq:LO-Gpol}.\
Then, in Section\,\ref{sec:3.3} we extended the analyses of
Section\,\ref{sec:3.2} to the case of the 5d orbifold compactification
of $S^1\!/\ZZ$.\ We demonstrated explicitly that all the four-point
double-copied KK graviton scattering amplitudes under orbifold compactification can be derived from the corresponding
KK graviton partial amplitudes under $S^1$ compactification
by transformations of the external KK state, and thus are expressed
in terms of proper combinations of the corresponding KK
graviton partial amplitudes of Section\,\ref{sec:3.2}.\

\vspace*{0.5mm}

In Section\,\ref{sec:3.4}, we further constructed explicitly
the massive five-point KK graviton scattering amplitudes by the
extended BCJ double-copy formulation using the shifting method,
as shown in Eqs.\eqref{eq:MS1-5pt}-\eqref{eq:5pt-Numerator}
and \eqref{eq:MS1-5pt-2}.\
In Section\,\ref{sec:3.5}, we studied the scattering amplitudes
of super massive KK states in the nonrelativistic limit.\
We first derived the nonrelativistic amplitudes of the KK gauge bosons
and of the KK gravitons in Sec.\,3.5.1.\  We found that
the LO elastic amplitudes of the KK gauge boson and of the KK graviton
exhibit a low energy behavior of $1/Q^2$ due to the exchange of
massless zero modes in the $t$ and $u$ channels, as shown in
Eqs.\eqref{eq:4AL-nnnn-NR-LO} and \eqref{eq:4hL-nnnn-NR-LO}.\
After Fourier transformation, this $1/Q^2$ behavior reproduces the
classical Coulomb potential $1/r$ (for the elastic KK gauge boson
scattering) and the classical Newtonian gravitational potential $1/r$
(for KK graviton scattering) in nonrelativistic limit.\
In Sec.\,3.5.2, we further considered a compactified 5d
Einstein-Scalar theory (with scalar field
coupled to gravity) in Eq.\eqref{eq:SEH} and a compactified
5d Einstein-YM theory or Einstein-Maxwell theory 
(with gauge fields coupled to gravity)
in Eq.\eqref{eq:EYM5}.\
We derived the four-point scattering amplitudes
of KK scalars and of KK gauge bosons through exchanges of the zero-mode graviton, the KK gravitons, and radion.\
We computed both the elastic and inelastic amplitudes and derived
their behaviors in the nonrelativistic limit.\
The nonrelativistic elastic KK scalar amplitudes are given in
Eq.\eqref{eq:4phi/h-nnnn}, whereas the nonrelativistic elastic KK
gauge boson amplitudes are given in
Eqs.\eqref{eq:4A-nnnn-NRexp} and \eqref{eq:4AL-nnnn-NRexp}.\
They exhibit different angular behaviors in the low energy
nonrelativistic limit.\  Their LO nonrelativistic
elastic amplitudes are given by Eqs.\eqref{eq:4phi/h-nnnn-LO},
\eqref{eq:4A-nnnn-NRexpLO}, and \eqref{eq:4AL-nnnn-NRexpLO},
respectively.\ We found that these LO nonrelativistic elastic amplitudes
have a massless pole due to exchanging massless zero modes in
the $t$ and $u$ channels.\ After Fourier transformation into the
coordinate space, they recover the classical Newtonian gravitational potential $1/r\hs$.\

\vspace*{1mm}

In Section\,\ref{sec:4},
by generalizing the conventional massless CHY method,
we presented an extended massive CHY formulation for
the KK scattering amplitudes under the 5d toroidal compactification.\
In Section\,\ref{sec:4.0},
we presented the extended massive KK CHY scattering equations in
Eqs.\eqref{eq:SEq-Ea}-\eqref{eq:Mab^2}.\
We derived the massive KK KLT-type double-copy relation
\eqref{eq:KK-CHY-DC} and formulated the massive KK graviton
scattering amplitude $\M_{\hsm N}^{}(\{\bn\}\hsm)$ as products of
the two color-ordered KK gauge boson amplitudes
$\A_N^{}([\al],\!\{\bn\}\hsm)$ and $\A_N^{}([\be],\!\{\bn\}\hsm)$,
together with a kernel $\hs\KK([\al|\be],\!\{\bn\}\hsm)\hs$
which is interpreted as the
inverse of the massive KK bi-adjoint scalar amplitude.\
In Section\,\ref{sec:4.1}, we presented the extended CHY construction for
massive KK bi-ajoint scalar amplitudes, which are given by the
formula \eqref{eq:Amp-BAS-1} and can be computed according to
the massive KK scattering equations \eqref{eq:SEq-Ea}-\eqref{eq:Mab^2}.\
We derived explicitly the four-point elastic KK bi-adjoint scalar
amplitudes in Eqs.\eqref{eq:BASamp4partial}-\eqref{eq:Amp-BAS-2}
and the inelastic KK bi-adjoint
scalar amplitudes in Eqs.\eqref{eq:BASamp4-00nn}-\eqref{eq:AmpS-00nn}
for illustrations.\
Then, we presented the five-point massive scattering amplitudes of
KK bi-adjoint scalars in Eq.\eqref{eq:BAS5pt}
which can form a rank-2 kernel matrix.\
In Section\,\ref{sec:4.2}, we presented the extended CHY construction of
the massive KK gauge boson amplitudes and KK graviton amplitudes
in the compactified 5d YM and GR theories.\
We derived in Eq.\eqref{eq:CHY-DC-GR} the BCJ-type double-copy
construction of the KK graviton amplitude by the product of
KK gauge boson amplitudes including a KK {bi-adjoint scalar} kernel
$\bar{\A}^{\hs\rm{BAS}}_{N}([\al|\be],\!\{\bn\}\hsm)$
given by Eq.\eqref{eq:BAS-kernel-PTPT}.\
We further converted this into the massive KLT formulation of the
KK graviton amplitude in Eq.\eqref{eq:CHY-DC-GR-BCJ}, in which
the KLT kernel $\hs\KK([\al|\be],\!\{\bn\}\hsm)\hs$ is given by
the inverse of the KK bi-adjoint scalar amplitude
$\bar{\A}^{\hs\rm{BAS}}_{N}([\al|\be],\!\{\bn\}\hsm)$
in Eq.\eqref{eq:BAS-kernel-PTPT}
and the color-ordered KK gauge boson amplitude
$\TT^{\rm{YM}}_{N}\big[\al,\hsmx\{\bn\}\big]$
is given by Eq.\eqref{eq:A=N/K}.\
These explicitly build up the connections of the massive KK CHY
formulation with the massive KK BCJ approach
(shown in Section\,\ref{sec:3})
and the massive KK KLT relations (shown in the current
Section\,\ref{sec:3} and in Ref.\,\cite{KKString}).\

\vspace*{1mm}

In Section\,\ref{sec:5}, we studied the possible solutions to the
mass spectral condition \eqref{eq:mass-spectrum}
of the four-point KK scattering amplitudes
and to the mass spectral condition \eqref{eq:mass-cond-N}
of the general $N$-point KK scattering amplitudes.\
For this we newly proposed a group theory approach to prove that this
four-point spectral condition uniquely determine the allowed mass
spectrum to be that of the KK theories under toroidal
compactification with the conservation of KK numbers.\
In our proof, we identified the group structure underlying the
spectral condition \eqref{eq:mass-spectrum} is a product of
integer groups $\hs\Z^r$ (with rank $\hs r\hs$)
in the finitely generated Abelian group of Eq.\eqref{eq:G-decompose}.\
Then, we proved that the unique
solution to the spectral condition \eqref{eq:mass-spectrum}
is given by Eq.\eqref{eq:Mk=k*M1f} for the case of
$\hs r\!=\!1\hs$ and by Eq.\eqref{eq:Mn2-Zr} for the general case of
$\hs r\!\geqq\!1\hs$.\ By inspecting all the known consistent
QFTs, we concluded that
{\it only the KK theories with $\delta\,(=\!r)$ extra dimensions
under the toroidal compactification could have a mass spectrum
behave exactly as in} Eq.\eqref{eq:Mn2-Zr}.\
Moreover, we found that the spectral condition \eqref{eq:mass-spectrum}
has imposed a nontrivial condition on the conservation of the
mass indices (associated with the group $\hs\Z^r$), as shown in
Eq.\eqref{eq:k1234} for the case of $\hs r\!=\!1\hs$
(corresponding to a single extra dimension $\hs \delta\!=\!1\hs$),
or as shown in Eq.\eqref{eq:k1234-r}
for the general case of $\hs r\!\geqq\!1\hs$
(corresponding to $\hs\delta\! =\hsmx r\!\hsm\geqq\hsm\!1\hs$
extra dimensions).\
These just single out the KK theories under the
toroidal compactifications without orbifold
(which can conserve the discretized extra-dimensional momenta and
thus KK numbers) as the unique realization of the spectral condition
\eqref{eq:mass-spectrum}.\
The general $N$-point spectral condition \eqref{eq:mass-cond-N}
also imposes the KK number conservation
as in Eq.\eqref{eq:k1234-1N} for the case of $\hs \delta\!=\!1\hs$
and in Eq.\eqref{eq:k1234-rN} for the case of
$\hs\delta\!\geqq\!1\hs$.\

\vspace*{1mm}

Finally, for the analyses in the main text, we also provided
in Appendix\,\ref{App:A}
the kinematics of the four-point scattering for both the massless
5d theories and the compactified massive 4d KK theories.\
Then, we presented the kinematic numerators for KK gauge
boson scattering amplitudes in Appendix\,\ref{App:B},\
and the double-copied full scattering amplitudes of KK gravitons
in Appendix\,\ref{App:C}.\

\vspace*{1mm}

\acknowledgments
\vspace*{-3mm}
We thank Song He for discussing the extension of
the massless CHY method\,\cite{CHY} and related issues.\
We thank Henry Tye for discussing the extended massive KLT relations
in compactified KK string theories.\ 
This research was supported in part
by National Natural Science Foundation
of China (under grants Nos.\,11835005 and 12175136) and
by National Key R\,\&\,D Program of China
(under grant No.\,2017YFA0402204).\

\vspace*{8mm}
\appendix

\noindent
{\Large\bf Appendix:}

\vspace*{-3mm}
\section{\hspace*{-2mm}Kinematics of Four-Point KK Scattering Amplitudes}
\label{App:A}

In this Appendix we present the kinematics of four-particle scattering processes of the KK states in the (4+1)d spacetime.\
The Minkowski metric tensor is chosen as
$\eta^{MN} \!\!=\! \eta_{MN}^{} \!\!=\! \diag(-1,1,1,1,1)\hs$.

\subsection{\hspace*{-2mm}Kinematics for Massless Scattering Amplitudes
in 5d Spacetime}
\vspace*{2mm}
\label{App:A1}

The 5d momenta in the center-of-mass (CM) frame are defined 
as follows:

%
\beq
\begin{alignedat}{3}
\label{eq:Momenta-massless}
p_1^M & = -E_1^{}\(1, 0, 0, 1 , 0\) \hsm,
&\hspace*{8mm}&
p_2^M = -E_2^{}\(1, 0, 0, -1 , 0\) \hsm,
\\[1mm]
p_3^M &= E_3^{}\(1, \st, 0, \ct , 0\) \hsm,
&\hspace*{8mm} &
p_4^M = E_4^{}\(1, -\st, 0, -\ct , 0\) \hsm,
\end{alignedat}
\eeq
In the above and hereafter, all the external momenta of each
scattering process are defined as out-going.\
The 5d Mandelstam variables are defined in terms of the momenta \eqref{eq:Momenta-massless}:
\begin{align}
\label{eq:stu-massless}
\sh = -(\hat{p}_1^{}\! + \hat{p}_2^{})^2 , \quad
\th = -(\hat{p}_1^{}\! + \hat{p}_4^{})^2 , \quad
\uh = -(\hat{p}_1^{}\! + \hat{p}_3^{})^2 .
\end{align}
We can decompose each 5d momentum of Eq.\eqref{eq:Momenta-massless}
into a 4d component plus a vanishing 5d component:
\begin{equation}
\hat{p}_j^M =\hs \big(p_j^\mu, \, 0\big) , \quad~~~
(j=1,2,3,4)\hs.
\end{equation}
Accordingly, the 4d Mandelstam variables are defined as usual:
\begin{align}
\label{eq:stu-massless-2}
s = -(p_1^{}\! + p_2^{})^2 , \quad
t = -(p_1^{}\! + p_4^{})^2 , \quad
u = -(p_1^{}\! + p_3^{})^2 ,
\end{align}
which equal the 5d Mandelstam variables $(\sh,\th,\uh)$\,.\
Then, we define the transverse polarization vectors for each
massless gauge boson in 5d as follows:
\begin{alignat}{3}
\label{eq:pols}
\hze_{1+}^M\! = \hze_{2-}^M\! &=
\Fr{1}{\sqrt{2\,}\,}(0, 1, \ii, 0, 0)  \hs,  \qquad
&&\hze_{3+}^M\!= \hze_{4-}^M\! =
\Fr{1}{\sqrt{2\,}\,}(0, -\ii \ct ,1,\ii \st,0) \hs,
\nn\\[1mm]
\hze_{1-}^M\! = \hze_{2+}^M\! &=
\Fr{1}{\sqrt{2\,}\,}(0, 1, -\ii, 0, 0) \hs,  \qquad
&&\hze_{3-}^M\! =\hze_{4+}^M\! =
\Fr{1}{\sqrt{2\,}\,}(0, \ii \ct ,1, -\ii \st,0)  \hs,
\\[1mm]
\hze_{10}^M\! = \hze_{20}^M\! &= (0,0,0,0,1) \hs,  \quad
&&\hze_{30}^M\! = \hze_{40}^M\! = (0,0,0,0,1)\hs.
\nn
\end{alignat}

\vspace*{2mm}
\subsection{\hspace*{-2mm}Kinematics for KK Scattering Amplitudes
under 5d Compactification}
\vspace*{2mm}
\label{App:A2}

With the compactified 5d space, the momenta of a given four-particle
scattering process in the center-of-mass frame are defined as follows:
\beq
\begin{alignedat}{3}
\label{eq:Momenta-massive}
\hat{p}_1^{\pp M} & =  -(E_1^{}, 0, 0, q, -M_{\bn_1}\hsm ) \hs,
&\hspace*{8mm}&
\hat{p}_2^{\pp M} = -(E_2^{}, 0, 0, -q, -M_{\bn_2}\hsm ) \hs,
\\[1mm]
\hat{p}_3^{\pp M} &= (E_3^{}, q'\st, 0, q'\hsm\ct, M_{\bn_3}\hsm ) \hs,
&\hspace*{8mm}&
\hat{p}_4^{\pp M} = (E_4^{}, -q'\st, 0, -q'\hsm\ct, M_{\bn_4}\hsm ) \hs,
\end{alignedat}
\eeq
where the fifth component of each 5d momentum is discretized as the KK mass-parameter $M_{\bn_i}\!\!=\!\bn_i^{}/\hsm R$\,.\
The 5d momentum conservation leads to the KK mass conservation
$M_{\bn_1}\!\!+\!M_{\bn_2}\!\!+\!M_{\bn_3}\!\!+\!M_{\bn_4}\!\!
 =\hsm 0\hs$
and the 4d energy conservation
$E_1^{}\hsm +\hsm E_2^{}\hsm =\hsm E_3^{}\hsm +\hsm E_4^{}
 \hsm =\!\hsm\sqrt{s\,}\hs$.\
The 4d energy conservation determines the sizes of
the spatial momenta $q$ and $q'$ as follows:
%
\beq
\begin{aligned}
q &\hs = \frac{1}{\,2\sqrt{s\,}\,}\!
\([s-(M_{\bn_1}\!\!+\!M_{\bn_2})^2][s-(M_{\bn_1}\!\!-\!
M_{\bn_2})^2]\)^{\!\frac{1}{2}} ,
\\ 
q' &\hs =\frac{1}{\,2\sqrt{s\,}\,}\!
\([s-(M_{\bn_3}\!\!+\!M_{\bn_4})^2][s-(M_{\bn_3}\!\!-\!
M_{\bn_4})^2]\)^{\!\frac{1}{2}} .
\end{aligned}
\eeq
Each 5d momentum in Eq.\eqref{eq:Momenta-massive} can be decomposed
into a 4d momentum and a discretized mass-component in the
compactified fifth dimension:
\begin{equation}
\label{eq:Momenta-massive-2}
p_j^{\hs\pp M} = (\hs p_j^{\hs\pp \mu}, \, M_{\bn_j})  \hs,
\quad~~~ (j=1,2,3,4)\hs.
\end{equation}
Using Eqs.\eqref{eq:stu-massless},\eqref{eq:stu-massless-2}
and Eqs.\eqref{eq:Momenta-massive},\eqref{eq:Momenta-massive-2},
we can connect the Mandelstam variables in 4d and 5d as follows:
\begin{align}
\label{eq:stu-comp}
\sh = s - (M_{\bn_1^{}}\!\!+\hsm M_{\bn_2^{}})^2 ,
\qquad
\th = t - (M_{\bn_1^{}}\!\!+\hsm M_{\bn_4^{}})^2 ,
\qquad
\uh = u - (M_{\bn_1^{}}\!\!+\hsm M_{\bn_3^{}})^2 ,
\end{align}
where the sum of each set of Mandelstam variables obeys
\beq
\label{eq:Sum-stu}
\sh+\th+\uh=0\hs, \hspace*{6mm}
s+t+u=\sum_{j=1}^4 \! M_{\bn_j^{}}^2\hs.\
\eeq
The transverse polarization verctors for the external KK gauge bosons
take the same forms as those defined in \eqrefe{eq:pols},
whereas the longitudinal polarization vectors are defined as follows:
\beq
\begin{alignedat}{3}
\hze_{1L}^M  &= \frac{1}{\,M_{\bn_1}}(q,0,0,E_1^{},0) \hs,  \qquad
&&\hze_{2L}^M = \frac{1}{\,M_{\bn_2}}(q,0,0,-E_2^{},0) \hs,
\\[1mm]
\hze_{3L}^M  &= \frac{1}{\,M_{\bn_3}}(q',E_3^{}\st,0,E_3^{}\ct,0) \hs,  \qquad
&&\hze_{4L}^M = \frac{1}{\,M_{\bn_4}}(q',-E_4^{}\st,0,-E_4^{}\ct,0)\hs.
\end{alignedat}
\eeq

\section{\hspace*{-2mm}Kinematic Numerators for KK Gauge Boson Scattering Amplitudes}
\label{App:B}
\label{App:B1}
\vspace*{1mm}

In this Appendix, we present systematically the
kinematic numerators of the various four-point 
scattering amplitudes of the massive KK gauge bosons at tree level
under the 5d toroidal compactification of $S^1$ without orbifold.\
(Some of the four-point KK gauge boson amplitudes under the 5d
orbifold compactification of $S^1\!/\ZZ$ were computed
previously\,\cite{Hang:2021fmp}\cite{5DYM2002}\cite{5DSM2003}.)

\vspace*{1mm}

We first summarize the
the kinematic numerators for the 
four-point scattering amplitudes 
of 5d massless gauge bosons as follows:
\beqs
\label{eq:Nstu-5d}
\begin{align}
\whN_s \,&=\,
- \Big[ ( \hat{p}^{}_1\!-\hat{p}^{}_2 )
(\hze^{}_1 \!\cdot \hze^{}_2) + 2 (\hat{p}^{}_2 \!\cdot \hze^{}_1)
\hze^{}_2 - 2 (\hat{p}^{}_1 \!\cdot \hze^{}_2 )\hze^{}_1 \Big]
\!\cdot\! \Big[ ( \hat{p}^{}_4\!-\hat{p}^{}_3) (\hze^{}_3 \!\cdot \hze^{}_4)
\nn\\
&\hspace{5mm}
+ 2 (\hat{p}^{}_3 \!\cdot \hze^{}_4) \hze^{}_3
-2 (\hat{p}^{}_4 \!\cdot \hze^{}_3 )\hze^{}_4\Big]
- \sh \hs \Big[(\hze^{}_1 \!\cdot \hze^{}_3)
(\hze^{}_2 \!\cdot \hze^{}_4)
\hsm-\hsm (\hze^{}_1 \!\cdot \hze^{}_4)
(\hze^{}_2 \!\cdot \hze^{}_3)\Big] ,
\\[1mm]
\whN_t \,&=\,
 \Big[ ( \hat{p}^{}_1 \!- \hat{p}^{}_4 )(\hze^{}_1 \!\cdot \hze^{}_4)
+ 2 (\hat{p}^{}_4 \!\cdot \hze^{}_1) \hze^{}_4
- 2 (\hat{p}^{}_1 \!\cdot \hze^{}_4 )\hze^{}_1 \Big]
\cdot \Big[ ( \hat{p}^{}_2 \!- \hat{p}^{}_3) (\hze^{}_2 \!\cdot \hze^{}_3)
\nn\\
&\hspace{5mm}
+ 2 (\hat{p}^{}_3 \!\cdot \hze^{}_2) \hze^{}_3
- 2 (\hat{p}^{}_2 \!\cdot \hze^{}_3 )\hze^{}_2 \Big]
+ \th \hs \Big[(\hze^{}_1 \!\cdot \hze^{}_3)
(\hze^{}_2 \!\cdot \hze^{}_4)
+(\hze^{}_1 \!\cdot \hze^{}_2 ) (\hze^{}_3 \!\cdot \hze^{}_4)\Big],
\\[1mm]
\whN_u \,&=\,
-\Big[(
\hat{p}^{}_1 \!- \hat{p}^{}_3 )(\hze^{}_1 \!\cdot \hze^{}_3)
+ 2 (\hat{p}^{}_3 \!\cdot \hze^{}_1) \hze^{}_3
- 2 (\hat{p}^{}_1 \!\cdot \hze^{}_3 )\hze^{}_1\Big]
\cdot\Big[(\hat{p}^{}_2 \!- \hat{p}^{}_4  ) (\hze^{}_2 \!\cdot \hze^{}_4)
\nn\\
&\hspace{5mm}
+ 2 (\hat{p}^{}_4 \!\cdot \hze^{}_2) \hze^{}_4
- 2 (\hat{p}^{}_2 \!\cdot \hze^{}_4 )\hze^{}_2\Big]
+\uh \hs\Big[
(\hze^{}_1 \!\cdot \hze^{}_4 ) (\hze^{}_2 \!\cdot \hze^{}_3)
- (\hze^{}_1 \!\cdot \hze^{}_2 ) (\hze^{}_3 \!\cdot \hze^{}_4)
\Big] .
\end{align}
\eeqs

\vspace*{1mm}
\subsubsection*{$\blacklozenge$~Inelastic KK Scattering Amplitudes of
\boldmath{$\,\{0, 0, n, n$\}}:}
\vspace*{2mm}


For the inelastic scattering channel $\{0, 0, n, n\}$,
there are two combinations of KK indices,
$\fA \!=\!\{0,0,\pm n,\mp n\}$,
which correspond to the same scattering amplitude.\
We derive the numerators of the KK scattering amplitude
$\TT\big[\hsm A^{a\hs0}_{\pm 1} A^{b\hs0}_{\mp 1}
A^{c \hs\pm n}_L\hsmx A^{d\hs\mp n}_L\big]\hs$
as follows:
\begin{equation}
\label{eq:NA-00nn}
\NN_s^\fA\! = 0 \,, \qquad
\NN_t^\fA\! = -\,\NN_u^\fA = \frac{~\Mn^2~}{4}
[\hs(\bs + 4)\!-\hsm (\bs +4)\ctt \hs] \hs,
\end{equation}
where $\hs\bs =s/M_n^2\hs$.\
Under high energy expansion, we derive the following
leading-order (LO) kinematic numerators:
\begin{equation}
\NN_t^{0,\fA}=-\NN_u^{0,\fA}=\frac{~\bs\hs\Mn^2~}{4}
(1\!-\hsm\ctt)\hs.
\end{equation}
These LO numerators will be needed for computing the LO
scattering amplitudes of KK gauge bosons and of KK gravitons
(via double-copy) and for verifying the identity
\eqref{eq:LO-Cond-N0L=N0}.

\vspace*{1mm}
\subsubsection*{$\blacklozenge$~Inelastic KK Scattering Amplitudes of
\boldmath{$\,\{n, 2n, 3n, 4n$\}}:}
\vspace*{2mm}

For the inelastic scattering channel $\{n, 2n, 3n, 4n\}$,
there are two combinations of KK indices,
$\fA \!=\! \{\pm n,\mp 2n,\mp 3n,\pm 4n\}$,
which correspond to the same scattering amplitude.\
We compute the numerators of the KK scattering amplitude
$\TT\big[\hsm A^{a\hs\pm n}_{L} A^{b\hs\mp 2n}_{L}
A^{c \hs\mp 3n}_L\hsmx A^{d\hs\pm 4n}_L\big]$
as follows:
\beqs
\label{eq:NA-1234}
\begin{align}
\NN_s^\fA &= -\frac{~\Mn^2(49\bs'\!-\!1)
~}{672 \hs \bs'}
[(7203 \bs'^2 \!-\! 4410 \bs' \!+\! 375)
+ \bar{\omega}_-^{} \hs(1519\bs'\!+\! 125) \ct\hs]\,,
\\[1mm]
\NN_t^\fA &= \frac{~\Mn^2(49\bs'\!-\!1)~}{1818816 \hs \bs'^2} [
(175273 \bs'^3 \!-\! 133427 \bs'^2 \!+\! 1915\bs' \!+\! 783 )
+ 4\bar{\omega}_-^{}(8918 \bs'^2 \!+\! 511\bs' \!+\! 87) \ct
\nn\\
&\hspace*{5mm}
+ (74431 \bs'^3 \!-\! 86681 \bs'^2 \!+\! 853 \bs' \!+\! 261) \ctt \hs ]\,,
\\[1mm]
\NN_u^\fA &= \frac{~\Mn^2(49\bs'\!-\!1)~}{1818816 \hs \bs'^2} [
(26411 \bs'^3 \!-\! 9947 \bs'^2 \!+\! 8585\bs' \!-\! 783 )
+ 4\bar{\omega}_-^{}(1715 \bs'^2 \!+\! 364\bs' \!-\! 87) \ct
\nn\\
&\hspace*{5mm}
- (74431 \bs'^3 \!-\! 86681 \bs'^2 \!+\! 853 \bs' \!+\! 261) \ctt \hs ]\,,
\end{align}
\eeqs
where we have defined the notations
$\,\bs\!=\!s/\Mn^2\hs$,
$\,\bs'\!=\!\bs/49\hs$, and
$\hs{\omega}_\pm \hsm\!=\!\hsm
 \sqrt{49\hs\bs^{\prime\hs 2}\!\pm 58\hs\bs'\! +\hsm 9\,}\hs$.\hs\
Under high energy expansion, we derive the following LO
kinematic numerators:
\beqs
\begin{align}
\NN_s^{0,\fA} &= \frac{~3\bs'\Mn^2~}{16}(35\hsm + 43\hs\ct) \hs,
\\
\NN_t^{0,\fA} &= -\frac{~\bs'\Mn^2~}{64}
(451\hsm + 468\hs\ct\! + 145\hs\ctt) \hs,
\\
\NN_u^{0,\fA} &= \frac{~\bs'\Mn^2~}{64}
(31\hsm -48\hs\ct\! + 145\hs\ctt) \hs.
\end{align}
\eeqs

\vspace*{1mm}
\subsubsection*{$\blacklozenge$~Inelastic KK Scattering Amplitudes of
\boldmath{$\,\{n, n, m, m$\}}:}
\vspace*{2mm}

For the inelastic scattering channel $\{n, n, m, m\}$,
there are two types of independent combinations of KK indices,
$\fA\!=\!\{\pm n,\mp n,\mp m,\pm m\}$
and
$\fB\!=\!\{\pm n,\mp n,\pm m,\mp m\}$.\
For the type-A scattering,
we compute the kinematic numerators of the KK scattering amplitude
$\TT\big[\hsm A^{a\hs\pm n}_{L}\hsm A^{b\hs\mp n}_{L}\hsm
A^{c \hs\mp m}_L\hsmx A^{d\hs\pm m}_L\big]$
as follows:
\beqs
\label{eq:NA-2n2m}
\begin{align}
\hspace*{-6mm}
\NN_s^\fA &=\,
-\frac{~\Mn^2~}{r^2}
\big[r\hs (\bs^2 \!-\! 2\hs\bs\hs r_+^2 \!+\!4\hs r^2 )
\!+\! 2\hs\qb\hs\qb' (\bs\hs r_+^2 \!+\!2\hs r^2)\hs\ct\hs
\big] \hs,
\\[1mm]
\hspace*{-6mm}
\NN_t^\fA &=\,
\frac{~M_n^2~}{\,8\hs r^2\,}\big\{\bs^2(r_+^2\!+\!4\hs r)
\!-\! 4\hs\bs\hs [r_+^2\hs r\hs (r\!+\!2)\!+\!1 ]
\hsm +\hsm 16\hs r^3
\!+\hsm 8\hs\qb\hs\qb'
[\bs\hs (r \!+\! 1)^2 \!+\!2\hs r^2 \hs]\hs\ct
\nn\\
&\hspace*{7mm}\!+\!\bs\hs r_+^2 (\bs \!-\!4\hs r_+^2)\hs\ctt
\big\}\hs,
\\[1mm]
\hspace*{-6mm}
\NN_u^\fA &=\, -\frac{~M_n^2~}{\,8\hs r^2\,}
\big\{\bs^2(r_+^2\!-\!4\hs r)
\!-\hsm 4\hs\bs\hs [\hs r_+^2\hs r\hs (r\!-\!2)\!+\!1 ]
\!-\! 16\hs r^3
\!-\hsm 8\hs\qb\hs\qb'
[\bs\hs (r \!-\! 1)^2 \!+\hsm 2\hs r^2 \hs]\hs\ct
\nn\\
&\hspace*{7mm}
\!+\! \bs\hs r_+^2 (\bs\!-\!4\hs r_+^2)\hs\ctt\big\} \hs,
\end{align}
\eeqs
where the mass ratios $(r,\,r_+^{})$
and the 3-momenta $(q,\,q')$ are given by
\begin{align}
\label{eq:qsr-AppB}
& r =\hsm M_m^{}/\Mn \hs ,
\hspace*{4mm}
r_+^2 = 1+r^2 ,
\hspace*{4.mm}
q \!=\hsm \sqrt{E^2 \!-\! M_n^2\,}\,,
\hspace*{4.mm}
q' \!=\hsm \sqrt{E^2 \!-\! M_m^2\,}\,,
\nn\\[-3.5mm]
\\[-2mm]
&\bar{q}^2\!= q^2/M_n^2 = \bs/4 \hsm -\hsm 1 \hs,
\hspace*{3mm}
\bar{q}^{\pp 2}\!= q^{\pp 2}\hsm / M_n^2= \bs/4\hsm -\hsm r^2 ,
\hspace*{3mm}
\qb^2\qqbp \hsm =\hsm (\bs\hsm -\hsm 4)
(\bs\hsm -\hsm 4\hs r^2)/16 \hs.~~~
\hspace*{3mm}
\nn
\end{align}
Then, making the high energy expansion, we derive the following LO
kinematic numerators:
\\[-5mm]
{\small
\beqs
\begin{align}
\NN_s^{0,\fA} &= \frac{~\bs\hs\Mn^2~}{r^2}
\big[2\hs r\hs r_+^2 + (1 + r^2 r_+^2)\hs\ct\big]\hs,
\\[1mm]
\NN_t^{0,\fA} &= -\frac{~\bs\hs\Mn^2~}{4\hs r^2}
\big[(1\!+\hsm 4\hs r\!+\hsm 8\hs r^2\!+\hsm 4\hs r^3\!+\hsm r^4)\!
+ 2\hs (1\!+\hsm 2\hs r\!+\hsm r^2\!+\hsm 2\hs r^3\!+\hsm r^4)\hs\ct
\!+\hsm r_+^4\ctt\big]\hs,
\\[1mm]
\NN_u^{0,\fA} &= \frac{~\bs\hs\Mn^2~}{4\hs r^2}
\big[(1 - 4 r + 8 r^2 - 4 r^3 + r^4)
- 2 (1 - 2 r + r^2 - 2 r^3 + r^4) \ct + r_+^4 \ctt\big]\hs.
\end{align}
\eeqs

For the type-B scattering,
we compute the kinematic numerators of the KK scattering amplitude
$\TT\big[\hsm A^{a\hs\pm n}_{L}\hsm A^{b\hs\mp n}_{L}\hsm
A^{c \hs\pm m}_L\hsmx A^{d\hs\mp m}_L\big]$
as follows:
\beqs
\label{eq:NB-2n2m}
\begin{align}
\hspace*{-6mm}
\NN_s^\fB \,&=\,
\frac{~\Mn^2~}{r^2}
\big[\hs r (\bs^2 \!-\! 2\hs\bs\hs r_+^2 \!+\!4\hs r^2 )
\hsm -\hsm 2\hs\qb\hs\qb' (\bs r_+^2 \!+\! 2\hs r^2)\hs\ct\big]\hs,
\\
\hspace*{-6mm}
\NN_t^\fB \,&=\,
\frac{~\Mn^2~}{8\hs r^2}\big\{\bs^2(r_+^2\!-\!4\hs r)
\hsm -\hsm 4\hs\bs\hs\big[r_+^2\hs r\hs (r\!-\!2)\!+\!1 \big]
\!-\! 16r^3
\!+\! 8 \qb \qb'\big[\bs\hs (r \!-\! 1)^2\hsm +\hsm
2\hs r^2 \hs\big] \ct
\nn\\
&\hspace*{6mm}
+\hsm \bs\hs r_+^2 (\bs \hsm -\hsm 4\hs r_+^2)\hs\ctt \big\}\hs,
\\
\hspace*{-6mm}
\NN_u^\fB \,&=\, -\frac{~\Mn^2~}{8\hs r^2}
\big\{\bs^2(r_+^2\!+\hsm 4\hs r)
\!-\hsm 4\hs\bs \big[r_+^2\hs r\hs (r\hsm +\hsm 2)\hsm +\!1\big]
\hsm +\! 16\hs r^3
\!-\hsm 8\hs\qb\hs\qb'\big[\hs\bs\hs (r \!+\! 1)^2
\!+\hsm 2\hs r^2 \hs]\hs\ct
\nn\\
&\hspace*{6mm}\!+\! \bs\hs r_+^2\hs (\bs\hsm -\hsm 4\hs r_+^2)
\hs\ctt \big\}\hs.
\end{align}
\eeqs
}
Then, under high energy expansion, we derive the following LO
kinematic numerators:
{\small
\beqs
\begin{align}
\NN_s^{0,\fB} &=-
\frac{~\bs\Mn^2~}{r^2}\big[2\hs r\hs r_+^2\hsm -
(1\hsm + r^2 r_+^2)\hs\ct \big],
\\
\NN_t^{0,\fB} &= -\frac{~\bs\hs\Mn^2~}{4\hs r^2}
\big[(1\hsm -\hsm 4\hs r\hsm +\hsm 8\hs r^2 \!- 4\hs r^3 \!+ r^4)
\!+ 2\hs (1\hsm -\hsm 2\hs r\hsm +\hsm r^2\!-\hsm 2\hs r^3\!
+\hsm r^4)\hs\ct \!+ r_+^4 \ctt\big],
\\
\NN_u^{0,\fB} &= \frac{~\bs\hs\Mn^2~}{4\hs r^2}
\big[(1\hsm +\hsm 4\hs r\hsm +\hsm 8\hs r^2\! +\hsm 4\hs r^3
\!+\hsm r^4)\hsm -\hsm 2\hs (1\hsm +\hsm 2\hs r\hsm +\hsm r^2\!
+\hsm 2\hs r^3\!+\hsm r^4)\hs \ct\! +\hsm r_+^4 \ctt\big].
\end{align}
\eeqs
}

\vspace*{-4mm}
\subsubsection*{$\blacklozenge$~Elastic KK Scattering Amplitudes of
\boldmath{$\,\{n, n, n, n$\}}:}
\vspace*{2mm}

For the elastic scattering channel $\{n, n, n, n\}$,
there are three types of independent combinations of KK indices,
$\fA\!=\!\{\pm n,\pm n,\mp n,\mp n\}$,
$\fB\!=\!\{\pm n,\mp n,\mp n,\pm n\}$, and
$\fC\!=\!\{\pm n,\mp n,\pm n,\mp n\}$.\
For the type-A scattering,
we compute the kinematic numerators of the
elastic KK scattering amplitude
$\TT\big[\hsm A^{a\hs\pm n}_{L} A^{b\hs\pm n}_{L}
A^{c \hs\mp m}_L\hsmx A^{d\hs\mp m}_L\big]$
as follows:
\beqs
\label{eq:NA-4n}
\begin{align}
\NN_s^\fA \,&=\, -(2\hs\bs^2\hsm -\hsm 7\hs\bs\hsm
-\hsm 4)\hs\ct\hs\Mn^2 \,,
\\[1mm]
\NN_t^\fA \,&=\, \Fr{1}{\,2\,} (\bs\hsm -\hsm 4)
\big[\hs (\bs\hsm -\hsm 3)\hsm +\hsm (2\hs\bs\hsm +\!1)\hs\ct
\hsm +\hsm \bs\hs\ctt\hs \big] \Mn^2 \,,
\\[1mm]
\NN_u^\fA \,&=\, -\Fr{1}{\,2\,} (\bs \hsm -\hsm 4)
\big[\hs (\bs\hsm -\hsm 3)\hsm -\hsm (2\hs\bs\hsm +\!1)\hs\ct
\hsm +\hsm \bs\hs\ctt\hs \big]\Mn^2 \,.
\end{align}
\eeqs
Then, making the high energy expansion, we derive the following
expressions for the LO kinematic numerators of type-A:
\begin{align}
\label{eq:NA-4n-LO}
\NN_s^{0,\fA} = -s\hs\ct \hs,\qquad
\NN_t^{0,\fA} = -\Fr{1}{\,2\,} s\hs (3-\ct) \hs,\qquad
\NN_u^{0,\fA} = \Fr{1}{\,2\,} s\hs (3+\ct) \hs.
\end{align}
%

Next, for the type-B elastic KK scattering,
we compute the kinematic numerators of the
elastic KK scattering amplitude
$\TT\big[\hsm A^{a\hs\pm n}_{L} A^{b\hs\mp n}_{L}
A^{c \hs\mp n}_L\hsmx A^{d\hs\pm n}_L\big]$
as follows:
\beqs
\label{eq:NB-4n}
\begin{align}
\hspace*{-6mm}
\NN_s^\fB \,&=\,
- [ (\bs \!-\! 2)^2\hsm +\hsm (\bs^2\!-\hsm 3\hs\bs\!-\!4)\hs\ct]
\Mn^2 \,,
\\[1mm]
\hspace*{-6mm}
\NN_t^\fB \,&=\,
\Fr{1}{\,4\,}\big[(3\hs\bs^2 \!-\!14\hs\bs\hsm +\hsm 8)
\hsm +\hsm 2\hs (2\hs\bs^2 \!-\hsm 7\hs\bs\!-\hsm 4)\hs\ct
\hsm +\hsm \bs\hs (\bs\hsm -\hsm 8)\hs\ctt\hs\big] \Mn^2 \,,
\\[1mm]
\hspace*{-6mm}
\NN_u^\fB \,&=\, \Fr{1}{\,4\,}\big[(\bs^2 \!-\hsm 2\hs\bs\hsm +\hsm 8) \hsm +\hsm 2\hs (\bs\hsm -\hsm 4)\hs\ct\hsm -\hsm
\bs\hs (\bs\hsm -\hsm 8)\hs\ctt\hs\big] \Mn^2 \,.
\end{align}
\eeqs
Then, under high energy expansion, we derive the following formulas
for the LO kinematic numerators of type-B:
\begin{equation}
\label{eq:NB-4n-LO}
\dis
\NN_s^{0,\fB}\! = s\hs (4+\hsm 3\hs\ct)\hs, \quad
\NN_t^{0,\fB}\! = -\Fr{1}{\,2\,}\hs s\hs
(9\hsm +\hsm 7\hs\ct\hsm +\hsm 2\hs\ctt)\hs, \quad
\NN_u^{0,\fB}\! =
\Fr{1}{\,2\,}s (1\hsm +\hsm\ct\hsm +\hsm 2\hs\ctt) \hs.
\end{equation}

Finally, for the type-C elastic KK scattering,
we compute the kinematic numerators of the
elastic KK scattering amplitude
$\TT\big[\hsm A^{a\hs\pm n}_{L} A^{b\hs\mp n}_{L}
A^{c \hs\pm n}_L\hsmx A^{d\hs\mp n}_L\big]$
as follows:
\beqs
\label{eq:NC-4n}
\begin{align}
\NN_s^\fC &=\,\big[(\bs\hsm -\hsm 2)^2\hsm -\hsm
(\bs^2\!-\hsm 3\hs\bs\hsm -\hsm 4 )\hs\ct\hs\big]\Mn^2 \hs,
\\[1mm]
\NN_t^\fC &=\, -\Fr{1}{\,4\,}\big[(\bs^2 \!-\hsm 2\hs\bs
\hsm +\hsm 8) \!-\! 2\hs (\bs\hsm -\hsm 4)\hs\ct\hsm -\hsm  \bs\hs (\bs\hsm -\hsm 8)\hs\ctt\hs\big] \Mn^2 \hs,
\\[1mm]
\NN_u^\fC &=\, -\fr{1}{4} [(3 \bs^2 \!-\!14 \bs \!+\! 8)
\!-\! 2 (2 \bs^2 \!-\! 7 \bs \!-\! 4 ) \ct \!+\! \bs(\bs\!-\! 8) \ctt] \Mn^2 \,.
\end{align}
\eeqs
Then, making the high energy expansion, we derive the following
expressions for the LO kinematic numerators of type-C:
\begin{equation}
\label{eq:NC-4n-LO}
\NN_s^{0,\fC} \!= -s\hs (4\! -\hsm 3\hs\ct) \hs,\quad
\NN_t^{0,\fC} \!= -\Fr{1}{\,2\,}s\hs
(1\!-\hsm\ct\hsm +\hsm 2\hs\ctt)\hs,\quad
\NN_u^{0,\fC} \!= \Fr{1}{\,2\,} s\hs
(9\! -\hsm 7\hs\ct\!+\hsm 2\hs\ctt) \hs.
\end{equation}
%

\section{\hspace*{-2mm}Full Scattering Amplitudes of KK Gravitons by Double-Copy}
\label{App:B2}
\label{App:C}

In this Appendix, we present the full massive scattering amplitudes
of KK gravitons for various four-particle elastic and inelastic
scattering processes at the tree level.\
The scattering amplitudes for these processes have been studied
in Sec.\,\ref{sec:3.2} by using the leading-order longitudinal
polarization tensor \eqref{eq:LO-Gpol}
for each external KK graviton state.\
In the following, we derive the full scattering amplitudes
of four KK gravitons by using the exact
polarization tensors in Eq.\eqref{eq:zeta-munu-4d}.

\subsubsection*{$\blacklozenge$~Inelastic KK Scattering Amplitudes of
\boldmath{$\,\{0,0,n,n$\}}:}
\vspace*{1mm}

As the first eaxmple, we present the four-point full scattering amplitude
of zero-mode and KK gauge bosons, by using the exact KK graviton polarization tensors \eqref{eq:zeta-munu-4d}.\
The amplitudes under $S^1$ and $S^1\!/\ZZ$ compactification
are equal and take the following form:
\begin{equation}
\label{eq:M00nn-full}
\M\!\[h^0_{\pm 2} h^0_{\mp 2} h^{\pm n}_L h^{\mp n}_L\] =
\M\!\[h^0_{\pm 2} h^0_{\mp 2} h^n_L h^n_L\] =
\frac{~\ka^2\hsm M_n^2\hs (\bs^2\!+\!16\hs\bs\hsm +\! 16) \hs s^4_\theta~}{{8} \hs
[(\bs\hsm +\hsm 4)\hsmx -\hsmx (\bs \hsm -\hsm 4)\ctt]}\hs,
\end{equation}
where we have defined
$\,\bs\hsm =\hsm s/\Mn^2\hs$.
Then, making the high energy expansion, we derive its LO and NLO
amplitudes as follows:
\beqs
\label{eq:h00nnFull}
\begin{align}
\label{eq:h00nnFull-LO}
{\M}_0^{}\big[h_{\pm 2}^{0} h_{\mp 2}^{0}
\bar{h}_L^{\mp n}\bar{h}_L^{\pm n}\big] \hs &=\hs
\frac{\,\ka^2 }{~{16}~}s \hs \sst  \,,
\\
\label{eq:h00nnFull-NLO}
{\dM}\big[h_{\pm 2}^{0} h_{\mp 2}^{0}
\bar{h}_L^{\mp n}\bar{h}_L^{\pm n}\big] \hs &=\hs
\frac{~\ka^2\Mn^2\,}{{8}}  (3- 5\hs\ctt) \,.
\end{align}
\eeqs
%

\vspace*{1.5mm}
\subsubsection*{$\blacklozenge$~Inelastic KK Scattering Amplitudes of
\boldmath{$\{n, 2n, 3n, 4n\}$}:}
\vspace*{1mm}

We note that the double-copied inelastic KK graviton
scattering amplitudes of
$\{\pm n, \mp 2n,$ $\mp 3n, \pm 4n\}$
under the $S^1$ compactification and of
$\{n,2n,3n,4n\}$
under the $S^1\!/\ZZ$ compactification are connected in the same way
as Eq.\eqref{eq:S1xZ2-1234n}.\
Then, we use the exact KK graviton polarization tensors
\eqref{eq:zeta-munu-4d}. and
derive the full inelastic KK scattering amplitudes as follows:
\begin{align}
\label{eq:M1234-full}
& \M\!\[h^{\pm n}_L h^{\mp 2n}_L h^{\mp 3n}_L h^{\pm 4n}_L\] =
2\M\!\[h^{n}_L h^{2n}_L h^{3n}_L h^{4n}_L\]
\nn\\
&~~~=\frac{~\ka^2 M_n^2\hs (49\hs\bs'\hsm -\! 1)~}
{~12544 \hs \bs^{\prime 2} \hs Q_-^2 \hs Q_+^{}\,}\!
\(X_0^{}\!+\!X_1^{}\ct \!+\! X_2^{}\hs\ctt
\!+\!X_3^{}\hs\cttt\!+\!X_4^{}\hs\ctf\!+\!X_5^{}\hs\ctfv\)\hsm
,
\end{align}
where we have denoted
$\,\bs\!=\!s/\Mn^2\hs$ and
$\,\bs'\!=\!\bs/49\hs$.\
The polynomials $Q_\pm$ and $\{X_j^{}\}$ are given by
\\[-3mm]
\begin{equation}
\begin{aligned}
Q_\pm^{} &= (\pm 7\bs'\! + 3)\hsm + \omega_-^{} \ct \,,
\\
X_0^{} &=
6\hs (554631\hs\bs^{\prime\hs 5}\!
-\hsm 909979\hs\bs^{\prime\hs 4}\!+\hsm 747894\hs\bs^{\prime\hs 3}\!
-\hsm 317850\hs\bs^{\prime\hs 2}\!+\hsm 65079\hs\bs'\! -\hsm 5103)
\hs,~~~
\\
X_1^{} &= -2\hs\omega_-^{} (271313\hs\bs^{\prime\hs 4}\!-\hsm
314776\hs\bs^{\prime\hs 3}\!+\hsm 210026\hs\bs^{\prime\hs 2}\!
-\hsm 69048\hs\bs'\! +\hsm 8505 ) \hs,
\\
X_2^{} &= 8\hs (117649\hs \bs^{\prime\hs 5}\!-\hsm
86093\hs\bs^{\prime\hs 4}\!+\hsm 61894\hs\bs^{\prime\hs 3}\!
-\hsm 47490 \bs^{\prime\hs 2}\!+\hsm 25893\hs\bs'\! -\hsm 3645)
\hs,
\\
X_3^{} &= -\omega_-^{}(69629\hs\bs^{\prime\hs 4}\!-\hsm 9016\hs\bs^3
\!-\hsm 43798\hs\bs^{\prime\hs 2}\!+\hsm 11592\hs\bs'\! +\hsm 3645)  \hs,
\\
X_4^{} &= 33614\hs\bs^{\prime\hs 5}\!-\hsm 85750\hs\bs^{\prime\hs 4} \!+\hsm 201740\hs\bs^{\prime\hs 3}\!+\hsm 80556\hs\bs^{\prime\hs 2}\!
-\hsm 23922\hs\bs'\! -\hsm 2430 \hs,
\\
X_5^{} &= -\omega_-^{}(2401\hs\bs^{\prime\hs 4}\!+\hsm 11368\hs\bs^{\prime\hs 3}\!+\hsm 9538\hs\bs^{\prime\hs 2}\!+\hsm 2088\hs\bs'\! +\hsm 81)
\hs.
\end{aligned}
\end{equation}
where we have defined $\hs{\omega}_\pm \hsm\!=\!
\sqrt{49\hs\bs^{\prime 2}\!\pm 58\hs\bs'\! +\hsm 9\,}\hs$.\
Making the high energy expansion, we derive the
following LO and NLO scattering amplitudes
of Eq.\eqref{eq:M1234-full}:
\beqs
\label{eq:M1234-full-LONLO}
\begin{align}
\label{eq:M1234-full-LO}
\M_0^{}\big[h^{\pm n}_L h^{\mp 2n}_L h^{\mp 3n}_L h^{\pm 4n}_L\big]
&=\frac{~\ka^2\hs s~}{\,64\,}\hs (7\hsmx +\hsm\ctt)^2\csc^2\!\theta
\hs,
\\[1mm]
\label{eq:M1234-full-NLO}
\dM^{}\big[h^{\pm n}_L h^{\mp 2n}_L h^{\mp 3n}_L h^{\pm 4n}_L\big]
&=
- \frac{~\ka^2\Mn^2\,}{512}
(23950 - 20328 \ct + 3279 \ctt - 1764 \cttt
\nn\\
& \hspace*{5mm}
+ 2178\hs\ctf  + 588\hs\ctfv + 289\hs\cts)
\csc^4\!\theta  \,.
\end{align}
\eeqs
%

\vspace*{1.5mm}
\subsubsection*{$\blacklozenge$~Inelastic KK Scattering Amplitudes of
\boldmath{$\{n, n, m, m\}$}:}
\vspace*{1mm}

In the following, we use the exact KK graviton polarization tensors \eqref{eq:zeta-munu-4d} and derive the full inelastic
KK graviton scattering amplitudes under $S^1$ compactification:
\beqs
\label{eq:Amp-4hL-nnmm}
\begin{align}
\label{eq:Amp-4hL-nnmm+--+}
\M\big[h_L^{\pm n}h_L^{\mp n}h_L^{\mp m}h_L^{\pm m}\big]
&= \frac{~\ka^2 M_n^2\hs (X_0^{} +\!X_1^{} \ct +\!X_2^{}\hs\ctt +\!
X_3^{}\hs \cttt +\!X_4^{}\hs\ctf )\,}
{128\hs\bs\hs (\bs^2\hsm -\! 16\hs\qb\hs\qb'\cct
- 32\hs\qb\hs\qb' r\hs\ct\!-\!16\hs r^2 )} \hs,
\\[1.5mm]
\label{eq:Amp-4hL-nnmm+-+-}
\M\big[h_L^{\pm n}h_L^{\mp n}h_L^{\pm m}h_L^{\mp m}\big]
&= \frac{~\ka^2M_n^2\hs
(X_0^{}\hsm -\! X_1^{}\hs\ct +\!X_2^{}\hs\ctt\hsm -\!
X_3^{}\hs\cttt\hsm +\!X_4^{}\hs\ctf )\,}
{128\hs\bs\hs (\bs^2\!-\! 16\hs\qb\hs\qb'\cct\!
+\hsm 32\hs\qb\hs\qb'r\hs\ct\hsmx -\hsmx 16\hs r^2 )} \,,
\end{align}
\eeqs
where the coefficients $\{X_j^{}\}$ take the following forms:
\begin{equation}
\begin{aligned}
\label{eq:Xj-nnmm}
X_0^{} &= 99\hs\bs^4\!-\hsm 336\hs \bs^3 r_+^2\!
+ 36\hs\bs^2(12\hs r^4 \!+\! 64\hs r^2 \!+\! 12)
\!-\hsm 3840\hs\bs\hs r^2 r_+^2\!+\hsm 8960\hs r^4 \hs,
\\
X_1^{} &= 512\hs\qb\hs\qb'r\hs (3\hs\bs^2\!-6\hs\bs\hs r_+^2\!+\hsm
28\hs r^2)\hs,
\\
X_2^{} &= 4\hs [\hs 7\hs \bs^4\!+\hsmx 16\hs\bs^3 r_+^2\!
-\hsmx 16\hs \bs^2(5\hs r^4 \hsm\!-\! 4\hs r^2 \hsm\!+\! 5)
\hsm -\hsm 128 \bs r^2r_+^2
\!+\hsm 1792 r^4 \hs]\,,
\\
X_3^{} &= 512\hs (\bs^2 \qb\hs \qb' r
+ 2 \hs\bs\hs\qb\hs\qb' r\hs r_+^2\!
+4\hs\qb\hs\qb' r^3) \,,
\\
X_4^{} &= \bs^4\! +\!16\hs\bs^3 r_+^2\hsmx
+\hsm 16\hs\bs^2 (r^4 \hsm\!+\!16\hs r^2\hsm\!+\! 1)\!
+256\hs\bs\hs r^2 r_+^2\! +256\hs r^4  \,.
\end{aligned}
\end{equation}
Under the high energy expansion,
we derive the LO and NLO amplitudes
of Eqs.\eqref{eq:Amp-4hL-nnmm+--+}-\eqref{eq:Amp-4hL-nnmm+-+-}
as follows:
\beqs
\label{eq:Mnnmm-LONLO}
\begin{align}
\label{eq:Mnnmm-LO}
{\M}_0^{}\big[{h}_L^{\pm n}{h}_L^{\mp n}{h}_L^{\mp m}{h}_L^{\pm m}\big] &=
{\M}_0^{}\big[{h}_L^{\pm n}{h}_L^{\mp n}{h}_L^{\pm m}{h}_L^{\mp m}\big]
= \frac{~\ka^2\hs s~}{\,64\,}\hs (7\hsmx +\hsm\ctt)^2\csc^2\!\theta \,,
\hspace*{8mm}
\\
\label{eq:Mnnmm-NLO+--+}
{\dM}\big[{h}_L^{\pm n}{h}_L^{\mp n} {h}_L^{\mp m}{h}_L^{\pm m}\big]
&=
-\frac{\,\ka^2\Mn^2\,}{128}
(410\hs r_+^2 - 968\hs r\hs\ct + 59\hs r_+^2\hs\ctt
- 84\hs r\hs\cttt
\nn\\
& \hspace*{5mm}
+ 38\hs r_+^2 \ctf+ 28\hs r\hs\ctfv
+ 5\hs r_+^2 \cts)\hsm\csc^4\!\theta \,,
\\[1mm]
\label{eq:Mnnmm-NLO+-+-}
{\dM}\big[{h}_L^{\pm n}{h}_L^{\mp n}{h}_L^{\pm m}{h}_L^{\mp m}\big]
&=
-\frac{\,\ka^2\Mn^2\,}{128}
(410\hs r_+^2 \!+ 968\hs r\hs\ct + 59\hs r_+^2 \ctt
+ 84\hs r\hs \cttt
\nn\\
& \hspace*{5mm}
+ 38\hs r_+^2\hs \ctf - 28\hs r\hs \ctfv + 5\hs r_+^2 \cts)\hsm\csc^4\!\theta \,.
\end{align}
\eeqs
The scattering amplitudes under $S^1/\ZZ$ compactification is
derived as follows:
\begin{align}
\M\hsmx\[h^n_L \hs h^n_L \hs h^m_L \hs h^m_L\] &=
\frac{1}{2}\Big\{\M\big[h_L^{\pm n}h_L^{\mp n}h_L^{\mp m}h_L^{\pm m}\big]\hsmx +
\M\big[h_L^{\pm n}h_L^{\mp n}h_L^{\pm m}h_L^{\mp m}\big]
\!\Big\}
\nn\\
& = -\frac{~\ka^2 M_n^2~}
{~2048\hs\bs\hsx Q_+^{}\hs Q_-^2 ~}
\sum_{k=0}^{8} X_k^{} \cos(k\hs\theta)
\,,	
\label{eq:Mnnmm-Z2}
\end{align}
where the coefficients $(Q_+^{},\hs Q_-^{},\hs X_k^{})$
are polynomial functions given by
\begin{align}
Q_\pm^{} =&\  \bs^2 + 8 (\qb\hs\qb')^2 -\hsm 16\hs r^2
\pm 8\hs\bs\hs\qb\hs\qb'\ct + 8 (\qb\hs\qb')^2 \ctt \,,
\nn\\
X_0^{} =&\ 877 \bs^8\!+\!2024 \bs^7 r_+^2\!-\!96 \bs^6 (205 r^4\!+\!594 r^2\!+\!205)\!+\!128 \bs^5 (301 r^6\!+\!1194 r^2r_+^2\!+\!301)
\nn\\
& \!-\!256 \bs^4 (83 r^8\!+\!716 r^6\!+\!428 r^4\!+\!716 r^2\!+\!83)\!+\!2048 \bs^3 r^2 (21 r^6\!-\!118 r^2r_+^2\!+\!21)
\nn\\
&\!+\!24576 \bs^2 r^4 (35 r^4\!+\!78 r^2\!+\!35)\!-\!2981888 \bs r^6r_+^2 \!+\!2949120 r^8 \,,
\nn\\
X_1^{} =&\ 16 \bs \qb\qb' [197 \bs^6\!+\!2140 \bs^5 r_+^2\!-\!16 \bs^4 (383 r^4\!+\!931 r^2\!+\!383)\!+\!64 \bs^3 (83 r^6\!+\!155 r^2r_+^2\!+\!83)
\nn\\
&-\!256 \bs^2 r^2 (95 r^4\!-\!141 r^2\!+\!95)\!+\!7168 \bs r^4r_+^2\!-\!45056 r^6]\,,
\nn\\
X_2^{} =&\ 32 [17 \bs^8\!-\!348 \bs^7 r_+^2\!+\!8 \bs^6 (155 r^4\!+\!259 r^2\!+\!155)\!-\!16 \bs^5 (88 r^6\!+\!53 r^2r_+^2 \!+\!88)
\nn\\
&+\!128 \bs^4 (5 r^8\!-\!28 r^6\!-\!138 r^4\!-\!28 r^2\!+\!5)\!-\!256 \bs^3 r^2 (13 r^6\!-\!31 r^2r_+^2 \!+\!13)
\nn\\
&+\!2048 \bs^2 r^4 (18 r^4\!+\!53 r^2\!+\!18)\!-\!167936 \bs r^6r_+^2\!+\!131072 r^8]\,,
\nn\\
X_3^{} =& -16 \bs \qb\qb' [169 \bs^6\!-\!1892 \bs^5 r_+^2\!+\!16 \bs^4 (181 r^4\!+\!233 r^2\!+\!181)
\nn\\
&+\!64 \bs^3 (3 r^6\!+\!179r^2r_+^2\!+\!3)\!-\!256 \bs^2 r^2 (11 r^4\!+\!239 r^2\!+\!11)
\!+\!48128 \bs r^4r_+^2\!-\!61440 r^6]\,,
\nn\\
X_4^{} =&\ 4 (75 \bs^8\!-\!776 \bs^7 r_+^2\!+\!352 \bs^6 (5 r^4\!+\!6 r^2\!+\!5)\!-\!640 \bs^5 (r^6\!+\!4 r^4\!+\!4 r^2\!+\!1)
\nn\\
&-\!256 \bs^4 (5 r^8\!-\!76 r^6\!-\!140 r^4\!-\!76 r^2\!+\!5)\!-\!2048 \bs^3 r^2 (11 r^6\!+\!12 r^2r_+^2\!+\!11)
\nn\\
&-\!57344 \bs^2 r^4 (3 r^4\!+\!10 r^2\!+\!3)\!+\!819200 \bs r^6r_+^2\!-\!327680 r^8)\,,
\nn\\
X_5^{} =&-16 \bs \qb\qb' [27 \bs^6\!-\!76 \bs^5 r_+^2\!-\!16 \bs^4 (49 r^4\!+\!149 r^2\!+\!49)\!+\!64 \bs^3 (17 r^6\!+\!49 r^2r_+^2\!+\!17)\,,
\nn\\
&+\!256 \bs^2 r^2 (47 r^4\!+\!115 r^2\!+\!47)\!-\!44032 \bs r^4r_+^2\!+\!12288 r^6]\,,
\nn\\
X_6^{} =&\ 32 \bs (\bs^2\!-\!4 \bs r_+^2\!+\!16 r^2)^2 [\bs^3\!+\!4 \bs^2 r_+^2\!-\!8 \bs (r^4\!+\!3 r^2\!+\!1)\!-\!112 r^2r_+^2]\,,
\nn\\
X_7^{} =&-16 \bs \qb\qb' [\bs^6\!+\!12 \bs^5 r_+^2\!-\!48 \bs^4 (r^4\!-\!3 r^2\!+\!1)\!-\!64 \bs^3 (r^6\!+\!9 r^2r_+^2\!+\!1)
\nn\\
&-\!768 \bs^2 (r^6\!-\!3 r^4\!+\!r^2)\!+\!3072 \bs r^4r_+^2\!+\!4096 r^6]\,,
\nn\\
X_8^{} =&\ (\bs^2\!-\!4 \bs r_+^2\!+\!16 r^2)^2 [\bs^4\!+\!16 \bs^3 r_+^2\!+\!16 \bs^2 (r^4\!+\!16 r^2\!+\!1)\!+\!256 \bs r^2r_+^2 \!+\!256 r^4] \,.
\end{align}
Under the high energy expansion, we derive the LO and NLO
amplitudes from Eq.\eqref{eq:Mnnmm-Z2} as follows:
%
\beqs
\begin{align}
\M_0^{}\[h^n_L \hs h^n_L \hs h^m_L \hs h^m_L\] &=
{\M}_0^{}\big[\bar{h}_L^{\pm n}\bar{h}_L^{\mp n} \bar{h}_L^{\mp m}\bar{h}_L^{\pm m}\big]
= \frac{~\ka^2\hs s~}{\,64\,}\hs (7\hsmx +\hsm\ctt)^2\hsm\csc^2\!\theta \hs,
\hspace*{5mm}
\\
\dM\[h^n_L \hs h^n_L \hs h^m_L \hs h^m_L\] &=
-\frac{\,\kappa^2 M_n^2\,}{128} r_+^2
(410+59\hs\ctt\!+38\hs \ctf\!+5\hs \cts)\hsm
\csc^4\!\theta \hs.
\end{align}
\eeqs
%

\vspace*{1.5mm}
\subsubsection*{$\blacklozenge$~Elastic KK Scattering Amplitudes of
\boldmath{$\{n, n, n, n\}$}:}
\vspace*{1.5mm}

Finally, we consider the elastic KK graviton scattering amplitudes.\
Under the $S^1$ compactification, we derive the full elastic KK graviton
scattering amplitudes as follows:
\beqs
\label{eq:Amp-4hL-nnnn}
\begin{align}
\M\big[ h_L^{\pm n}h_L^{\pm n} h_L^{\mp n}h_L^{\mp n}\big]
&=\frac{\,\ka^2\Mn^2\,}{{64}} (\bs-4) (7+\ctt)^2\hsm
\csc^2\hsmx\theta \,,
\\[1mm]
\M\big[h_L^{\pm n}h_L^{\mp n} h_L^{\mp n}h_L^{\pm n}\big]
&=\frac{~\ka^2 M_n^2(X_0 \!+\! X_1 \ct \!+\! X_2 \ctt \!+\! X_3 \cttt \!+\! X_4 \ctf)\csc^2\! \frac{\theta}{2}~}
{{256}\hs\bs\hs (\bs-4) [(\bs+4)\!+\! (\bs-4)\ct]} \,,
\\[1mm]
\M\big[h_L^{\pm n}h_L^{\mp n} h_L^{\pm n}h_L^{\mp n}\big]
&= \frac{~\ka^2 M_n^2(X_0 \!-\! X_1 \ct \!+\! X_2 \ctt \!-\! X_3 \cttt \!+\! X_4 \ctf)\sec^2\!\frac{\theta}{2}~}
{{256}\hs\bs\hs (\bs-4) [(\bs+4)\!-\! (\bs-4)\ct]} \,,
\end{align}
\eeqs
where
\begin{equation}
\label{eq:Xj-nnnn}
\begin{aligned}
X_0 &= 99 \bs^4-672 \bs^3+3168 \bs^2-7680 \bs+8960  \,,
\\
X_1 &= 128 (3 \bs^3-24 \bs^2+76 \bs-112 ) \,,
\\
X_2 &= 4 (7 \bs^4+32 \bs^3-96 \bs^2-256 \bs+1792 ) \,,
\\
X_3 &= 128 \bs^3 -1536 \bs-2048  \,,
\\
X_4 &= \bs^4 +32 \bs^3+288 \bs^2 +512 \bs+256 \,.
\end{aligned}
\end{equation}
Under the high energy expansion, we derive the LO and NLO
elastic KK graviton scattering amplitudes as follows:
\beqs
\begin{align}
\M_0^{}\big[h_L^{\pm n}h_L^{\pm n} h_L^{\mp n}h_L^{\mp n}\big]
&=
\M_0^{}\big[h_L^{\pm n}h_L^{\mp n} h_L^{\mp n}h_L^{\pm n}\big]
=
\M_0^{}\big[h_L^{\pm n}h_L^{\mp n} h_L^{\pm n}h_L^{\mp n}\big]
\hspace*{12mm}
\nn\\
\hspace*{6mm}
&= \frac{~\ka^2\hs s~}{\,64\,}\hs (7\hsmx +\hsm\ctt)^2\hsm\csc^2\!\theta \,,
\label{eq:MnnnnFull-LO}
\\[1mm]
\dM\big[h_L^{\pm n}h_L^{\pm n} h_L^{\mp n}h_L^{\mp n}\big]
&=
-\frac{~\ka^2\hs \Mn^2~}{\,16\,}\hs (7+\hsm\ctt)^2\!
\csc^2\!\theta \,,
\\[1mm]
\dM\big[h_L^{\pm n}h_L^{\mp n} h_L^{\mp n}h_L^{\pm n}\big]
&=
\frac{~\ka^2\Mn^2\hs
(-70+\!196\hs\ct+\hsm 443\hs\ctt\hsm
+\hsm 282\hs\cttt\!+\hsm 134\hs\ctf\!+\hsm
34\hs\ctfv\!+\hsm 5\hs\cts)}
{8\hs (5\hsm +\hsm 4\hs\ct\!-\hsm 4\hs\ctt\!-4\hs\cttt\!
-\hsm\ctf)} \hs,
\\[1mm]
\dM\big[h_L^{\pm n}h_L^{\mp n} h_L^{\pm n}h_L^{\mp n}\big]
&=
\frac{~\ka^2\Mn^2\hs
(-70\hsm -\hsm 196\hs\ct\!+443\hs\ctt\!-\hsm 282\hs\cttt
\!+\!134\hs\ctf\!-\hsm 34\hs\ctfv\!+\hsm 5\hs\cts)~}
{8\hs (5\hsm -\hsm 4\hs\ct\!-\hsm 4\hs\ctt\!+\hsm
4\hs\cttt\!-\hsm\ctf)} \hs.
\end{align}
\eeqs

\vspace*{1mm}

Next, we consider the $S^1\!/\ZZ$ compactification and
compute the elastic scattering amplitude of
longitudinal KK gravitons as follows:
\begin{align}
\label{eq:M4hLZ2-nnnn}
\M\big[h^n_L\hs h^n_L \hs h^n_L \hs h^n_L\big]
&=\frac{1}{2}
\LB\M\big[h_L^{\pm n}h_L^{\pm n} h_L^{\mp n}h_L^{\mp n}\big] \!+\! \M\big[h_L^{\pm n}h_L^{\mp n} h_L^{\mp n}h_L^{\pm n}\big] \!+\! \M\big[h_L^{\pm n}h_L^{\mp n} h_L^{\pm n}h_L^{\mp n}\big] \RB
\nn\\[1mm]
&=- \frac{~\ka^2 M_n^2\hs (X_0^{} \hsm +\! X_2^{}\ctt
\hsm +\! X_4^{}\ctf \hsm +\! X_6^{}\cts)\hsm\csc^2\!\theta~}
{~512\hs\bs\hs (\bs\hsmx -\hsmx 4)
[\hs(\bs^2  \!+\hsm 24\hs\bs\hsm +\!16) \!-\hsm (\bs\hsm -\hsm 4)^2\ctt ]~}  \,,
\end{align}
where the coefficients $\{X_j^{}\}$
are polynomial functions given by
\begin{equation}
\label{eq:Xj4hLZ2-nnnn}
\begin{aligned}
X_0^{} &= -2\hs (255\hs\bs^5\! +\hsm 2824\hs\bs^4\!-\! 19936\hs\bs^3
\!+\hsm 39936\hs\bs^2 \!-\hsm 256\hs\bs \hsm +\!14336)\hs ,
\\[1mm]
X_2^{} &= 429\hs\bs^5\!-\hsm 10152\hs\bs^4\!+\hsm 30816\hs\bs^3
\!-\hsm 27136\hs\bs^2\!-\hsm 49920\hsm\bs\hsm +\hsm 34816 \hs ,
\\[1mm]
X_4^{} &= 2\hs (39\hs\bs^5\!-\hsm 312\hs\bs^4\!-\hsm 2784\hs\bs^3\!
-\hsm 11264\hs\bs^2\!+\hsm 26368\hsm\bs \hsm-\! 2048) \hs ,
\\[1mm]
X_6^{} &= 3\bs^5\!+\hsm 40\hs\bs^4\!+\hsm 416\hs\bs^3
\!-\hsm 1536\hs\bs^2 \!-\hsm 3328\hs\bs \hsm-\hsm 2048 \hs .
\end{aligned}
\end{equation}
Then, making the high energy expansion,
we derive the following LO and NLO elastic
scattering amplitudes of KK gravitons:
\beqs
\begin{align}
\label{eq:MnnnFull-LO-Z2}
&
\M_0^{}\big[h_L^{n}h_L^{n}h_L^{n}h_L^{n}\big] =
\frac{~3\hs\ka^2\,}{\,128\,}\hs s^{}\hs
( 7\hsm + \ctt )^2 \!\csc^2\!\theta  \,,
\\[1mm]
&
\dM\big[h_L^{n}h_L^{n}h_L^{n}h_L^{n}\big] =
-\frac{\,\ka^2 \Mn^2~}{256}
(1810+93\hs \ctt\hsmx +\hsm 126\hs\ctf\hsmx +\hsm
19\hs \cts)\hsm\csc^4\!\theta   \hs.
\end{align}
\eeqs


\newpage 

\end{document}